\documentclass[12pt,twoside]{report}

\usepackage{fancyheadings}
\usepackage{times}
\usepackage{graphics,cite,amssymb,epsfig,float,psfrag}
\RequirePackage{epsfig}
\RequirePackage{subfigure}
\catcode`@=11

\def\@citex[#1]#2{\if@filesw\immediate\write\@auxout{\string\citation{#2}}\fi
  \def\@citea{}\@cite{\@for\@citeb:=#2\do
    {\@citea\def\@citea{,\penalty\@m}\@ifundefined
      {b@\@citeb}{{\bf ?}\@warning
       {Citation `\@citeb' on page \thepage \space undefined}}%
\hbox{\csname b@\@citeb\endcsname}}}{#1}}

\def\citer{\@ifnextchar [{\@tempswatrue\@citexr}{\@tempswafalse\@citexr[]}}

\def\@citexr[#1]#2{\if@filesw\immediate\write\@auxout{\string\citation{#2}}\fi
  \def\@citea{}\@cite{\@for\@citeb:=#2\do
    {\@citea\def\@citea{--\penalty\@m}\@ifundefined
       {b@\@citeb}{{\bf ?}\@warning
       {Citation `\@citeb' on page \thepage \space undefined}}%
\hbox{\csname b@\@citeb\endcsname}}}{#1}}
\catcode`@=12
\def\cseff{C_7^{\mbox{eff}}}

\def\bxsll{$B \rightarrow X_s \ell^+ \ell^- $ }

\def\mco{\multicolumn}
\def\bsll{$b \rightarrow s \ell^+ \ell^- $ }

\def\bxsg{$B \rightarrow X_s \gamma $ }
\def\bsg{$b \rightarrow s \gamma $}

\newcommand{\gev}{\; \hbox{GeV}}

\newcommand{\ds}{\displaystyle}

\renewcommand{\a}{\alpha}

\renewcommand{\c}{\chi}
\renewcommand{\d}{\delta}

\newcommand{\g}{\gamma}

\newcommand{\G}{\Gamma}

\newcommand{\bea}{\begin{eqnarray}}
\newcommand{\eea}{\end{eqnarray}}
\newcommand{\beq}{\begin{eqnarray}}
\newcommand{\eeq}{\end{eqnarray}}
\newcommand{\nn}{\nonumber}

\newcommand{\noi}{\noindent}
\def\A{{\scriptscriptstyle A}}
\def\BB{{\scriptscriptstyle B}}

\def\D{{\scriptscriptstyle D}}
\def\E{{\scriptscriptstyle E}}

\def\G{{\scriptscriptstyle G}}

\def\I{{\scriptscriptstyle I}}
\def\J{{\scriptscriptstyle J}}

\def\M{{\scriptscriptstyle M}}

\def\T{{\scriptscriptstyle T}}
\def\U{{\scriptscriptstyle U}}

\def\W{{\scriptscriptstyle W}}

\def\CL{{\cal L}}

\def\a{\alpha}
\def\d{\delta}

\def\g{\gamma}
\def\n{\eta}

\def\th{\theta}
\def\u{\mu}
\def\v{\nu}

\def\half{{1 \over 2}}
\def\quarter{{1 \over 4}}

\def\third{{1 \over 3}}

\def\twothirds{{2 \over 3}}

\def\Athree{{\scriptscriptstyle A3}}
\def\bar#1{\overline{#1}}

\def\Ai{{\scriptscriptstyle Ai}}
\def\Bi{{\scriptscriptstyle Bi}}

\def\bsll{b \to s \ell^+ \ell^-}

\def\ccdot{\hbox{\kern-.1em$\cdot$\kern-.1em}}
\def\dAB{\delta_{\A\BB}}

\def\dIJ{\delta_{\I\J}}
\def\DL{{D_L}}
\def\DR{{D_R}}
\def\Dslash{D\hskip-0.65 em / \hskip+0.30 em} 

\def\gtap{\raise.3ex\hbox{$>$\kern-.75em\lower1ex\hbox{$\sim$}}}
\def\hc{{\rm h.c.}}

\def\KMangles{V^*_{ts} V_{tb}}
\def\KManglesb{V^*_{ts} V_{tb}}

\def\ltap{\raise.3ex\hbox{$<$\kern-.75em\lower1ex\hbox{$\sim$}}}

\def\mc{{m_c}}
\def\mchI{m_{\tilde{\chi}^\pm_\I}}
\def\mchJ{m_{\tilde{\chi}^\pm_\J}}
\def\mdsqA{m_{\tilde{d}_A}}
\def\mdsqB{m_{\tilde{d}_B}}
\def\meslone{m_{\tilde{e}_1}}
\def\meslfour{m_{\tilde{e}_4}}

\def\mgluino{m_{\tilde{g}}}

\def\mhsq{m^2_{h^\pm}}
\def\mneI{m_{\tilde{\chi}^0_\I}}
\def\mneJ{m_{\tilde{\chi}^0_\J}}

\def\msneutrinoone{m_{\tilde{\nu}_1}}

\def\mtsq{m_t^2}
\def\musqA{m_{\tilde{u}_A}}
\def\musqB{m_{\tilde{u}_B}}
\def\mW{m_W}
\def\mWsq{m_W^2}
\def\mZ{m_Z}

\def\slash#1{#1\hskip-0.5em /}  

\def\therefore{{\hbox{..}\kern-.43em \raise.5ex \hbox{.}}\>\>}
\def\twoA{{\scriptscriptstyle 2A}}
\def\xt{x_t}

\def\UL{{U_L}}
\def\UR{{U_R}}
\def \cl#1{{#1\%\ \mathrm{C.L.}}}

\def\QL{(p'\cdot q)}
\def\PL{(p\cdot q)}
\def\QN{(p'\cdot q')}
\def\PN{(p\cdot q')}
\def\NPLQ{(\bar{q'pqp'})}
\def\LQN{(\bar{qp'q'})}
\def\QPL{(\bar{p'pq})}
\def\PLQ{(\bar{pqp'})}
\def\LPN{(\bar{qpq'})}

\def\ABC{(\bar{ABC})}
\def\ABCD{(\bar{ABCD})}
\def\journal#1#2#3#4{{\it #1} {\bf #2} (#3) #4}

\def\np{Nucl. Phys.}

\def\pr{Phys. Rev.}

\newcommand{\tfrac}[2]{{\frac{#1}{#2}}}
\newcommand{\beqa}{\begin{eqnarray}}
\newcommand{\eeqa}{\end{eqnarray}}
\def\ifmath#1{\relax\ifmmode#1\else$#1$\fi}
\def\to{\rightarrow}

\def\trans{_\perp}
\def\ml{{\hat{m_{\ell}}}}
\def\mc{{\hat{m_c}}}

\def\c{C}
\def\cs{{\c_7}}
\def\cn{{\c_9}}
\def\ct{{\c_{10}}}
\def\cne{\cn^{\rm eff}}
\def\cse{\cs^{\rm eff}}
\def\m{{\cal M}}
\def\gl{\Gamma}
\def\g{\gamma}
\def\l{\ell}

\def\bsg{B \rightarrow X_s \, \g}

\def\ph{{p_h}}

\def\d{{\rm d}}

\def\mh{\hat{m}}
\def\mbh{\mh_b}

\def\mvh{\mh_{K^*}}

\def\mlh{\mh_\l}

\def\ph{\hat{p}}
\def\sh{\hat{s}}

\def\a{{\cal A}}

\def\uh{{\hat{u}}}

\def\be{\begin{equation}}
\def\ee{\end{equation}}
\def\ba{\begin{eqnarray}}
\def\ea{\end{eqnarray}}

\def\thetakstar {\ensuremath{\theta_{K^*}}}
\def\phitr      {\ensuremath{\phi_{tr}}}   
\def\thetatr    {\ensuremath{\theta_{tr}}}
\def\cthetatr   {\ensuremath{\cos{\thetatr}}}

\def\cthetakstar{\ensuremath{\cos{\thetakstar}}}

\def\sphitr     {\ensuremath{\sin{\phitr}}}

\def\jpsi     {\ensuremath{ J/\psi }}
\def\dd          {\ensuremath{d }}   

\newcommand{\cq}[1]{\cos^{2}{#1}}
\newcommand{\sq}[1]{\sin^{2}{#1}}
\newcommand{\gfrac}[2]{\displaystyle\frac{#1}{#2}}
\newcommand{\text}[1]{\mbox{\rm #1}}
\def \m{\mu}
\def \g{\gamma}
\def \gev{{\hbox{GeV}}}
\def \cl#1{{#1\%\ \mathrm{C.L.}}}

\begin{document}
\setlength{\baselineskip}{24pt}
\setlength{\baselineskip}{6mm}

\input{qm.cmd}

\begin{titlepage}
\begin{flushright}\normalsize
\vspace{-2cm}
DESY-THESIS-2002-040 \\ 
hep-ph/0211103\\
October 2002
\vspace{0.cm}
\end{flushright}
\begin{center}

\vspace{0.5cm}
\huge {
\sc 

Theoretical studies of exclusive rare $B$ decays in the
Standard Model and Supersymmetric theories}
\vspace{2.cm}
\vfill


\vfill
{\LARGE \bf
 Dissertation \\\vspace{-0.4cm}
 zur Erlangung des Doktorgrades\\\vspace{-0.4cm}
 des Fachbereichs Physik\\\vspace{-0.4cm}
 der Universit\"at Hamburg\\\vspace{-0.4cm}
}
\vspace{2.5cm}
\vfill

\vfill

 {\large \bf vorgelegt von} \\\vspace{-0.4cm}
 {\Large \sc Abdelkader Salim Safir} \\\vspace{-0.4cm}
  {\large \bf aus} {\Large Algier (ALGERIEN)}\\
\vspace{1.5cm}

{\large \bf Hamburg}\\\vspace{-0.4cm}
{\large \bf 2002}

\vspace*{1cm}

\end{center}
\newpage
\thispagestyle{empty}

\vspace*{14cm}
\vfill
\small 

\begin{tabular}{ll}
\begin{minipage}[t]{6.0cm}
Gutachter der Dissertation:
\end{minipage} 
& 
\begin{minipage}[t]{5.0cm}
Prof.~Dr.~Ahmed~Ali\\
Prof.~Dr.~Jochen~Bartels\\
\end{minipage} \\
\bigskip

\begin{minipage}[t]{6.0cm}
Gutachter der Disputation:
\end{minipage} 
& 
\begin{minipage}[t]{8.0cm}
Prof.~Dr.~Ahmed~Ali\\
Prof.~Dr.~Bernd~Andreas~ Kniehl \\
\end{minipage} \\
\bigskip

\begin{minipage}[t]{6.0cm}
Datum der Disputation:
\end{minipage} 
& 
\begin{minipage}[t]{5.0cm}
21.~Oktober 2002
\end{minipage} \\
\bigskip

\begin{minipage}[t]{6.0cm}
Vorsitzender des Pr\"ufungsausschusses:
\end{minipage} 
& 
\begin{minipage}[t]{5.0cm}
Prof.~Dr.~Gerhard~Mack
\end{minipage} \\
\bigskip

\begin{minipage}[t]{8.0cm}
Vorsitzender des Promotionsausschusses:\\
Dekan des Fachbereichs Physik:
\end{minipage} 
& 
\begin{minipage}[t]{8.0cm}
Prof.~Dr.~G\"unter ~Huber  \\
Prof.~Dr.~Friedrich-Wilhelm ~B\"usser
\end{minipage} 
\end{tabular}

\newpage
\clearpage
\thispagestyle{empty}
\vspace*{4cm}
\begin{flushright}
\emph{Existe-t-il au monde une connaissance dont la certitude soit\\
telle qu'aucun homme raisonnable ne puisse la mettre en doute?}\\
{\scriptsize Bertrand Russell}
\end{flushright}
\clearpage
\newpage
\thispagestyle{empty}
\vspace*{4cm}
\clearpage

\pagenumbering{roman}

\newpage
\begin{minipage}[t]{12cm}
{\begin{center}\bf{Abstract}\end{center}}
We calculate the independent helicity amplitudes in the decays $B \to K^*
\ell^+ \ell^-$ and $B \to \rho \ell \nu_\ell$ in the so-called
Large-Energy-Effective-Theory (LEET). Taking into account the dominant
$O(\alpha_s)$ and $SU(3)$ symmetry-breaking effects, we calculate 
various Dalitz distributions in these decays making use of the presently
available data and decay form factors calculated in the QCD sum
rule approach. Differential decay rates in the dilepton 
invariant mass and the Forward-Backward asymmetry in $B \to K^* \ell^+ \ell^-$
are worked out. We also present the decay amplitudes 
in the transversity basis which has been used in the analysis of data on
the resonant decay $B \to K^* J/\psi (\to \ell^+ \ell^-)$. 
Measurements of the ratios $R_i(s) \equiv d \Gamma_{H_i}(s)(B \to K^*
\ell^+ \ell^-)/ d \Gamma_{H_i}(s)(B \to \rho \ell \nu_\ell)$, involving
the helicity amplitudes $H_i(s)$, $i=0,+1, -1$, as precision tests of
the standard model in semileptonic rare $B$-decays are emphasized. We
argue that $R_0(s)$ and $R_{-}(s)$ can be used to determine the CKM ratio
$\vert V_{ub}\vert/\vert V_{ts} \vert$ and search for new physics, where
the later is illustrated by supersymmetry.
  \\
{\begin{center}\bf{Zusammenfassung}\end{center}}
Wir berechnen die unabh\"angigen Helizit\"atsamplituden der Zerf\"alle
$B\to K^*\ell^+\ell^-$ und $B\to\rho \ell\nu_\ell$ in der sogenannten
Large-Energy-Effective-Theory (LEET). Unter Be\-r\"uck\-sich\-ti\-gung der
dominierenden $O(\alpha_s)$ und $SU(3)$ Symmetrie-brechenden Effekte
berechnen wir verschiedene Dalitz Distributionen in diesen Zerf\"allen
unter Einbeziehung der z.~Zt.\ verf\"ugbaren Daten und Formfaktoren,
die mittels QCD Summenregeln berechnet wurden.
Differentielle Zerfallsraten in der Dilepton-invarianten Masse und der
Vorw\"arts-R\"uckw\"arts Asymmetrie in $B\to K^*\ell^+\ell^-$ wurden
ausgearbeitet. Au\ss erdem pr\"asentieren wir die Zerfallsamplituden in der
Transversalit\"atsbasis, die bei der Analyse der Daten des resonanten
Zerfalls $B\to K^*J/\psi(\to\ell^+\ell^-)$ benutzt wurde.
Messungen der Verh\"altnisse $R_i(s)\equiv d\Gamma_{H_i}(s)
(B\to K^*\ell^+\ell^-)/d\Gamma_{H_i}(s)(B\to\rho\ell\nu_\ell)$ mit
den Helizit\"atsamplituden $H_i(s), i=0, +1, -1$ werden als
Pr\"azisionstests des Standarsmodells in semileptonischen seltenen
$B$-Zerf\"allen hervorgehoben.
Wir diskutieren, da\ss\ $R_0(s)$ und $R_-(s)$ benutzt werden k\"onnen, um das
CKM Verh\"altnis $|V_{ub}|/|V_{ts}|$ zu bestimmen, und um nach neuer
Physik zu suchen. Letzteres wird mittels Supersymmetrie illustriert.
\end{minipage}

\vfill
\mbox{ }

\newpage
\pagestyle{empty}
\mbox{}

\newpage
\pagestyle{empty}
\end{titlepage}

\tableofcontents
\cleardoublepage
\pagestyle{fancyplain}
\pagenumbering{arabic}
\chapter{Introduction}
Rare $B$ decays have always played a crucial role in shaping the flavour
structure of the Standard Model~\cite{GSW} and particle physics in general. Since the first measurement
of rare radiative $B \to K^* \g$ decays by the CLEO collaboration
\cite{CLEOkstar} this area of particle physics has received much experimental 
\cite{skwarnicki97} and theoretical \cite{aliapctp97} attention.
In particular, flavour-changing neutral current (FCNC) $B$-decays,
involving  the $b$-quark transition $b\to (s,d)+\gamma$ and $b\to
(s,d)+\ell^+ \ell^-$ ($\ell= e,\ \mu,\tau,\nu$), provide a crucial testing grounds for the standard model at the quantum level, since such transitions are forbidden in the Born approximation. Hence, these rare $B$-decays are characterized by their high sensitivity to New physics.

In the standard model, the short distance contribution to rare
$B$-decays is dominated by the top quark, and long distance
contributions by form factors. Precise measurements of these
transition will not only provide a good estimate of the top quark mass
and the Cabibbo Kobayashi Maskawa (CKM) matrix elements \cite{CKM}
$V_{td},\ V_{ts},\ V_{tb}$, but also of the hadronic properties of $B$-mesons, namely form factors which in turn would provide a good knowledge of the corresponding dynamics and more hint for the non-perturbative regime of QCD.

Via the machinery of operator product expansion (OPE) and the
renormalization group equations (RGE) in the framework of an effective
 Hamiltonian formalism \citer{effhamali,grinstein} (see section
\ref{sec:effham} for a discussion) ${\cal H}_{eff} \sim \sum C_i O_i$,
one can factorize low energy weak processes in terms of
perturbatively short-distance Wilson coefficients  $C_i$ from the long-distance operator matrix elements $<{\cal O}_i>$. 
The (new) vertices ${\cal O}_i$, which are absent in the full Lagrangian, 
are obtained by integrating out the heavy particles, namely the
$W$ and the $top$ in the SM, from the full theory. 
Their effective coupling  is given by the $C_i$, which characterize the short-distance dynamics of the underlying theory.

The Wilson coefficient $\cseff$, reflecting  the $b \to s \g$
transitions, is actually well constrained\footnote{The modulus of the
effective coefficient of the electromagnetic penguin operator in the
SM agrees well
with the experimental bounds, but there is no experimental information
on the phase of $\cseff(m_B)$.} by the current precise measurement
of the inclusive radiative $B \to X_s \g$ decays at the $B$-factories. 
The current world average based on the improved measurements by the
BABAR~\cite{Aubert:2002pd}, CLEO~\cite{Chen:2001fj}, ALEPH~\cite{alephbsg} and BELLE~\cite{bellebsg}
collaborations, 
\begin{eqnarray}
{\cal B}(B \to X_s \gamma)=(3.43^{+0.42}_{-0.37} ) \times 10^{-4},
\label{eq:bsgexp} 
\end{eqnarray}
is in good agreement with the estimates of the standard model (SM) 
\citer{Chetyrkin:1997vx,Gambino:2001ew}, which we shall take
as ${\cal B}(B \to X_s \gamma)=(3.50 \pm 0.50) \times 10^{-4}$,
and moreover, can exclude large parameter spaces of non-standard models.

The $b \to s~ \ell^+ \ell^{-}$ $(\ell^{\pm}=e^{\pm},
\mu^{\pm}, \tau^{\pm})$ transition involves
besides the electromagnetic penguin $b \to s \g^* \to s~ \ell^+ \ell^{-}$ also
electroweak penguins $b \to s Z^{0 *} \to s~ \ell^+ \ell^{-}$ and boxes.
They give rise to two additional Wilson coefficients in the semileptonic 
\bxsll decays, $C_9$ and $C_{10}$. 
%

\begin{table}[t]
\renewcommand{\arraystretch}{}
\begin{center}
\begin{tabular}{|c||c|c|}
\hline\hline
 & \multicolumn{2}{|c|}{${\cal B}~ (\times 10^{-6})$}\\ 
\hline
Decay Mode& {\bf BELLE \cite{bellebsg}} & {\bf BABAR \cite{babarbsll}}\\
\hline\hline
$B\to K e^+ e^-$ & $0.38^{+0.21}_{-0.17}\pm 0.06$ & 
$0.24^{+0.49+0.14}_{-0.32-0.15}$ \\
$B\to K \mu^+ \mu^-$& $0.8^{+0.28}_{-0.23}\pm 0.08$ & 
$1.33^{+1.07}_{-0.78}\pm0.26$  \\
$B\to K \ell^+ \ell^-$ &$ 0.58^{+0.17}_{-0.15}\pm 0.06$ & 
$0.78^{+0.24+0.11}_{-0.20-0.18}$\\
\hline
$B\to K^* e^+ e^- $       & $\leq 2.4$  &   
$1.78^{+0.87+0.48}_{-0.72-0.49}$ \\ 
$B\to K^* \mu^+ \mu^-$    & $\leq 1.2$  &  
$0.99^{+1.14}_{-0.82}\pm 0.39$ \\
$B\to K^* \ell^+ \ell^-$  & $\leq 1.4$  &   $1.68^{+0.68}_{-0.58}\pm 0.28$\\
\hline
$B\to X_s e^+ e^-$       & $5.0 \pm 2.3 ^{+1.2}_{-1.1}$   &  \\
$B\to X_s \mu^+ \mu^-$   & $7.9\pm 2.1^{+2.0}_{-1.5}$   &    \\
$B\to X_s \ell^+ \ell^-$ & $6.1 \pm 1.4 ^{+1.3}_{-1.1}$ &   \\
\hline\hline
\end{tabular}
\end{center}
\caption{\it Experimental results of the semileptonic rare
$B$-decays \cite{bellebsg,babarbsll}.
\label{tab:bsllexp}}
\end{table}
A model independent fit of the short-distance coefficients 
$\cseff$, $C_8$, $C_9$ and $C_{10}$ can be obtained, using the
exclusive as well as the inclusive semileptonic (and radiative) rare
$B$-decays experimental constraints on the corresponding branching
ratio $\Big(B\to (X_s, K, K^*) \gamma\Big)$, the various
angular distributions and the Forward-Backward (FB) asymmetry
\cite{amm91} in $B\to (X_s, K, K^*) \ell^+ \ell^-$ decays. 
They involve independent combinations of the Wilson coefficients,
which allows the determination of sign and magnitude of $\cseff,C_9$
and $C_{10}$ from data. On the other hand, these measurements are also
of great help in studying that part of strong interaction physics
which is least understood, the non-perturbative confinement interactions.

Using the recent $\bsg$ experimental result
(\ref{eq:bsgexp}), with the new measurements of the semileptonic rare
$B$-decays recently reported by BELLE \cite{bellebsg} and BABAR
\cite{babarbsll} (see Table \ref{tab:bsllexp}), it has been shown
that the bounds on these Wilson coefficients are
consistent with the SM, but considerable room for new physics effects
are not excluded \cite{Ali:2002jg}. 

With the advent of the Fermilab booster BTeV (Fermilab)
and LHCb (CERN) experiments at hadron colliders, and also the ongoing
experiments at CLEO and the $B$-factories, the semileptonic rare
decays of  $B
\to (X_s,~K,~K^*) \ell^+ \ell^-$ will be precisely measured. 
On the theoretical side, partial
results in next-to-next-to-leading logarithmic (NNLO) accuracy are now
available in the inclusive decays $B \to X_s \ell^+ \ell^-$
\cite{BMU,AAGW}. What concerns the exclusive
decays, some theoretical progress in calculating their decay rates
to NLO accuracy in the $B \to (K^*,\rho) \gamma$
\citer{Ali:2001ez,Beneke:2001at}, and to NNLO accuracy in $B
\to K^* \ell^+ \ell^-$ \cite{Beneke:2001at} decays, including the leading
$\Lambda_{\rm QCD}/m_B$, has been reported.

Making use of the exclusive semileptonic  $B\to V~\ell^+\ell^-$ (with
$V$ stands for a vector) theoretical improvements, obtained within the
large-energy-effective-theory \cite{Dugan:1990de,Charles:1998dr}, we
explore here a detailed phenomenological analysis of the exclusive $B\to
K^*~\ell^+\ell^-$ and $B\to \rho~\ell\nu_{\ell}$ decays  with
$\ell=e,\mu$ (since we neglected lepton masses in our calculation the
result cannot be used in the $\tau$ case) in the SM and supersymmetric
theories .

This thesis contains the following points \cite{Ali:2002qc,Safir:SUSY02}:
\begin{itemize}
\item Using the effective Hamiltonian approach, and incorporating the 
perturbative improvements\cite{Beneke:2001at}, we expressed the
various helicity components in the decays $B \to K^* \ell^+ \ell^-$ in
the context of the Large-Energy-Effective-Theory.

\item As this framework does not predict the
decay form  factors, which have to be supplied from outside, 
we combined the $B \to K^* \gamma$ experimental constrains obtained in
the NLO-LEET  approach with the light cone QCD sum
rule\cite{Ball:1998kk,Ali:1999mm} 
to extract the LEET form factor $\xi_{\perp}^{(K^*)}$ and
$\xi_{||}^{(K^*)}$ in the large $E_{K^*}$-region $s < 8$ GeV$^2$ 
(namely the LEET validity range).

\item We calculate a number of Dalitz distributions and 
the dilepton invariant mass distribution for the individual helicity 
amplitudes (and the sum) in $B \to K^* \ell^+ \ell^-$. We find that
the NLO order corrections to latter distribution is significant in the
low dilepton mass region $(s\leq 2\mathrm{GeV}^2)$.

\item We show the $O(\alpha_s)$ effects on the forward-backward
asymmetry, confirming essentially the earlier work of {\it Beneke, Feldmann
and Seidel}~\cite{Beneke:2001at}. 

\item We have compared the LEET-based transversity amplitudes in this
basis with the data \citer{ref:cleo,ref:belle} currently available on $B \to K^* J/\psi(\to \ell^+
\ell^-)$ and find that the short-distance based transversity amplitudes
are very similar to their long-distance counterparts.  

\item Using the $SU(3)$-breaking corrections, we relate the $B\to
\rho$ LEET form factors namely $\xi_{\perp}^{(\rho)}$ and
$\xi_{||}^{(\rho)}$ with the $B\to K^*$ corresponding ones, and we  
implement the $O(\alpha_s)$-improved analysis of the various
helicity components in the decays $B \to \rho \ell \nu_\ell$. We 
carry out in the context of the LEET a number of Dalitz distributions,
the dilepton invariant mass distribution for the individual helicity 
amplitudes (and the sum).

\item Combining the analysis of the decay modes 
$B \to K^* \ell^+ \ell^-$ and $B \to \rho \ell \nu_\ell$, we show that
the ratios of differential decay rates involving definite helicity states, 
$R_{-}(s)$ and $R_{0}(s)$, can be used for extracting the CKM matrix elements
$\vert V_{ub}\vert/\vert V_{ts}\vert$.

\item We investigate possible effects on these
ratios from New Physics contributions, exemplified by representative
values for the effective Wilson coefficients in the large-$\tan \beta$ SUGRA
models.

\end{itemize}

\subsubsection{Organization of the work}
An introduction to rare $B$ decays and the methods used is given in
Chapter \ref{chap:rare}, where we discuss the effective Hamiltonian
theory, rare radiative \bxsg decays and long-distance method in
exclusive $B$ decays. 
In Chapter \ref{chap:btoK*} we investigate in details an helicity
analysis of  $B \to K^* \ell^+ \ell^-$ and $B \to \rho~ \ell \nu_\ell$
in the SM. 
We present in the context of the NLO-LEET approach, various 
angular distributions and their uncertainties are worked out. Furthermore,
for the $B \to K^* \ell^+ \ell^-$ decays, we project out the forward-backward
asymmetry  and the corresponding transversity basis.
Chapter~\ref{chap:susy} is devoted to the semileptonic rare $B \to \
K^{*} \ell^{+} \ell^{-}$ decay, by contrasting its anticipated phenomenological
profile in some variants of supersymmetric models. After a review on
the $b \to s~ \ell^{+} \ell^{-}$ decay in the MSSM, we propose to study
the ratios $R_0(s)$ and $R_-(s)$ as probes of new physics effects in
$B \to \ K^{*} \ell^{+} \ell^{-}$, using some generic SUSY
effects. Finally, Chapter \ref{chap:out} contains a summary and an
outlook. Input parameters, Feynman rules and utilities are collected in 
Appendix~\ref{app:generalities}. The large energy expansion with its
machinery is presented in
Appendix~\ref{app:leet}. Appendix~\ref{app:susy} contains various loop
functions, introduced in SUSY.

\clearpage
\chapter{Rare $B$ Decays: Motivation and Methods 
\label{chap:rare}}

In this chapter we outline the flavour structure of the standard model (SM).
We discuss the CKM mixing matrix and motivate the importance of studying 
flavour changing neutral current (FCNC) $b \to s$ transitions. 
We introduce the necessary tool to include
QCD perturbative corrections in weak decays, the effective 
Hamiltonian theory. 
As an application of the former we discuss the $b \to s \gamma$ decay as the 
most prominent example of a rare $B$ decay.
Finally we discuss in details the large energy quark expansion
technique as the appropriate non-perturbative approach for the
heavy-to-light transitions.
\section{The Flavour Sector in the Standard Model}

In the quark sector of the SM, there are six quarks organized in 3 families.
The left-handed quarks are put into weak isospin $SU(2)_L$ doublets 
\begin{eqnarray}
{q_{up} \choose q_{down}^{\prime}}_{i=1,2,3}=
{u \choose d^{\prime} }_L, \; {c \choose s^{\prime} }_L, \; 
{t \choose b^{\prime} }_L \; ,
\end{eqnarray}
and the corresponding 
right-handed fields transform as singlets under $SU(2)_L$.
Under the weak interaction an up-quark (with $Q_{u}=2/3 e$) can decay into a
down-quark (with $Q_{d}=-1/3 e$) and a $W^+$ boson.
This charged current is given as
\begin{equation}
J_{\mu}^{CC}=
 \frac{e}{ \sqrt{2} \sin \theta_W }
 \left( \bar{u}, \bar{c}, \bar{t} \right)_L
\gamma_\mu V_{\mbox{\footnotesize{ CKM}}}
\left( \begin{array}{c} d\\ s \\ b \end{array} \right)_L \; ,
\end{equation}
where the subscript $L =(1-\gamma_5)/2$ denotes the left-handed projector and 
reflects the $V-A$ structure of $J_{\mu}^{CC}$ in the SM.
Here the weak mixing (Weinberg-)angle $\theta_W$ is a 
parameter of the SM, which is measured with high accuracy \cite{PDG2000}.
The so-called Cabibbo Kobayashi Maskawa (CKM) matrix $V_{CKM}$ \cite{CKM} 
describes the mixing between different quark flavours. 
It contains the angles describing the rotation between the eigen vectors of 
the weak interaction $(q^{\prime})$ and the mass eigen states $(q)$
\begin{equation}
\left( \begin{array}{c} d^{\prime}\\ s^{\prime} \\ b^{\prime} \end{array} 
\right)=
V_{CKM}  \left( \begin{array}{c} d\\ s \\ b \end{array} \right)\; .
\end{equation}
Symbolically, $V_{CKM}$ can be written as
\begin{equation}
\label{eq:ckm}
V_{\mbox{\footnotesize CKM}} \equiv \left( \begin{array}{lll}
V_{ud} & V_{us} & V_{ub}\\
V_{cd} & V_{cs} & V_{cb}\\
V_{td} & V_{ts} & V_{tb}
\end{array} \right).
\end{equation}
In general all the entries are complex numbers, only restricted by
unitarity $V_{CKM} V_{CKM}^{\dagger}=1$.
They are parameters of the SM and can only be obtained from an 
experiment.
Note that only three independent real parameters and one phase 
are left after imposing the unitarity condition. 
Some parametrizations of $V_{CKM}$ can be seen in ref.~\cite{PDG2000}.

A useful parametrization of the CKM matrix has been proposed by Wolfenstein 
\cite{Wolfenstein}
 \begin{equation}
V_{\mbox{\footnotesize Wolfenstein}} = \left( \begin{array}{lll}
1-\frac{1}{2} \lambda^2 & \lambda &
A\lambda^3 (\rho - i \eta) \\
- \lambda & 1-\frac{1}{2} \lambda^2 
 & A \lambda^2 \\
A\lambda^3 (1-\rho-i \eta) & -A \lambda^2 & 1
\end{array} \right) +{\cal{O}}(\lambda^4) \; .
\label{eq:wolfenstein}
\end{equation}
The parameters $A,\lambda,\rho$ and the phase $\eta$ are real numbers.
$\lambda$ is related to the Cabibbo angle through
$\lambda=\sin \theta_C$ \cite{PDG2000}, 
which describes the quark mixing with 4 quark flavours. 
Since $\lambda \simeq 0.221$,
the relative sizes of the matrix elements in eq.~(\ref{eq:ckm})
can be read off from eq.~(\ref{eq:wolfenstein}). 
As can be seen, the diagonal entries are close to unity and the more 
off-diagonal they are, the smaller is the value of their corresponding 
matrix elements.
The parameter $A$ has been determined from the decays $b \to c \ell \nu_\ell$
and $B \to D^* \ell \nu_\ell$, yielding
$A=0.81 \pm 0.04$. The measurement of the ratio $|V_{ub}/V_{cb}|=0.08 \pm 0.02$ yields $\sqrt{\rho^2+\eta^2}=0.36 \pm 0.09$.
Likewise the mass difference $\triangle M_d \equiv M(B_d^{(1)})-M(B_d^{(2)})
\simeq 0.50 \, (ps)^{-1}$ constrains the combination
$\sqrt{(1-\rho)^2+\eta^2}$.
The observed CP-asymmetry parameter $\epsilon_K=2.26 \cdot 10^{-3}$ constrains
$\rho$ and $\eta$.
The precise determination of the parameters $\rho$ and $\eta$ is a high and 
important 
goal, since it corresponds to two important questions:
\begin{itemize}
\item Does CP hold in the SM ??
A non zero phase $\eta\not=0$ in the CKM matrix directly leads to CP 
violating effects.
\item The unitarity of the CKM matrix can be used to write down relations 
among its elements 
$\sum_{j=1}^{3} V_{ij} V^{\dagger}_{j k}$ $ =\delta_{ik}, \; i,k=1,2,3$. 
There are 6 orthogonality equations possible ($i\not=k$), and each
can be represented graphically as a triangle, a so-called unitarity 
triangle (UT) \cite{jarlskog88}.
The sides and angles of such an UT can be constrained by different types of 
experiments.
For {\it the} UT given by the relation
\begin{eqnarray}
V_{ud} V_{td}^* + V_{us} V_{ts}^* + V_{ub} V_{tb}^* = 0 \; ,
\end{eqnarray}
there are 3 scenarios possible, which at present are not excluded 
experimentally and are a sign for {\it new physics}:
1. the UT does not close, i. e., $\sum_{i=1}^3 \alpha_i \not=0$, where
$\alpha_i$ denotes the three angles of the triangle.
2. $\sum_{i=1}^3 \alpha_i =0$, but the values of the $\alpha_i$ are outside of 
their SM ranges determined by another type of experiment
3. $\sum_{i=1}^3 \alpha_i =0$, but the values of the angles are inconsistent 
with the measured sides of the triangle.
\end{itemize}
In the literature special unitarity triangles are discussed.
A recent review over the present status on the CKM matrix and {\it the} 
unitarity triangle is given in~\cite{aliapctp97}.

\subsection{Flavour changing neutral currents}

In the SM, the neutral currents mediated through the gauge bosons
$Z^0, \gamma, g$ do not change flavour. Therefore, the so-called 
Flavour changing neutral currents (FCNC) do not appear at tree level and
are described by loop effects.
The subject of the present work is an analysis of such rare 
(FCNC mediated) $B$ decays in the SM.
The quarks are grouped into {\it light} $(u,d,s)$ and
{\it heavy} $(c,b,t)$ ones in the sense, that the mass of a heavy quark is
 much larger than the typical scale of the strong interaction, 
$\Lambda_{QCD} \sim 200 \, {\mbox{MeV}}$.
The sixth quark, the top, is too heavy to build 
bound states because it decays too fast.
The special role of the $b$-quark is that it is the heaviest one 
building hadrons.
We will not discuss the ``double" heavy $B_c$ and concentrate on 
$B \equiv (\bar{b} q)$ meson transitions with light $q=u,d,s$.
Since the $b$-quark is heavy, the $B$-system is well suited for a clean 
extraction of the underlying short-distance dynamics.
Unlike the $K$-system, long-distance effects are expected to play 
a subdominant role in $B$ decays except where such effects are present in a 
resonant form.

The motivation to investigate
$b\to s (d)$ transitions is to improve the knowledge of the
CKM matrix elements and to study loop effects. 
For the latter the interest is large, since there is no tree level FCNC decay 
possible in the SM. 
The leading loops give the leading contribution and they are 
sensitive to the masses and other properties of the internal virtual 
particles like e.~g. the top. 
They can be heavy and therefore can be studied in a rare $B$ decay at energies 
which are much lower than the ones necessary for a direct production of such 
particles. 
The idea is to compare the SM based prediction for a rare $B$ decay with an
experiment. A possible deviation gives a hint not only for the existence, but
also for the structure of the ``new physics" beyond the SM.

Further the $B$-system can be used as a testing ground for QCD, to check 
perturbative and non-perturbative methods. One example is the decay 
$B \to X_s \gamma$, which can be described in the lowest order at parton level
through $b \to s \gamma$. As a 2-body decay, the photon energy in the $b$-quark
rest frame is fixed: $E_\gamma=(m_b^2-m_s^2)/2 m_b$ for an on-shell $\gamma$.
A possible non trivial spectrum 
can result from gluon bremsstrahlung $b \to s \gamma g$ and/or a 
non-perturbative mechanism, which is responsible for the motion of 
the $b$-quark inside the meson thus boosting the distribution.

Some exclusive rare $B$ decays have already been detected. The recent
experimental observations of the rare decay mode $B \to K^* \gamma$
have been determined by CLEO \cite{ref:cleo}, and more recently also by
BABAR \cite{ref:babar} and BELLE \cite{ref:belle} 
\begin{eqnarray}
{\cal B}(B^0 \to K^{*0} \gamma)=\cases{
(4.55\pm 0.70\pm 0.34)  \times 10^{-5}&\cite{ref:cleo}, \cr 
(4.39\pm 0.41\pm 0.27)  \times 10^{-5}&\cite{ref:babar},\cr 
(4.96\pm 0.67\pm 0.45)  \times 10^{-5}& \cite{ref:belle}, \cr}
\label{Brk*0}
\end{eqnarray}
and
\begin{eqnarray}
{\cal B}(B^+ \to K^{*+} \gamma)=\cases{
(3.76\pm 0.86\pm 0.28)  \times 10^{-5}&\cite{ref:cleo}, \cr 
(3.89\pm 0.93\pm 0.41)  \times 10^{-5}&\cite{ref:belle}.\cr}
\label{Brk*+} 
\end{eqnarray}

However, the first observation of the rare $B$-decay to the orbitally
excited strange mesons has been reported by CLEO~\cite{ref:cleo} and
recently confirmed by BELLE~\cite{bellebsg} with a branching fraction of  
\begin{eqnarray} 
{\cal B}(B \to K_2^*(1430) \gamma)= \cases{
(1.66 ^{+0.59}_{-0.53} \pm 0.13) \times 10^{-5}&\cite{ref:cleo}, \cr 
(1.5 ^{+0.6}_{-0.5} \pm 0.1) \times 10^{-5}&\cite{bellebsg},\cr}
\label{Brk2*}  
\end{eqnarray}
These important experimental measurements provides a crucial challenge
to the theory. Many theoretical approaches have been employed  to
predict the exclusive $B\to K^*(892) \gamma$ decay rate (for a review
see \cite{ali2} and references therein). On the other hand less
attention has been devoted to rare radiative $B$-decays to excited
strange mesons \citer{altomari,Faustov}. Most of these
theoretical approaches rely on non-relativistic quark
models\cite{altomari,mannel}, HQET \cite{Veseli}, relativistic
model\cite{Faustov} and light cone QCD sum rules~\cite{Safir:2001cd}. 
However there is a large spread between different results, due to a
different treatment of the long distance effects. Thus, the difficulty
with the exclusive mode is the large theoretical  
uncertainties due to the hadronic matrix elements, which has to be
controlled. A large section is devoted to this issue in end of this chapter. 

\begin{figure}[htb]
\vskip -0.8truein
\centerline{\epsfysize=10in
{\epsffile{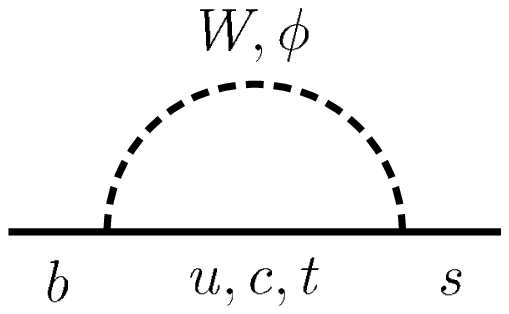}}}
\vskip -7.8truein
\caption{ \it A FCNC $b \to s$ diagram.}
\label{fig:bsg}
\end{figure}  

A typical diagram for a virtual $b \to s $ transition is displayed in Fig.~\ref{fig:bsg}
from where the CKM couplings can be directly read off.
The amplitude $T$ is the sum over all internal up-quarks
\begin{eqnarray}
T= \sum_{i=u,c,t} \lambda_i  T_i \; \; ; \; \;   
\lambda_i \equiv V_{i b} V_{i s}^{\ast} \; .
\end{eqnarray}
Using the CKM unitarity $\sum_{i=u,c,t} \lambda_i=0$ and the smallness of
$V_{ub}$  yielding $\lambda_u \ll \lambda_t$, we arrive at 
\begin{eqnarray}
T=\lambda_t (T_t-T_c)+\lambda_u (T_u-T_c) \simeq \lambda_t (T_t-T_c)
\end{eqnarray}
for a $b \to s$ amplitude in the SM.
In the $D$-system the FCNC transition rates ($c \to u$) are much more 
suppressed due to an inbuilt GIM mechanism \cite{GIM}. Here we have 
\begin{eqnarray}
T_{c \to u}& = &\sum_{i=d,s,b}   V_{c i} V_{u i}^{\ast} T_i \nonumber   \\
&=& V_{c b} V_{u b}^{\ast} (T_b-T_s)+ V_{c d} V_{u d}^{\ast}(T_d-T_s) \; ,
\end{eqnarray}
in which the first term is CKM suppressed and the second one GIM suppressed 
since $m_s^2-m_d^2 \ll m_W^2$.
The SM rates in the charm sector for decays such as 
$D^0 \to \gamma \gamma$, $D^0 \to \ell^+ \ell^-$ are out of reach for present 
experiments. If one nevertheless finds something in the rare charm sector, 
it would be a direct hint for the desired physics beyond the SM.

\section{The Effective Hamiltonian Theory \label{sec:effham}}

As a weak decay under the presence of the strong interaction, rare $B$ decays
require special techniques, to be treated economically.
The main tool to calculate such rare $B$ decays is the effective Hamiltonian
theory. It is a two step program, starting with an operator product 
expansion (OPE) and performing
a renormalization group equation (RGE) analysis afterwards.
The necessary machinery has been developed over the last years, see
\citer{effhamali,grinstein}, \cite{buchallaburasreview}
and references therein.

\begin{figure}[htb]
\vskip -1.0truein
\centerline{\epsfysize=10in
{\epsffile{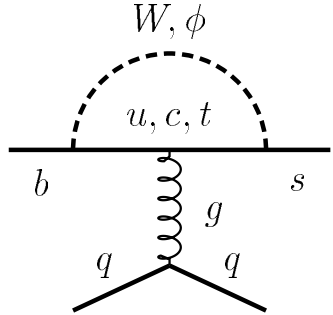}}}
\vskip -7.3truein
\caption{ \it A gluonic penguin diagram.}
\label{fig:penguin}
\end{figure}

The derivation starts as follows:
If the kinematics of the decay are of the kind that the masses of the
internal particles $m_i$ are much larger than the external momenta
$p$, $m_i^2 \gg p^2$, then 
the heavy particles 
can be {\it integrated out}. This concept takes a concrete form with 
the functional integral formalism. 
It means that the heavy particles  are removed
as dynamical degrees of freedom  from the theory, hence their fields do not 
appear in the (effective) Lagrangian anymore. 
Their residual effect lies in the generated effective vertices.
In this way an effective low energy theory can be constructed from a full 
theory like the SM.
A well known example is the four-Fermi interaction, where the 
$W$-boson propagator is made local for $q^2 \ll m_W^2$
($q$ denotes the momentum transfer through  the $W$):
\begin{eqnarray}
-i \frac{g_{\mu \nu}}{q^2-m_W^2} \to 
i g_{\mu \nu} (\frac{1}{m_W^2} + \frac{q^2}{m_W^4} +\dots \,) \; ,
\end{eqnarray}
where the ellipses denote terms of higher order in $1/m_W$.
\footnote{
We remark here that the original way was reversed:
The main historical step was to extrapolate the observed low energy 4-Fermi 
theory in nuclear $\beta$-decay to a dynamical theory of the weak interaction
with heavy particle exchange.}
Performing an OPE for QCD and electroweak 
interactions,
the effective Hamiltonian for a FCNC $b\to s \gamma$ transition in the SM
can be obtained by integrating out $W,t,\phi$.
Up to ${\cal{O}}(\frac{1}{m_W^4})$ it is given as:\\
\begin{eqnarray}
{\cal{H}}_{eff}(b\to s \gamma)=- \frac{G_{F}}{\sqrt{2}} \lambda_t 
\sum_{i=1}^{8} C_{i}(\mu) {\cal O}_{i}(\mu) \; ,
\label{eq:heff}
\end{eqnarray}
where the weak couplings $g_W=\frac{e}{ \sin{\theta_W}}$ are collected in the 
Fermi constant $G_F$
\begin{eqnarray}
\frac{G_F}{\sqrt{2}}&=&\frac{g_W^2}{8 m_W^2} \; , \\
G_F &=&1.16639 \cdot 10^{-5} \; {\mbox{GeV}}^{-2} \; .
\end{eqnarray}
The on-shell operator basis is chosen to be \cite{effhamali,effhamburas}
\begin{eqnarray}
\label{eq:operatorbasis}
 {\cal O}_1 &=&  4 (\bar{s}_{L \alpha} \gamma_\mu b_{L \alpha})
               (\bar{c}_{L \beta} \gamma^\mu c_{L \beta}), \nonumber   \\
{\cal O}_2 &=& 4 (\bar{s}_{L \alpha} \gamma_\mu b_{L \beta})
               (\bar{c}_{L \beta} \gamma^\mu c_{L \alpha}), \nonumber   \\
{\cal O}_3 &=& 4 (\bar{s}_{L \alpha} \gamma_\mu b_{L \alpha})
               \sum_{q=u,d,s,c,b}
               (\bar{q}_{L \beta} \gamma^\mu q_{L \beta}), \nonumber   \\
{\cal O}_4 &=& 4 (\bar{s}_{L \alpha} \gamma_\mu b_{L \beta})
                \sum_{q=u,d,s,c,b}
               (\bar{q}_{L \beta} \gamma^\mu q_{L \alpha}), \nonumber   \\
{\cal O}_5 &=& 4 (\bar{s}_{L \alpha} \gamma_\mu b_{L \alpha})
               \sum_{q=u,d,s,c,b}
               (\bar{q}_{R \beta} \gamma^\mu q_{R \beta}), \nonumber   \\
{\cal O}_6 &=& 4 (\bar{s}_{L \alpha} \gamma_\mu b_{L \beta})
                \sum_{q=u,d,s,c,b}
               (\bar{q}_{R \beta} \gamma^\mu q_{R \alpha}), \nonumber   \\  
{\cal O}_7 &=&-\frac{g_{em}}{4\pi^2}\,\bar{s}\ \sigma^{\mu\nu} (m_b R + m_s L) b\ F_{\mu\nu}, \nn \\ 
{\cal O}_8 &=& -\frac{g_s}{4\pi^2}\,\bar{s}_{\alpha}\sigma^{\mu\nu} (m_b R +
               m_s L) T^a_{\alpha \beta}b_{\beta} G^a_{\mu\nu}, 
\end{eqnarray}
where $\alpha_{em}=g^2_{em}/4 \pi$ is the
electromagnetic fine-structure constant, $L(R)=1/2(1\mp \gamma_5)$, 
$\sigma_{\mu \nu}=\frac{i}{2} [\gamma_{\mu}, \gamma_{\nu}]$ and 
$\alpha,\beta$ are $SU(3)$ colour indices. $T^a, \, a=1 \dots 8$ are the
generators of QCD, some of their identities can be seen in appendix
\ref{app:qcd}. 
Here $F^{\mu \nu}, \, G^{a \mu \nu}$ denote the electromagnetic and 
chromomagnetic field strength tensor, respectively.
As can be seen from the operator basis, only degrees of freedom which are 
light compared to the heavy integrated out fields ($W,t,\phi$), 
remain in the theory.
The basis given above contains four-quark operators ${\cal O}_{1 \dots 6}$, which 
differ by colour and left-right structure. 
Among them, the current-current operators ${\cal O}_1$ and ${\cal O}_2$ are the 
dominant four-Fermi operators.
A typical diagram generating the so-called gluonic penguins ${\cal O}_{3 \dots 6}$  
is displayed in Fig.~\ref{fig:penguin}.
The operators ${\cal O}_7$ and $O_8$
are effective $b \to s \gamma$, $b \to s g$ vertices, respectively.
All operators have dimension 6.
For $b \to s \ell^+ \ell^{-}$ transitions the basis 
eq.~(\ref{eq:operatorbasis})
should be complemented by two additional operators containing dileptons. They
are discussed together with their corresponding Wilson coefficients in 
chapter \ref{chap:btoK*}.

The coupling strength of the introduced effective vertices ${\cal O}_i$ is given by
the (c-numbers) Wilson coefficients $C_{i}(\mu)$.
Their values at a large scale $\mu=m_W$ are obtained from a
``matching" of the effective with the full theory.
In the SM, the $C_{i}(m_W)$ read as follows \cite{inamilim} 
\begin{eqnarray}
\label{eq:CimW}
C_{1,3 \dots 6}(m_W)&=&0  \; , \\
C_2(m_W)&=&1  \; , \\
C_7(m_W)&=&\frac{3 x^3-2 x^2}{4(x-1)^4} \ln x+
\frac{-8x^3-5 x^2+7 x}{24 (x-1)^3} \; , \\
C_8(m_W)&=&\frac{-3 x^2}{4(x-1)^4}\ln x+
\frac{-x^3+5 x^2+2 x}{8 (x-1)^3} \; ,
\label{eq:CimWend}
\end{eqnarray}
with $x=m_t^2/m_W^2$.
It is convenient to define {\it effective} coefficients
$C_{7,8}^{{\mbox{eff}}}(\mu)$ of the operators ${\cal O}_7$ and ${\cal O}_8$. They
contain renormalization scheme dependent contributions from the 
four-quark operators ${\cal O}_{1\ldots 6}$ in ${\cal H}_{eff}$ to the effective 
vertices in $b \rightarrow s \gamma$ and $b \rightarrow s g$, respectively.
In the NDR scheme
{\footnote{We recall that in the naive dimensional regularization (NDR) scheme
the $\gamma_5$ matrix is total anti-commuting, 
i. e. $\{\gamma_5,\gamma_\mu \}=0$, thus $L \gamma_\mu=\gamma_\mu R$.}}
, which will be used throughout this work, they are written as
\cite{effhamburas}
\begin{eqnarray}
\cseff(\mu)&=&C_7(\mu)+Q_d C_5(\mu)+Q_d N_c C_6(\mu) \; , \\
C_8^{{\mbox{eff}}}(\mu)&=&C_8(\mu)+ C_5(\mu)  \; .
\end{eqnarray}
Here $N_c$ denotes the number of colours, $N_c=3$ for QCD.
\begin{figure}[htb]
\vskip -1.0truein
\centerline{\epsfysize=10in
{\epsffile{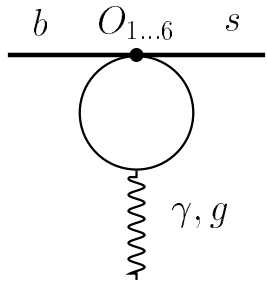}}}
\vskip -7.3truein
\caption{ \it The diagram contributing to the one loop $b \to s \gamma$,
$b \to s g$ matrix element, respectively.}
\label{fig:c78eff}
\end{figure}  
The above expressions can be found from evaluating the diagram shown in 
Fig.~\ref{fig:c78eff}.
Contributions from ${\cal O}_{1 \dots 4}$, which correspond to an
$\gamma_\mu L \otimes \gamma_\mu L $ like insertion, vanish for an on-shell 
photon, gluon, respectively.
The Feynman rules consistent with these definitions are given in 
appendix \ref{app:feynrules}.

\subsection{QCD improved $\alpha_s$ corrections}

Our aim is now to include perturbative QCD corrections in the framework of 
the effective Hamiltonian theory.
This can be done by writing down the
renormalization group equation for the Wilson coefficients
{\footnote{with $C_i=C_i(\mu,g)$ we have equivalently
$\mu \frac{{\mbox{d}}}{{\mbox{d}}\mu} C_i=
(\mu \frac{\partial}{\partial \mu} +\mu  \frac{{\mbox{d}} g}{{\mbox{d}}\mu}
\frac{\partial}{\partial g}) C_i $.}}
\begin{eqnarray}
\label{eq:rge}
\mu \frac{{\mbox{d}}}{{\mbox{d}}\mu} C_i(\mu)=\gamma_{j i} C_j(\mu) \; ,
\end{eqnarray}
where $\gamma$ denotes the anomalous dimension matrix, i.e., in general the 
operators mix under renormalization.
Solving this equation yields
the running of the couplings $C_i(\mu)$ under QCD 
from the large matching scale (here $\mu=m_W$) down to the low scale 
$\mu \approx m_b$, 
which is the relevant one for $b$-decays.
Eq.~(\ref{eq:rge}) can be solved in perturbation theory $g^2=4 \pi \alpha_s$:
\begin{eqnarray}
\gamma_{j i}&=&\frac{g^2}{16 \pi^2} \gamma_{j i}^{(0)} 
+ (\frac{g^2}{16 \pi^2})^2 \gamma_{j i}^{(1)} + \ldots \; , \\
C_i(\mu)&=&C_i(\mu)^{(0)}+\frac{g^2}{16 \pi^2} C_i(\mu)^{(1)}  + \ldots \; .
\end{eqnarray}
The initial values of the above RGE are the $C_i(m_W)$, which in 
the lowest order in the SM are given in eq.~(\ref{eq:CimW}-\ref{eq:CimWend}).

Let us for the moment concentrate on the special case that $\gamma$ is a 
number. Then the lowest order solution is given by
\begin{eqnarray}
\label{eq:LLOG}
C(\mu)&=&\eta^{\frac{\gamma^{(0)}}{2 \beta_0}} C(m_W) \; , \\
\eta&=& \frac{\alpha_s(m_W)}{\alpha_s(\mu)} \; ,
\end{eqnarray}
which can be easily checked by substituting it into eq.~(\ref{eq:rge}).
In the derivation we have used the QCD $\beta$ function, 
which describes the running of the strong coupling:
\begin{eqnarray}
\label{eq:qcdbeta}
\beta(g)= \mu \frac{d}{d \mu} g =-g (\frac{g^2}{16 \pi^2} \beta_0 
+ (\frac{g^2}{16 \pi^2})^2 \beta_1) + \ldots \; , 
\end{eqnarray}
with its lowest order solution
\begin{eqnarray}
\label{eq:alphaslo}
\frac{\alpha_s(m_W)}{\alpha_s(\mu)}=\frac{1}
{1+\beta_0 \frac{\alpha_s(\mu)}{4 \pi} \ln(\frac{m_W^2}{\mu^2})} \; .
\end{eqnarray}
We see that our obtained solution eq.~(\ref{eq:LLOG}) contains all powers of 
$\alpha_s(\mu) \ln(\frac{\mu}{m_W})$.
It is called leading logarithmic (LLog) approximation and
is an improvement of the conventional perturbation theory.
In general such a QCD improved solution contains all large 
logarithms like $n=0,1, \dots$ (here with $\mu = m_b$)
\begin{eqnarray}
\alpha^n_s(m_b) \ln^m(\frac{m_b}{m_W}) \; ,
\end{eqnarray} 
where  $m=n$ corresponds to LLog.
A calculation including the next to lowest order terms is called
next to leading order (NLO) and would result in a summation of all terms with
$m=n-1$ and so on.
In the following we use the 2-loop expression for $\alpha_s(\mu)$
which can be always cast into the form
\begin{eqnarray}
\label{eq:2loop}
\alpha_s(\mu)&=&\frac{4 \pi}{\beta_0 \ln(\mu^2/\Lambda_{QCD}^2)}
 \left[1- \frac{\beta_1 \ln \ln(\mu^2/\Lambda_{QCD}^2)}
{\beta_0^2 \ln(\mu^2/\Lambda_{QCD}^2)} \right]\; .
\end{eqnarray}
With $N_f=5$ {\it active} flavours (note that we integrated out the top)
and $SU(N_c=3)$ the values of the coefficients of the $\beta$ function are 
\begin{eqnarray}
 \beta_0 =\frac{23}{3} \; ,  \hspace{1cm} \beta_1 =\frac{116}{3} \; .
\end{eqnarray}
They are given in appendix \ref{app:qcd} for arbitrary $N_c$ and $N_f$.
The strong scale parameter $\Lambda_{QCD} \equiv \Lambda_{QCD}^{(N_f=5)}$ is 
chosen to reproduce the
measured value of $\alpha_s(\mu)$ at the $Z^0$ pole.

We recall that in LLog the calculation of the anomalous 
dimension and the matching conditions at lowest order, 
$\gamma^{(0)},C_i^{(0)}(m_W)$ is required.
In NLO a further evaluation of higher order diagrams yielding
$\gamma^{(1)},C_i^{(1)}(m_W)$ is necessary and in addition the hadronic 
matrix elements $\langle O_i \rangle$ have also to be known in 
${\cal{O}}(\alpha_s)$.

In a general theory and also in the one relevant for rare radiative $b$ 
decays given in eq.~(\ref{eq:heff}), the operators mix and the matrix $\gamma$ 
has to diagonalized.
In the latter case the $(8\times 8)$ matrix $\gamma^{(0)}$ has been obtained 
by \cite{Ciuchini,misiakE} and the running of the $C_i(\mu)$ in LLog 
approximation cannot be given analytically.
The LLog solution for the Wilson coefficients ready for numerical analysis 
can be taken from \cite{burasmuenz}. We display the $C_i$ for different
values of the scale $\mu$ in Table \ref{LOWilson}. As can be seen, there is a 
strong dependence on the renormalization scale $\mu$, especially for 
$C_1$ and $\cseff$.
Other sources of uncertainty in the short-distance coefficients $C_i$
are the top mass and the value of $\alpha_s(m_Z)$. We keep them fixed to their 
central values given in appendix \ref{app:input}.
\begin{table}[t]
        \begin{center}
        \begin{tabular}{|l|r|r|r|r|}
        \hline\hline
        \multicolumn{1}{|c|}{{\small{$C_i(\mu)$}}}&{{\small{$\mu=m_W$}}} & 
{{\small{$\mu=10$ GeV}}}&{{\small{$\mu=5$ GeV}}}&{{\small{$\mu=2.5$ GeV}}} \\
        \hline \hline
$C_1$       &  $0$     &$-0.161$  & $-0.240$ & $-0.347$  \\
$C_2$       &  $1$     &$1.064$   & $1.103$  & $1.161$\\
$C_3$       &  $0$     &$0.007$   & $0.011$  & $0.016$         \\
$C_4$       &  $0$     &$-0.017$  & $-0.025$ & $-0.035$ \\
$C_5$       &  $0$     &$0.005$   & $0.007$  & $0.010$  \\
$C_6$       &  $0$     &$-0.019$  & $-0.030$ & $-0.046$  \\   
$\cseff$ & $-0.196$ &$-0.277$  & $-0.311$ & $-0.353$  \\
$C^{{\mbox{eff}}}_8$ & $-0.098$ &$-0.134$  & $-0.148$ & $-0.164$\\
        \hline\hline
        \end{tabular}
        \end{center}
\caption{ \it Leading order Wilson coefficients in the Standard Model as a 
function of the renormalization scale $\mu$.}
\label{LOWilson}
\end{table}

Here a comment about power counting in our effective theory is in order:
As can be seen from Fig.~\ref{fig:c78eff} with an external photon, the 
insertion of four-Fermi operators generates a contribution to $b\to s \gamma$,
which is also called a ``penguin".
It is a 1-loop diagram, but unlike ``normal" perturbation theory, of order
$\alpha_s^0$. To get the $\alpha_s^1$ contribution, one has to perform already
2 loops and so on.
That means, the calculation of the LO(NLO) anomalous dimension matrix was a 
2(3)-loop task.

A comprehensive discussion of weak decays beyond leading logarithms can 
be seen in ref.~\cite{buchallaburasreview}.
The main results of the NLO calculation in $B \to X_s \gamma$ decay 
will be given in section \ref{sec:bsgamma}.

The advantages of the effective theory 
compared to the full theory can be summarized as follows:
\begin{itemize}
\item The effective theory is the more appropriate way to
include QCD corrections. Large logarithms like $\ln(\mu/m_W)$ multiplied by 
powers of the strong coupling $\alpha_s(\mu)$, which spoil the
perturbation series in the full theory, can be resummed with the help of the 
RGE.
\item On the level of a generic amplitude $A=\langle {\cal{H}}_{eff} \rangle 
\sim \sum_i C_i(\mu) \langle O_i\rangle (\mu)$
the problem can be factorized into two parts: 
The short-distance (SD) information,
which can be calculated perturbatively, is encoded in the $C_i$, and it is 
independent of the external states, i.e. quarks or hadrons.
The long-distance (LD) contribution lies in the hadronic matrix elements.
Both are separated
by the renormalization scale $\mu$. Of course the complete physical answer 
should not depend on the scale $\mu$, truncating the perturbation 
series causes such a remaining dependence, which can be reduced only after 
including higher order terms or a full resummation of the theory.
\item As long as the basis is complete, the effective Hamiltonian theory 
enables one to write down a model independent analysis in terms of the SD 
coefficients $C_i$. This is true for SM {\it near} extensions like the 
two Higgs doublet model (2HDM) 
and the minimal supersymmetric model (MSSM). 
Here one can try to fit the $C_i$ from the data \cite{agm94}.
However, new physics scenarios like, e.g., the 
left-right symmetric model (LRM) require an extended operator 
set \citer{chomisiak,hewett98}.
Wilson coefficients in the 2HDM and in supersymmetry (SUSY) can be seen in 
ref.~\cite{grinstein90} and ref.~\cite{bsgsusy}, respectively.
\end{itemize}

\section{$b \rightarrow s \gamma$ in the Effective Hamiltonian Theory
\label{sec:bsgamma}}

The effective Hamiltonian theory displayed in the previous section is applied 
to $b \rightarrow s \gamma$ transitions. Several groups have worked on the 
completion of the LLog calculation \cite{Ciuchini,misiakE}.
The anomalous dimension matrix at leading order $\gamma^{(0)}$
and the lowest order matching conditions 
(eq.~(\ref{eq:CimW}-\ref{eq:CimWend})) govern the 
running of the LLog Wilson coefficients, denoted in this and only this 
section by $C_i^{(0)}(\mu)$, to separate them from the NLO coefficients. 
We discuss the improvement of the theory in $B \rightarrow X_s \gamma$ 
obtained from NLO analysis. In the remainder of this work we treat the
Wilson coefficients $C_{i}, \, i=1, \dots 8$ in LLog approximation.

In the spectator model, the inclusive $B \rightarrow X_s \gamma$
branching ratio  
in LLog approximation can be written as 
\begin{eqnarray}
{\cal{B}}(B \to X_s \gamma) = 
{\cal{B}}_{sl} \frac{\Gamma(b \to s \gamma)}{\Gamma(b \to c e \bar{\nu}_e)}=
\frac{|\lambda_t|^2}{|V_{cb}|^2} \frac{6 \alpha_{em}}{\pi f(\mc)} 
|C_7^{(0) {\mbox{eff}}}(\mu)|^2 \; ,
\end{eqnarray}
where a normalization to the semileptonic decay $B \to X_c \ell \nu_{\ell}$
to reduce the uncertainty in the $b$-quark mass has been performed.
Here ${\cal B}_{sl}$ denotes the measured semileptonic branching ratio and
the phase space factor\footnote{For the semileptnic $B \rightarrow
X_c \ell \nu_{\ell}$ decay, the phase space factor read as:  $f(\mc) =
1 - 8  \mc^2 + 8 \mc^6 - \mc^8 - 24 \mc^4 \ln \mc$ $(\mc=mc/mb)$ .}
$f(\mc)$ for $\Gamma (B \rightarrow X_c \ell \nu_{\ell})$. 

As the branching ratio for $B \to X_s \gamma$ is mainly driven by 
$C_7^{(0) {\mbox{eff}}}(\mu)$, several effects can be deduced:
\begin{itemize}
\item Including LLog QCD corrections enhance the branching ratio for 
$B \to X_s \gamma$ about a factor $2-3$, as can be seen in 
Table \ref{LOWilson} (here denoted by $C_i(\mu)$) and changing the scale from 
$\mu=m_W$ down to $\mu \sim m_b$.
\item While the sign of $C_7^{(0) {\mbox{eff}}}$ is fixed within the SM, i.e. 
negative, it can be plus or minus in possible extensions of the SM. 
A measurement of ${\cal{B}}(B \to X_s \gamma)$
alone is not sufficient to determine the sign of $C_7^{(0) {\mbox{eff}}}$, or 
in general, the sign of $C_7^{{\mbox{eff}}}$ resulting from possible higher 
order calculations.
\item The strong scale dependence of the value of $C_7^{(0) {\mbox{eff}}}(\mu)$
causes serious problems in the accuracy of the LLog prediction.
Varying the scale between $\frac{m_b}{2} \leq \mu \leq 2 m_b$ , results in an 
error in the branching ratio of $\pm 25 \%$ 
\cite{effhamburas,aligreub93}.
\end{itemize}
Because of the last point the NLO calculation was required.
Several steps have been necessary for a complete NLO analysis.
Let us illustrate how the individual pieces look like:
At NLO, the matrix element for $b \to s \gamma$ renormalized around
$\mu=m_b$ can be written as \cite{effhamburas}:
\begin{eqnarray}
{\cal{M}}(b \to s \gamma)=-4 \frac{G_{F}}{\sqrt{2}} \lambda_t 
D \langle O_{7}(m_b) \rangle_{tree} \; ,
\label{eq:matbsg}
\end{eqnarray}
with
\begin{eqnarray}
D=\cseff(\mu)+\frac{\alpha_s(m_b)}{4 \pi} \sum_{i} \left(
C_i^{(0) \mbox{eff}}(\mu) \gamma_{i7}^{(0)} \ln \frac{m_b}{\mu}+
C_i^{(0)\mbox{eff}}(\mu) r_{i7} \right)  \; .
\label{eq:D}
\end{eqnarray}
The $r_{i7}$ are computed in ref.~\cite{nlogreub}.
They contain the $b \to s \gamma g$ bremsstrahlung
corrections~\cite{effhamali,Pott} and virtual corrections to the
${\cal O}_7$ matrix element~\cite{nlogreub}.
Especially the latter with an ${\cal O}_2$ operator insertion demands
an involved 2-loop calculation, see Figs.~1-4 in \cite{nlogreub}, where the 
corresponding diagrams are shown.
It is consistent to keep the pieces in the parentheses in eq.~(\ref{eq:D}), 
which are multiplied by $\alpha_s(m_b)$, in LLog approximation.

Now $\cseff(\mu)$ has be be known at NLO precision,
\begin{eqnarray}
\cseff(\mu)=C_7^{(0)\mbox{eff}}(\mu)+
\frac{\alpha_s(m_b)}{4 \pi}C_7^{(1)\mbox{eff}}(\mu)  \; .
\end{eqnarray}
As this job consists out of two parts, 
the work has been done by two groups: The ${\cal{O}}(\alpha_s^2)$ anomalous 
dimension matrix was obtained in ref.~\cite{nlomisiak}, which required the 
calculation of 
the residue of a large number of 3-loop diagrams, describing the mixing 
between the four-Fermi operators ${\cal O}_{1 \dots 6}$ and ${\cal O}_{7,8}$.
The second part, the NLO matching at $\mu=m_W$ has been done in 
ref.~\cite{adelyao94} and confirmed in ref.~\cite{greubhurth}.
The NLO calculation reduces the $\mu={\cal{O}}(m_b)$ scale uncertainty 
in varying $\mu$ in the range $\frac{m_b}{2} \leq \mu \leq 2 m_b$
drastically to $\pm 4.3 \%$ \cite{buraskwiatpott}
and suggests for $B \to X_s \gamma$ a scale $\mu=\frac{m_b}{2}$ as an 
``effective" NLO calculation through
\begin{equation}
\Gamma(B \rightarrow X_s \gamma)_{LO} (\mu=\frac{m_b}{2})
\approx \Gamma (B \rightarrow X_s \gamma)_{NLO} \; .
\end{equation}

As a final remark on scale uncertainties it should be noted that in the 
foregoing the top quark and the $W$ have been integrated out at the same scale
$\mu = m_W$, which is an approximation to be tested.
It is justified by the fact that the difference between
$\alpha_s(m_W)$ and $\alpha_s(m_t)$ is much smaller than the one
between $\alpha_s(m_W)$ and $\alpha_s(m_b)$
\footnote{Using eq.~(\ref{eq:2loop}) and the input parameters in 
Table~\ref{parameters}, we have
$\alpha_s(m_W)=0.12,\alpha_s(m_t)=0.11$ and $\alpha_s(m_b)=0.21$.}.
The authors of \cite{buraskwiatpott} analysed the dependencies on both
the $W$ matching scale $\mu_{W}={\cal{O}}(m_W)$ and the one at which the 
running top mass is defined: $\bar{m}_{t}(\mu_t)$ and 
$\mu_{W} \not=\mu_{t}$.
Similar to the $m_b$ scale they allowed for $\mu_W,\mu_t$ a possible range:
$\frac{m_x}{2} \leq \mu_x \leq 2 m_{x}$  where $x=W,t$.
Their findings are that the $\mu_{W},\mu_{t}$ uncertainty is
much smaller 
(namely $\pm 1.1 \%,\pm 0.4 \%$  at $\mu \sim m_b$ in NLO, respectively) than 
the uncertainty in the scale around $m_b$ and therefore negligible.

Concerning the exclusive $b\to s \gamma$ transitions, the situation is
more complicated. For a generic radiative decay $B \to F \gamma$,
where $F=P, V, S, A, T$ and $T_A$ stands for pseudoscalar, vector,
scalar, axial-vector, tensor and pseudo-tensor respectively, one 
defines the corresponding exclusive branching ratio as following:
\begin{eqnarray}
{\cal B}(B \to P(S)\gamma)& = &0\\
{\cal B}(B \to V(A)\gamma)& = & \tau_B {\alpha_{em}\ G_{F}^2 \over 32 \pi^4}|\lambda_t|^2 \ m_{b}^5\ |C_7^{\mbox{eff}}(m_b)|^2\ F^{V(A)}_1(0)^2\nn\\ 
&\times &\left (1-{m_{V(A)}^2\over {m_B}^2}\right )^3 \left (1+ {m_{V(A)}^2\over {m_B}^2}\right )\ \\
{\cal B}(B \to T(T_A)\gamma)& = & \tau_B {\alpha_{em}\ G_{F}^2 \over 256 \pi^4}|\lambda_t|^2 \ m_{b}^5\ |C_7^{\mbox{eff}}(m_b)|^2\ F^{T(T_A)}_1(0)^2\ {m_{B}^2 \over m_{T(T_A)}^2} \nn\\
&\times &\left (1-{m_{T(T_A)}^2 \over m_{B}^2}\right )^5 \left (1+ {m_{T(T_A)}^2 \over m_{B}^2}\right )
\end{eqnarray}
where $m_{B}$ ($m_{F}$) and $\tau_B$ are the $B$-meson mass (the
generic $F$-meson) and life time respectively, whereas $F^{F}_{i}(q^2)$ is the
so-called transition form factor, which will be given in
section~\ref{sec:ff}.

A good quantity to test the model dependence of the form factors for the exclusive decay is the ratio of the exclusive-to-inclusive radiative decay branching ratio:
\begin{eqnarray}
{\cal R}_{V(A)} &\equiv& {{\cal B}(B \to V(A)\ \gamma)\over {\cal B}(B \to X_s\ \gamma)} 
\nn\\
&=&   F^{V(A)}_1(0)^2 {\left (1-\mh_{V(A)}^2 \right )^3 \left (1+ \mh_{V(A)}^2 \right ) \over \left (1-m_{s}^2/ {m_b}^2\right )^3 \left(1+m_{s}^2/ {m_b}^2\right )},
\label{eqRV}  
\end{eqnarray}
and
\begin{eqnarray}
{\cal R}_{T(T_A)} &\equiv& {{\cal B}(B \to T(T_A)\ \gamma)\over {\cal B}(B \to X_s\ \gamma)}
\nn\\ 
&=&{ F^{T(T_A)}_1(0)^2\over 8 \mh_{T(T_A)}^2}
{\left (1-\mh_{T(T_A)}^2 \right )^5 \left (1+ \mh_{T(T_A)}^2 \right )
\over \left (1-m_{s}^2/ {m_b}^2\right )^3 \left (1+m_{s}^2/{m_b}^2\right )},
\label{eqRT} 
\end{eqnarray}

\noi where $\mh_{F}= m_{F}/ m_{B}$. With this normalization, one
eliminates the uncertainties from the CKM factor $\lambda_t$ and the
short distance Wilson coefficient $C_7^{\mbox{eff}}(m_b)$. Thus, we
are left in eqs.(\ref{eqRV}) and (\ref{eqRT}) with unknown form
factors $F^{F}_1(0)$, which have to be computed using some
non-perturbative methods, which will be presented in the forthcoming section.
\section{Long-Distance Effects in Exclusive $B$-decays 
\label{sec:ff}}
After having witnessed in explicit terms the short-distance
(coefficients) of the OPE, we will now turn to the long-distance
(operator matrix elements) contributions in {\it exclusive} $B$ Decays.

In general, $B$-mesons transitions can be measured inclusively over the
hadronic final state or exclusively by tagging a particular light
hadron (typically a Kaon for $B \to K^* \gamma^*$ transition). The
{\it inclusive} measurement is experimentally more difficult but
theoretically simpler to interpret, since the decay rate is well and
systematically approximated by the calculation of quark level
processes. However, the theoretical difficuly with exclusive decay
modes is usually due to their nonperturbative nature encoded in their 
hadronic form factors.

For a $B$-meson decay into a pseudoscalar meson $(P)$, the corresponding
form factors are defined by the following Lorentz decomposition of
bilinear quark current matrix elements:
\begin{eqnarray}
\langle P(p)|\bar q \, \gamma_\mu b |B(p_B)\rangle &=&
f_+(q^2)\left[(p_B+p)_{\mu}-\frac{m_B^2-m_P^2}{q^2}\,q_\mu\right]
+f_0(q^2)\,\frac{m_B^2-m_P^2}{q^2}\,q_\mu,\;\;\;\;\;\;\;\;\label{fvector}
\end{eqnarray}
\begin{eqnarray}
\langle P(p)|\bar q \, \sigma_{\mu\nu} q^\nu b|B(p_B) \rangle &=&
\frac{i f_T(q^2)}{m_B+m_P}\left[q^2(p_B+p)_{\mu}-
(m_B^2-m_P^2)\,q_\mu\right],\;\;\;\;\;\;\;\;\;\;\;\;\;\;\;\;\;\;\;\;\;\;\;
\label{ftensor}
\end{eqnarray}

\noi where $m_B$ is the $B$ meson mass, $m_P$ the mass of the pseudoscalar 
meson and $q^\mu=p_B^\mu-p^\mu$ is the four-momentum transfer . The
relevant form factors for $B$ decays into vector meson $(V)$ are defined as 

\begin{eqnarray}
\langle V(p,\varepsilon^\ast)| \bar q \gamma_\mu b | B(p_B) \rangle &=&
 \frac{2V(q^2)}{m_B+m_V} \,\epsilon_{\mu\nu\rho\sigma}
 \varepsilon^{\ast}_\nu \, p^\rho p_B^{\sigma},
\label{V}\\
\langle V(p,\varepsilon^\ast)| \bar q \gamma_\mu\gamma_5 b | B(p_B) 
\rangle &=&
  2\,i\,m_VA_0(q^2)\,\frac{\varepsilon^\ast\cdot q}{q^2}\,q_\mu + 
  i\,(m_B+m_V)\,A_1(q^2)\left[\varepsilon^{\ast}_{\mu}-
  \frac{\varepsilon^\ast\cdot q}{q^2}\,q_\mu\right]
\nonumber\\
&&
-i\,A_2(q^2)\,\frac{\varepsilon^\ast\cdot q}{m_B+m_V}
 \left[(p_{B}+p)_{\mu} -\frac{m_B^2-m_V^2}{q^2}\,q_\mu\right],\\
\langle V(p,\varepsilon^\ast)| \bar q \sigma_{\mu\nu}q^\nu b | B(p_B)
\rangle &=&
  2\,i\,T_1(q^2)\,\epsilon_{\mu\nu\rho\sigma}\varepsilon^{\ast \nu}\, 
  p_B^{\rho} p^\sigma,\\
\langle V(p,\varepsilon^\ast)| \bar q \sigma_{\mu\nu} \gamma_5 q^\nu b | 
B(p_B) \rangle &=&
T_2(q^2)\left[(m_B^2-m_V^2)\,\varepsilon_{\ast\mu}-(\varepsilon^\ast\cdot
q)\,(p_B+p)_\mu\right]
\nonumber\\[0.0cm]
 && 
+\,T_3(q^2)\,(\varepsilon^\ast\cdot
q)\left[q_\mu-\frac{q^2}{m_B^2-m_V^2}(p_B+p)_\mu\right],
\label{ffdef}
\end{eqnarray}
where $m_V$ ($\varepsilon$) is the mass (polarisation vector) of the
vector meson, $A_{0,1,2}(q^2)~ \textrm{and}~ T_{1,2,3}(q^2)$ are
defined respectively as the semileptonic and the penguin form factors.

Clearly, in order to compute these form factors $\Big(f_{+,0,T}(q^2),
A_{0,1,2}(q^2)~ \textrm{and}~ T_{1,2,3}(q^2) \Big)$ one is forced to use
some theoretical methods such as the Heavy Quark Effective Theory
(HQET), the Large Energy Effective Theory (LEET), QCD sum rules,
Lattice QCD or Quark Models. Needless to say, all these
non-perturbative methods have some limitations. Consequently the
dominant theoretical uncertainties in the exclusive modes reside in
these form factors. 

Among all these theoretical approaches, it has been shown recently,
that an adequate tool to describe heavy-to-light $B$-transitions is
the so-called Large Energy Effective Theory (LEET). Since we are
dealing in this analysis with $B \to (K^*, \rho)$ transition, we will
focus in the following on the LEET approach, more appropriate for our work.
%
%
\subsection{The Large Energy Effective Theory (LEET)\label{sec:leet}}
Although the HQET~\cite{falklecture96} has permitted a great succes
in the description of heavy-to heavy semileptonic decays such as $B\to
D^{(*)} \ell \nu_{\ell}$, it fails unfortunately in its description of the
heavy-to-light decays, such as  $B\to M$ transitions, where $M$ stands for a light
meson\footnote{a bound states of light quarks: $u,~d~ \textrm{and}~ s$.}. 

The LEET was first introduced by {\it Dugan and Grinstein}~\cite{Dugan:1990de} to study factorization of non-leptonic matrix elements in decays like
$B\to D^{(*)}\pi, D^{(*)}\rho...$, where the light meson is emitted by
the $W$-boson. Later on, {\it Charles et al.} \cite{Charles:1998dr}
have established this formalism for semileptonic and radiative rare
$B$-decays, such as $B \to \pi \ell \nu_{\ell}$, $B \to K \ell^+ \ell^-$
and $B \to K^* \gamma$. They have shown that to leading order all the
weak current $P\to P(V)$ matrix elements can be expressed in terms of
only three universal form factors. However, the LEET symmetries are
broken by QCD interactions and the leading $O(\alpha_s)$ corrections
in perturbation theory are known~\cite{Beneke:2001at,Beneke:2000wa}.
\begin{figure}[t]
   \vspace{-3.7cm}
   \epsfysize=27cm
   \epsfxsize=18cm
   \centerline{\epsffile{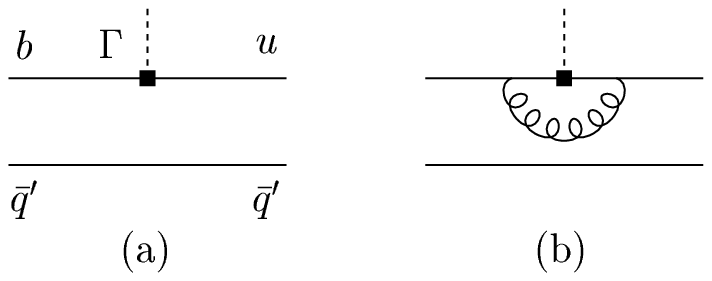}}
   \vspace*{-22.7cm}
   \epsfysize=27cm
   \epsfxsize=18cm
   \centerline{\epsffile{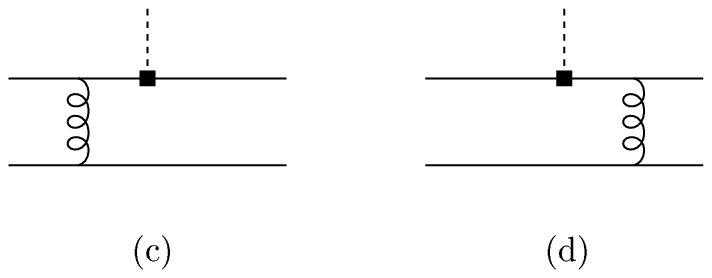}}
   \vspace*{-21.5cm}
\caption{\label{fig01}\it 
Different contributions to the 
$B\to P(V)$ transition, where $u$ stands for light quark $(u,~d,~s)$. (a) Soft contribution (soft interactions with 
the spectator antiquark $\bar{q}'$ are not drawn). (b) Hard vertex 
renormalisation. (c,d) Hard spectator interaction.}
\end{figure}

\subsubsection{Heavy-light form factors at large recoil
\label{sec:leetL}}
Let us switch to the system under consideration, the $B$-mesons. Since
the $b$-quark (inside the $B$-meson) is heavy, i .e. $m_b \sim
4.6~\textrm{GeV} >> \Lambda_{QCD}\sim 200 \textrm{MeV}$, it will
transmit all its momentum to the light quark (inside the
final light meson $P(V)$) with a large energy $E_{P(V)}$, in almost the whole
physical phase space except the vicinity of the zero-recoil point:
\begin{eqnarray}
\label{eq:E}
E_{P(V)}={m_B\over 2}\Big(1- {q^2\over m_B^2} + {m_{P(V)}^2\over
m_B^2}\Big). 
\end{eqnarray}
This assumption holds in the limit where such transitions are
dominated by soft gluon exchange, i. e. the $b$-quark and the light one
must interact with the spectator quark (and other soft degrees of
freedom) exclusively via soft exchange, as it is shown in Fig.~\ref{fig01}a.

To work out the large-recoil symmetry constraints 
on the soft form factor, one uses a technique familiar from HQET
\citer{Eichten:1990zv,Neubert:1994mb}, in writing the heavy-to-light current
$[\bar{q}\Gamma b]_{\rm QCD} = [\bar{q}_n \Gamma b_v]_{\rm eff}$. The
form factors at large recoil can be calculated within the following
set up of the LEET: 
\begin{itemize}
\item The heavy $b$-quark momentum is written as $p_b=m_b
v +k$, where $k$ is a small residual momentum of order $\Lambda_{QCD}$
and $v$ denotes the velocity of the meson with momentum $p_B=m_B v$
which at rest is $v=(1,0,0,0)$.

\item The heavy $b$-quark is then described by its large spinor field
$Q(x)$ component $b_v(x)=e^{i~ m_b v \cdot x} {1+\slash{v}\over 2}Q(x)$.

\item The light $q$-quark momentum carries a $p_q=E_{P(V)} n_{-}+ k'$
momentum, where $k' \sim \Lambda_{QCD}$  and the light-like vector
$n_{-}$ $(n_{-}^2=0)$ is parallel to the four-momentum $p=E_{P(V)} n_{-}$ of the
light meson $P(V)$.

\item The light $q$-quark is described by the large components $q_n(x)=e^{i E_q
n_- \cdot x} \, \frac{\slash{n}_- \slash {n}_+}{4} \,q(x)$ of its
quark spinor field $q(x)$. Here $n_+=2 v-n_-$ is another 
light-like vector with $n_+ \cdot n_- = 2$ and $E_q\approx E_{P(V)}$
is the energy of the light quark. 
\end{itemize}
Following the above description, the form factors at large recoil are then 
represented by (defining  $L=P,V$)
\begin{eqnarray}
\;\;\;\;\;\;\;\;\;\;\;\;\;\;\;\;\;
  \langle L(E_L n_-)|~ \bar{q}_n \Gamma \, b_v~| B(m_B v)\rangle
  &=& {\rm tr} \left[A_L(E_L)\, \overline {\mathcal M}_{L} 
    \, \Gamma \, {\mathcal M}_{\rm B} \right],
\;\;\;\;\;\;\;\;\;\;\;\;\;\;\;\;\;\;\;\;\;\;\;\;\;\;\;\;\;\;\;
\label{trace-form} 
\end{eqnarray}
where 
\begin{eqnarray}
\overline{\mathcal M}_{\rm P(V)}&=&  -\gamma_5 \frac{\slash n_+ \slash n_-}{4}
\Big(\slash \varepsilon^*  \frac{\slash n_+ \slash n_-}{4}\Big),\nn\\
{\mathcal M}_{\rm B} \,&=&\,  \frac{1+\slash v}{2} \,(-\gamma_5),
\label{IW-pi}
\end{eqnarray}
with $\varepsilon$ the polarisation vector of the vector meson. 
The function $A_L(E)$ contains the 
long-distance dynamics, but it is independent on the Dirac structure 
$\Gamma$ of the current, because the effective lagrangians (see
Appendix \ref{app:leetrules}) does not contain a Dirac matrix. The
most general form $A_L(E)$ can take is therefore 
\begin{equation}
A_L(E_L) = a_{1L}(E_L) +a_{2L}(E_L)\,\slash v+a_{3L}(E_L)\,\slash n_-+
a_{4L}(E_L)\,\slash n_-\slash v,
\end{equation}
but the projectors $\overline{\mathcal M}_{\rm L}$, 
${\mathcal M}_{\rm B}$ imply that not all the $a_{iL}(E_L)$ are 
independent. Accounting for these projectors, the most general 
form is 
\begin{eqnarray}
A_P(E_P) &=& 2 E_P \,\xi^{(P)}(E_P), 
\\
A_V(E_V) &=& E_V \,\slash n_-\left\{\xi^{(V)}_\perp(E_V)-\frac{\slash
v}{2} \,\xi^{(V)}_{||}(E_V)\right\},
\end{eqnarray}
with a conveniently chosen overall normalisation. It follows that 
the three pseudoscalar meson form factors are all related to a single
function $\xi^{(P)}(E_P)$ and the seven vector meson form factors are all 
related to two unknown functions, $\xi^{(V)}_\perp(E_V)$ and
$\xi^{(V)}_{||}(E_V)$. The latter two functions are chosen such that 
only $\xi^{(V)}_\perp(E_V)$ contributes the form factors for a transversely 
polarised vector meson and only $\xi^{(V)}_{||}(E_V)$ contributes the 
production of a longitudinally polarised vector meson. Performing 
the trace in Eq.~(\ref{trace-form}), we obtain 
\begin{eqnarray}
&&\hspace*{-0.3cm}
\langle P(p)|\bar q \, \gamma^\mu b |B(p_B)\rangle =
2 E_P \,\xi^{(P)}(q^2)\,n_-^\mu,
\nn\\
&&\hspace*{-0.3cm}
\langle P(p)|\bar q \, \sigma^{\mu\nu} q_\nu b|B(p_B) \rangle =
2i E_P\,\xi^{(P)}(q^2)\,\Big\{(m_B-E_P) \,n_-^\mu-m_B v^\mu\Big\},
\label{fvector1}
\end{eqnarray}
for pseudoscalar mesons, and 

\begin{eqnarray}
\left\langle V(p,\varepsilon^\ast)\left|\bar q \gamma^\mu b\right|B\right\rangle&=&i2E_{V}\,\xi^{(V)}\trans(q^2)\,\epsilon^{\mu\nu\rho\sigma}v_\nu
n_{-\rho} \epsilon^\ast_\sigma\,,\nn
\\
\left\langle V(p,\varepsilon^\ast)\left|\bar q \gamma^\mu\gamma_5 b\right|B(p_B)\right\rangle&=&
2E_{V}\left\{\xi^{(V)}\trans(q^2)\left[\epsilon^{\ast\mu}-(\epsilon^\ast\cdot
v)n_-^\mu\right] +\,\xi^{(V)}_{||}(q^2)\frac{m_V}{E_{V}}(\epsilon^\ast\cdot
v)n_-^\mu\right\}\,,\nn
\\
\langle V(p,\varepsilon^\ast)| \bar q \sigma^{\mu\nu}q_\nu b | B(p_B)
\rangle &=& 2 E_{V} m_B \,\xi^{(V)}_\perp(q^2)\,
\epsilon^{\mu\nu\rho\sigma}\varepsilon^{\ast}_\nu\, 
  v_\rho n_{-\sigma},\nn
\\
\left\langle V(p,\varepsilon^\ast)\left|\bar q \sigma^{\mu\nu} \gamma_5 q_\nu b\right|B(p_B)\right\rangle&=&
-i2E_{V}\xi^{(V)}\trans(q^2)\left(\epsilon^{\ast\mu}n_-^\nu-
\epsilon^{\ast\nu}n_-^\mu\right)\nn
\\
&&-i2E_{V}\xi^{(V)}_{||}(q^2)\frac{m_V}{E_{V}}(\epsilon^\ast\cdot v)\left(n_-^\mu
v^\nu-n_-^\nu v^\mu\right)\,,\label{ffdef1}
\end{eqnarray}

\noi for vector mesons, in agreement with Ref.~\cite{Charles:1998dr}. Comparing 
Eqs.~(\ref{fvector})-(\ref{ffdef}) with Eqs.~(\ref{fvector1})-(\ref{ffdef1}), we find the following form
factor relations:
\begin{eqnarray}
f_+(q^2)&=&\xi^{(P)}(q^2)\,,\label{f+}\\
f_0(q^2)&=&\left(1-\frac{q^2}{m_B^2-m_P^2}\right)\xi^{(P)}(q^2)\,,\\
f_T(q^2)&=&\left(1+\frac{m_P}{m_B}\right)\xi^{(P)}(q^2)\,,
\label{pirelation}
\end{eqnarray}
for pseudoscalar mesons and

\begin{eqnarray}
A_{0}(q^2) &=& (1 - {m_{V}^2\over m_{B} E_V })\ \xi^{(V)}_{||}(q^2) +
 {m_{V}\over m_B}\ \xi^{(V)}_{\perp}(q^2)~,\label{A0} \\
A_{1}(q^2) &=&{2 E_V \over m_{B}+m_{V}}\ \xi^{(V)}_{\perp}(q^2)~,\label{A1} \\
A_{2}(q^2)&=& (1 + {m_{V}\over m_B})\ [\ \xi^{(V)}_{\perp}(q^2) - {m_{V}\over E_V}\
 \xi^{(V)}_{||}(q^2)\ ]~,\label{A2} \\
V(q^2)&=&(1 + {m_{V}\over m_B})\ \xi^{(V)}_{\perp}(q^2)\label{V} \\
T_{1}(q^2)&=& \xi^{(V)}_{\perp}(q^2)~, \label{T1} \\
T_{2}(q^2)&=&(1- {q^2\over m_{B}^2-m_{V}^2}) \ \xi^{(V)}_{\perp}(q^2) ~,\label{T2} \\
T_{3}(q^2)&=& \xi^{(V)}_{\perp}(q^2)- {m_{V}\over E_V}\ (1- {m_{V}^2\over m_{B}^2}) \
 \xi^{(V)}_{||}(q^2)~,\label{T3} 
\end{eqnarray}
for vector mesons. These relations are valid for the {\it soft 
contribution}\/ to the form factors at large recoil, neglecting
corrections of order $1/m_b$ and $\alpha_s$. 
\subsubsection{Symmetry-breaking corrections to the LEET form factors 
\label{sec:leetL}}
We have just seen the  LEET effect in
describing the exclusive heavy-to light semileptonic decays by
reducing the number of independent form factors from ten to
three. However theses symmetries are broken by factorizable and
non-factorizable QCD corrections, worked out by Beneke {\it et al.}
\cite{Beneke:2001at,Beneke:2000wa}.

While the form factors obtained in Eqs.(\ref{f+})-(\ref{T3}) are a
straighforward evaluation of the soft contributions in
Fig.~\ref{fig01}a, their ${\cal O}(\alpha_s)$-corrections are
originated from the two following processes:
\begin{itemize}
\item the vertex corrections (Fig.~\ref{fig01}b)
\item the hard scattering corrections (Fig.~\ref{fig01}c-d)
\end{itemize}
The vertex corrections are a straightforward calculations using
standard techniques, in contrast to the hard scattering ones where one
makes use of the two-particle light-cone distribution amplitudes of the $B$ meson and
the light meson (more details can be found in
Ref.\cite{Beneke:2000wa}).

Finally, having these ${\cal O}(\alpha_s)$-corrections at hand, the form factors
defined in Eqs.(\ref{f+})-(\ref{T3}) get modified as follows\cite{Beneke:2000wa}:
\begin{eqnarray}
f_+(q^2) &=&  \xi^{(P)}(q^2)+ \frac{\alpha_s \, C_F}{4\pi} \, \Delta f_+,
\label{add+}\\ 
f_0(q^2) &=& \frac{2E_P}{m_B} \, \xi^{(P)}(q^2) \, \left\{1 +
    \frac{\alpha_s \, C_F}{4\pi} \,
    \left[ 2 - 2 \, {\cal Y}  \right] \right\}
  + \frac{\alpha_s \, C_F}{4\pi} \, \Delta f_0, 
\label{add1}\\
  f_T(q^2) &=& \frac{m_B+m_P}{m_B} \, \xi^{(P)}(q^2) \, \left\{1 +
    \frac{\alpha_s \, C_F}{4\pi} \,
    \left[ \ln \frac{m_b^2}{\mu^2} + 2 \, {\cal Y} \right]
  \right\}+\frac{\alpha_s \, C_F}{4\pi} \, \Delta f_T,
\label{vertexcorr1}
\end{eqnarray}
for the form factors of pseudoscalar mesons and
%
\begin{eqnarray}
A_{0}(q^2) &=& (1 - {m_{V}^2\over m_{B} E_V })\ \xi^{(V)}_{||}(q^2) +
 {m_{V}\over m_B}\ \xi^{(V)}_{\perp}(q^2)~+\frac{\alpha_s \, C_F}{4\pi} \, \Delta A_{0}~,\label{A0cor} \\
V(q^2)&=&(1 + {m_{V}\over m_B})\ \xi^{(V)}_{\perp}(q^2)+\frac{\alpha_s \, C_F}{4\pi} \, \Delta V~,\label{Vcor} \\
A_{1}(q^2)&=& {2 E_{V}\over m_{B}+m_{V}}\
\xi^{(V)}_{\perp}(q^2)+\frac{\alpha_s \, C_F}{4\pi} \, \Delta A_{1}~,\label{A1cor}
\\
A_{2}(q^2)&=& {m_{B} \over m_{B} - m_{V}} \left\{ \xi^{(V)}_{\perp}(q^2) - {m_{V}\over
E_{V}} \ \xi^{(V)}_{||}(q^2) (1 +  {\alpha_{s} C_{F}\over 4 \pi} [-2
+2 \, {\cal Y}])\right\}+ \frac{\alpha_s \, C_F}{4\pi} \, \Delta A_{2}~,\;\;\;\;\;\;\;\;\;
\label{A2cor}\\
T_1(q^2) &=& \xi^{(V)}_\perp(q^2) \, \left\{1 +
    \frac{\alpha_s \, C_F}{4\pi} \,
    \left[  \ln \frac{m_b^2}{\mu^2} - {\cal Y} \right]
  \right\}+\frac{\alpha_s \, C_F}{4\pi} \, \Delta T_1,
\\
T_2(q^2) &=& \xi^{(V)}_\perp(q^2) \Big(1- {q^2\over m_{B}^2-m_{V}^2}\Big)  \, \left\{1 +
    \frac{\alpha_s \, C_F}{4\pi} \,
    \left[ \ln \frac{m_b^2}{\mu^2} - {\cal Y} \right]
  \right\}+\frac{\alpha_s \, C_F}{4\pi} \, \Delta T_2,
\;\;\;\;\;\;\\
T_3(q^2) &=& \xi^{(V)}_\perp(q^2) \,  \left\{1 +
    \frac{\alpha_s \, C_F}{4\pi} \,
    \left[ \ln \frac{m_b^2}{\mu^2} - {\cal Y} \right]
  \right\} -\xi^{(V)}_{||}(q^2) {m_{V}\over E_{V}} \Big(1- {m_{V}^2\over m_{B}^2}\Big)\nn
\\&& 
    \times \left\{1 +
    \frac{\alpha_s \, C_F}{4\pi} \,
    \left[ \ln \frac{m_b^2}{\mu^2} -2  + 4 \, {\cal Y} \right]
  \right\} +\frac{\alpha_s \, C_F}{4\pi} \, \Delta T_3
\label{vertexcorr2}
\end{eqnarray}
\noi for the form factors of vector mesons. The abbreviation
    ${\cal Y}$ stands for 
\begin{eqnarray}
 {\cal Y} &=& - \frac{2E_{P(V)}}{m_B-2E_{P(V)}} \, \ln\frac{2E_{P(V)}}{m_B}
\label{Labbrev}
\end{eqnarray}
\begin{figure}[t]
   \vspace{-3.2cm}
   \epsfysize=25.2cm
   \epsfxsize=18cm
   \centerline{\epsffile{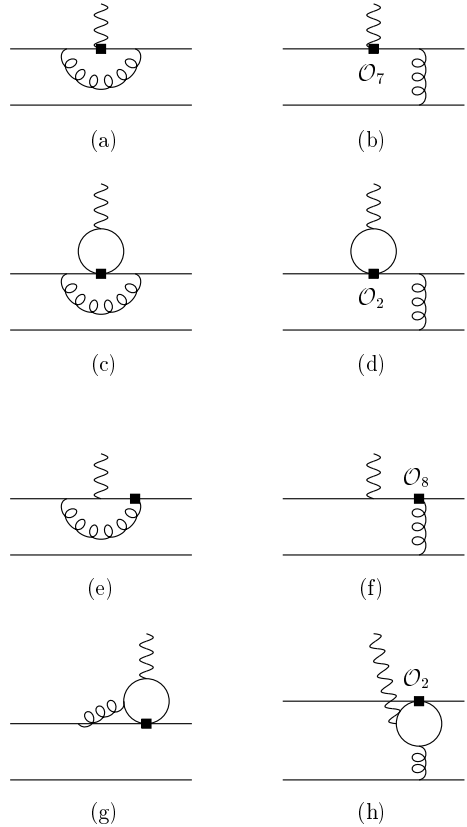}}
   \vspace*{-10.8cm}
\caption[dummy]{\label{fig3}\small Various next-to-leading 
order contributions to the $B\to V\gamma^*$ matrix elements.}
\end{figure}
Moreover in Eqs.~(\ref{add+})-(\ref{vertexcorr2}),  the form factors
receive a further additive correction from the interaction with the
spectator quark, indicated by $\Delta F_i$ (and can be found in Appendix
\ref{app:leetcorr}). Its general form reads as   
\begin{eqnarray}
\Delta F_i \approx \Phi_B \otimes T_i \otimes \Phi_L\nn
\end{eqnarray}
where $T_i$ is a hard-scattering kernel convoluted with the light-cone
distribution amplitudes of the $B$ meson and the light meson $L$. 
Thus, we can summarize the $O(\alpha_s)$-LEET corrections by the following, tentative, factorization formula for a heavy-light form factor at large recoil, and at leading order in $1/m_B$:
\begin{eqnarray}
\label{fff}
f_i(q^2)  &=& C_i \, \xi^{L}(E_L) + \Phi_B \otimes T_i \otimes \Phi_L,
\end{eqnarray}
where $\xi^{L}(E_L)$ is the soft part of the form factor, to which the 
LEET symmetries discussed above apply and $C_i=1+O(\alpha_s)$ 
is the hard vertex corrections. At this stage, we have seen that the
$O(\alpha_s)$-LEET corrections can be absorbed into the redefinition
of the corresponding LEET form factor $\xi$. 

However, concerning the $B\to V$-transition,  there exist
further corrections at order $\alpha_s$, originating from  four quark operators and the
chromomagnetic dipole operator in the weak effective hamiltonian,
which cannot be expressed in terms of form factors, i.e.\ matrix
elements of the type $\langle V|\bar{s}\Gamma b|B\rangle$, and will be
presented in section~\ref{sec:matrixNLO}. Sample Feynman diagrams are shown in Fig.~\ref{fig3}e-g, compared to the diagrams in Fig.~\ref{fig3}a-d, which do assume 
the structure of form factor matrix elements and it will be discussed
in section 3.2.
%
\subsubsection{Soft-collinear contributions to the LEET form factors}
\begin{figure}[t]
   \vspace{0.2cm}
   \epsfxsize=12cm
   \centerline{\epsffile{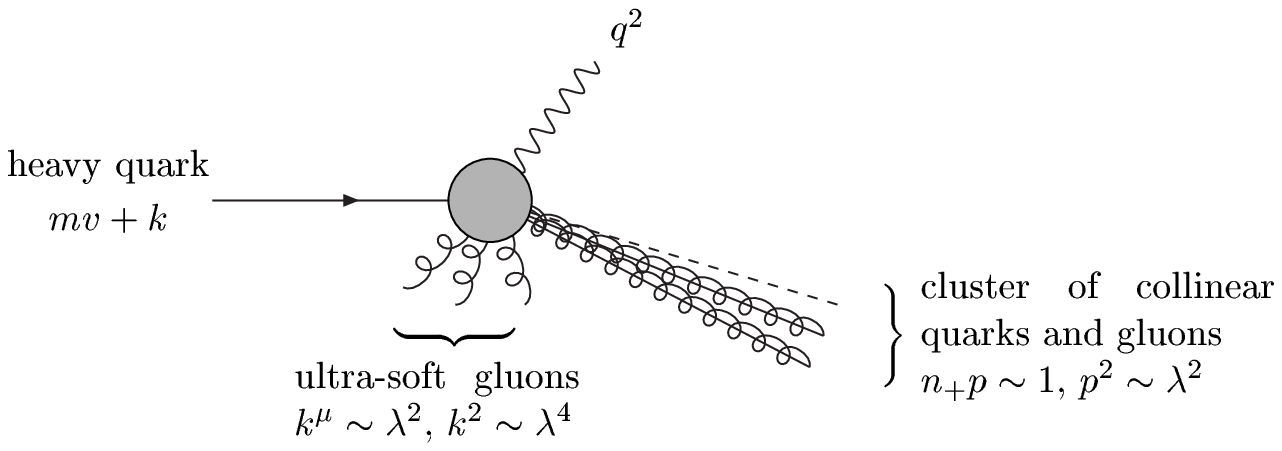}}
   \vspace*{-0.3cm}
\caption{\label{setup} \it Kinematics of heavy quark decay into 
a single cluster of collinear and ultrasoft particles.}
\end{figure}
An important unresolved question in strong interaction physics
concerns the parameterization of power-suppressed long-distance
effects to hard processes that do not admit an operator product
expansion (OPE). For a large class of processes of this type the
principal difficulty arises from the presence of collinear modes,
i.e. highly energetic, but nearly massless particles.

Recently {\it Bauer et al.}\cite{Bauer:2000yr} have claimed that the
missing collinear gluons in the LEET do not allowed this effective
theory to reproduce the Infrared (IR) physics of
QCD\cite{Aglietti:1997zk}. Thus an effective theory is able to
reproduce correctly the infrared physics of QCD at one loop, only by
including 
both collinear and soft gluons\cite{Bauer:2000ew}.

Happily, {\it Bauer et al.}\cite{Bauer:2000yr} and {\it Beneke et
al.}\cite{Beneke:2002ph} have formulated separately the heavy-to-light
soft-collinear effective theory by taking into account the collinear
and ultrasoft particles (see Figure~\ref{setup}), missing in the LEET approach.
They found independently that the presence of collinear gluons does not spoil the
relations among the soft form factors, therefore establishing the
corresponding results in the large energy limit of QCD \cite{Beneke:2000wa}.

Since the numerical effect of the collinear gluons contributions
are negligeable\cite{Beneke:2002ph} on the LEET form factors (defined
in section~\ref{sec:leetL}), we will not consider them in our work.

\chapter{Exclusive $B \to \ K^{*} \ell^{+}\ell^{-}$ Decay in the SM
\label{chap:btoK*}}

This chapter contains a comprehensive helicity analysis of the $B \to K^*
\ell^+ \ell^-$ and  the $B \rightarrow \rho  \ell \bar{\nu}$ decays in
the so-called Large-Energy-Effective-Theory (LEET). Taking into
account the dominant $O(\alpha_s)$ and $SU(3)$ symmetry-breaking
effects, we calculate various double and single distributions in these decays
making use of the presently available data and decay form factors
calculated in the QCD sum rule approach. As precision tests of
the standard model in semileptonic rare $B$-decays, we propose a model
independent extraction of the CKM matrix elements
$\vert V_{ub}\vert/\vert V_{ts}\vert$.
\section{Introduction}             
Flavour changing neutral current (FCNC) decays $ b\to s~\gamma$ and $ b\to s~\l^+\l^-$ are governed in the SM by loop effects.
They provide a sensitive probe of the flavour sector in the SM and search for 
physics beyond the SM. In the context of rare $B$ decays the radiative mode
$ b\to s~\gamma$ has been extensively discussed in chapter~\ref{chap:rare}.

In this chapter we address the exclusive semileptonic $B \to K^*
\ell^+ \ell^-$ and the $B \rightarrow \rho \ell \bar{\nu}$ decays with
$\ell^\pm =e^\pm,\mu^\pm$ in the LEET framework. Since we are 
neglecting finite lepton masses we cannot apply our results to the $\tau$-case.
The theoretical study of the exclusive rare decays proceeds
in two steps. First, the effective Hamiltonian  for such
transitions is derived by calculating the leading and next-to-leading
loop diagrams in the SM and by using the operator product expansion
and renormalization group techniques (for a review see \cite{Buras:1995} and references therein). Second, one needs to evaluate the matrix elements of the 
effective Hamiltonian between  hadronic states. This part of 
the calculation is model dependent since it involves  nonperturbative QCD.
Many theoretical approaches have been employed to predict the
exclusive radiative $B\to (K,K^\ast)\,\ell^+\ell^-\,(\ell=e,\mu,\tau)$
decays. Most of them rely on QCD sum rules
\cite{Colangelo:1995jv,Ali:1999mm}, quark model \cite{Greub:1994pi},
lattice-constrained dispersion quark model in \cite{Melikhov:1997wp}
and  perturbative QCD \cite{Chen:2002}.
Recently, an updated analysis of  these decays has been
done in \cite{Ali:2002jg} by including  explicit $O(\alpha_s)$
and $\Lambda_{\rm QCD}/m_b$ corrections. Concerning, the lepton polarizations
in the $B\to K^\ast(\to K \pi)\,\ell^+\ell^-$ decay in terms of the
helicity amplitudes $H_{0,\pm}^{L,R}(q^2)$ without the explicit
$O(\alpha_s)$ corrections, was undertaken in a number of
papers~\citer{Melikhov:1998cd,Faessler:2002ut}.
In particular, Kim et al.~\cite{Kim:2000dq,Kim:2001xu}  
emphasized the role of the azimuthal angle distribution as a
precision test of the SM. Following closely the earlier analyses, we now
calculate the $O(\alpha_s)$ corrections in the LEET framework.

Concentrating on the decay $B \to K^* \ell^+ \ell^-$, the main theoretical
tool is the factorization Ansatz which enables one to relate the form 
factors in full QCD \big(called in the literature $A_{i}(q^2)~~(i=0,1,2)$, 
$V(q^2)$, $T_{i}(q^2)~~(i=1,2,3)$\big) and the two LEET form factors
$\xi^{(V)}_\perp (q^2)$ and $\xi^{(V)}_{||}(q^2)$~\cite{Beneke:2001at,Beneke:2000wa};
\begin{equation}
f_k(q^2)= C_{\perp}\xi^{(V)}_\perp (q^2) + C_{||}\xi^{(V)}_{||}(q^2) + \Phi_B \otimes 
T_k \otimes \Phi_V~,
\label{eq:fact}
\end{equation}
where the quantities $C_i$ $(i=\perp, \parallel)$ encode the perturbative 
improvements of the factorized part
\begin{equation}
C_i=C_i^{(0)} + \frac{\alpha_s}{\pi} C_i^{(1)} + ... ,
\nonumber
\end{equation}
and $T_k$ is the hard spectator kernel (regulated so as to be free of 
the end-point singularities), representing the 
non-factorizable perturbative corrections, with the direct product
understood as a convolution of $T_k$ with the light-cone distribution 
amplitudes of the $B$ meson ($\Phi_B$) and the vector meson ($\Phi_V$). 
With this Ansatz, it is a straightforward exercise to implement the
$O(\alpha_s)$-improvements in the various helicity amplitudes. 
The non-perturbative information is encoded in the LEET-form factors,
which are {\it a priori} unknown, and the various parameters which enter 
in the description of the non-factorizing hard spectator contribution,
which we shall discuss at some length. The normalization of the LEET form
factor $\xi_\perp^{(K^*)} (q^2)$ at $q^2=0$ is determined by the
$B \to K^* \gamma$ decay rate; the other form factor $\xi_{||}^{(K^*)}(q^2)$
has to be modeled entirely for which we
use the light cone QCD sum rules. This input, which
for sure is model-dependent, is being used to illustrate the various
distributions and should be replaced as
more precise data on the decay $B \to \rho \ell \nu_\ell$ becomes
available, which then can be used directly to determine 
the form factors $\xi_\perp^{(K^*)}(q^2)$ and $\xi_{||}^{(K^*)}(q^2)$,
taking into account the $SU(3)$-breaking effects.

This chapter is divided roughly into three parts. The first one (this
section up to and including section \ref{sec:BtoK}), contains an
introduction to $B \to K^* \ell^+ \ell^-$ decay, basic definitions and
the  $O(\alpha_s)$ 
improvements to the $b \to s~ \ell^+ \ell^-$ matrix elements in the
LEET framework. It is mainly devoted to the analysis of the double and
single angular distributions for the individual helicity amplitudes,
and their sum, and the Forward-Backward (FB) asymmetry. In doing that
we have shown the systematic improvement in $O(\alpha_s)$ and $1/E$ in
the exclusive radiative $B \to K^* \ell^+ \ell^-$ decay, using the
large energy expansion (LEET). Further, we carry out in the so-called
transversity basis, the LEET-based transversity amplitudes (both in the
LO and NLO accuracy), and compare them to the  currently available data.

The second part discussed in section \ref{sec:Btorho}, describes a
helicity distributions analysis of the exclusive semileptonic $B
\rightarrow \rho  \ell \nu_\ell$ decay in the LEET. We display the
various helicity components, Dalitz distributions, and the dilepton
($\nu_\ell \ell$) invariant mass, making explicit the $O(\alpha_s)$ corrections. The estimates of the $B \to \rho$ LEET form factors
$\xi^{(\rho)}_{\perp}(s)$ and $\xi^{(\rho)}_{\parallel}(s)$, which are scaled from their $B \to K^*$ counterparts incorporating SU(3)-breaking,
are also displayed here.

Finally the last part deals with subsection \ref{sec:Vub}, is devoted to
the determination of the ratio of the CKM matrix elements $\vert
V_{ub}\vert /\vert V_{ts}\vert$ from the ratio of the dilepton mass spectra in $B \to \rho \ell \nu_\ell$ and $B \to K^* \ell^+ \ell^-$ decays involving definite helicity states. In particular, we show the dependence of the ratio 
$R_{i}(s) ={{d\Gamma_{H_i}^{B \rightarrow  K^{*} \ \l^{+}  \l^{-}}/ds}
\over{d\Gamma_{H_i}^{B \rightarrow  \rho \ \l \nu_\l}/ds}}$ $(i=0,-1)$ involving
the helicity-0 and -1 components, on the CKM matrix elements $\vert
V_{ub} \vert/\vert V_{ts} \vert$.  
%
\subsection{Kinematics \label{sec:kin}}
We start with the definition of the kinematics of the exclusive
semileptonic $B \to V (\ell^+ \ell^-, \ell~ \nu_{\ell})$ decays,
\begin{equation}
B (p_B) \to V (p_{V}, \epsilon_{V}) +\ell (p_{+})+(\ell, \nu_{\ell}) (p_{-}) ~,
\end{equation} 
where the index $V$ stands for the corresponding vector meson.
We define the momentum transfer to the lepton pair and the invariant 
mass of the dilepton system, respectively, as
\begin{eqnarray}
q &\equiv & p_{+}+p_{-} \; , \\
s &\equiv & q^2 \; .
\end{eqnarray}
The dimensionless variables with a hat denotes normalization in terms
of the $B$-meson mass, e.g.,
\begin{eqnarray}
\sh & = & \frac{q^2}{m_B^2} = (\ph_+ + \ph_-)^2 \; , \\
\hat{m}_{V} & = & \frac{m_{V}}{m_B}, \\
\ml & = & \frac{m_{\ell}}{m_B},
\end{eqnarray}
etc., where $m_{V}$ $(m_{\ell})$  is the corresponding vector meson mass (lepton mass). 
Since we are dealing in our analysis with the two lepton generations,
namely $\ell^\pm =e^\pm,\mu^\pm$, one can neglect their finite
masses. Thus, the scaled variable $\hat{s}$ in the $B\to V
(\ell^+ \ell^-, \ell~ \nu_{\ell})$  decays is bounded as follows,
\begin{eqnarray}
  (\mlh^2\approx 0) \leq & \sh & \leq (1 - \hat{m}_{V})^2  \; .
  \label{eq:sbound}
\end{eqnarray}

\subsection{NLO-corrected amplitude for $b \to s ~\ell^+ \ell^- $ 
\label{sec:NLO}}

Next, the  explicit 
expressions for the matrix element and (partial) branching ratios in
the decays  $b \to s ~\ell^+ \ell^- $ are presented in terms of the Wilson coefficients of the 
effective 
Hamiltonian obtained by integrating out the top quark and the $W^\pm$ bosons,
\begin{equation}\label{heffbsll}
{\cal H}_{eff}(b \to s + \ell^+ \ell^-)
  = {\cal H}_{eff} (b \to s + \gamma) -\frac{G_F}{\sqrt{2}} V_{ts}^* V_{tb}
\left[ C_9(\mu)  {\cal O}_9 +C_{10} {\cal O}_{10} \right],
\end{equation}
where ${\cal H}_{eff}(b \to s +\gamma)$ together with the operators
$O_{1 \dots 8}$ and their corresponding Wilson coefficients $C_i(\mu)$ 
\cite{effhamburas,effhamali} can be seen in section \ref{sec:effham}.
The two additional operators involving the dileptons ${\cal O}_9$ and ${\cal O}_{10}$ 
are defined as:
\begin{eqnarray}
{\cal O}_{9} &=& \frac{\alpha_{em}}{\pi}\;
\bar{s}_\alpha \gamma^{\mu} L b_\alpha
\bar{\ell} \gamma_{\mu} \ell \;,\nonumber\\
{\cal O}_{10} &=& \frac{\alpha_{em}}{\pi}\;
\bar{s}_\alpha \gamma^{\mu} Lb_\alpha 
\bar{\ell} \gamma_{\mu}\gamma_5 \ell  \; .
\end{eqnarray}
A usual, CKM unitarity has been used in factoring out the product $V_{ts}^\ast
V_{tb}$. 
The Wilson coefficients are given in the
literature (see, for example, \cite{misiakE,burasmuenz}).
They depend, in general, on the renormalization scale $\mu$,
except for $C_{10}$. 
With the help of the effective Hamiltonian in eq.~(\ref{heffbsll})
 the matrix element for the decay $\bsll$  can be factorized 
into a leptonic and a hadronic part as,
\begin{eqnarray}
{\cal M}(b\to s\ell^+\ell^-) & = & \frac{G_F \alpha_{em}}{\sqrt{2}
\pi} \ V_{t s}^{*} V_{tb} \,  
\Big\{ C_{9}  \left[ \bar{s}  \gamma_{\mu}  L  b \right] \left[
\bar{\l}  \gamma^{\mu}  \l \right] 
+ C_{10} \left[ \bar{s} \gamma_{\mu}  L  b \right] \left[ \bar{\l}
\gamma^\mu  \gamma_5  \l \right] 
\nn\\
& & \; \; \; \; \; \; \; \; \; \; \; \; \; \; 
\; \; \; \; \; \; \; \; \; \left. 
- 2 \hat{m_{b}}  C_{7}^{\bf eff}  \left[ \bar{s}  i  \sigma_{\mu \nu}
  {\hat{q^{\nu}}\over\hat{s}} R  b \right]  
\left[ \bar{\l}  \gamma^{\mu}  \l \right] \right \} \; .
\label{eqn:hamiltonian}
\end{eqnarray}
where we neglect the $m_s$ mass. Here  and in the remainder of this work we shall
denote by $m_b \equiv m_b(\mu)$ the $\overline{MS}$ mass evaluated at a
scale $\mu$, and by $m_{b,pole}$ the pole mass of the $b$-quark. To
next-to-leading order the pole and $\overline{\rm MS}$ masses are related by 
\begin{equation}
\label{polerel}
m_b(\mu) = m_{b,pole}\left(1+\frac{\alpha_s(\mu) C_F}{4\pi}\left[
3\ln\frac{m_b^2}{\mu^2}-4\right]+O(\alpha_s^2)\right)~.
\end{equation}
Since we are including the next-to-leading corrections into our
analysis, we will take the Wilson coefficients in
next-to-leading-logarithmic order (NLL) given in Table \ref{tab1}.
\begin{table}[t]
\vspace{0.1cm}
\begin{center}
\begin{tabular}{|l|c|c|c|c|c|c|}
\hline\hline
\rule[-2mm]{0mm}{7mm}
 & $\bar{C_1}$ & $\bar{C_2}$ & $\bar{C_3}$ & $\bar{C_4}$ & $\bar{C_5}$ & $\bar{C_6}$ \\
\hline
\rule[-0mm]{0mm}{4mm}
LL    & $-0.257$ & $1.112$ & $0.012$ & $-0.026$ & $0.008$ & $-0.033$ \\

NLL   & $-0.151$ & $1.059$ & $0.012$ & $-0.034$ & $0.010$ & $-0.040$ \\
\hline
\rule[-2mm]{0mm}{7mm}
 & $C_7^{\rm eff}$ & $C_8^{\rm eff}$ & $C_9$ & $C_{10}$
 & $C_9^{\rm NNLL}$ &  $C_{10}^{\rm NNLL}$ \\
\hline
\rule[-0mm]{0mm}{4mm}
LL  & $-0.314$ & $-0.149$ & $2.007$ & 0
 & & \\
NLL & $-0.308$ & $-0.169$ & $4.154$ & $-4.261$
 & \raisebox{2.5mm}[-2.5mm]{$4.214$} & \raisebox{2.5mm}[-2.5mm]{$-4.312$} \\
\hline\hline
\end{tabular}
\end{center}
\centerline{\parbox{14cm}{\caption{\label{tab1}
 \it Wilson coefficients at the scale $\mu=4.6\,$GeV in
leading-logarithmic (LL) and next-to-leading-logarithmic order
(NLL) \cite{Beneke:2001at}.}}}
\end{table}

The $K^*$ meson subsequently decays to $K$ and $\pi$, with effective
Hamiltonian
\begin{equation}
{\cal H}_{eff} = g_{K^*K\pi} ( p_K -p_\pi) \cdot \epsilon_{K^*},
\end{equation}
with $g_{K^*K\pi}$ denotes the strong coupling of $K$-mesons, to
$P$-wave pion. In the following analysis, we neglect the masses of
leptons, kaon and pion. Then the final 4-body decay amplitude can be written as the sum of two amplitudes,
\begin{equation}
{\mathcal M}^{\it 4-body} = {\mathcal{M}_R}+{\mathcal{M}_L}~, 
\label{Amp:4-body}
\end{equation}
where ${\mathcal{M}_R}$ and ${\mathcal{M}_L}$ denote respectively the
left and right helicity amplitudes in the dilepton system; and they 
can be written in a compact form,
\begin{eqnarray}
{\mathcal{M}_R}=&&
\frac{G_F}{\sqrt{2}} V_{tb}V_{ts}^* g_{K^*K\pi}
 \frac{\alpha_{em} m_b}{ \pi s} (\bar \ell_R \gamma^\mu \ell_R)
\left( a_R g_{\mu\nu} -b_R {p_\mu q_{\nu}} +
ic_R ~\varepsilon _{\mu \nu \alpha \beta} p^\alpha q^\beta \right ) \\
\nonumber
&&
\frac{g^{\nu\alpha}-p^\nu p^\alpha/m_{K^*}^2}
{ p^2-m_{K^*}^2 +i m_{K^*} \Gamma_{K^*}} (p_K-p_\pi)_\alpha,\nonumber \\
{\mathcal{M}_L}=&&
\frac{G_F}{\sqrt{2}} V_{tb}V_{ts}^* g_{K^*K\pi}
 \frac{\alpha_{em} m_b}{\pi s} (\bar \ell_L \gamma^\mu \ell_L)
\left( a_L g_{\mu\nu} -b_L {p_\mu q_{\nu}} +
ic_L ~\varepsilon _{\mu \nu \alpha \beta} p^\alpha q^\beta \right ) \\
\nonumber&&
\frac{g^{\nu\alpha}-p^\nu p^\alpha/m_{K^*}^2}
{ p^2-m_{K^*}^2 +i m_{K^*} \Gamma_{K^*}} (p_K-p_\pi)_\alpha,
\end{eqnarray}
with $p \equiv p_K + p_{\pi}$, and the auxiliary functions $a_R,b_R,c_R$ and $a_L,b_L,c_L$ can be expressed as
\begin{eqnarray}
a_{L/R} &=&  {i(m_B +  m_{K^*})\over {2\ m_b\ m_B \sqrt{s}}} \Big[s\ m_B
(\pm C_{10} - C_{9})  A_1(s) + 4 {\cal T}_1(s)\ m_b (m_{K^*} -  m_B) E_{K^*}\Big],\label{aL}\\
b_{L/R} &=& {i\over m_b\ m_B\ (m_B^2 -m_{K^*}^2) \ \sqrt{s}} 
\Big[ 4\ {\cal T}_1(s) \ m_b \ (-m_B^2 + m_{K^*}^2) \ E_{K^*} 
\label{bL}\\
& & + m_B s \Big(-2\ m_b  \left\{ {\cal T}_1(s)+ {\cal T}_3(s)-\frac{m_B}{2E_{K^*}}\,{\cal T}_2(s)\right \}
+ A_2(s)\ (\pm C_{10} - C_{9})(m_B -m_{K^*}) \Big) \Big ]\nn\\
c_{L/R} &=&  {i\over {m_b\ (m_B + m_{K^*})\ \sqrt{s}}} 
\Big[2 {\cal T}_1 (s)\ m_b \ (m_B + m_{K^*}) + (\mp C_{10} + C_9)\ s \
V(s)\Big].
\label{cL}
\end{eqnarray}
Note that our conventions for $a_{L,R}$, $b_{L,R}$ and $c_{L,R}$ are
slightly differents from those defined by {\it Kim et al.} in 
ref.~\cite{Kim:2000dq} by a factor of $1/\sqrt{s}$. 
The form factors $V,~A_{0},~A_{1},~A_{2},~T_{1},~T_{2}$ and $T_{3}$
have already been introduced in section 2.4 and their
numerical values will be presented below.
The functions ${\cal T}_1,~{\cal T}_2$ and ${\cal T}_3$ are related to the so-called penguin form factors, and will be defined in the next section.

With the help of the above expressions, the differential decay width
becomes,
\begin{eqnarray}
\frac{d^5 \Gamma }{dp^2~ ds~ d\cos\theta_K~ d\cos\theta_+~ d \phi}=
\frac{2\sqrt{\lambda}}{128 \times 256 \pi^6 m_B^3}(|{\cal M}_R|^2+|{\cal M}_L|^2),
\end{eqnarray}
with 
\begin{eqnarray}
\lambda=\left[{1\over 4}(m_{B}^2-m_{K^*}^2-s)^2 - m_{K^*}^2\ s\right].
\label{eq:lambda }
\end{eqnarray}
\noi Here we introduced the various angles as: $\theta_{K}$ is the
polar angle of the K meson momentum in the rest system of the $K^*$
meson with respect to the helicity axis, {\it{i.e.}} the outgoing
direction of $K^*$. Similarly $\theta_{+}$ is the polar angle of the
positron in the dilepton rest system with respect to the helicity axis
of the dilepton. And $\phi$ is the azimuthal angle between the planes
of the two decays   $K^* \rightarrow K  \pi$ and $\gamma^*
\rightarrow\ \l^{+} \l^{-}$. And then,
\begin{eqnarray}
|{\cal M}_R|^2&=&\left|\frac{G_F}{\sqrt{2}} V_{tb}V_{ts}^* g_{K^*K\pi}
 \frac{\alpha_{em} m_b}{\pi s}\right|^2
 \frac{1}{ (p^2-m_{K^*}^2)^2 + (m_{K^*} \Gamma_{K^*})^2}\nonumber \\ \nonumber
&&\left[|a_R|^2 \left\{ \QL^2-\QN^2-\frac{(s-q'^2) p'^{2}}{2} \right\}
\right. \\ \nonumber
&+& 2 Re(a_R b_R^*) \left\{-\QL^2 \PL + \QN \PN \QL \right\} \\ \nonumber
&+& |b_R|^2 \left\{ \PL^2 \QL^2 - \PN^2 \QL^2-\frac{s-q'^{2}}{2} p^2 \QL^2 
\right\} \\
\nonumber
&+& |c_R|^2 \left\{- \NPLQ  \NPLQ -\frac{s-q'^2}{2} \PLQ \cdot \PLQ 
\right\}\\ \nonumber
&+& 2 Im(b_R c_R^*) \PN \QL \NPLQ + 2 Im(c_R a_R^*) \QN \NPLQ \\ \nonumber
&-& 2 Im(b_R a_R^*) \QL \NPLQ - 2 Re(c_R a_R^*) \LQN \cdot \QPL \\
&+& \left.2 Re(b_R c_R^*) \QL \LPN \cdot \QPL \right],\label{ar}
\end{eqnarray}
and
\begin{eqnarray}
|{\cal M}_L|^2&=&\left|\frac{G_F}{\sqrt{2}} V_{tb}V_{ts}^* g_{K^*K\pi}
 \frac{\alpha_{em} m_b}{\pi s}\right|^2
 \frac{1}{ (p^2-m_{K^*}^2)^2 + (m_{K^*} \Gamma_{K^*})^2} \nonumber \\ \nonumber
&&\left[|a_L|^2 \left\{ \QL^2-\QN^2-\frac{(s-q'^2) s}{2} \right\} 
\right.\\ \nonumber
&+& 2 Re(a_L b_L^*) \left\{-\QL^2 \PL + \QN \PN \QL \right\} \\ \nonumber
&+& |b_L|^2 \left\{ \PL^2 \QL^2 - \PN^2 \QL^2-\frac{s-q'^2}{2} p^2 \QL^2 
\right\} \\
\nonumber
&+& |c_L|^2 \left\{- \NPLQ  \NPLQ -\frac{s-q'^2}{2} \PLQ \cdot \PLQ 
\right\}\\ \nonumber
&+& 2 Im(b_L c_L^*) \PN \QL \NPLQ + 2 Im(c_L a_L^*) \QN \NPLQ \\ \nonumber
&+& 2 Im(b_L a_L^*) \QL \NPLQ + 2 Re(c_L a_L^*) \LQN \cdot \QPL \\
&-& \left.2 Re(b_L c_L^*) \QL \LPN \cdot \QPL \right] , \label{al}
\end{eqnarray}
where\footnote{We use $Tr(\g^{\alpha}\g^{\beta}\g^{\gamma}\g^{\delta}\g_5)=+4 i
\varepsilon^{{\alpha}{\beta}{\gamma}{\delta}} $.} $\ABC_{\mu}=
\varepsilon_{{\mu}{\alpha}{\beta}{\gamma}}A^{\alpha}B^{\beta}C^{\gamma}$,
$\ABCD=
\varepsilon_{{\alpha}{\beta}{\gamma}{\delta}}A^{\alpha}B^{\beta}C^{\gamma}D^{
\delta}$, 
$p'=p_K-p_\pi$, and $q'=p_+-p_-$.
Comparing $|{\cal M}_L|^2$ with $|{\cal M}_R|^2$, 
we see that the signs of the corresponding last three terms
are opposite to each other.
We can simplify the expression by introducing the
helicity amplitudes, defined as,
\begin{eqnarray}
H_{(\pm1,0)}^L &=& -{\epsilon_{K^*}^{(\pm,0)}}^{\nu *}
 {\epsilon_{\gamma}^{(\pm,0)}}^{\mu *} 
 \left( a_L g_{\mu\nu} -b_L {p_\mu q_{\nu}} +
ic_L ~\varepsilon _{\mu \nu \alpha \beta}~ p^\alpha q^\beta \right )~, \nn\\
H_{(\pm1,0)}^R &=& -{\epsilon_{K^*}^{(\pm,0)}}^{\nu *}
 {\epsilon_{\gamma}^{(\pm,0)}}^{\mu *} 
 \left( a_R g_{\mu\nu} -b_R {p_\mu q_{\nu}} +
ic_R ~\varepsilon _{\mu \nu \alpha \beta}~ p^\alpha q^\beta \right )~,
\label{hel}
\end{eqnarray}
where the auxiliary functions $a_{L,R}$, $b_{L,R}$, $c_{L,R}$ are
given in Eqs.~(\ref{aL})-~(\ref{cL}), and we define the following
polarization vectors:
\begin{equation}
\begin{array}{lclccrcl}
\epsilon_{K^*}^+ &=& (&0,&1,&i,&0&)/\sqrt{2}~,\\
\epsilon_{K^*}^- &=& (&0,&1,&-i,&0&)/\sqrt{2}~,\\
\epsilon_{K^*}^0 &=& (&\frac{\sqrt{\lambda}}{m_B},&0,&0,&
\sqrt{\frac{\lambda}{m_B^2}+m_{K^*}^2}&)/m_{K^*}~,\\
\epsilon_{\gamma}^+ &=& (&0,&1,&-i,&0&)/\sqrt{2}~,\\
\epsilon_{\gamma}^- &=& (&0,&1,&+i,&0&)/\sqrt{2}~,\\
\epsilon_{\gamma}^0 &=& (&\frac{\sqrt{\lambda}}{m_B},&0,&0,&
-\sqrt{\frac{\lambda}{m_B^2}+s}&)/\sqrt{s}~.
\label{pol}
\end{array}
\end{equation}
Substituting them into Eq. (\ref{hel}), we obtain the 
following helicity amplitudes,
\begin{eqnarray}
H_{+1}^{L,R}(s)&=& (a_{L,R} + c_{L,R} \sqrt{\lambda})\nn \\
H_{-1}^{L,R}(s)&=& (a_{L,R} - c_{L,R}\sqrt{\lambda}) \nn \\  
H_{0}^{L,R}(s) &=& -a_{L,R} {p.q\over m_{K^*} \sqrt{s}} +
{b_{L,R} \lambda \over m_{K^*} \sqrt{s}}\label{Hs}
\end{eqnarray}
 
\noi where $p.q = (m_B^2-m_{K^*}^2 - s)/2$. From now on, we will omit for
simplicity the index $1$ in the helicity amplitudes\footnote{ We will
refer to $H_{+1}^{L,R}(s)$ and $H_{-1}^{L,R}(s)$ respectively as
$H_{+}^{L,R}(s)$ and $H_{-}^{L,R}(s)$.} in Eqs.~(\ref{Hs}).
Using these equations, we can get the results for Eqs.~(\ref{ar},\ref{al}),
whose sum makes the decay angular distribution of 
$B\to K^*(\to K\pi) \ell^+ \ell^-$,

\begin{eqnarray}
&&\frac{d^5 \Gamma }{dp^2\,ds\, d\cos \theta_ K\, d\cos \theta_+\, d \phi} = 
\frac{
\alpha_{em}^2 G_F^2 g_{K^*K\pi}^2 \sqrt{\lambda}p^2 m_b^2 |V_{tb} V_{ts}^*|^2}
{64 \times 
8 (2\pi)^8 m_B^3 [(p^2-m_{K^*}^2)^2 + m_{K^*}^2 \Gamma_{K^*}^2]} \nn\\
&&\,\,\,\,\,\times \left\{
4 \cos ^2 \theta_K  \sin^2 \theta_+  (|H_{0}^R(s)|^2+|H_{0}^L(s)|^2) \right.
\nonumber \\
&&\,\,\,\,\,+\sin^2 \theta_K  (1+ \cos^2 \theta_+ ) ( |H_{+}^L(s)|^2+  |H_{-}^L(s)|^2
+|H_{+}^R(s)|^2+  |H_{-}^R(s)|^2) \nonumber \\
&&\,\,\,\,\,-2\sin ^2\theta_K \sin^2 \theta_+ \left[ \cos 2\phi Re (H_{+}^R(s) H^{R*}_{-}(s)+
H_{+}^L(s) H^{L*}_{-}(s) ) \right.
\nonumber \\
&&\,\,\,\,\,- \sin 2\phi  \left.Im (H_{+}^R(s) H^{R*}_{-}(s)+H_{+}^L(s) H^{L*}_{-}(s) ) \right]\nonumber \\
&&\,\,\,\,\,-\sin 2\theta_K \sin 2\theta_+ \left[\cos \phi Re (H_{+}^R(s) H^{R*}_{0}(s)
+H_{-}^R(s)H_0^{R*}(s)+
H_{+}^L(s) H^{L*}_{0}(s)+H_{-}^L(s)H_0^{L*}(s))\right.
\nonumber\\
&&\,\,\,\,\,-\sin \phi \left. Im (H_{+}^R(s) H^{R*}_{0}(s)-H_{-}^R(s)H_0^{R*}(s)
+H_{+}^L(s) H^{L*}_{0}(s)-H_{-}^L(s)H_0^{L*}(s) ) \right] \nonumber \\
&&\,\,\,\,\, -2 \sin^2\theta_K \cos\theta_+ 
(|H_{+}^R(s)|^2-|H_{-}^R(s)|^2- |H_{+}^L(s)|^2+|H_{-}^L(s)|^2) \nn \\
&&\,\,\,\,\,+2 \sin\theta_+ \sin 2\theta_K 
 \left [\cos \phi Re (H_{+}^R(s) H^{R*}_{0}(s)
-H_{-}^R(s)H_0^{R*}(s)-H_{+}^L(s) H^{L*}_{0}(s)
+H_{-}^L(s)H_0^{L*}(s)) \right. \nn \\
&&\,\,\,\,\,-  \sin \phi  \left.\left.
Im (H_{+}^R(s) H^{R*}_{0}(s)+ H_{-}^R(s) H^{R*}_{0}(s)
- H_{+}^L(s) H^{L*}_{0}(s)-H_{-}^L(s) H^{L*}_{0}(s))
\right] \right\} .
\label{eq:d5}
\end{eqnarray}

From the decay angular distribution presented above, it turns out that
the main theoretical difficulty in evaluating this quantity is the
estimate of non-perturbative part located in the helicity
amplitudes (see Eq~.\ref{Hs}). Henceforth, we will see in the next 
section our estimate of the related hadronic matrix elements.
\section{Hadronic matrix elements for $B\to K^* \ell^+ \ell^-$
\label{sec:matrixNLO}}
\hspace*{\parindent}

In this section, we present our estimates of the non-perturbative
effects on the exclusive $B\to K^* \ell^+ \ell^-$ decay, which are
described by the matrix elements of the quark operators in
Eq.~(\ref{eqn:hamiltonian}) over meson states, and can be parameterized in terms of form
factors.

For the vector meson $K^*$ with polarization vector $\epsilon_\mu$,
the semileptonic form factors of the $(V-A)$ current are defined as
\begin{eqnarray}
&&\langle{K^* (p,\epsilon^*)} | (V-A)_\mu | B(p_B)\rangle  = -i\
\epsilon^*_\mu 
(m_B+m_{K^*}) A_1(s) + i\ (p_B+p)_\mu (\epsilon^* p_B)\,
\frac{A_2(s)}{m_B+m_{K^*}}\nn\\
&& +  i\ q_\mu (\epsilon^* p_B) \,\frac{2m_{K^*}}{s}\,
\Big(A_3(s)-A_0(s) \Big) +\epsilon_{\mu\nu\rho\sigma}\ \epsilon^{*\nu} p_B^\rho p^\sigma\,
\frac{2V(s)}{m_B+m_{K^*}} ~.
\hspace*{2cm}\label{eq:ff3}
\end{eqnarray}
Note the exact relations:
\begin{eqnarray}
 A_3(s) & = & \frac{m_B+m_{K^*}}{2m_{K^*}}\, A_1(s) -
\frac{m_B-m_{K^*}}{2m_{K^*}}\, A_2(s),\nonumber\\
A_0(0) & = & A_3(0), \nonumber\\
\langle K^* |\partial_\mu A^\mu | B\rangle & = & 2 m_{K^*}
(\epsilon^* p_B) A_0(s).
 \label{eq:A30}
\end{eqnarray}
The second relation in (\ref{eq:A30}) ensures that there is no kinematical
singularity in the matrix element at $s=0$. The decay $B\to
K^*\ell^+\ell^-$ is described by the above semileptonic form factors and the
 following penguin form factors:
\begin{eqnarray}
\langle { K^* (p,\epsilon^*)} | C_7^{\,\rm eff}
\bar{s}\sigma_{\mu\nu}q^{\nu}(1+\gamma_5)b |B(p_B)\rangle & = & i\ \epsilon_{\mu\nu\rho\sigma} \epsilon^{*\nu}
p_B^\rho p^\sigma \, 2 {\cal T}_1(s)\label{eq:T}\\
&+&  {\cal T}_2(s)  \Big\{ \epsilon^*_\mu (m_B^2-m_{{K^*}}^2) 
- (\epsilon^* p_B) \,(p_B+p)_\mu  \Big\}\nn\\
&+ &  {\cal T}_3(s)(\epsilon^* p_B) 
\left \{ q_\mu - \frac{s}{m_B^2-m_{{K^*}}^2}\ (p_B+p)_\mu \right \} ~.\nn
\end{eqnarray}
The matrix element decomposition is defined such that the leading
order contribution from the electromagnetic dipole operator ${\cal
O}_7$ reads ${\cal T}_i(s) = C^{\bf eff}_7 T_i(s)+\ldots$, where $T_i(s)$ denote the tensor form factors. Including also the four-quark operators (but neglecting for the moment annihilation contributions), the leading logarithmic expressions are \cite{Grinstein:1989me}

\begin{eqnarray}
\label{calT1}
{\cal T}_1(s) &=& C_7^{\,\rm eff} \,T_1(s) + Y(s) \,\frac{s}
{2 m_b (m_B+m_{K^*})}\,V(s), \\
\label{calT2}
{\cal T}_2(s) &=& C_7^{\,\rm eff} \,T_2(s) + Y(s) \,\frac{s}
{2 m_b (m_B-m_{K^*})}\,A_1(s), \\
{\cal T}_3(s) &=& C_7^{\,\rm eff} \,T_3(s) + Y(s) \,\left[\frac{m_B-m_{K^*}}{2 m_b} \,A_2(s)- \frac{m_B+m_{K^*}}{2 m_b}\,A_1(s)\right], 
\label{calT3}
\end{eqnarray}

\begin{figure}[htb]
\vskip -0.4truein
\centerline{\epsfysize=7in
{\epsffile{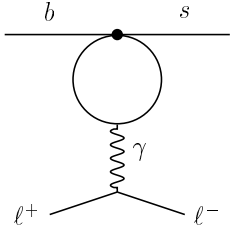}}}
\vskip -5.0truein
\caption[]{ \it The Feynman diagram responsible for the 
four-Fermi-operator contribution (depicted by the blob) to the
operator ${\cal O}_9$.}
\label{fig:ycharm}
\end{figure}

\noi with $C_7^{\,\rm eff} = C_7-C_3/3-4 C_4/9-20 C_5/3-80 C_6/9
=C_7-(4 \bar{C}_3-\bar{C}_5)/9-(4 \bar{C}_4-\bar{C}_6)/3$, and the
function $Y(s)$ represents the one-loop matrix element of the
four-Fermi operators \cite{burasmuenz,misiakE}, see
Fig.~\ref{fig:ycharm}. 
It is written as:
\begin{eqnarray}
\label{yy}
Y(s) &=& h(s,m_c) \left(3 \bar{C}_1+\bar{C}_2+3 \bar{C}_3+\bar{C}_4+3
 \bar{C}_5+\bar{C}_6\right) \nonumber\\
&&-\,\frac{1}{2}\,h(s,m_b) \left(4 \,(\bar{C}_3+\bar{C}_4)+3 \bar{C}_5+
\bar{C}_6\right) -\frac{1}{2}\,h(s,0) \left(\bar{C}_3+3 \bar{C}_4\right)
\nonumber\\
&&+\,\frac{2}{9}\,\left(\frac{2}{3}\bar{C}_3+2 \bar{C}_4+\frac{16}{3} 
 \bar{C}_5\right)~,
\end{eqnarray}
\noi where the ``barred'' coefficients $\bar{C}_i$ ( for i=1,...,6) are
defined as certain linear combinations of the $C_i$, such that the
$\bar{C}_i$ coincide  {\it at leading logarithmic order} with the
Wilson coefficients in the standard basis \cite{Buchalla:1996vs}.
Following Ref. \cite{Beneke:2001at}, they are expressed as~:
 \begin{eqnarray}
\bar{C}_1 &=& \frac{1}{2} \,C_1,\nonumber\\
\bar{C}_2 &=& C_2-\frac{1}{6}\,C_1,\nonumber\\
\bar{C}_3 &=& C_3-\frac{1}{6}\,C_4+16\,C_5-\frac{8}{3}\,C_6,\nonumber\\
\bar{C}_4 &=& \frac{1}{2}\,C_4+8\,C_6,\nonumber\\
\bar{C}_5 &=& C_3-\frac{1}{6}\,C_4+4\,C_5-\frac{2}{3}\,C_6,\nonumber\\
\bar{C}_6 &=& \frac{1}{2}\,C_4+2\,C_6.
\end{eqnarray}
\noi The functions
\begin{eqnarray}
h(s,m_q) &=& -\frac{4}{9}\left(\ln\frac{m_q^2}{\mu^2} - \frac{2}{3} - z \right)-
\frac{4}{9} \,(2+z) \,\sqrt{\,|z-1|} \,
\left\{\begin{array}{l}\,\arctan\displaystyle{\frac{1}{\sqrt{z-1}}}\qquad\quad 
z>1~,\\
[0.4cm]\,\ln\displaystyle{\frac{1+\sqrt{1-z}}{\sqrt{z}}} - \frac{i\pi}{2}
\quad z\leq 1~,\end{array}\right.
\end{eqnarray}
and
\begin{eqnarray}\hspace*{-10cm}
h(s,0) &=& \frac{8}{27} -\frac{4}{9} {\mathrm{ln}}\frac{s}{\mu^2}
 +\frac{4}{9}~i\pi,
\end{eqnarray}
\noi are related to the basic fermion loop. (Here $z$ is defined as
$4 m_q^2/s$.) $Y(s)$ is given in the NDR scheme with anticommuting
$\gamma_5$ and with respect to the operator basis of
\cite{Chetyrkin:1997vx}. As can be seen from the above equations, 
internal $b$-quarks $\sim h(s,m_b)$, $c$-quarks $\sim h(s,m_c)$ and 
light quarks $q$, (with $m_q=0$ for $q=u,d,s$) $\sim h(s,0)$
contribute to the function $Y(s)$; only the charm loop involves the dominant 
``current-current" operators $O_1$ and $O_2$.

Since $C_9$ is basis-dependent starting from next-to-leading logarithmic
order, the terms not proportional to $h(s,m_q)$ differ from those given in
\cite{Buchalla:1996vs}.
The contributions from the four-quark operators ${\cal O}_{1-6}$ are
usually combined with the coefficient $C_9$ into an ``effective''
(basis- and scheme-independent) Wilson coefficient
\begin{eqnarray}
C_9^{\,\rm eff}(s)=C_9+Y(s). 
\end{eqnarray}
The effective Wilson coefficient $C_9^{\mbox{eff}}(s)$ 
receives contributions from various pieces especially from the
$c\bar{c}$ states\footnote{ 
This effect will not be treated here since the LEET symmetry is
restricted to the kinematic region in which the energy of the final
state meson scales with the heavy quark mass. For $B\to K^* \ell^+
\ell^-$ decay this region is identified as $s\simeq
8~\mathrm{GeV}^2$.}. However the contribution given below is just the
perturbative part. 
%

We have seen in the previous chapter that in the Large Energy
Effective Theory framework, one can relate the seven $B\to K^*$ form
factors to only two universal
quantities~\cite{Charles:1998dr,Beneke:2000wa}, namely
$\xi^{(K^*)}_{\perp}$ and $\xi^{(K^*)}_{||}$. Adopting this formalism,  the
various form factors 
appearing in (\ref{calT1})-(\ref{calT3}) simplify to 
\begin{eqnarray}
\label{tperpdef}
&& {\cal T}_1(s) \equiv {\cal T}_\perp(s) = 
\xi^{(K^*)}_\perp(s) \left[C_7^{\,\rm eff} \,\delta_1 + 
\frac{s}{2 m_b m_B} \,Y(s)\right], \\
&&{\cal T}_2(s) = \frac{2 E_{K^*}}{m_B}\, {\cal T}_\perp(s), \\
&&{\cal T}_3(s) - \frac{m_B}{2 E_{K^*}}\,{\cal T}_2(s) \equiv  
{\cal T}_\parallel(s) = - \xi^{(K^*)}_\parallel(s) 
\left[C_7^{\,\rm eff} \,\delta_2 + \frac{m_B}{2 m_b}\, Y(s)
\,\delta_3\right],
\label{tpardef}
\end{eqnarray}
where $E_{K^*}$ refers to the energy of the final state $K^*$-meson
(see Eq.~(\ref{eq:E}) in section \ref{sec:leetL}). 
The factors $\delta_i$ are defined such that $\delta_i=1+O(\alpha_s)$.  
The $O(\alpha_s)$-corrections represent the next-to-leading terms related
to these form factors in the LEET and they can seen from Eqs.~(\ref{A1cor})-(\ref{vertexcorr2}).

At next-to-leading order, the invariant amplitudes ${\cal
T}_{\perp,\,\parallel}(s)$, which refer to the decay into a
transversely and longitudinally polarized vector meson (virtual
photon), get contributions both from factorizable corrections as well
from the non-factorizable ones, and they read respectively \cite{Beneke:2001at}
\begin{eqnarray}
\label{nlodef}
{\cal T}_{\perp} &=& \xi_{\perp} \left(C_{\perp}^{(0)}+\frac{\alpha_s C_F}
{4\pi} \,C_{\perp}^{(1)}\right) \nonumber\\
&&\hspace*{0cm}+ \,\frac{\pi^2}{N_c}
\,\frac{f_B f_{K^*,\,{\perp}}}{m_B} \,\,\Xi_{\perp}\,\sum_{\pm}
\int\frac{d\omega}{\omega}\,\Phi_{B,\,\pm}(\omega)\int_0^1
\!du\,\Phi_{K^*,\,{\perp}}(u) \,T_{{\perp},\,\pm}(u,\omega),\nn\\
{\cal T}_{||} &=& \xi_{||} {m_K^*\over E_K^*} \left(C_{||}^{(0)}+\frac{\alpha_s C_F}
{4\pi} \,C_{||}^{(1)}\right) \nonumber\\
&&\hspace*{0cm}+ \,\frac{\pi^2}{N_c}
\,\frac{f_B f_{K^*,\,{||}}}{m_B} \,\,\Xi_{||}\,\sum_{\pm}
\int\frac{d\omega}{\omega}\,\Phi_{B,\,\pm}(\omega)\int_0^1
\!du\,\Phi_{K^*,\,{||}}(u) \,T_{{||},\,\pm}(u,\omega)~.
\end{eqnarray}
\noi Here $C_F=4/3$, $N_c=3$, $\Xi_\perp\equiv 1$, $\Xi_{||}
\equiv m_{K^*}/E_K^*$, $f_{K^*,\,||}$ denotes the usual $K^*$ decay
constant $f_{K^*}$ and $f_{K^*,\,\perp}$ refers to the
(scale-dependent) transverse decay constant defined by the matrix
element of the tensor current. The leading-ordercoefficient
$C^{(0)}_a$ follows by comparison with Eqs.~(\ref{tperpdef}) and (\ref{tpardef}) setting $\delta_i=1$.
\begin{figure}[t]
   \vspace{-6.9cm}
   \epsfysize=30cm
   \epsfxsize=22cm
   \centerline{\epsffile{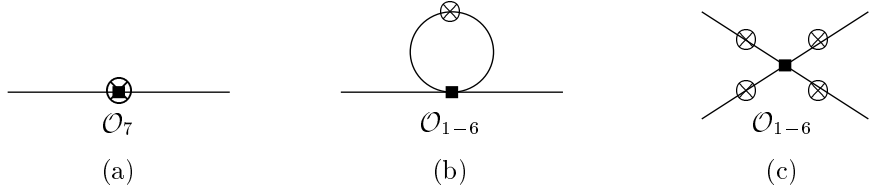}}
   \vspace*{-20.1cm}
\centerline{\parbox{14cm}{\caption{\label{fig2}
\it Leading contributions to 
$\langle \gamma^*  K^*| H_{\rm eff} | B \rangle$. 
The circled cross marks the possible insertions of the virtual 
photon line.}}}
\end{figure}
The term $T_{a,\,\pm}(u,\omega)$ and $C^{(1)}_a$ $(a=\perp,||)$ represent
respectively the hard-scattering and the form factor corrections
which will be discussed below.
\subsection{hard-spectator corrections}
\begin{figure}[t]
   \vspace{-3.5cm}
   \epsfysize=30cm
   \epsfxsize=22cm
   \centerline{\epsffile{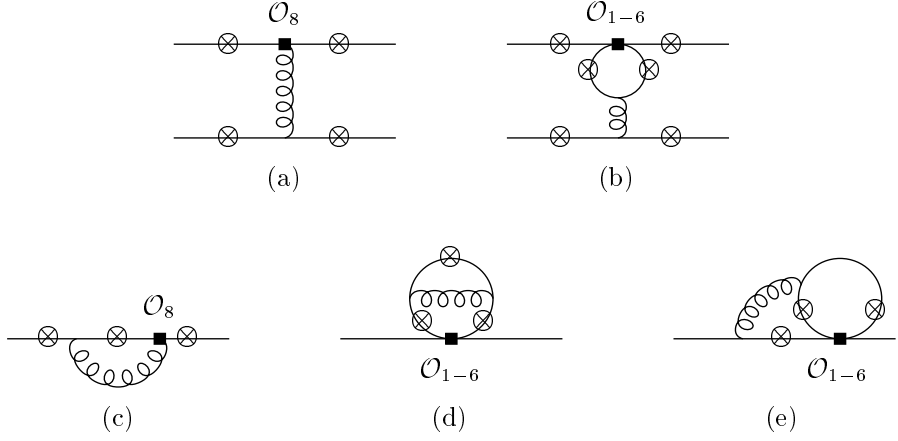}}
   \vspace*{-20.3cm}
\caption[]{ \it 
Non-factorizable contributions to 
$\langle \gamma^* K^*| H_{\rm eff} | B\rangle$. 
The circled cross marks the possible insertions of the virtual 
photon line. Diagrams 
that follow from (c) and (e) by symmetry are not shown. Upper line: 
hard spectator scattering. Lower line: diagrams involving a 
$B\to K^*$ form factor (the spectator quark line is not drawn for 
these diagrams).}
\label{fig1}
\end{figure}

The hard scattering functions $T_{a,\,\pm}(u,\omega)$ $(a=\perp,||)$
in Eq. (\ref{nlodef}) is expanded as~:
\begin{equation}\label{eq:Tadef}
T_{a,\,\pm}(u,\omega)= T^{(0)}_{a,\,\pm}(u,\omega)+\frac{\alpha_s C_F}{4\pi} \,
T^{(1)}_{a,\,\pm}(u,\omega)~,
\end{equation}
To compute the leading-order term $T^{(0)}_{a,\,\pm}$ we have to compute the weak 
annihilation amplitude of Figure~(\ref{fig2}-c), which has no analogue 
in the inclusive decay and generates the hard-scattering term 
$T^{(0)}_{a,\,\pm}(u,\omega)$ in (\ref{nlodef}). To compute this 
term we perform the projection of the amplitude on the $B$ meson 
and $K^*$ meson distribution amplitude as explained in 
 \cite{Beneke:2000wa}. The four diagrams in Figure~(\ref{fig2}-c) 
contribute at different powers in the $1/m_b$ expansion. It turns out 
that the leading contribution comes from the single diagram with the 
photon emitted from the spectator quark in the $B$ meson, because 
this allows the quark propagator to be off-shell by an amount 
of order $m_b\Lambda_{\rm QCD}$, the off-shellness being of order 
$m_b^2$ for the other three diagrams. Hence the result of the
leading-order term $T^{(0)}_{a,\,\pm}(u,\omega)$ reads \cite{Beneke:2001at}
\begin{eqnarray}
\label{loann}
T^{(0)}_{\perp,\,+}(u,\omega) &=& T^{(0)}_{\perp,\,-}(u,\omega) = 
 T^{(0)}_{\parallel,\,+}(u,\omega) = 0\\
T^{(0)}_{\parallel,\,-}(u,\omega) &=& -e_q\,\frac{m_B\omega}{m_B \omega
-s-i\epsilon}\,\frac{4 m_B}{m_b} \,(\bar{C}_3+3 \bar{C}_4).
\end{eqnarray}
The hard scattering functions $T^{(1)}_{a,\,\pm}$ in (\ref{eq:Tadef}) 
contain a factorizable term from expressing the full QCD form factors 
in terms of $\xi^{(K^*)}_a$, related to the $\alpha_s$-correction to the 
$\delta_i$ in Eqs.~(\ref{tperpdef}), (\ref{tpardef}) above. 
We write $T^{(1)}_{a,\,\pm} = T_{a,\,\pm}^{(\rm f)}+ 
T_{a,\,\pm}^{(\rm nf)}$. The factorizable correction 
reads \cite{Beneke:2000wa}:
\begin{eqnarray}
\label{start1}
&& \hspace*{-0.5cm}
T^{(\rm f)}_{\perp,\,+}(u,\omega) = C_7^{\,\rm eff} \,
\frac{2 m_B}{\bar{u}E_{K^*}},\\
&& \hspace*{-0.5cm}
T^{(\rm f)}_{\parallel,\,+}(u,\omega) = \left[C_7^{\,\rm eff}  + 
\frac{s}{2 m_b m_B} \,Y(s)\right]\,
\frac{2 m_B^2}{\bar{u}E_{K^*}^2}
\\
&& \hspace*{-0.5cm}
T^{(\rm f)}_{\perp,\,-}(u,\omega) = 
T^{(\rm f)}_{\parallel,\,-}(u,\omega) = 0
\end{eqnarray}
The non-factorizable correction is obtained by computing matrix
elements of four-quark operators and the chromomagnetic dipole
operator represented by diagrams (a) and (b) in Figure~(\ref{fig1}),
using the projection on the meson distribution amplitudes.
They read as~\cite{Beneke:2000wa}:
\begin{eqnarray}
T^{(\rm nf)}_{\perp,\,+}(u,\omega) &=&
-\frac{4 e_d \,C_8^{\,\rm eff}}{u+\bar{u} s/m_B^2} 
+\frac{m_B}{2 m_b}\,\Big[e_u t_\perp(u,m_c) \, (\bar{C}_2+\bar{C}_4-\bar{C}_6)
\nonumber\\
&& +\,e_d \,t_\perp(u,m_b)\, 
(\bar{C}_3+\bar{C}_4-\bar{C}_6-4 m_b/m_B\,\bar{C}_5)+ e_d
\,t_\perp(u,0) 
\,\bar{C}_3\Big]
\label{Tnfperp}
\\
T^{(\rm nf)}_{\perp,\,-}(u,\omega) &=& 0
\\
T^{(\rm nf)}_{\parallel,\,+}(u,\omega) &=& \frac{m_B}{m_b}\,
\Big[e_u t_\parallel(u,m_c) \, 
(\bar{C}_2+\bar{C}_4-\bar{C}_6)+ e_d \,t_\parallel(u,m_b)\, 
(\bar{C}_3+\bar{C}_4-\bar{C}_6)
\nonumber\\
&& + \,e_d \,t_\parallel(u,0) \,\bar{C}_3\Big]
\\
T^{(\rm nf)}_{\parallel,\,-}(u,\omega) &=& e_q\,\frac{m_B\omega}{m_B \omega
-s-i\epsilon}\, \bigg[\frac{8 \,C_8^{\,\rm eff}}{\bar{u}+u s/m_B^2} 
\nonumber\\
&& 
+\,\frac{6 m_B}{m_b}\, \Big(h(\bar{u}m_B^2+u s,m_c) 
\,(\bar{C}_2+\bar{C}_4+\bar{C}_6) 
+ h(\bar{u}m_B^2+u s,m_b)\,
(\bar{C}_3+\bar{C}_4+\bar{C}_6)
\nonumber\\
&& 
+ \,h(\bar{u}m_B^2+u s,0)\,
(\bar{C}_3+3 \bar{C}_4+3
\bar{C}_6)
-\frac{8}{27}\,(\bar{C}_3-\bar{C}_5-15\bar{C}_6)\Big)\bigg].
\label{tnf4}
\end{eqnarray}
Here $C_8^{\,\rm eff}=C_8+C_3-C_4/6+20 C_5-10 C_6/3 = 
C_8+(4 \bar{C}_3-\bar{C}_5)/3$, $e_u=2/3$ ($e_d=-1/3$), $e_q$ is 
the electric charge of the spectator quark in the $B$
meson and $h(s,m_q)$ has been defined above. The functions  
$t_a(u,m_q)$ arise from the two diagrams of Figure~(\ref{fig1}-b) in which 
the photon attaches to the internal quark loop. They are given
by~\cite{Beneke:2000wa}:  
\begin{eqnarray}
t_\perp(u,m_q) &=& \frac{2 m_B}{\bar{u} E_{K^*}} \,I_1(m_q) + 
\frac{s}{\bar{u}^2 E_{K^*}^2}\,\Big(B_0(\bar{u} m_B^2+u s,m_q)-B_0(s,m_q)\Big),
\\
t_\parallel(u,m_q) &=& \frac{2 m_B}{\bar{u} E_{K^*}}\,I_1(m_q) + 
\frac{\bar{u} m_B^2+u s}{\bar{u}^2 E_{K^*}^2}\,
\Big(B_0(\bar{u} m_B^2+u s,m_q)-B_0(s,m_q)\Big),
\end{eqnarray}
where $B_0$ and $I_1$ are defined as
\begin{eqnarray}
\label{b0def}
&& \hspace*{-0.5cm}B_0(x,m_q)=-2\,\sqrt{4 m_s/x-1}\,\arctan\frac{1}
{\sqrt{4 m_s/x-1}},
\\
&& \hspace*{-0.5cm}I_1(m_q) = 1+\frac{2 m_s}{\bar{u} (m_B^2-s)}\,
\Big[L_1(x_+)+L_1(x_-)-L_1(y_+)-L_1(y_-)\Big],
\end{eqnarray}
and
\begin{eqnarray}
&& \hspace*{-0.5cm}x_\pm=\frac{1}{2}\pm\left(\frac{1}{4}-\frac{m_s}
{\bar{u} m_B^2+u s} \right)^{\!1/2},
\qquad
y_\pm=\frac{1}{2}\pm\left(\frac{1}{4}-\frac{m_q^2}{s}\right)^{\!1/2},
\\
&& \hspace*{-0.5cm}
L_1(x)=\ln\frac{x-1}{x}\,\ln(1-x)-\frac{\pi^2}{6}+\mbox{Li}_2\left(
\frac{x}{x-1}\right).
\end{eqnarray}
The correct imaginary parts are obtained by interpreting $m_q^2$ as 
$m_q^2-i\epsilon$. Closer inspection shows that contrary to 
appearance none of the hard-scattering functions is more singular than 
$1/\bar{u}$ as $u\to 1$. It follows that the convolution integrals 
with the kaon light-cone distribution are convergent at the endpoints. 

The limit $s\to 0$ ($E_{K^*}\to m_B/2$) 
of the transverse amplitude is relevant to the 
decay $B\to  K^*\gamma$. The corresponding limiting function 
reads
\begin{equation}
t_\perp(u,m_q)_{|s=0} = \frac{4}{\bar{u}} \,\left(1+\frac{2 m_q^2}
{\bar{u} m_B^2}\,\Big[L_1(x_+)+L_1(x_-)\Big]_{|s=0}\right)
\end{equation}
In the 
same limit the longitudinal amplitude develops a logarithmic
singularity, which is of no consequence, because the longitudinal 
contribution to the $B\to  K^*\ell^+\ell^-$ decay rate 
is suppressed by a power of $s$ relative to the transverse 
contribution in this limit. 
%
        \subsection{vertex corrections} 
The next-to-leading order coefficients 
$C^{(1)}_{a}$ in (\ref{nlodef}) 
contain a factorizable term from expressing the full QCD form factors 
in terms of $\xi^{(K^*)}_a$, related to the $\alpha_s$-correction to the 
$\delta_i$ in Eqs.~(\ref{tperpdef}), (\ref{tpardef}).  
In writing $C^{(1)}_{a} = C_{a}^{(\rm f)}+ 
C_{a}^{(\rm nf)}$, the factorizable correction 
reads \cite{Beneke:2001at}:
\begin{eqnarray}
\label{start2}
&& \hspace*{-0.5cm}
C^{(\rm f)}_{\perp} = C_7^{\,\rm eff} \left(4\ln\frac{m_b^2}{\mu^2} 
-4-L \right),\\
&& \hspace*{-0.5cm}
C^{(\rm f)}_{\parallel} = - C_7^{\,\rm eff} \left(4\ln\frac{m_b^2}{\mu^2} 
-6+4 L\right)
+ \frac{m_B}{2 m_b}\,Y(s)\,\Big(2-2 L\Big)
\label{end2}
\end{eqnarray}
with
\begin{equation}
\label{Ldef}
L\equiv -\frac{m_b^2-s}{s}\ln\left(1-\frac{s}{m_b^2}\right).
\end{equation}
Note that the brackets multiplying $ C_7^{\,\rm eff}$ include the term 
$3\ln(m_b^2/\mu^2)-4$ from expressing the $\overline{\rm MS}$ quark 
mass in the definition of the operator ${\cal O}_7$ 
in terms of the $b$ quark pole
mass according to (\ref{polerel}). 
The non-factorizable correction is obtained by computing matrix
elements of four-quark operators and the chromomagnetic dipole
operator represented by diagrams (c) through (e) in 
Figure~(\ref{fig1}). 

\begin{figure}[t]
    \begin{center}
    \leavevmode
    \includegraphics[height=4cm]{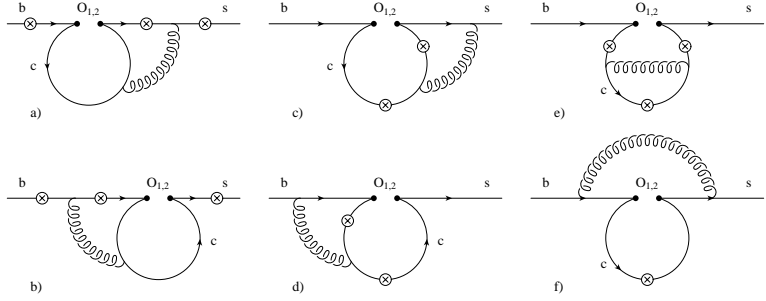}
    \vspace{2ex}
    \caption[f1]{\it Complete list of two-loop Feynman 
    diagrams for $b \to s \gamma^*$
    associated with the operators $O_1$ and $O_2$.
    The fermions ($b$, $s$ and $c$ quarks) are represented by solid 
    lines;
    the curly lines represent gluons.
    The circle-crosses denote the possible locations for emission 
    of a virtual photon.}
    \label{figG:1}
    \end{center}
\end{figure}

The matrix elements of four-quark operators require the calculation of
two-loop diagrams with several different mass scales. The result for
the current-current operators ${\cal O}_{1,2}$ is 
presented in  \cite{AAGW} as a double expansion in 
$s/m_b^2$ and $m_c/m_b$. Since we are only interested in small 
$s$, this result is adequate for our purposes. For that note that only the 
result corresponding to Figure~(\ref{figG:1}a-e) of is needed for 
this calculation. 
%
The 2-loop 
matrix elements of penguin operators have not yet been computed and 
will hence be neglected. Due to the small Wilson coefficients of 
the penguin operators, this should be a very good approximation. 
The matrix element of the chromomagnetic dipole operator
[Figure~(\ref{fig1}-c)] is also given in  \cite{AAGW} in 
expanded form. The exact result is given in
Appendix~\ref{app:leetcorr2}. 
All this combined, we obtain
\begin{eqnarray}
\label{start3}
C_F C^{(\rm nf)}_{\perp} &=& -\,\bar{C}_2 F_2^{(7)}- C_8^{\rm eff}
F_8^{(7)}\nonumber\\
&&\hspace*{0.0cm}
-\,\frac{s}{2 m_b m_B} \left[\bar{C}_2 F_2^{(9)}+2\bar{C}_1\left(F_1^{(9)}+
\frac{1}{6}\,F_2^{(9)}\right)+C_8^{\rm eff} F_8^{(9)}\right] ,
\\
C_F C^{(\rm nf)}_{\parallel} &=& \bar{C}_2 F_2^{(7)}+C_8^{\rm eff}
F_8^{(7)}\nonumber\\
&&\hspace*{0.0cm}
+\,\frac{m_B}{2 m_b} 
\left[\bar{C}_2 F_2^{(9)}+2\bar{C}_1\left(F_1^{(9)}+
\frac{1}{6}\,F_2^{(9)}\right)+C_8^{\rm eff} F_8^{(9)}\right].
\label{end3}
\end{eqnarray}
The quantities $F_{1,2}^{(7,9)}$ and $F_{8}^{(7,9)}$  are 
given in Appendix~\ref{app:leetcorr2}, or can also be extracted from  
 \cite{AAGW} in expanded form. 
In expressing the result in terms of the coefficients $\bar{C}_{1,2}$,
we have made use of $F_1^{(7)}+F_2^{(7)}/6=0$. We also substituted 
$C_8$ by $C_8^{\rm eff}$, taking into  account a subset of penguin
contributions.
        \subsection{form factors values \label{sec:ff}}
\begin{table}
\addtolength{\arraycolsep}{3pt}
\renewcommand{\arraystretch}{1.4}
$$
\begin{array}{|l|llll|lll|}
\hline\hline
& A_1 & A_2 & A_0 & V & T_1 & T_2 & T_3\\ \hline
F(0) & 0.294 & 0.246 & 0.412 &0.399 & 0.334 & 0.334 & 0.234\\

c_1 & 0.656 & 1.237 & 1.543 & 1.537 & 1.575 & 0.562 & 1.230\\

c_2 & 0.456 & 0.822 & 0.954 & 1.123 & 1.140 & 0.481 & 1.089\\
\hline\hline
\end{array}
$$
\caption[] {\it Input values for the parameterization 
  (\protect{\ref{eq:para}}) of the $B\to K^*$ form factors. Renormalization
   scale for
  the penguin form factors $T_i$ is $\mu = m_b$ 
\cite{Ali:1999mm}.} \label{tab:p1}
 \end{table}
In the description of exclusive $B$-decays hadronic matrix elements
$<X|{\cal O}_i|B>$ are involved\footnote{ where $X$ is any meson (with mass
$m_X$)}. However, in order for these quantities to become available, it
is necessary to confront the fact these hadrons are color bound state
objects. While understood {\it in principle}, the non-perturbative
nature of these bound states makes problematic the extraction of
precision information about the exclusive $B$-physics. To explore them
one faces a daunting theoretical challenge to evaluate first the
corresponding form factors.

This is not a problem which has been solved in its entirely, nor is
it likely ever to be. Rather, what is available is a variety of
theoretical approaches and techniques, appropriate to a variety of
specific problems and with varying levels of reliability. While
approaches which are based directly on QCD, and which allow for
quantitative error estimates, are clearly to be preferred, more
model-dependent methods are often all that are available and thus have
an important role to play as well.

\begin{figure}[H]
\psfrag{a}{\hskip 0cm $\mathrm{s\ (GeV^2)}$}
\psfrag{c}{\hskip -1.cm $\xi^{(K^*)}_{||}(s)$}
\psfrag{b}{\hskip -1.cm $\xi^{(K^*)}_{\perp}(s)$}
\psfrag{d}{\hskip 0cm [AS]}
\psfrag{e}{\hskip 0cm [BFS]}
\begin{center}
\includegraphics[width=18cm,height=16cm]{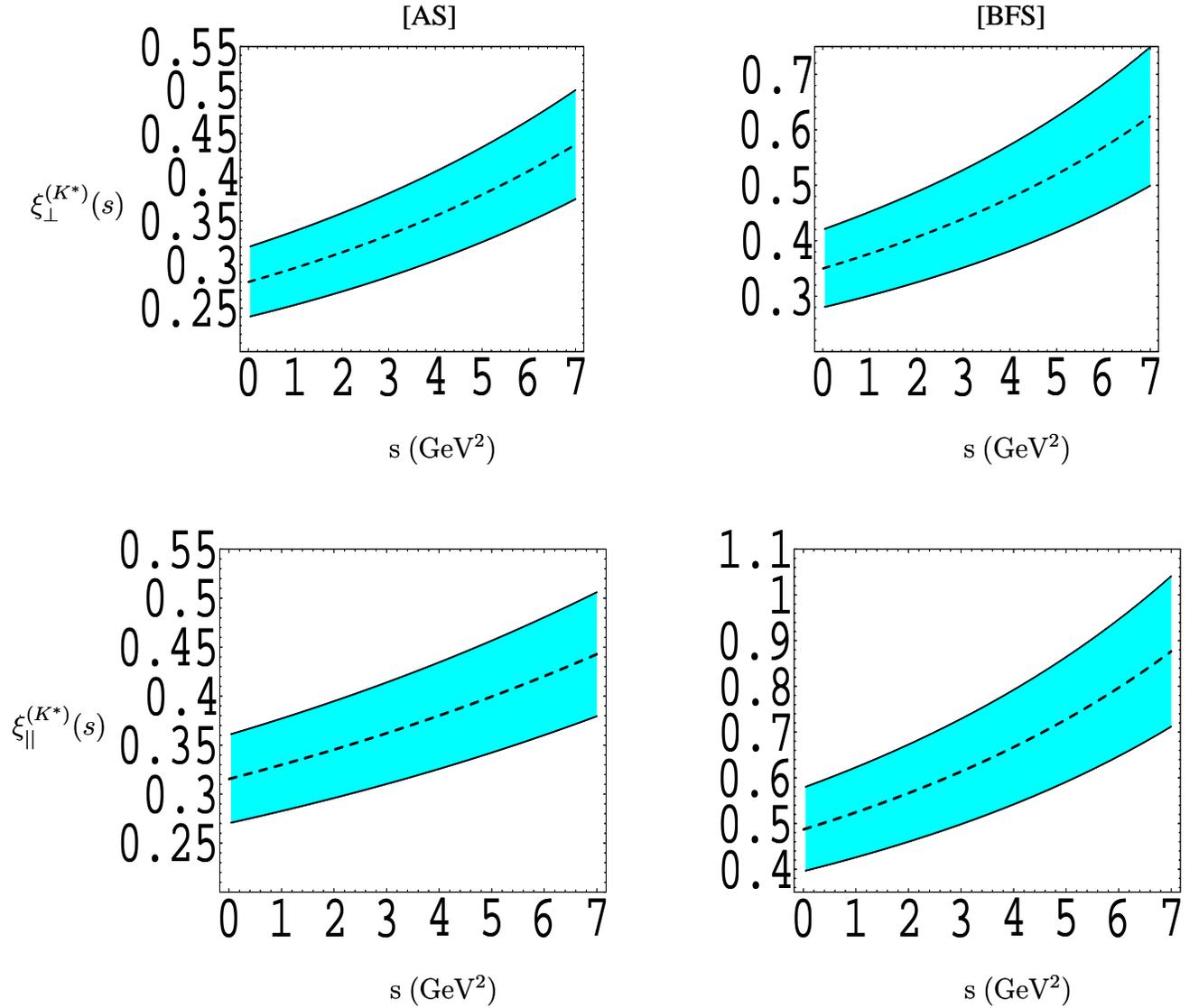}
\caption{\it LEET form factors $\xi^{(K^*)}_{{\perp},||}(s)$ 
for  $B \rightarrow  K^*\ l^+\ l^-$. 
The two columns denoted by [AS] and [BFS] represent, respectively, our
$\xi^{(K^*)}_{{\perp},||}(s)$ and the ones used by Beneke {\it et al.} in
ref\cite{Beneke:2001at}. The central values are represented by the dashed
curves, while the bands reflect the uncertainties on the form
factors~\cite{Ali:2002qc}.} 
\label{ffASetBF}
\end{center}
\end{figure}

Concerning our guess on the corresponding $B\to K^*$ form factors
(see Eqs.~(\ref{eq:ff3}) and (\ref{eq:T}))
we have combined roughly two theoretical approaches to compute them:
\begin{itemize}  
\item
First we have 
used the LEET symmetry to reduce the number of independent form
factors from seven to two universal ones, namely $\xi^{(K^*)}_{\perp}(s)$ and
$\xi^{(K^*)}_{||}(s)$ (see Eqs.~(\ref{A0})-(\ref{T3})). 
\item
At large recoil, namely $s=0$, the normalization of the LEET form
factor $\xi_\perp^{(K^*)}(0)$ is determined using the $B \to K^*
\gamma$ experimental constraint on the corresponding NLO-LEET
estimates. Thus, the magnetic moment form factors turns out to be in
the range\citer{Ali:2001ez,Beneke:2001at} 
\begin{equation}
\label{eq:T1(0)}
T_1^{(K^*)}(0)=0.28\pm 0.04.
\end{equation}
Thus, from Eqs.~(\ref{eq:T1(0)}) and (\ref{T1}), the numerical value
for the $\xi_\perp^{(K^*)}$ at large recoil momentum is defined as
\begin{equation}
\xi_\perp^{(K^*)}(0)=0.28\pm 0.04.
\end{equation}
\item
The second universal LEET form factor $\xi_{||}^{(K^*)}(s)$ has to be
modeled entirely from some approximate methods. For that we have used
a non-perturbative approach, the so-called Light-cone sum-rule
approach\footnote{The method of light-cone sum-rules was first
suggested for the study of weak baryon decays in~\cite{Balitsky} and
later extended to heavy-meson decays in \cite{Chernyak}. It combines
the traditional QCD sum rule method~\cite{Shifman} with the twist expansion
characteristic of hard exclusive processes in
QCD~\cite{Brodsky}.}~\cite{Balitsky,Chernyak}, based on the
approximate conformal invariance of QCD. While
in principle this technique is rigorous, it suffers in its current
practical implementations from a degree of uncontrolled model
dependence (for a review see~\cite{RuckletKhodja} and reference
therein). Using the result of \cite{Ball:1998kk} for $B\to K^*$ form
factors, presented in Table~\ref{tab:p1}, which include NLO radiative
corrections and higher twist 
corrections up to twist four. The result turns out to be:
\begin{equation}
\xi_{||}^{K^*}(0)=0.31\pm 0.04.
\end{equation}
\item
To extrapolate these form factors, namely $\xi_{\perp, ||}^{(K^*)}$, at small recoil (large values of $s$), we use the
following parametrization (with its coefficients listed in Table \ref{tab:p1}):
\begin{equation}
\label{eq:para}
F(\hat{s}) = F(0) \exp ( c_1 \hat{s} + c_2 \hat{s}^2 ).
\end{equation}
\end{itemize}   
Using the ingredients described above, we show in  Fig.~\ref{ffASetBF}
the corresponding LEET form factors  $\xi^{(K^*)}_{\perp}(s)$ and
$\xi^{(K^*)}_{||}(s)$.
Note that the values used by {\it Beneke el al.} in
ref.\cite{Beneke:2001at} are very different of the ones used by
us. This descripency is related to the fact that their choice is based
on the QCD sum rules estimates for $\xi^{(K^*)}_{\perp}=0.35 \pm 0.07$
and $\xi^{(K^*)}_{||}=0.49 \pm 0.09$. On the other hand, our values are somewhat
lower than the corresponding estimates in the lattice-QCD framework,
yielding \cite{DelDebbio:1997kr} $T_1^{K^*}(0)=0.32^{+0.04}_{-0.02}$, and in 
the light cone QCD sum rule approach, which give
typically $T_1^{K^*}(0)=0.38 \pm 0.05$ \cite{Ali:1999mm}. (Earlier 
lattice-QCD results on $B \to K^* \gamma$ form factors are reviewed in
\cite{Soni:1995qq}.).

Finally, we have to keep in mind that such a descripency reflect after
all our poor knowledge
of this part of QCD, namely the non-perturbative QCD. Consequently, we
can anticipate the fact that the long distance uncertainty in our analysis will be
the  dominant one.
%
%

\newpage
\section{Decay Distributions in $B \rightarrow \ K^{*} \ell^{+}\ell^{-}$ 
\label{sec:BtoK}}
Using all the machinery presented until this point we are able now to
do the corresponding analysis for specific transitions. In this
section, we present our Helicty analysis of the $B \rightarrow 
\ K^{*} \ell^{+} \ell^{-}$ decay. 
In the LEET Limit, the helicity amplitudes (\ref{Hs}) are expressed as:

\begin{eqnarray}
H^{L/R}_{+}(s) &=& {i\over\ {2\ m_b\ m_B\ (m_B + m_{K^*}) \sqrt{s}}}
\Big[ -4\ {\cal T}_1(s)\ m_b\ (m_B - m_{K^*})\ (m_B + m_{K^*})^2\ E_{K^*}\nn\\
&& +(\pm C_{10}-C_{9})\ m_B\ (m_B + m_{K^*})^2 \  s \ A_1(s)\nn\\ 
&& +2 m_B \sqrt{\lambda} \Big \{ 2 {\cal T}_1(s)\ m_b\ (m_B+ m_{K^*}) +
(\mp C_{10}+C_{9})\ s\ V(s) \} \Big \} \Big], 
\label{HK+}
\end{eqnarray}
\begin{eqnarray}
H^{L/R}_{-}(s) &=& {i\over\ {2\ m_b\ m_B\ (m_B + m_{K^*})}\sqrt{s}} 
\Big[ -4\ {\cal T}_1(s)\ m_b\ (m_B - m_{K^*})\ (m_B + m_{K^*})^2\ E_{K^*}\nn\\
&& +(\pm C_{10} - C_{9})m_B (m_B + m_{K^*})^2\  s \ A_1(s)\nn\\ 
&& -2 m_B \sqrt{\lambda} \Big \{2 {\cal T}_1(s)\ m_b (m_B + m_{K^*})
+ (\mp C_{10} + C_{9})s V(s) \} \Big \} \Big],
\label{HK-} 
\end{eqnarray}
\begin{eqnarray}
H^{L/R}_{0}(s)&=& {i\over {4\ m_b \ m_B \ m_{K^*} (-m_B^2 + m_{K^*}^2)s}}
\Big[ 8\  \lambda\  m_b\ {\cal T}_1(s) \Big \{2 (m_B^2 - m_{K^*}^2) E_{K^*} + m_B\ s\Big \}\nn\\
&& + 4\ \lambda\  m_B\ s \Big \{ 2\  m_b\ ({\cal T}_3(s)-\frac{m_B}{2\
E_{K^*}}\,{\cal T}_2(s))-A_2(s)(\pm C_{10} - C_{9}) (m_B - m_{K^*})\Big \}\nn\\
&& +(m_B - m_{K^*}) (m_B + m_{K^*})^2 (m_B^2 - m_{K^*}^2 -s)\Big \{4\ {\cal T}_1(s)\
m_b\ E_{K^*}  (-m_B + m_{K^*})\nn\\
&&+ s\  m_B\ A_1(s)(\pm C_{10} - C_{9})\Big \} \Big].
\label{HK0}
\end{eqnarray}
It is interesting to observe that in the Large Energy Effective Theory
(LEET), both helicity amplitudes $\vert H_{+}^{ L,R}(s) \vert^2$ and
$\vert H_{-}^{ L,R}(s) \vert^2$ have essentially one dependence on the
universal form factor $\xi_{\perp}^{(K^*)}$. However the helicity
amplitude $\vert H_{0}^{ L,R}(s) \vert^2$ is more model-dependent,
since it depends on the two form factors $\xi_{\perp}^{(K^*)}$ and
$\xi_{||}^{(K^*)}$.  

\begin{figure}[H]
\begin{center}
\psfrag{a}{\hskip 0.cm $\mathrm{s\ (GeV^2)}$}
\psfrag{b}{\hskip 2.cm $ |\mathit{H}_{+}^{L}(s)|^2$}
\epsfig{file=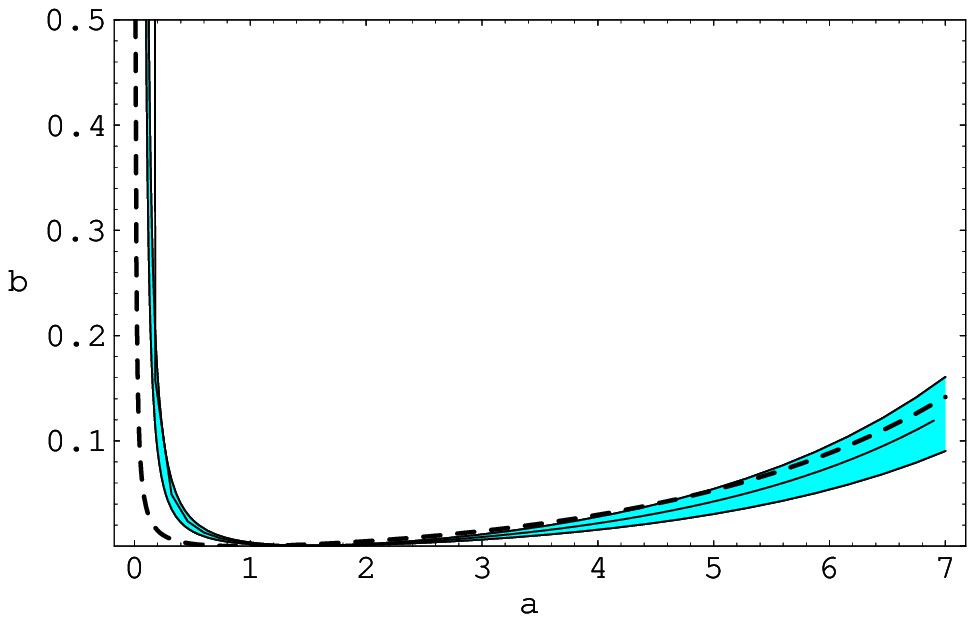,width=8cm,height=8cm}
\hspace*{0cm}
\psfrag{b}{\hskip 2.cm $ |\mathit{H}_{-}^{L}(s)|^2$}
\epsfig{file=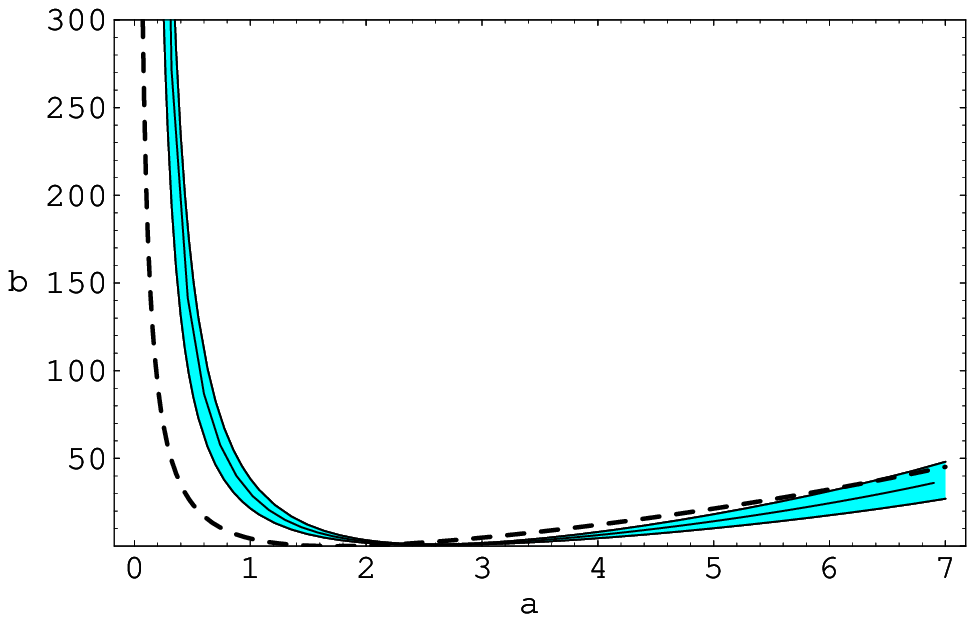,width=8cm,height=8cm}
\caption{\it The helicity amplitude $ |H_{+}^{L}(s)|^2$ (left-hand
plot) and $ |H_{-}^{L}(s)|^2$ (right-hand plot) at NLO (solid center line) and LO (dashed). The band reflects
theoretical uncertainties from the input parameters~\cite{Ali:2002qc}.}
\label{figHL2}
\vspace*{2cm}
\psfrag{HRp}{\hskip 2.cm$ |\mathit{H}_{+}^{R}(s)|^2$}
\epsfig{file=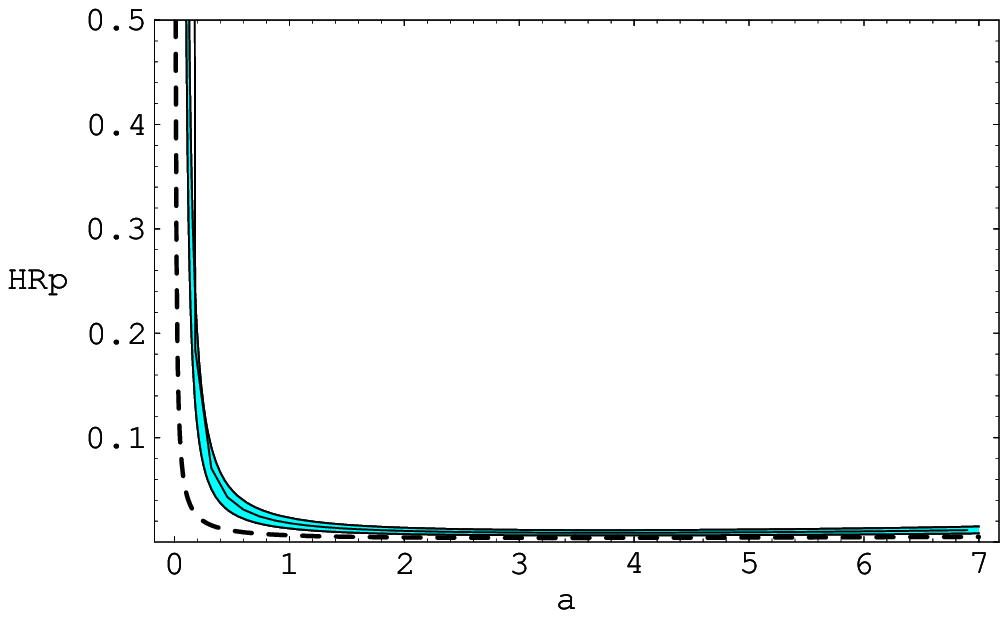,width=8cm,height=8cm}
\hspace*{0cm}
\psfrag{HRm}{\hskip 2.cm$ |\mathit{H}_{-}^{R}(s)|^2$}
\epsfig{file=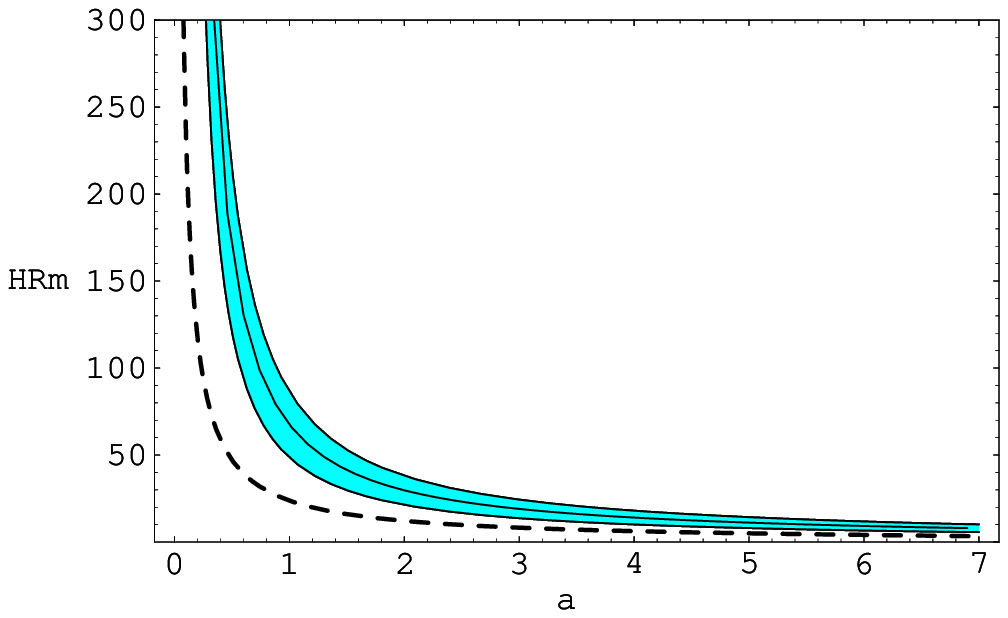,width=8cm,height=8cm}
\caption{\it  The helicity amplitude $ |H_{+}^{R}(s)|^2$ (left-hand
plot) and $ |H_{-}^{R}(s)|^2$ (right-hand plot) at NLO (solid center line) and LO (dashed). The band reflects
theoretical uncertainties from the input parameters~\cite{Ali:2002qc}.}
\label{figHR2}
\end{center}
\end{figure}

In Figs.~(\ref{figHL2}) and (\ref{figHR2}) we show respectively the
helicity amplitudes $\vert H_{+,-}^{ L}(s) \vert^2$ and $\vert
H_{+,-}^{R}(s) \vert^2$ at leading and at next-to-leading order. We
remark that both helicity amplitudes $\vert H_{+,-}^{R}(s) \vert^2$
are completely negligeable comparing to their left-helicity components.  
Moreover, the impact of the NLO corrections on the the amplitudes $\vert
H_{+,-}^{ L}(s) \vert^2$ and $\vert H_{+,-}^{R}(s) \vert^2$ increase
considerably their magnitude up to $\sim 100 \%$ in the small lepton
invariant mass ($s < 2~ \textrm{GeV}^2$). The NLO uncertaities are 
dominated mainly by the $B$-meson light-cone distribution amplitudes
$\lambda_{B,+}$, the $B$-decay constant $f_B$ and the form
factor $\xi_{\perp}^{(K^*)}$. The corresponding errors were calculated by
varing mainly these parameters in the indicated range, one at a time,
and adding the individual arrors in quadrature.
%

Using the above helicity amplitudes and taking the narrow resonance
limit of $K^*$ meson, {\it i.e.}, using the  equations
\begin{eqnarray}
&&\Gamma_{K^*}=\frac{g_{K^*K\pi}^2 m_{K^*}}{48 \pi}~, \nn \\
&&\lim_{\Gamma_{K^*} \to 0} 
\frac{\Gamma_{K^*} m_{K^*}}{(p^2-m_{K^*}^2)^2 + m_{K^*}^2 \Gamma_{K^*}^2}=
\pi \delta(p^2-m_{K^*}^2)~,
\end{eqnarray}
we can perform the integration over $p^2$ in Eq.~(\ref{eq:d5}) and obtain the
fourth differential angular distribution of  $B \rightarrow
K^{*}(\rightarrow K \pi )\ \ell^{+} \ell^{-}$with respect to dilepton mass
squared $s$, the azimuthal angle $\phi$, the polar angles $\theta_{K}$
and $\theta_{+}$, 
\begin{eqnarray}
\label{5diff} 
\hspace*{-3cm}
&&{d^4\Gamma  \over  ds\ d\cos\theta_{K}\ d\cos\theta_{+}\ d\phi} = {3\
\alpha_{em}^2 G_{F}^2  \sqrt{\lambda}  m_{b}^2 |V_{tb}V_{ts}^*|^2 \over 128
(2\pi)^6 m_{B}^3 }\label{eq:angdist}  \\ 
&& \times \Big\{4\ cos^2\theta_{K}\ sin^2\theta_{+}\ 
\Big(|H^R_{0}(s)|^2+|H^L_{0}(s)|^2 \Big)\nn  \\ 
&& + sin^2\theta_{K}\ (1+ cos^2\theta_{+})\
\Big( |H^L_{+}(s)|^2+|H^L_{-}(s)|^2 +|H^R_{+}(s)|^2 +|H^R_{-}(s)|^2 \Big)\nn \\ 
&& - 2\  sin^2\theta_{K}\  sin^2\theta_{+}\ \Big[\
cos2\phi \ Re \Big(H^R_{+}(s) H^{R*}_{-}(s)+H^L_{+}(s) H^{L*}_{-}(s)\Big)\nn \\ 
&& - sin2\phi \ Im\Big(H^R_{+}(s) H^{R*}_{-}(s) + H^L_{+}(s) H^{L*}_{-}(s)\Big)\ \Big]\nn \\
&& - sin2\theta_{K}\ sin2\theta_{+}\ \Big[\ cos\phi\ Re\Big(H^R_{+}(s)H^{R*}_{0}(s) 
+H^R_{-}(s) H^{R*}_{0}(s)+H^L_{+}(s)H^{L*}_{0}(s)+H^L_{-}(s) H^{L*}_{0}(s) \Big)\nn \\
&& -sin\phi\ Im \Big(H^R_{+}(s) H^{R*}_{0}(s) - H^R_{-}(s) H^{R*}_{0}(s)
+ H^L_{+}(s) H^{L*}_{0}(s) -H^L_{-}(s)H^{L*}_{0}(s)\Big)\ \Big]\nn \\
&& -2\ sin^2\theta_{K}\  cos\theta_{+}\
\Big(|H^R_{+}(s)|^2-|H^R_{-}(s)|^2 -|H^L_{+}(s)|^2+ |H^L_{-}(s)|^2 \Big)\nn \\
&& +2\ sin\theta_{+}\  sin2\theta_{K}\ 
\Big[\ cos\phi\ Re\Big(H^R_{+}(s) H^{R*}_{0}(s)- H^R_{-}(s)H^{R*}_{0}(s)
- H^L_{+}(s) H^{L*}_{0}(s) + H^L_{-}(s)H^{L*}_{0}(s) \Big)\nn \\
&& - sin\phi\ Im \Big(  H^R_{+}(s) H^{R*}_{0}(s) +
H^R_{-}(s) H^{R*}_{0}(s)  
- H^L_{+}(s) H^{L*}_{0}(s)- H^{L}_{-}(s) H^{L*}_{0}(s)  \Big) \Big]\Big\}.
\nn
\end{eqnarray}
\subsection{Dalitz distributions}
If we integrate out the angle  $\theta_{K}$ and $\theta_{+}$ from
Eq.(\ref{5diff}), we get the double $\phi$ angular distribution:
\begin{eqnarray} 
\label{dBrdl2dphi}
&&{d^2{\cal B} \over d\phi\ ds} = \tau_B {\alpha_{em}^2 G_{F}^2\over 384 \pi^5}
\sqrt{\lambda} {m_{b}^2\over m_{B}^3 }  |V_{tb}V_{ts}^*|^2 \ {1\over
2\pi}\ \Big\{ |H_{0}(s)|^2 + |H_{+}(s)|^2+ |H_{-}(s)|^2 \\ 
&& -  cos2\phi \ Re\Big(H_{+}^R(s) H_{-}^{R*}(s)+H_{+}^L(s)
H_{-}^{L*}(s)\Big) + sin2\phi \ Im\Big(H_{+}^R(s) H_{-}^{R*}(s)+H_{+}^L(s) H_{-}^{L*}(s)\Big)\Big\}, \nn
\end{eqnarray}

\noi where  $ \tau_B$ is the life time of the $B$-meson, and the
various terms in the expansion above can be specified , as follows~:
\begin{eqnarray} 
|H_{0}(s)|^2 & = &|H^L_{0}(s)|^2+|H^R_{0}(s)|^2,\nn\\
|H_{+}(s)|^2 &=&  |H^L_{+}(s)|^2+|H^R_{+}(s)|^2,\nn\\
|H_{-}(s)|^2 &=& |H^L_{-}(s)|^2 +|H^R_{-}(s)|^2.\label{H2def}
\end{eqnarray}

%
\begin{figure}[t]
\psfrag{c}{\hskip 0.3cm $s\ (\textrm{GeV}^2)$}
\psfrag{a}{\hskip -3.cm \Large {${d^2{\cal B}\over ds\ d\cos\theta_{+}\ 10^{-8}}$}}
\psfrag{b}{$ \textrm{cos}\theta_{+}$}
\begin{center}
\includegraphics[width=12cm,height=9cm]{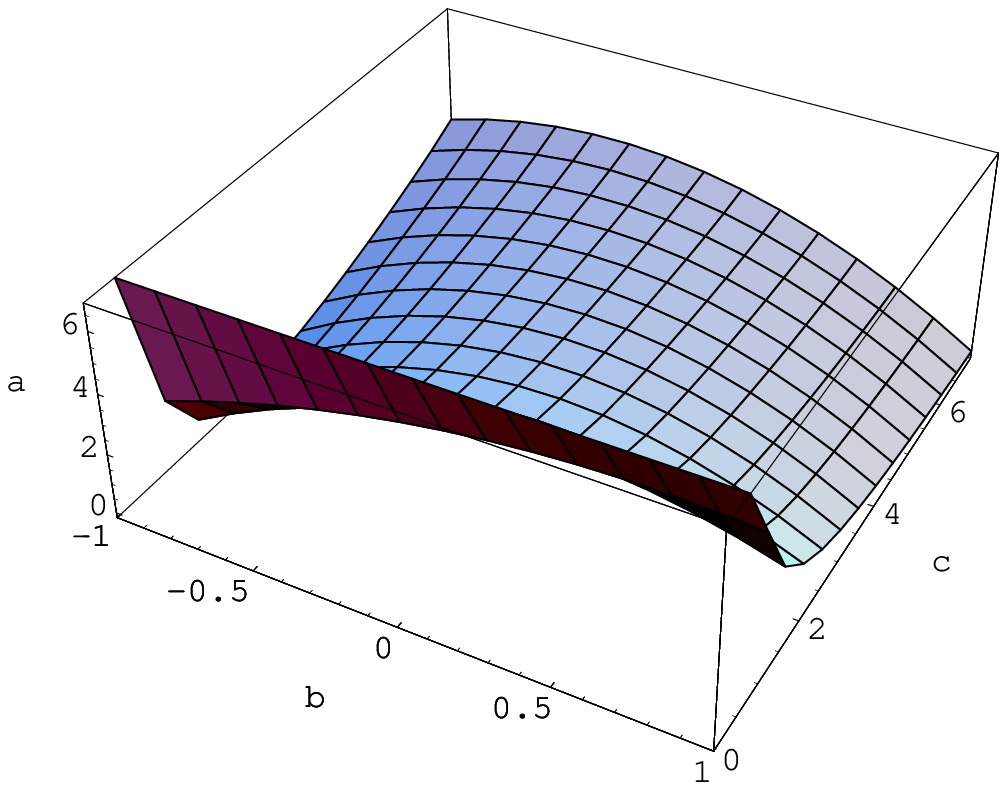}
\caption{\it Dalitz distribution {\Large ${d^2{\cal B} \over
d\cos\theta_{+}\ ds}$} for $ B \rightarrow K^*\  \ell^+ \  \ell^-
$~\cite{Ali:2002qc} .}
\label{dalitzdBrdcostheta+}
\end{center}
\end{figure}
%

\noi Similarly, we can get respectively the $\theta_{K}$ and $\theta_{+}$ angular distributions as following:
\begin{eqnarray} 
\label{dBrdl2dcostetaK}
{d^2{\cal B} \over d\cos\theta_{K}\ ds} &=& \tau_B {\alpha_{em}^2 G_{F}^2\over 384
\pi^5}  \sqrt{\lambda} {m_{b}^2\over m_{B}^3 }  |V_{tb}V_{ts}^*|^2 \
{3\over 4}\ \Big\{ 2\ cos^2\theta_{K} \ |H_{0}(s)|^2\cr \nn \\  
&& + sin^2\theta_{K}\ \Big( |H_{+}(s)|^2+ |H_{-}(s)|^2\Big) \Big\},
\end{eqnarray}
and 
\begin{eqnarray} 
{d^2{\cal B} \over d\cos\theta_{+}\ ds} &=& \tau_B {\alpha_{em}^2 G_{F}^2\over 384
\pi^5}  \sqrt{\lambda} {m_{b}^2\over m_{B}^3 }  |V_{tb}V_{ts}^*|^2 \
{3\over 8}\ \Big\{ 2\ sin^2\theta_{+} \ |H_{0}(s)|^2  \nn\\ 
&& +(1+ cos\theta_{+})^2\ |H_{+}^L(s)|^2 +(1- cos\theta_{+})^2\ |H_{+}^R(s)|^2\nn\\
&&+(1- cos\theta_{+})^2\ |H_{-}^L(s)|^2 +(1+ cos\theta_{+})^2\
|H_{-}^R(s)|^2 \Big\}, 
\label{dBrdl2dcosteta+}\\
&& ={d^2{\cal B}_{|H_{0}|^2} \over d\cos\theta_{+}\ ds}+ {d^2{\cal
B}_{|H_{-}|^2} \over 
d\cos\theta_{+}\ ds}+{d^2{\cal B}_{|H_{+}|^2} \over d\cos\theta_{+}\ds},
\label{dBrdl2dcosteta+Hs}
\end{eqnarray}
where the various terms in Eq.~(\ref{dBrdl2dcosteta+Hs}), namely
$d^2{\cal B}_{|H_{0}|^2}/ d\cos\theta_{+}\ ds$, $d^2{\cal
B}_{|H_{-}|^2}/ d\cos\theta_{+}\ ds$ and $d^2{\cal B}_{|H_{+}|^2}/
d\cos\theta_{+}\ ds$ can be defined respectively,as follows: 
\begin{eqnarray} 
{d^2{\cal B}_{|H_{0}|^2} \over d\cos\theta_{+}\ ds}&=&
\tau_B {\alpha_{em}^2 G_{F}^2\over 384 \pi^5}  \sqrt{\lambda}
{m_{b}^2\over m_{B}^3 }  |V_{tb}V_{ts}^*|^2 \ {3\over 8}\ \Big\{ 2\
sin^2\theta_{+} \ |H_{0}(s)|^2 \Big\}, 
\label{dBrdsdcosteta+Hs} \\ 
{d^2{\cal B}_{|H_{-}|^2} \over d\cos\theta_{+}\ ds} & = & \tau_B
{\alpha_{em}^2 G_{F}^2\over 384 \pi^5}  \sqrt{\lambda} {m_{b}^2\over
m_{B}^3 }  |V_{tb}V_{ts}^*|^2 \ {3\over 8}\ \Big\{(1-
cos\theta_{+})^2\ |H_{-}^L(s)|^2 +(1+ cos\theta_{+})^2\ 
|H_{-}^R(s)|^2 \Big\},
\nn\\
{d^2{\cal B}_{|H_{+}|^2} \over d\cos\theta_{+}\ ds} &=& \tau_B
{\alpha_{em}^2 G_{F}^2\over 384 \pi^5}  \sqrt{\lambda} {m_{b}^2\over
m_{B}^3 }  |V_{tb}V_{ts}^*|^2 \ {3\over 8}\ \Big\{(1+
cos\theta_{+})^2\ |H_{+}^L(s)|^2 +(1- cos\theta_{+})^2\ |H_{+}^R(s)|^2\Big\}.
\nn
\end{eqnarray}

%
%
%
%
\begin{figure}[H]\vspace{1.cm}
\psfrag{c}{\hskip 0.3cm $s\ (\textrm{GeV}^2)$}
\psfrag{b}{$ \textrm{cos}\theta_{+}$}
\begin{center}
\psfrag{a}{\hskip -3.cm \Large {${d^2{\cal B}_{|H_{-}|^2}\over ds\ d\cos\theta_{+}\ 10^{-8}}$}}
\includegraphics[width=12cm,height=9cm]{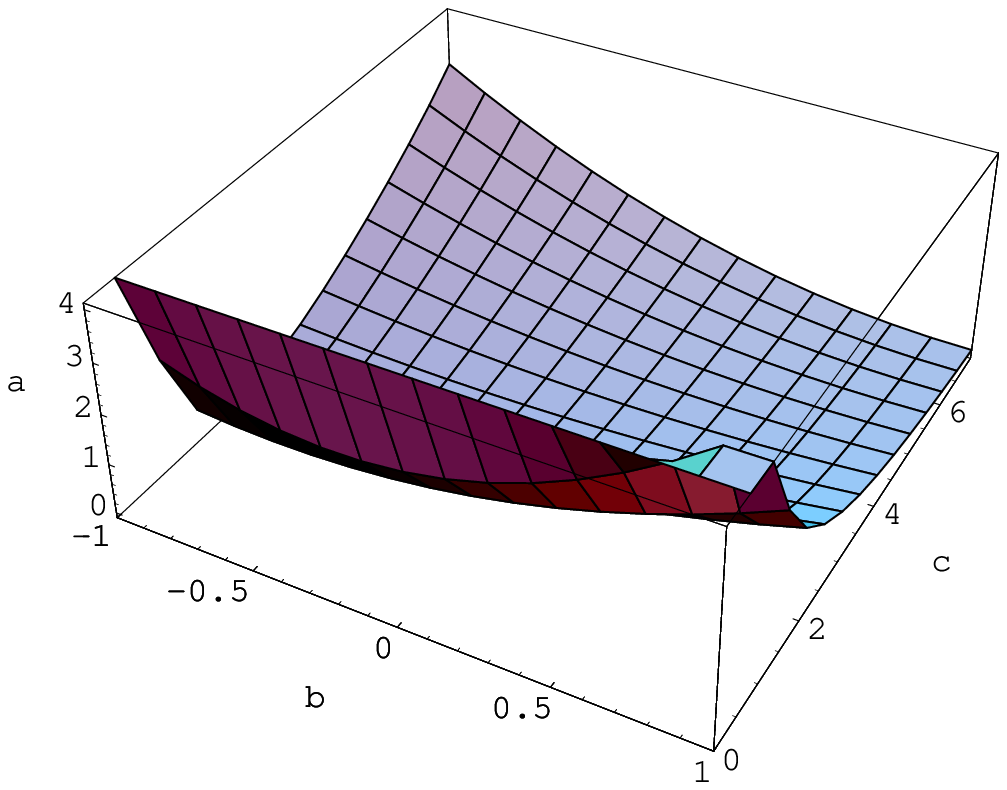}
\caption{\it Partial Dalitz distribution {\Large ${d^2{\cal
B}_{|H_{-}|^2} \over  
d\cos\theta_{+}\ ds}$} for $ B \rightarrow  K^*\  \ell^+ \  \ell^-
 $~\cite{Ali:2002qc} .}  
 \label{dalitzdBrH-1dcostheta+}
\vspace{2cm}
\psfrag{a}{\hskip -3.cm \Large {${d^2{\cal B}_{|H_{0}|^2}\over ds\ d\cos\theta_{+}\ 10^{-8}}$}}
\includegraphics[width=12cm,height=9cm]{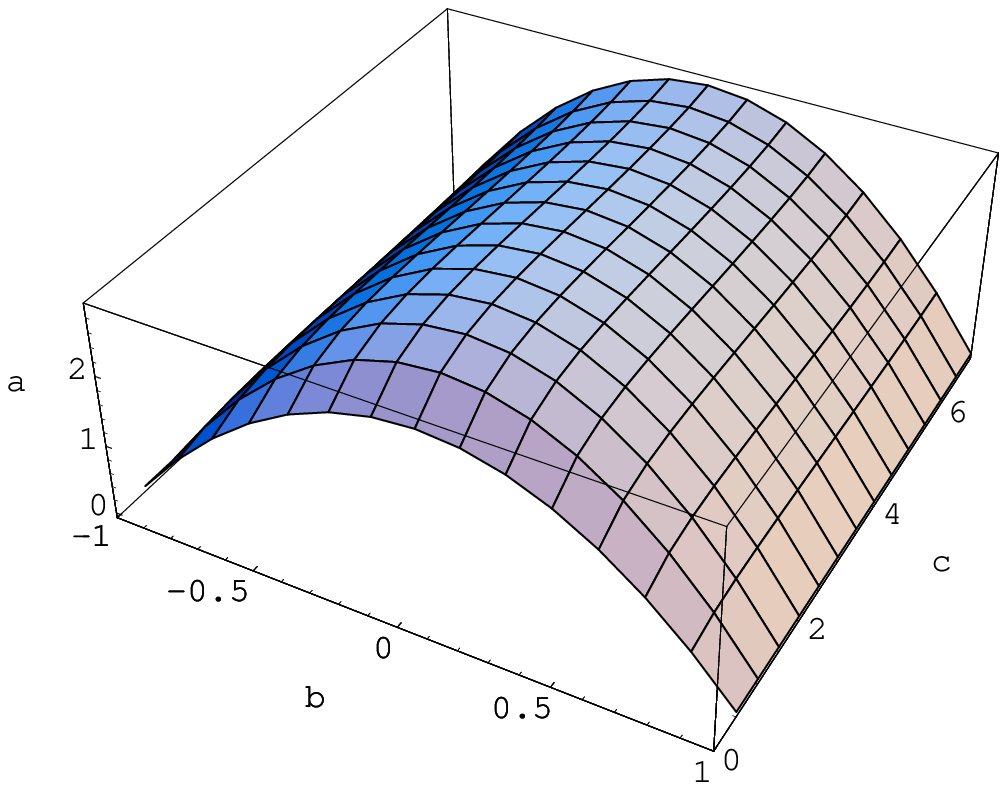}
\caption{\it Partial Dalitz distribution {\Large ${d^2{\cal
B}_{|H_{0}|^2} \over  
d\cos\theta_{+}\ ds}$} for $ B \rightarrow  K^* \ell^+ \ell^-
$~\cite{Ali:2002qc} .}
\label{dalitzdBrH0dcostheta+}
\end{center}
\end{figure}
%
%
Note that the polar angle distribution functions in
Eqs.~(\ref{dBrdl2dcostetaK}), (\ref{dBrdl2dcosteta+}) and
(\ref{dBrdsdcosteta+Hs}) depend only on the modular square terms of
the helicity amplitudes, which give the decay width of the
semileptonic decay (see next section).

Using our central input parameters given in Tables~(\ref{tab1}) and
(\ref{parameters}) (see Apendix~ \ref{app:input}), we show in
Figs.~(\ref{dalitzdBrdcostheta+}), (\ref{dalitzdBrH-1dcostheta+}) and
(\ref{dalitzdBrH0dcostheta+}) at the NLO accuracy the total Dalitz
distribution $d^2{\cal B}/ d\cos\theta_{+}\ ds$, the two angular partial
distributions $d^2{\cal B}_{|H_{-}|^2}/ d\cos\theta_{+}\ ds$ and $d^2{\cal
B}_{|H_{0}|^2}/ d\cos\theta_{+}\ ds$, respectively. 
From the experimental point of view these Dalitz distribution can
serve as a double check of whether the branching fraction is different
from the SM predictions. 
\subsection{ Dilepton mass spectrum and Forward-backward asymmetry}
Finally, after integrating over the polar angles $\phi$, $\theta_{+}$
and $\theta_{K}$, we derive the total differential branching ratio in
the scaled dilepton invariant mass for $B \rightarrow  K^* \ell^+
\ell^-$,

\begin{eqnarray} 
\label{dBrdl2}
{d{\cal B} \over ds} &=& \tau_B {\alpha_{em}^2 G_{F}^2\over 384 \pi^5}
\sqrt{\lambda} {m_{b}^2\over m_{B}^3 }  |V_{tb}V_{ts}^*|^2
\Big\{|H_{+}(s)|^2 + |H_{-}(s)|^2+|H_{0}(s)|^2 \Big\}
\cr\nn\\
 &=& {d{\cal B}_{|H_{+}|^2} \over ds}+{d{\cal B}_{|H_{-}|^2} \over ds}+{d{\cal B}_{|H_{0}|^2} \over ds},
\end{eqnarray}
where the various terms in the Eq.~(\ref{dBrdl2}), read as
\begin{eqnarray} 
{d{\cal B}_{|H_{+}|^2} \over ds} &=& \tau_B {\alpha_{em}^2
G_{F}^2\over 384 \pi^5} \sqrt{\lambda} {m_{b}^2\over m_{B}^3 }
|V_{tb}V_{ts}^*|^2\,\, |H_{+}(s)|^2,
\cr\nn\\
{d{\cal B}_{|H_{-}|^2} \over ds}&=& \tau_B {\alpha_{em}^2
G_{F}^2\over 384 \pi^5} \sqrt{\lambda} {m_{b}^2\over m_{B}^3 }
|V_{tb}V_{ts}^*|^2\,\, |H_{-}(s)|^2,
\cr\nn\\
{d{\cal B}_{|H_{0}|^2} \over ds}&=& \tau_B {\alpha_{em}^2
G_{F}^2\over 384 \pi^5} \sqrt{\lambda} {m_{b}^2\over m_{B}^3 }  |V_{tb}V_{ts}^*|^2\,\, |H_{0}(s)|^2.\label{dBrds}
\end{eqnarray}

\begin{figure}[H]
\vspace{-0.5cm}
\begin{center}
\psfrag{b}{\hskip -3.cm \Large{${d{\cal B}_{|H_{-}|^2} \over ds\ 10^{-7}}$}}
\psfrag{c}{}
\psfrag{a}{}
\epsfig{file=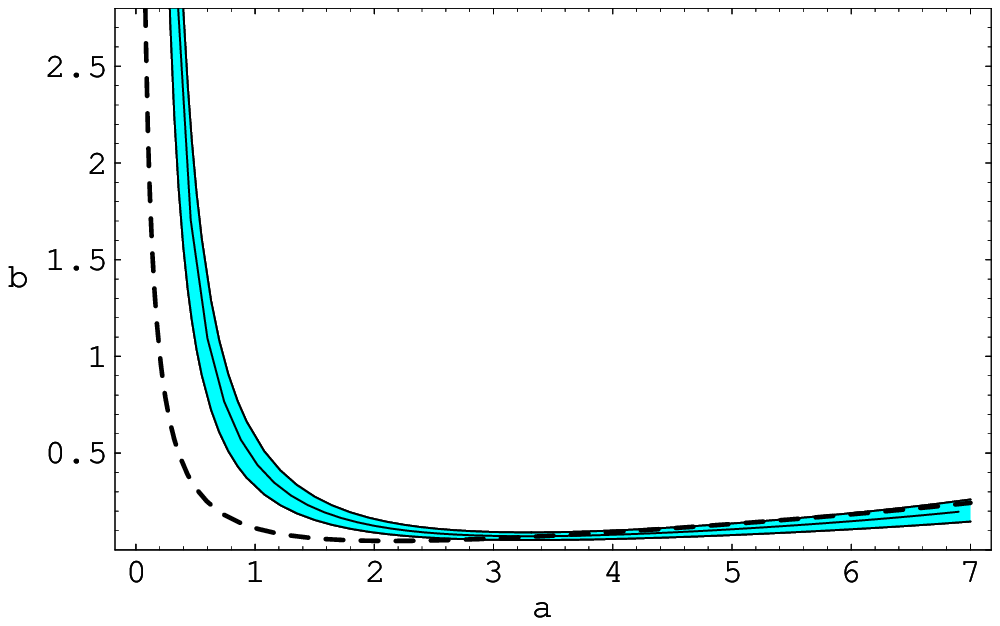,width=10cm,height=7.5cm}
\\
\psfrag{b}{\hskip -3.cm \Large{${d{\cal B}_{|H_{+}|^2}\over ds\ 10^{-10}}$}}
\epsfig{file=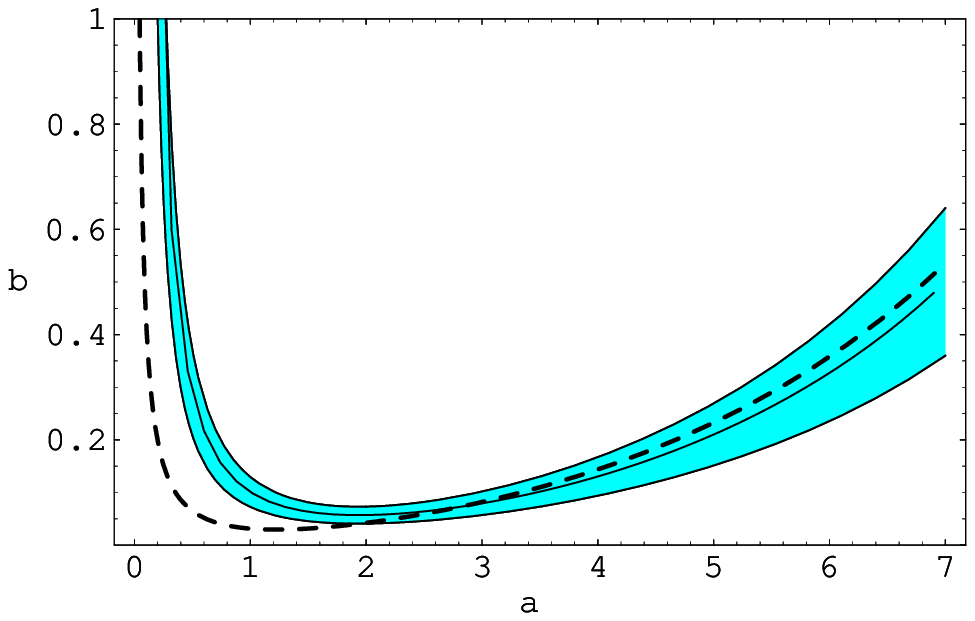,width=10cm,height=7.5cm}
\\
\psfrag{a}{\hskip 0.cm $\mathrm{s\ (GeV^2)}$}
\psfrag{b}{\hskip -3.cm \Large{${d{\cal B}_{|H_{0}|^2}\over ds\ 10^{-7}}$}}
\epsfig{file=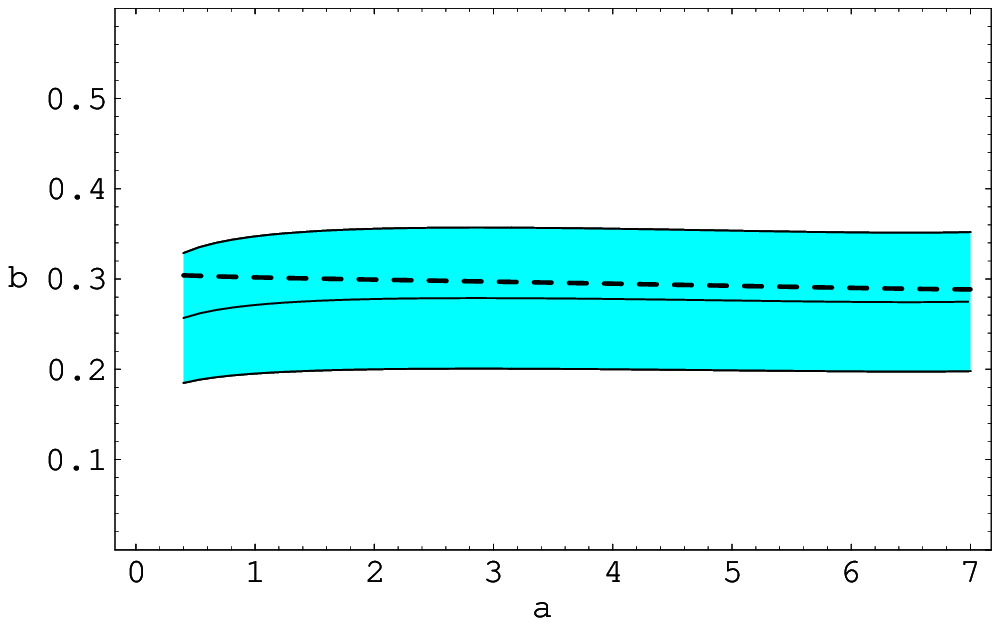,width=10cm,height=7.5cm}
\caption{\it The dilepton invariant mass distributions
{\large${d{\cal B}_{|H_{-}|^2}\over ds}$} (upper-plot),
{\large${dBr_{|H_{+}|^2}\over ds}$} (middle-plot) and {\large${d{\cal
B}_{|H_{0}|^2}\over ds}$} (lower-plot) for $ B\rightarrow K^*  \ell^+  \ell^-$
at NLO (solid center line) and LO
(dashed).The band reflects the theoretical uncertainties from input
parameters~\cite{Ali:2002qc,Safir:SUSY02}.}
\label{figdBrH}
\end{center}
\end{figure}

The partial lepton invariant mass spectrum $d{\cal B}_{|H_{-}|^2}/
ds$, $d{\cal B}_{|H_{+}|^2}/ ds$ and $d{\cal B}_{|H_{0}|^2}/ ds$
are shown in Fig.~(\ref{figdBrH}) showing in each case the
next-to-leading order and leading order results.
We remark that the partial single distribution $d{\cal B}_{|H_{+}|^2}/
ds$ is completely negligeable comparing to the others. In fact this is
due to the smallness of the helicity amplitude $|H_{+}|^2$, as it is
shown in Figs.~(\ref{figHL2}) and (\ref{figHR2}). 

\begin{figure}[t]
\psfrag{a}{\hskip 0.cm $\mathrm{s\ (GeV^2)}$}
\psfrag{b}{\hskip -2.5cm \Large{${d{\cal B}\over ds\ 10^{-7}}$}}
\psfrag{c}{}
\begin{center}
\includegraphics[width=14cm,height=10cm]{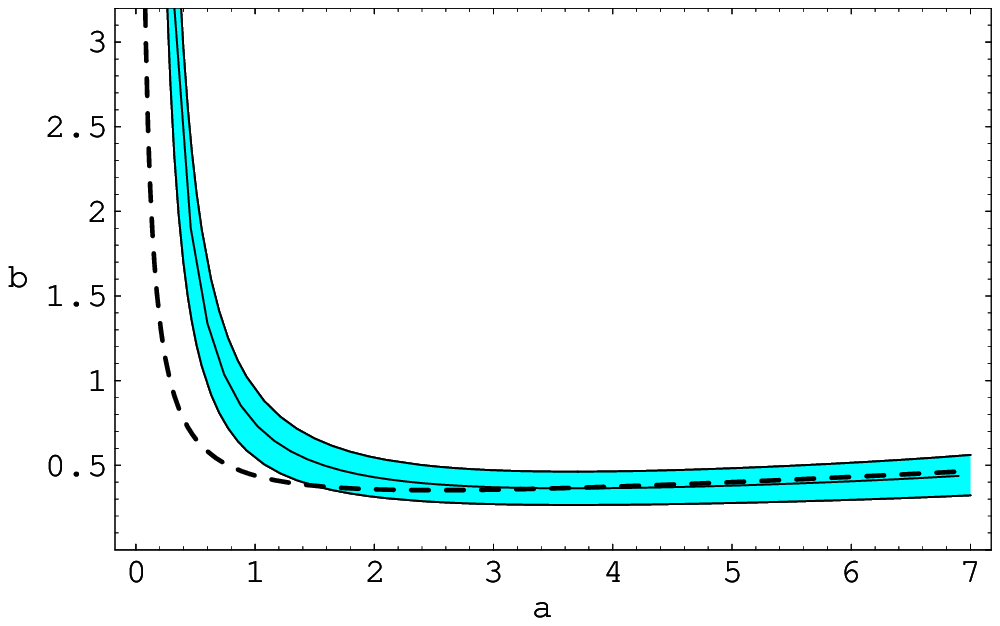}
\caption{\it The total dilepton invariant mass distribution for $ B \rightarrow
K^*  \ell^+  \ell^-$ at NLO (solid center line) and LO (dashed). The band reflects theoretical uncertainties from
the input parameters~\cite{Ali:2002qc}.}
\label{figdBr}
\end{center}
\end{figure}

In Fig.~(\ref{figdBr}), we plot the total dilepton invariant
mass $d{\cal B}/ ds$ at next-to-leading order and leading order. As it
is shown in Figs.~(\ref{figdBrH}--upper plot) and (\ref{figdBr}) the
total decay rate is dominated by the contribution of the helicity
$|H_{-}|$ component. 

We Note that the next-to-leading order correction to the
lepton invariant mass spectrum in $ B \rightarrow K^*  \ell^+  \ell^-$
is significant in the low dilepton mass region $(s \leq 2~\textrm{GeV}^2)$
but small beyond that shown for the anticipated validity of the LEET
theory $(s \leq 8~\textrm{GeV}^2)$.
Apparently rather large uncertainty of our prediction is mainly due to
the form factors with their current large uncertainty and to a lesser
extent respectively due to $\lambda_{B,+}^{-1}$ and the $B$-decay
constant. 

Besides the differential branching ratio, $B \to K^* \ell^+ \ell^-$
decay offers other distributions (with different combinations of
Wilson coefficients) to be measured.
An interesting quantity is the Forward-Backward (FB) asymmetry
defined in \cite{amm91,agm94}

\begin{eqnarray}
  \frac{\d \a_{\rm FB}}{\d \sh} = 
        -\int_0^{\uh(\sh)} \d\uh \frac{\d^2\gl}{\d\uh\, \d\sh}
              + \int_{-\uh(\sh)}^0 \d\uh \frac{\d^2\gl}{\d\uh\, \d\sh} \; ,
  \label{eq:dfba}
\end{eqnarray}
where the variable $\hat{u}$ corresponds to $\theta_{+}$, the angle 
between the momentum of the $B$-meson and the positively charged lepton 
$\ell^+$  in the dilepton CMS frame, through the relation $\hat{u} =
-\hat{u}(\hat{s}) \cos \theta_{+}$, and bounded as 
\begin{eqnarray}
-\hat{u}(\hat{s}) \leq  \hat{u}  \leq \hat{u}(\hat{s}) \; ,
\label{eq:ubound}
\end{eqnarray}
with 
\begin{eqnarray}
 \hat{u}(\hat{s}) = {2\over m_{B}^2} \sqrt{\lambda (1-4 {\hat{m}_{\ell}^2\over \hat{s}})}. 
\end{eqnarray}
%
\begin{figure}[t]
\psfrag{a}{$s\ (GeV^2)$}
\psfrag{b}{\hskip -2.5cm $dA_{FB}/ ds$}
\psfrag{c}{\hskip 0cm }
\begin{center}
\includegraphics[width=14cm,height=10cm]{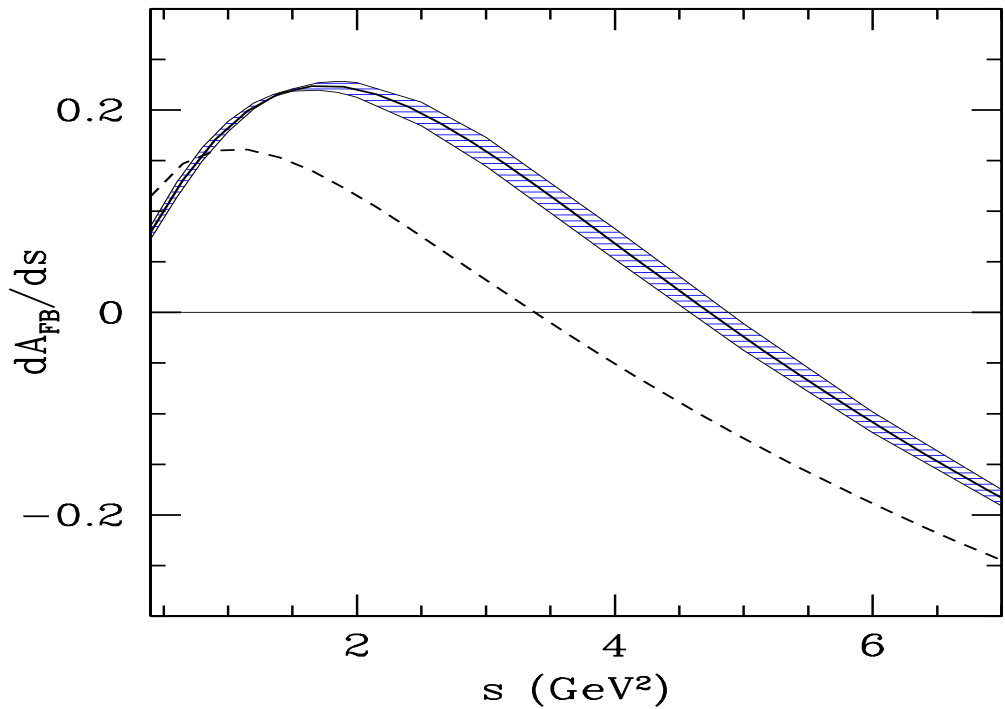}
\caption{\it The forward-backward asymmetry in $B \rightarrow K^*
\ell^+\ell^-$ decay at NLO (solid center line) and LO (dashed). The band reflects the theoretical uncertainties from
the input parameters~\cite{Ali:2002qc}.}
\label{FigFBA}
\end{center}
\end{figure}

It is interesting to observe that at the leading order in the LEET
approach, the FB-asymmetry  in $B\to 
K^*\ell^+\ell^-$ decays depends on one universal form factor
$\xi^{(K^*)}_{\perp}(s)$, and reads as follows
\begin{eqnarray}\cr
  \frac{\d \a_{\rm FB}}{\d \sh}& =& 
\frac{G_F^2 \, \alpha_{em}^2 \, m_B^5}{2^{8} \pi^5} 
      \left| V_{ts}^\ast  V_{tb} \right|^2 \, \sh\, \uh(\sh)^2 \label{eq:dfbabvllex} \\
& & \times  \ct 
\left[ (-\cse) {\mbh\over\ \sh}(-1 + \mvh^2 + \sh) + 
2\ {E_{K^*}\over m_B}\Big(\cse\ {\mbh\over\ \sh} +  {\rm Re}[\cne]\Big) \right] \xi^{(K^*)}_{\perp}(s)^2 \nn \; .
\cr
\end{eqnarray}

It has been noted in \cite{Burdman:1998mk} that the 
location of the forward-backward asymmetry zero $\sh_0$ is nearly 
independent of particular form factor models. An explanation 
of this fact was given in \cite{Ali:1999mm}, 
where it has been noted that the form factor ratios on which the asymmetry 
zero depends are predicted free of hadronic uncertainties in 
the combined heavy quark and large energy limit. Thus the position of
the zero $\sh_0$ is given by
\begin{eqnarray}
{\rm Re}\Big(\cne(\sh_0)\Big) =- \frac{\mbh}{\sh_0} \cse 
\left\{\frac{1-\mvh^2-\sh}{1+\mvh^2-\sh} +1 \right\} \; ,
\label{eq:fbzero}
\end{eqnarray}
which depends on the value of $m_b$ and the ratio of the effective
coefficients $\cse/{\rm Re}\Big(\cne(\sh_0)\Big)$. 

Thus, the precision on the zero-point of the FB-asymmetry in $B \to K^* \ell^+
\ell^-$ is determined essentially by the precision of the ratio of the
effective coefficients and $m_b$\footnote{the corresponding quantity
in the inclusive decays $B \to X_s \ell^+ \ell^-$, for which the
zero-point is given by the solution of the equation ${\rm
Re}\Big(\cne(\sh_0)\Big)=-\frac{2}{\sh_0} \cse$.}. We find  
the insensitivity of $\sh_0$ to the decay form factors in $B \to K^* 
\ell^+ \ell^-$ a remarkable result, which  has also been discussed in
\cite{Burdman:1998mk}. However, the LEET-based
result in Eq.~(\ref{eq:fbzero}) stands theoretically on more rigorous grounds
than the arguments based on scanning a number of form factor models.
Our result for FBA is shown in Fig.~(\ref{FigFBA}) to LO and NLO accuracy.
With the coefficients given in Table~(\ref{tab1}) and $m_b=4.6$ GeV,
we find the LO location of the FB-asymmetry zero is $s_0\simeq3.4\, \mbox{GeV}^2$.

In~\cite{Beneke:2000wa} the effect of the (factorizable) radiative
corrections to the form factor  has been studied and has been
found to shift the position of the asymmetry zero about 5\% towards
larger values. However the effect of both, factorizable {\em and}\/
non-factorizable radiative corrections modify considerably the
location of the FB-asymmetry zero $s_0$. As it is shown in Fig.~(\ref{FigFBA}), the numerical effect of NLO
corrections amounts to a substantial enhancement of the FB asymmetry for  
intermediate lepton invariant mass ($s= 1.5-6$~GeV$^2$) and a
significant shift of the location of the FB-asymmetry zero to $s_0\simeq 4.7\, \mbox{GeV}^2$.
The dominant uncertainty (between $5\%$  and up  $55\%$) is shared
mainly between the $B$-meson light-cone distribution amplitudes
$\lambda_{B,+}$, the $B$-decay constant $f_B$ and the form
factor $\xi_{\perp}^{(K^*)}$.

\subsection{Transversity Amplitudes for $B \to K^* \ell^+ \ell^-$ }
The decay $B \to J/\psi K^*$ is described
by three amplitudes\footnote{they should not be confused with the form
factors $A_0(s)$, $A_1(s)$ etc.} $({\cal A}_i; i=0,\parallel, \perp)$ in the
transversity basis, where ${\cal A}_0(s)$, ${\cal A}_{||}(s)$ and
${\cal A}_{\perp}(s)$  have CP eigenvalues $+1, +1$ and $-1$, respectively
\citer{ref:dunietz,ref:dighe}.
Here, ${\cal A}_0(s)$ corresponds to the longitudinal polarization of the vector meson
$K^*$ and ${\cal A}_{||}(s)$ and ${\cal A}_{\perp}(s)$ correspond to parallel and 
transverse polarizations, respectively. 
The relative phase between the parallel (transverse)
amplitude and the longitudinal amplitude is given by
$\phi_{\parallel(\perp)}(s)\equiv\mbox{arg}\Big({\cal A}_{\parallel(\perp)}/{\cal A}_0(s)  \Big)$.

\begin{figure}[!t]
\begin{center}
\epsfxsize8.6cm
\includegraphics[width=12cm,height=9cm]{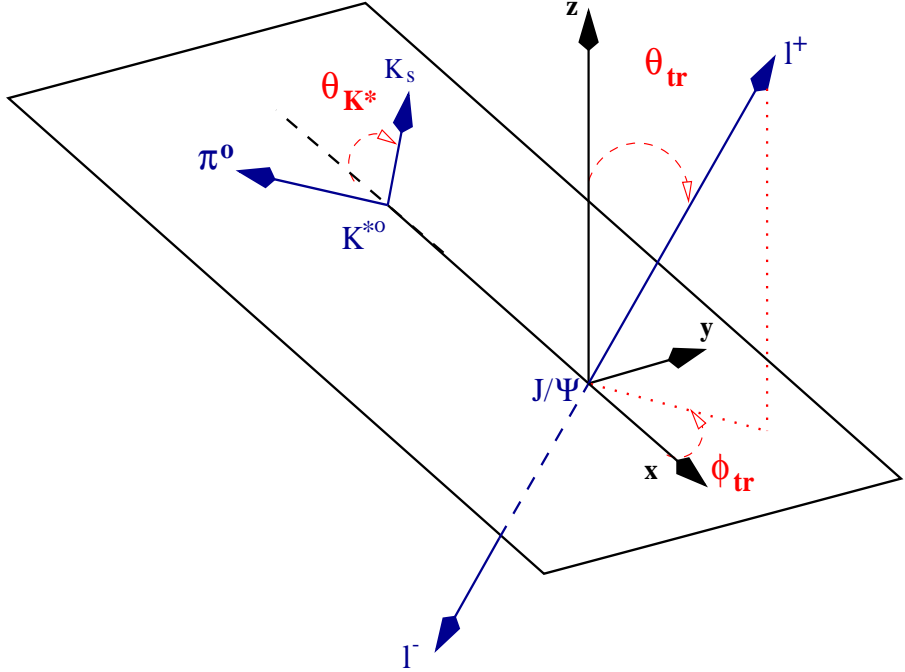}
\caption{ \it Definitions of the transversity angles \thetatr , \phitr, and
\thetakstar . The angles \thetatr~and \phitr~are determined in the
\jpsi\ rest frame. The angle  \thetakstar~is determined in the $K^*$ rest
 frame.}
\label{fig:jpsiksttrans}
\end{center}
\end{figure}

The transversity frame is defined as the \jpsi\ rest frame \big(see
Fig.~(\ref{fig:jpsiksttrans})\big). 
The $K^*$ direction defines the negative $x$ axis. The $K\pi$ decay plane
defines the $(x,y)$ plane, with $y$ oriented such that $p_y(K) > 0$. The
$z$ axis is the normal to this plane, and the coordinate system is
right-handed.  The transversity angles 
\thetatr\ and \phitr\ are defined as the polar and azimuthal angles of the
positively charged  lepton from the \jpsi\ decay; \thetakstar\ is the $K^*$
helicity
angle defined in the $K^*$ rest frame as the angle between the $K$
direction and the direction opposite to the \jpsi. This basis
has been used by the CLEO \cite{ref:cleo},
CDF \cite{ref:cdf}, BABAR \cite{ref:babar}, and the BELLE \cite{ref:belle}
collaborations to project out the amplitudes in the decay $B \to J/\psi
K^*$ with well-defined CP eigenvalues in their measurements of the 
quantity $\sin 2 \beta$, where $\beta$ is an inner angle of the unitarity
triangle. 

We also adopt this basis and analyze the various amplitudes
from the non-resonant (equivalently short-distance) decay $B \to K^*
\ell^+ \ell^-$. In this basis, both the
resonant $B \to K^* J/\psi \to K^* \ell^+ \ell^-$ (already measured) and
the non-resonant ($B \to K^* \ell^+ \ell^-$) amplitudes turn out to be
very similar, as we show here.
\par

The angular distribution is given in terms of the linear
polarization basis (${\cal A}_{\pm1}(s)=({\cal A}_{\parallel}(s)\pm
{\cal A}_{\perp}(s))/\sqrt{2}$) and ${\cal A}_0(s)$ by

\begin{eqnarray}\label{eqn:distrib}
\gfrac{\dd^{4} \Gamma}{ds\;\,\dd\cthetatr\;\,\dd\cthetakstar\;\,\dd\phitr} 
& = & f_1(w)\cdot |{\cal A}_{0}(s)|^2  + f_2(w)\cdot |{\cal
A}_{\parallel}(s)|^2  + f_3(w)\cdot |{\cal A}_{\perp}(s)|^2   \nonumber \\ 
& + &  \eta f_4(w)\cdot {\rm Im}({\cal A}^*_{\parallel}(s){\cal A}_{\perp}(s))
+ f_5(w)\cdot {\rm Re}({\cal A}^*_{0}(s){\cal A}_{\parallel}(s))  \nonumber\\
& + &   \eta f_6(w)\cdot {\rm Im}({\cal A}^*_{0}(s){\cal
A}_{\perp}(s))~, \nonumber 
\end{eqnarray}
\noi where $\eta=+1(-1)$ for $B^0$ and $B^+$ ($\bar{B}^0$ and $B^-$),
and the coefficients $f_{i=1,...,6}$, which depend on
the transversity angles $w=(\thetakstar, \thetatr, \phitr)$, are given by:

\begin{eqnarray*}
f_1(w) & = & ~ ~ \ \, 9/(32\pi)\cdot\, 
 2\cq{\thetakstar}(1-\sq{\thetatr}\cq{\phitr}), \\
f_2(w) & = & ~ ~ \ \, 9/(32\pi)\cdot\,
 \sq{\thetakstar}(1-\sq{\thetatr}\sq{\phitr}), \\
f_3(w) & = & ~ ~ \ \, 9/(32\pi)\cdot\,
 \sq{\thetakstar}\sq{\thetatr}, \\
f_4(w) & = & ~ ~ \ \, 9/(32\pi)\cdot\,
 \sq{\thetakstar}\sin{2\thetatr}\sphitr,\\
f_5(w) & = & - \ \, 9/(32\pi)\cdot\,
  1/\sqrt{2}\cdot\sin{2\thetakstar}\sq{\thetatr}\sin{2\phitr}, \\
f_6(w) & = & ~ ~ \ \, 9/(32\pi)\cdot\,
  1/\sqrt{2}\cdot\sin{2\thetakstar}\sin{2\thetatr}\cos{\phitr}~.
\end{eqnarray*}
%
\renewcommand{\arraystretch}{1}
\begin{table}[t]
\begin{center}
\begin{tabular}{|c|c|c|c|c|c|}   
\hline\hline
Group & $|\hat {\cal A}_0|^2$ & $|\hat {\cal A}_{\perp}|^2$ & $|\hat
{\cal A}_{||}|^2$ & 
$\phi_{\perp}$ & $\phi_{\parallel}$
\\ \hline
CLEO\cite{ref:cleo} &
$0.52\pm0.08$ & $0.16\pm0.09$ & $0.32\pm0.12$ &
$-3.03\pm0.46$ & $-3.00\pm 0.37$ \\
CDF\cite{ref:cdf} &    
$0.59\pm0.06$ & $0.13^{+0.13}_{-0.11}$ &$0.28\pm0.12$ &
$-2.58\pm0.54$ & $-2.20\pm 0.47$ \\
BaBar\cite{ref:babar} &
$0.60\pm0.04$ & $0.16\pm0.03$ & $0.24\pm0.04 $& 
$-2.97\pm0.17$ &$-2.50\pm 0.22$ \\
Belle \cite{ref:belle}&
$0.60\pm0.05$ & $0.19\pm0.06$ & $0.21\pm0.08 $&
$-3.15\pm0.21$ & $-2.86\pm0.25$ \\
AS \cite{Ali:2002qc} &$0.51$ &$0.21$ & $0.28$ &$-3.25$ &$-3.04$ \\
\hline\hline
\end{tabular}
\caption{\it Current measurements of the decay amplitudes in the transversity
basis for the decay $B\rightarrow J/\psi   K^*$ .
The corresponding amplitudes for the
non-resonant decay $B \to K^* \ell^+ \ell^-$ worked out
in this paper in the LO approximation at $m_{\ell^+
\ell^-}^2=m_{J/\psi}^2$ are given in the last
row~\cite{Ali:2002qc}. \label{others}} 
\end{center} \end{table}
%

\noi In terms of the helicity amplitudes
$H^{L/R}_{\pm 1,0}$, introduced earlier, the amplitudes in the linear
polarization basis, ${\cal A}_{0,\perp,\parallel}$, can be calculated  
from the relation:
\begin{eqnarray*}
{\cal A}_0(s)  &=& \kappa \Big(H^L_{0}(s)+H^R_{0}(s)\Big )~,\\
{\cal A}_{\pm1}(s)&=&\kappa \Big(H^L_{\pm}(s)+H^R_{\pm}(s)\Big)~,
\end{eqnarray*}
with
$\kappa^2= {\alpha_{em}^2 G_{F}^2\over 384 \pi^5}
\sqrt{\lambda} {m_{b}^2\over m_{B}^3 }  |V_{tb}V_{ts}^*|^2~$.

\begin{figure}[!hbtp]
\vspace{-1cm}
\begin{center}
\includegraphics[width=12cm,height=7.5cm]{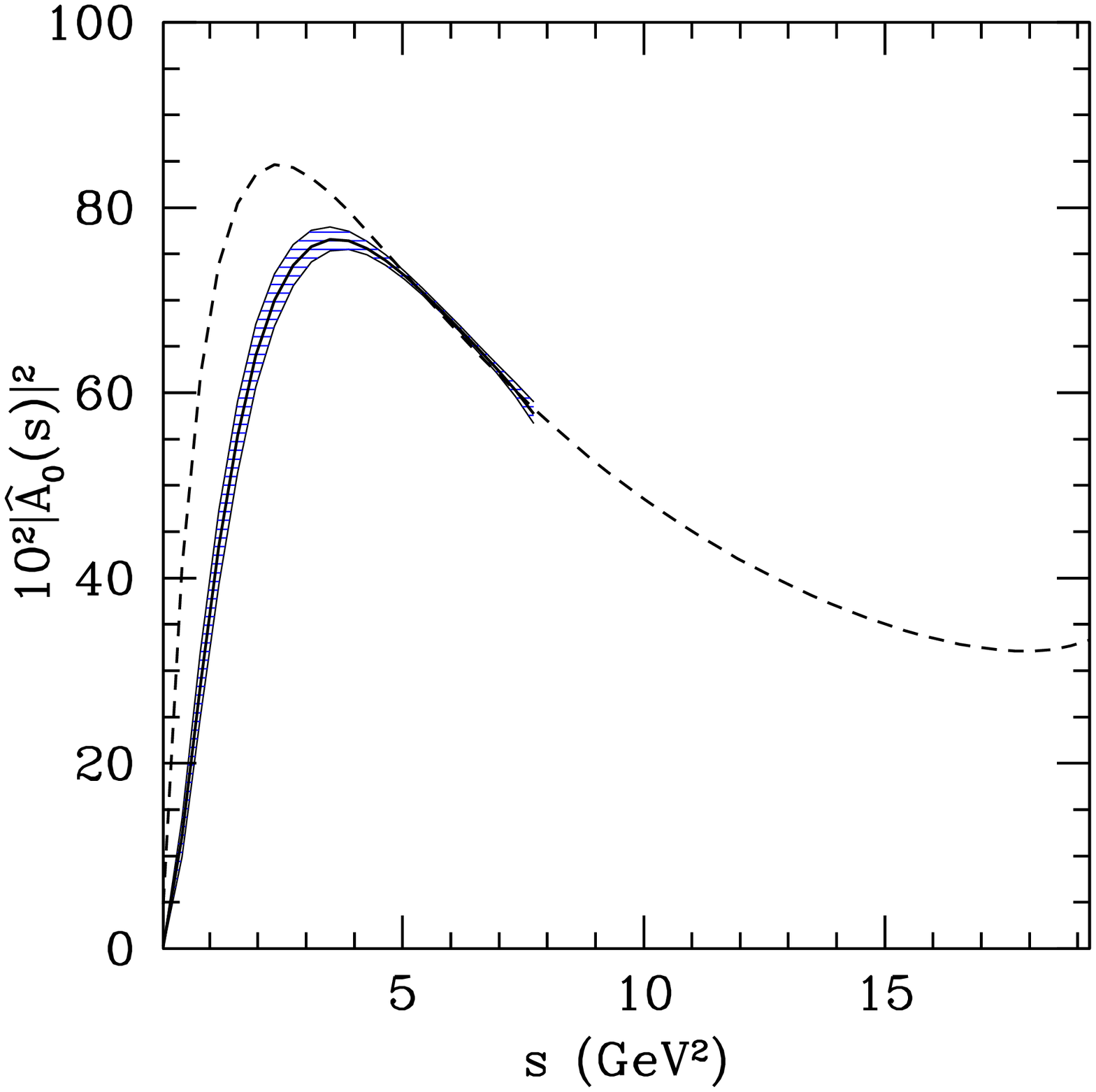}
\includegraphics[width=12cm,height=7.5cm]{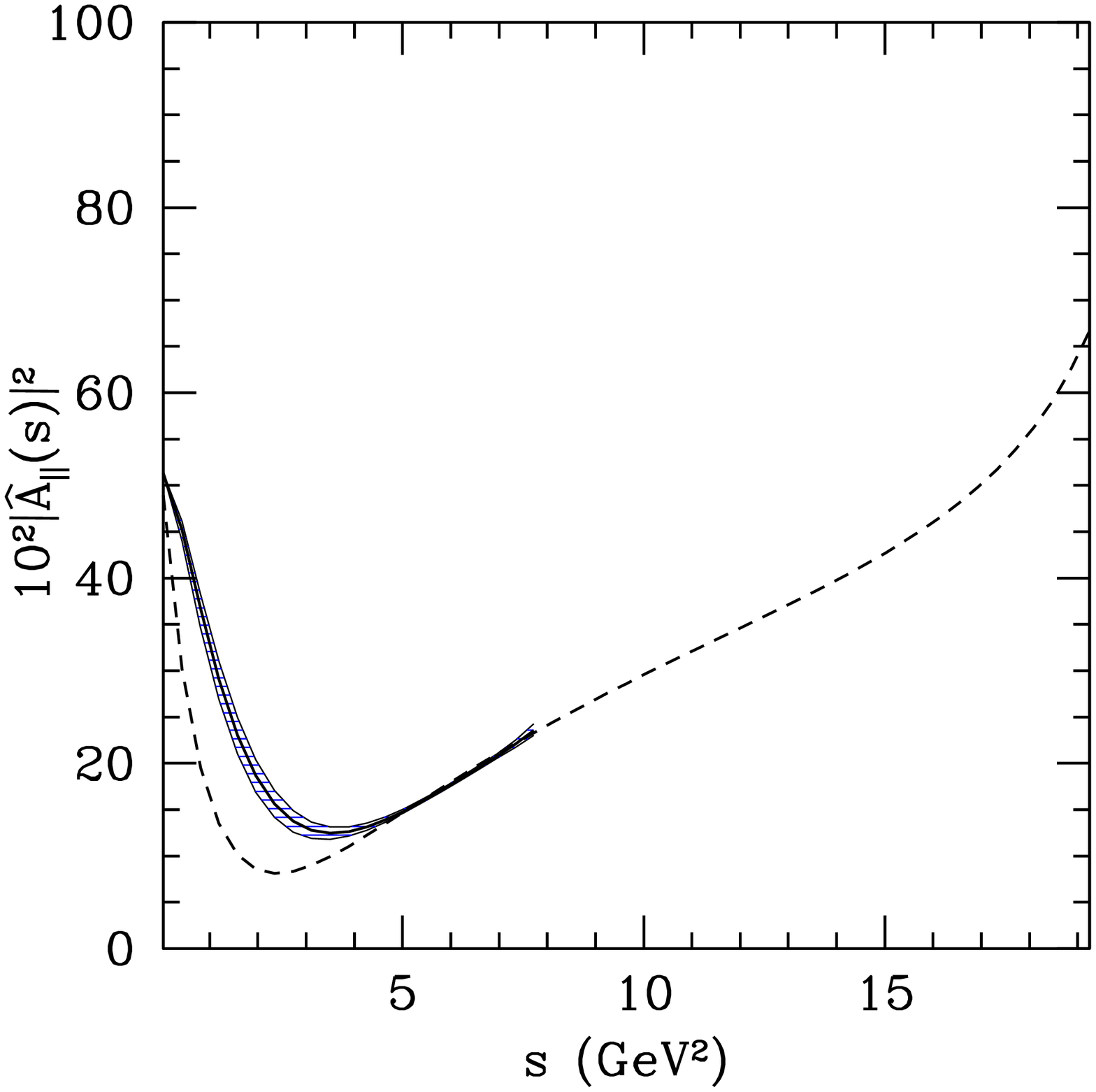}
\includegraphics[width=12cm,height=7.5cm]{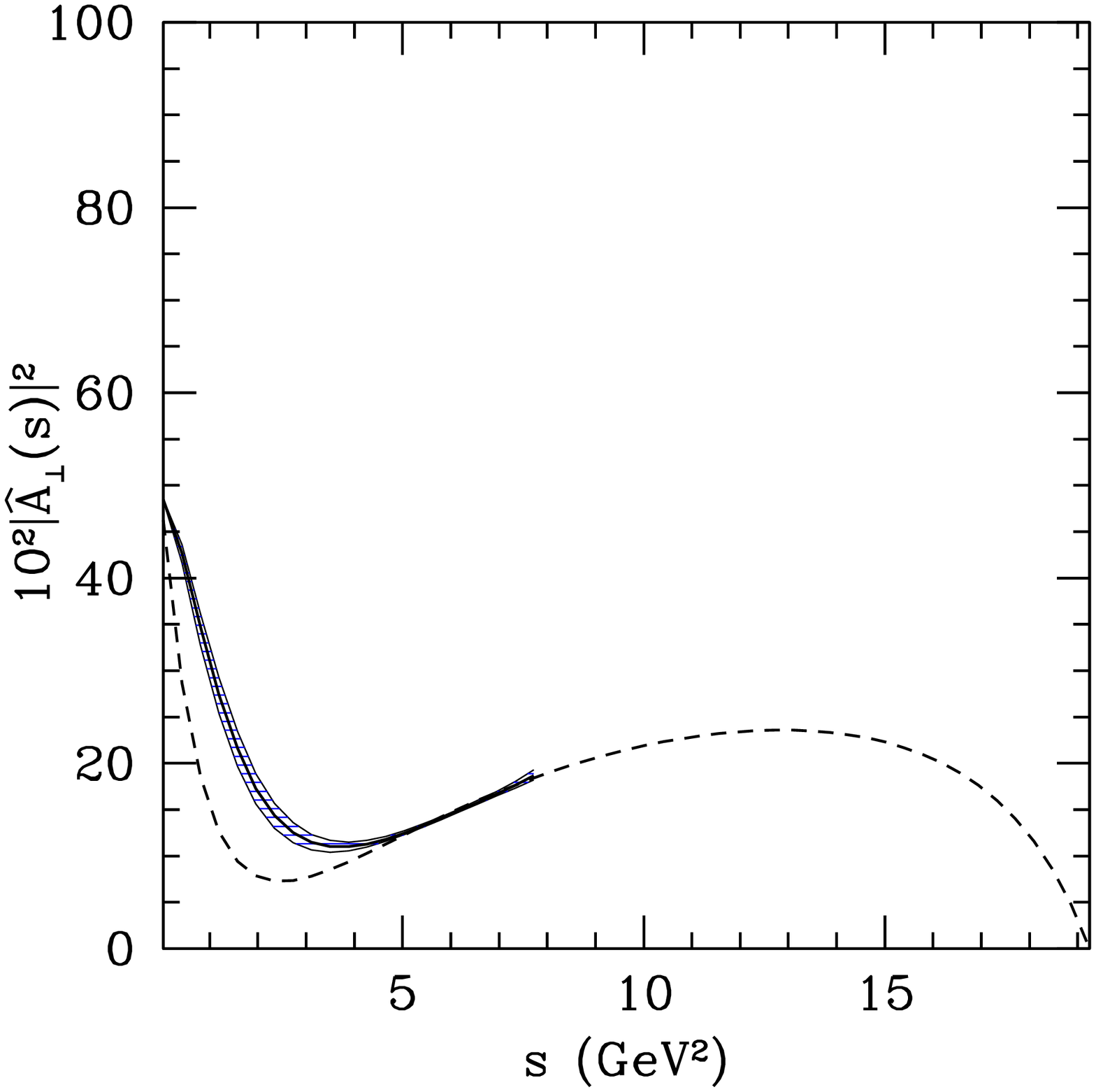}
\caption{ \it The helicity amplitudes $|\hat {\cal A}_0(s)|^2$
(upper-plot), $|\hat {\cal
A}_{\parallel}(s)|^2$ (middle-plot) and $|\hat {\cal A}_{\perp}(s)|^2$
(lower-plot) in $B \to K^* \ell^+ \ell^-$ at NLO
(solid center line) and LO (dashed). The band for NLO reflects 
theoretical uncertainties from input parameters~\cite{Ali:2002qc}.}
\label{figAperp2}
\end{center}
\end{figure}

Experimental results are conventionally expressed in terms of the spin
amplitudes $\hat {\cal A}_{0,\bot,\|}$ normalized to unity, with $|\hat
{\cal A}_0|^2+|\hat {\cal A}_\bot|^2+|\hat {\cal A}_\||^2=1$. We show the
polarization fractions, $\Gamma_0/\Gamma= |\hat {\cal A}_0(s)|^2$,
$\Gamma_{\parallel}/\Gamma= |\hat {\cal A}_{\parallel}(s)|^2$ and 
$\Gamma_{\perp}/\Gamma= |\hat {\cal A}_{\perp}(s)|^2$ in the
leading and next-to-leading order for the decay $B \to K^* \ell^+ \ell^-$
in Fig.~(\ref{figAperp2}),
respectively. Since the interference terms in the angular distribution are
limited to  \textrm{Re(${\cal A}_{||} {\cal A}_0^*$)},
\textrm{Im(${\cal A}_{\perp}
{\cal A}_0^*$)} and \textrm{Im(${\cal A}_{\perp} {\cal A}_{||}^*$)}, there
exists a phase ambiguity:
\begin{eqnarray}
\phi_{||} &\to& -\phi_{||}~,\\ 
\phi_{\perp} &\to& \pm \pi-\phi_{\perp}~,\\
\phi_{\perp}-\phi_{||} &\to& \pm \pi-(\phi_{\perp}-\phi_{||})~.
\label{eq:phidef}
\end{eqnarray}

\begin{figure}[H]
\begin{center}
\psfrag{s} { $$}
\psfrag{bCfi2}{\hskip -0.2cm $\cos \phi_{||}(s)$}
\psfrag{bSfi2}{\hskip -0.2cm  $\sin \phi_{||}(s)$}
\epsfig{file=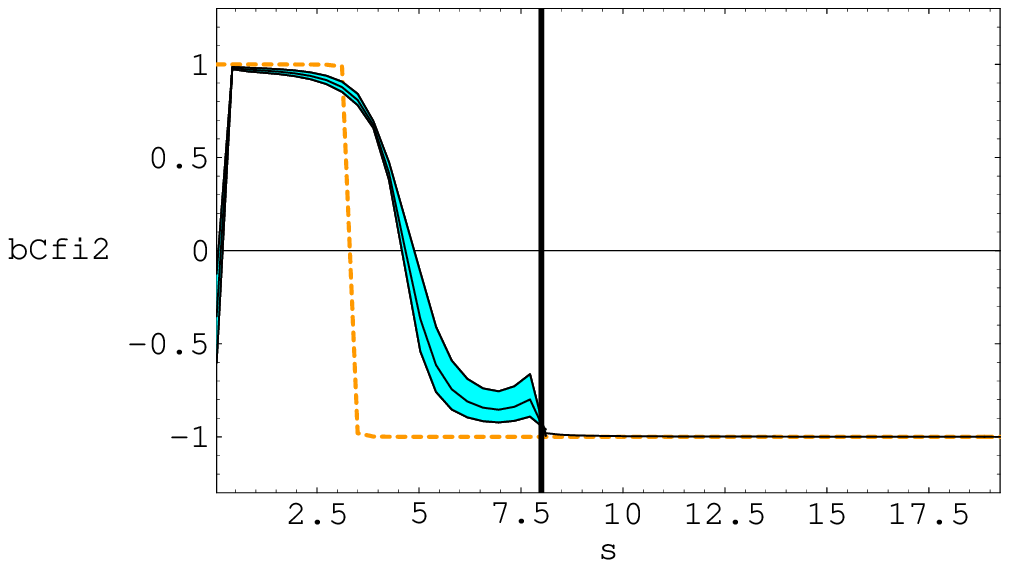,width=0.48\linewidth}
\hspace*{.2cm}
\epsfig{file=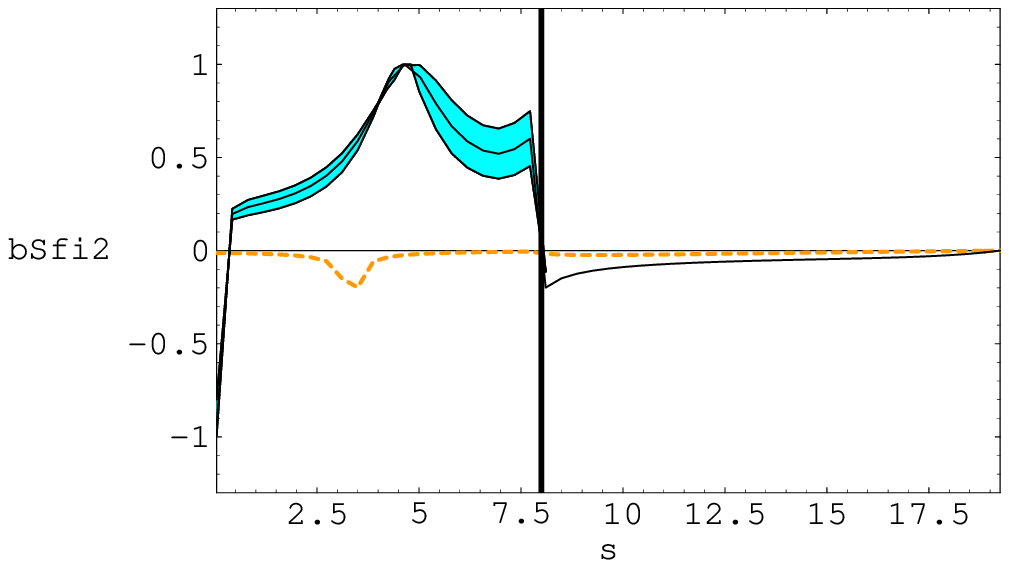,width=0.48\linewidth}
\vspace*{1cm}
\psfrag{s}{\hskip -0.3cm $s\ (\textrm{GeV}^2)$}
\psfrag{bCfiperp}{\hskip 0.cm $\cos \phi_{\perp}(s)$}
\psfrag{bSfiperp}{\hskip 0.2cm  $\sin \phi_{\perp}(s)$}
\epsfig{file=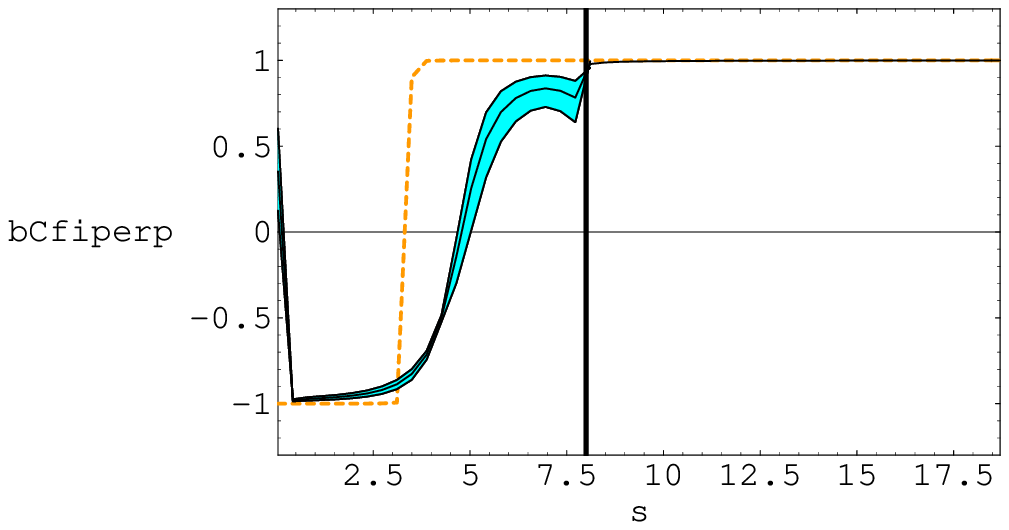,width=0.48\linewidth}
\hspace*{.2cm}
\epsfig{file=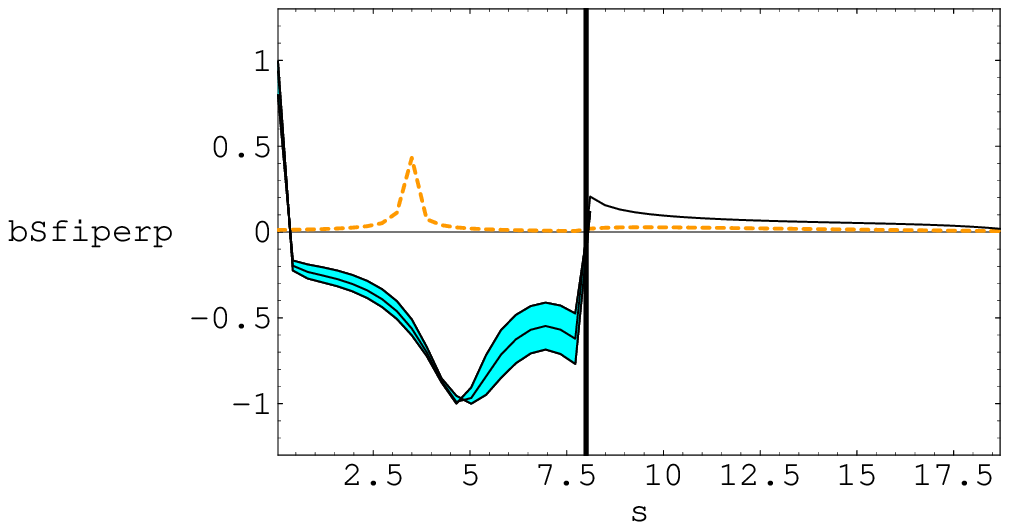,width=0.48\linewidth}
\vspace*{-1cm}
\caption{ \it
The functions $\cos \phi_{||,\perp}(s)$ and $\sin \phi_{||,\perp}(s)$ at
NLO (solid center line) and LO
(dashed). The band reflects all theoretical uncertainties from
parameters with most of the uncertainty due to the form factors
$\xi^{(K^*)}_{i}(0)$.The vertical line at \textrm{ s = 8
$\mbox{GeV}^2$} represents the 
domain of validity of the LEET approach in our case~\cite{Ali:2002qc}.}
\label{figphi-perp}
\end{center}
\end{figure}

\noi To avoid this, we have plotted in Fig.~(\ref{figphi-perp}) the
functions $\cos \phi_{\parallel,\perp}(s)$ and $\sin
\phi_{\parallel,\perp}(s)$, showing their behaviour at the leading and
next-to-leading order.  
The dashed lines in these figures correspond to using the LO amplitudes,
calculated in the LEET approach. In this order, the bulk of the parametric
uncertainty resulting from the form factors cancels. Although, strictly
speaking, the domain of validity
of the LEET-based distributions is limited by the requirement of large
energy of the $K^*$ (which we have translated into approximately $s < 8~
\mbox{GeV}^2$), we show this distribution for the entire $s$-region allowed
kinematically in $B \to K^* \ell^+ \ell^-$.
The shaded  curves correspond to using the NLO contributions in the LEET
approach. 

%

We compare the resulting amplitudes $\vert \hat {\cal A}_0
\vert^2$, $\vert
\hat {\cal A}_\perp \vert^2$, $\vert \hat {\cal A}_{\parallel} \vert^2$,
$\phi_{\parallel}(s)$, and $\phi_{\perp}(s)$ at the value $s=m_{J/\psi}^2$
with the corresponding results from the four experiments in
Table~(\ref{others}). In comparing these results for the phases, we had to
make a choice between the two phase conventions shown in
Eq.~(\ref{eq:phidef}) and the phases shown in the last row of this table 
correspond to adopting the lower signs in these equations. 
We note that the short-distance amplitudes from the decay $B \to K^*
\ell^+ \ell^-$ are similar to their resonant counterparts measured in the
decay $B \to J/\psi K^*$. We also note that a helicity analysis of the
decay $B \to J/\psi K^*$ has been performed in the QCD factorization
approach by {\it Cheng et al.}~\cite{Cheng:2001ez}.  

\begin{figure}[t]
\psfrag{s}{\hskip -0.5cm $s\ (\mathrm{GeV}^2)$}
\psfrag{b}{\hskip - 1.cm$\phi_{0}(s) $}
\begin{center}
\epsfig{file=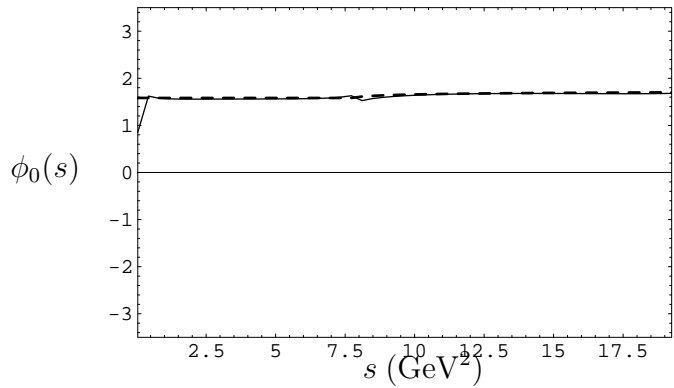,width=0.5\linewidth}
\caption{\it The phase $\phi_{0}(s) $ at NLO (solid line) and LO (dashed)~\cite{Ali:2002qc}.}
\label{figphi-0}
\end{center}
\end{figure}

The structures in the phases shown in  
Fig.~(\ref{figphi-perp}) deserve a closer look. We note that at
the leading order, the phases $\phi_{\perp}(s)$ and $\phi_{\parallel}(s)$
are given by the following expressions:
\begin{eqnarray}
\phi_{\perp}(s)&=& \textrm{Arg}\Big[{i\,\sqrt{\lambda}\over m_b\,
m_B\, \sqrt{s}}\Big\{s\, \cne+ 2\, m_b\, m_B\, \cse \Big\}\,
\xi^{(K^*)}_{\perp}(s)\Big] - \textrm{Arg}[{\cal A}_0(s)]~,\label{eq:phiperp}\\
\phi_{||}(s)&=& \textrm{Arg}\Big[{-i\,E_{K^*}\, \xi^{(K^*)}_{\perp}(s)\over m_b  
\,\sqrt{s}}\Big\{ \Big(s\, \cne+ 2\, m_b\, m_B\, \cse \Big)\nn\\
&&- 2\, m_b\, m_B\, \Big(\cse +{s\over  2\, m_b\, m_B} Y(s)\Big)({m_{K^*}^2
\over m_B^2}) \Big\} \Big] 
- \textrm{Arg}[{\cal A}_0(s)],\label{eq:phiII}
\end{eqnarray}
\noi where we can neglect the term proportional to $(m_V^2/ m_B^2)$ in
the latter equation. The phase
$\phi_{0}(s)\equiv\textrm{Arg}[{\cal A}_0(s)]$ is
constant in the entire phase space, as shown in Fig.~(\ref{figphi-0}).
The functions in the square brackets in Eqs.~(\ref{eq:phiperp}) and
(\ref{eq:phiII}) are purely imaginary. However, due to the fact that in
the SM the coefficients $C_9^{\rm eff}$ and $C_7^{\rm eff}$ have opposite
signs, these phases become zero at a definite value of $s$, beyond which
they change sign, yielding a step-function behaviour, shown by
the dotted curves in the
functions $\cos \phi_{\parallel}(s)$ and $\cos \phi_{\perp}(s)$ in
Fig.~(\ref{figphi-perp}), respectively.
The position of the zero of the two functions, denoted, respectively,
 by $s^{\perp}_0$ and $s^{||}_0$, are given by solving the following
equations:
\begin{eqnarray}
 \textrm{Arg}\Big[{i\,\sqrt{\lambda}\over m_b\,
m_B\, \sqrt{s^{\perp}_0}}\Big\{s^{\perp}_0\, \cne(s^{\perp}_0)+ 2\,
m_b\, m_B\, \cse \Big\}\, \xi^{(K^*)}_{\perp}(s^{\perp}_0)\Big] &=& \phi_0(s^{\perp}_0)~,\label{eq:solphiperp}\\
 \textrm{Arg}\Big[{-i\,E_{K^*}\ \over m_b 
\,\sqrt{s^{||}_0}}\Big\{ s^{||}_0\, \cne(s^{||}_0)+ 2\, m_b\, m_B\,
\cse \Big\}\, \xi^{(K^*)}_{\perp}(s^{||}_0) \Big] &=&\phi_0(s^{||}_0).
\label{eq:solphiII}
\end{eqnarray}
\noi For the assumed values of the Wislon coefficients and other
parameters, the zeroes of the two functions, namely
$s^{||}_0$ and $s^{\perp}_0$, occur at around $s\simeq 3\, \mbox{GeV}^2$, in the
lowest order, as can be seen in Figs.~(\ref{figphi-perp}), respectively. 

The LO contributions in $\sin\phi_{\parallel}(s)$ and $\sin
\phi_{\perp}(s)$ are constant, with a value 
around 0, with a small structure around $s\simeq 3\, \mbox{GeV}^2$,
reflecting the sign flip of the imaginary part in $ {\cal A}_{||}(s)$
(${\cal A}_{\perp}(s)$). At the NLO, the phases are influenced 
by the explicit $O(\alpha_s)$ contributions from the factorizable and
non-factorizable QCD corrections (see section 3), which also bring in
parametric uncertainties with them. The most important effect is that the
zeroes of the phases as shown for $\cos \phi_{\perp}(s)$ and
$\cos \phi_{\parallel}(s)$ are
shifted to the right, and the step-function type bahaviour of these
phases in the LO  gets a non-trivial shape.  Note
that in both figures a shoulder around $s\simeq 8\, \mbox{GeV}^2$ reflects charm
production whose threshold lies at $s = 4\, m_c^2$.
\section{ Decay Distributions in $B \rightarrow \rho\,\, \ell\nu_{\ell}$
\label{sec:Btorho}}
After a complete analysis of the $B \rightarrow K^* \ell^+ \ell^-$, we
turn now to the semileptonic $B \rightarrow \rho\,\, \ell\nu_{\ell}$
one. In this section, we present general spectra analysis in exclusive
$B \rightarrow \rho \ell \nu_{\ell}$ decay in terms of the corresponding
helicity amplitudes. Further, we calculate the different
dilepton invariant mass distributions and the corresponding Dalitz
distributions. We include the $O(\alpha_s)$-corrections, $1/E$ power
corrections by means of the large energy expansion technique (LEET)
and using the light-cone QCD sum rules approach.

First, let us describe the apropriate matrix elements for $B
\rightarrow \rho\,\ell \nu_{\ell}$. Since this decay is purely a
$(V-A)$ transition, one could get the corresponding matrix element
from the $B \to K^* \ell^+ \ell^-$ ones, using the following replacements :
\begin{eqnarray} \label{eq:c9CC}
        C_9 & = & - C_{10} = \frac{1}{2}
                \, , \\
        C_7^{\mbox{eff}} & = & 0 
                \, , \\
        \left( \frac{G_F \, \alpha_{em}}{\sqrt{2} \, \pi}
                V_{ts}^\ast V_{tb}\right)
                & \rightarrow &
        \left( - \frac{4 \, G_F}{\sqrt{2}} V_{ub} \right)
                \, .  
\label{eq:CKMCC} 
\end{eqnarray}

\noi This amounts to keeping only the charged current $(V-A)$ contribution in
$B \to K^* \ell^+ \ell^-$ decay. Thus, the corresponding amplitude for
the semileptonic $b \rightarrow u\,\ell \nu_{\ell}$ decay, can be
factorized into a leptonic and a partonic part as,
\begin{eqnarray}
{\cal M}(b \rightarrow u\,\ell \nu_{\ell})= -i \frac{G_F}{\sqrt{2}}
V_{ub} \Big\{[\,\bar{u}\, \gamma_{\mu} (1-\gamma_5)\, b\,]\,
[\,\bar{\ell} \gamma^{\mu} (1-\gamma_5) \nu_{\ell}\,]\Big\}. 
\label{eq:Mrho} 
\end{eqnarray}

From the semileptonic amplitude given in Eq.~(\ref{eq:Mrho}), we
notice that the exclusive $b \rightarrow u\,\ell \nu_{\ell}$ decays is a good
candidate for a clean determination of the modulus of $V_{ub}$, one of
the smallest and least well known CKM matrix elements. Experimentally, the main difficulty of the
observation of $b \rightarrow u\,\ell \nu_{\ell}$ signal events is the
large background from $b \rightarrow c\,\ell \nu_{\ell}$
events\footnote{Because $|V_{ub}/V_{cb}|\approx 0.1$, the branching
fractions of the exclusive $b \rightarrow u\,\ell \nu_{\ell}$ decays
$(\sim 10^{-4})$ are small compared to those of the charmed semileptonic
decays, which are of the order of some percent.}. For that, different
experimental distribution analysis are in order to overcome this
prblem. In this spirit,  we propose many angular distributions
studies of $B \rightarrow \rho \,\ell \nu_{\ell}$, where the vector
meson decays to two pseudoscalars\footnote{this is due to the fact
that the non-leptonic 
$\rho\rightarrow \pi^+ \pi^-$ decay is by far the dominant branching
ratio~\cite{PDG2000}, with  ${\cal B}(\rho\rightarrow \pi^+ \pi^-)\sim
100\%$.}, $\rho \rightarrow \pi^+ \pi^-$ .

Four independent kinematic variables completely describe the
semileptonic decay $B \rightarrow \rho(\to \pi^+ \pi^-) \,\ell \nu_{\ell}$
the four variables most commonly used are the invariant
dilepton mass distributions and three polar angles. Thus, the
differential decay rate for $B\rightarrow \rho (\rightarrow \pi^+ \pi^-) \ell
\nu_{\ell}$ is expressed in terms of the helicity amplitudes $H_{\pm,0}$ by \cite{Richman:wm}:
\begin{eqnarray}
\label{4diff}
{d^4\Gamma  \over ds\ d\cos\theta_{\rho}\ d\cos\theta_{+}\ d\phi} &=& {3
\over 8 (4\pi)^4} G_{F}^2 |V_{ub}|^2 {\sqrt{\lambda} s \over  m_{B}^3}
{\it{{\cal B}(\rho \rightarrow \pi^+ \pi^-)}}\nn \\
&\times&  \Big \{(1 - \ cos\theta_{+})^2 sin^2\theta_{\rho}\ |H_{+}(s)|^2\nn \\  
&& + (1 +\ cos\theta_{+})^2  sin^2\theta_{\rho}\  |H_{-}(s)|^2\nn \\
&& + 4\ sin^2\theta_{+}\  cos^2\theta_{\rho}\  |H_{0}(s)|^2\nn  \\
&& - 4\ \  sin\theta_{+}\ (1 -  cos\theta_{+})\ sin\theta_{\rho}\ cos\theta_{\rho}\ cos\phi\  H_{+}(s)  H_{0}(s)\nn \\
&& + 4\ \  sin\theta_{+}\ (1 + cos\theta_{+})\  sin\theta_{\rho}\ cos\theta_{\rho}\ cos\phi\ H_{-}(s) H_{0}(s)\nn \\
&& -2\ sin^2\theta_{+}\ sin^2\theta_{\rho}\ cos2\phi\ H_{+}(s)
H_{-}(s)\Big \}. 
\label{diffrho}
\end{eqnarray}
The function $\lambda$ can be found in subsection \ref{sec:NLO}.
The angles here $\theta_+$, $\theta_{\rho}$ and $\phi$  are defined
respectively as~:  the direction between the charged lepton and the
recoiling vector meson measured in the $W$ rest frame, the polar angle
between $\pi^{+}$ (or $\pi^{-}$) and the direction of the vector meson
in the parent meson's rest frame, and the azimuthal angle between the
planes of the two decays   $B \rightarrow \rho(\rightarrow \pi^{+}
\pi^{-}) \ell \nu_{\ell}$.  

The helicity  amplitudes can in turn be related to the two
axial-vector form factors, $A_{1}(s)$ and $A_{2}(s)$, and the vector
form factor, $V(s)$, which appear in the hadronic current 
\cite{Richman:wm}:
\begin{eqnarray}
H_{-}(s) &=& (m_{B} + m_{\rho})\ A_{1}(s) + 2\ { \sqrt{\lambda}
\over m_{B} + m_{\rho}}\ V(s), \label{H-} \\
H_{+}(s) &=& (m_{B} + m_{\rho})\ A_{1}(s) - 2\ { \sqrt{\lambda}
\over m_{B} + m_{\rho}}\ V(s), \label{H+} \\
H_{0}(s) &=& {1 \over 2 m_{\rho} \sqrt{s}}\ \Big[(m_{B}^2 - m_{\rho}^2-
s)(m_{B} + m_{\rho})\ A_{1}(s) - 4\ {\lambda \over m_{B} + m_{\rho}}\
A_{2}(s) \Big], 
\label{H0}
\end{eqnarray}
where $m_{\rho}$ stands for the $\rho$-meson mass. Using
Eqs.~(\ref{A1}), (\ref{A2}) and (\ref{V}) in Eqs.~(\ref{H-})-(\ref{H0}), we obtain at the LO accuracy in the large Energy Limit,
the following helicity  amplitudes:
\begin{eqnarray}
H_{-}(s) &=& 2 \left[ E_{\rho} + {\sqrt{\lambda} \over m_B} \right] \ \xi^{(\rho)}_{\perp}(s),\label{HsemiLET-}\\
H_{+}(s) &=& 2 \left[ E_{\rho} - {\sqrt{\lambda} \over m_B} \right] \ \xi^{(\rho)}_{\perp}(s),\label{HsemiLET+}\\
H_{0}(s) &=& {1 \over m_{B}\ m_{\rho}\ \sqrt{s}}  \Big[ m_B\ E_{\rho}\ (m_{B}^2
 - m_{\rho}^2 - s)  - 2\ \lambda  \Big]\ \xi^{(\rho)}_{\perp}(s) 
\nn \\
&& + {2\ \lambda \over m_B E_{\rho} \sqrt{s}}\ \xi^{(\rho)}_{||}(s),\label{HsemiLET0} 
\end{eqnarray}
where $E_{\rho}$ represents the $\rho$-meson energy, defined in
Eq.~(\ref{eq:E}) (see subsection \ref{sec:leetL}), $\xi^{(\rho)}_{\perp}$
and $\xi^{(\rho)}_{||}$ denote the two universal form factors for the
$B \rightarrow \rho \ell \nu_{\ell}$ transition in the LEET theory.

\begin{figure}[t]
\psfrag{a}{\hskip 0.3cm $s\ (\textrm{GeV}^2)$}
\psfrag{b}{\hskip -1.cm $\xi^{(\rho)}_{\perp} (s)$}
\psfrag{c}{\hskip -0.7cm $\xi^{(\rho)}_{||} (s)$}
\begin{center}
\includegraphics[width=18cm,height=8cm]{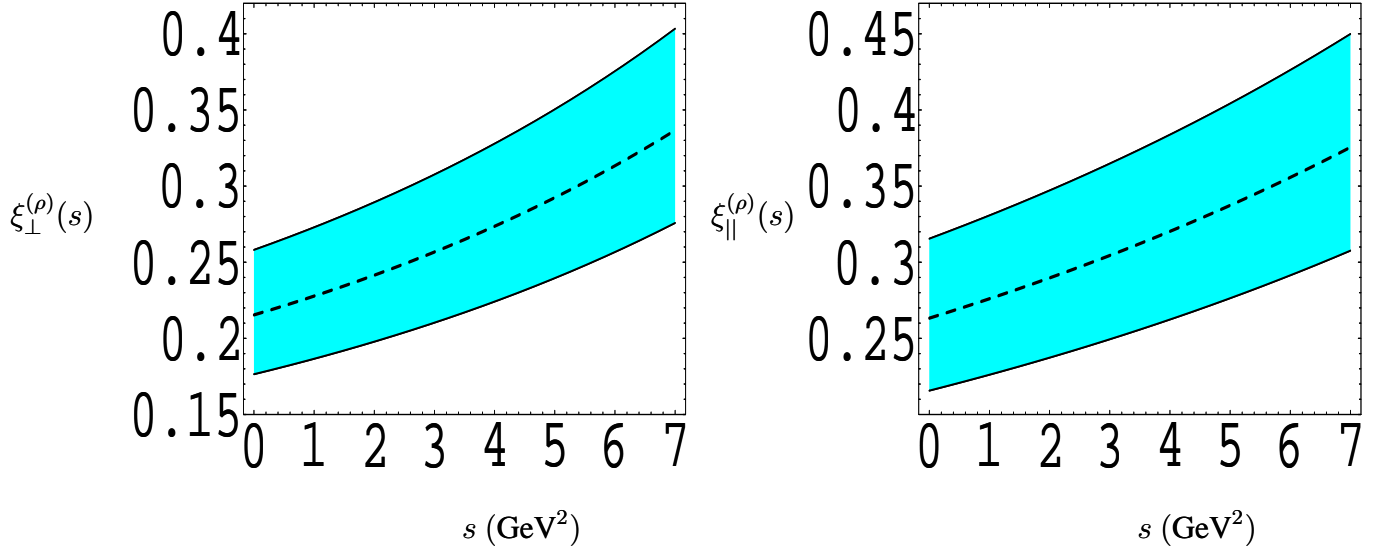}
\caption{\it 
LEET form factors $\xi^{(\rho)}_{{\perp},||}(s)$ 
in $B \rightarrow  \rho\ \ell \nu_{\ell}$. The central values are represented
by the dashed curves and  the bands reflect the uncertainties on the form
factors~\cite{Ali:2002qc}.}
\label{formfactorSemilep}
\end{center}
\end{figure}
As this framework does not predict the corresponding decay form
factors, they have to be supplied from outside. For that we have suggested
the following: 
\begin{itemize}  
\item
Having at hand the 
appropriate $B \rightarrow K^* \ell^+ \ell^-$ LEET form factors,
namely  $\xi^{(K^*)}_{\perp}$ and $\xi^{(K^*)}_{||}$, one can relate
them easily to the $B \rightarrow \rho \ell \nu_{\ell}$ ones in the
SU(3)-symmetry limit. Following this statement, then the semileptonic
$B \rightarrow \rho \ell \nu_{\ell}$  LEET form factors can be defined 
as ~$\xi^{(K^*)}_{\perp/||}(0)=\xi^{(\rho)}_{\perp/||}(0)$.
Unhappily, one has to consider the SU(3)-breaking effects in the
corresponding form factors, which have been evaluated within the QCD
sum-rules \cite{Ali:vd}. 
%
\item
Thus, we take the SU(3)-symmetry breaking factor, as  
\begin{eqnarray}
\zeta_{SU(3)}={\xi^{(K^*)}_{\perp,||}(0)\over
\xi^{(\rho)}_{\perp,||}(0)}= 1.3 \pm 0.06.
\end{eqnarray}
\item
Taking this and $\xi^{(K^*)}_{\perp}(0)$ from Table~(\ref{parameters})
into account, we obtain the corresponding one for the $B \to \rho$-transition:
\begin{eqnarray}
\xi^{(\rho)}_\perp(0)=0.22 \pm 0.04. 
\end{eqnarray}
\item
To extrapolate the $B \rightarrow  \rho \ell \nu_{\ell}$ form factors at $s
\neq 0$, we use the same extrapolation function as for $B \rightarrow
K^* \ell^+ \ell^-$ form factors \big(see Eq.~(\ref{eq:para}) in subsection \ref{sec:ff}\big):
\begin{eqnarray}
\xi^{(\rho)}_{\perp,||}(s)={\xi^{(K^*)}_{\perp,||}(s) \over
\zeta_{SU(3)}}.
\label{eq:paraball}
\end{eqnarray}
\end{itemize}
Using Eq.~(\ref{eq:paraball}), we have plotted in 
Fig.~(\ref{formfactorSemilep}) the corresponding $B\to \rho$ LEET from
factors, namely $\xi^{(\rho)}_{\perp}(s)$ and $\xi^{(\rho)}_{||}(s)$. To check the
consistency of the corresponding form factors, we have compared the
behavior of $\xi^{(\rho)}_{\perp,||}(s)$, with the one used by
\cite{Ball:1998kk} and surprisingly it turns out that the agreement is
reasonable.

We notice that the apparently rather large uncertainty of our prediction,
typically $\sim \pm 25\%$, is mainly due the the $B \to K^*$
form factor $\xi^{(K^*)}_{\perp,||}(0)$ form factors with their
current large uncertainty and to a lesser extent due to the
SU(3)-symmetry breaking factor $\zeta_{SU(3)}$. It may be hoped that in the 
longer term future the form factors could be known with much greater 
confidence.
\subsection{ Dalitz distributions}
Integrating out the angles $(\theta_{\rho},~\theta_{+})$ and
$(\phi,~\theta_{+})$ we obtain respectively the double $\phi$ and
$\theta_{\rho}$ angular distributions:
\begin{eqnarray}
{d^2{\cal B} \over d\phi\ ds} &=&\tau_B {G_F^2\ s\  \sqrt{\lambda}
 \over 192 m_{B}^3 \pi^4} |V_{ub}|^2 ({\it{{\cal B}(\rho \rightarrow \pi^+ \pi^-)}})\nn \\ 
&& \Big \{ |H_0(s)|^2 +  |H_-(s)|^2 + |H_+(s)|^2 -   cos 2\phi \ H_-(s)\
  H_+(s)\Big\},
\label{dgamadsdchi}
\end{eqnarray}
and
\begin{eqnarray}
{d^2{\cal B} \over d\cos\theta_{\rho}\ ds} &=&  \tau_B {G_{F}^2\ s\
\sqrt{\lambda}\over 128 m_{B}^3 \pi^3} |V_{ub}|^2 ({\it{{\cal B}(\rho
\rightarrow \pi^+ \pi^-)}})\nn \\ 
&&  \Big \{ 2\ cos^2\theta_{\rho}\ |H_0(s)|^2 +\   sin^2\theta_{\rho}\
 \Big( |H_{+}(s)|^2 + |H_{-}(s)|^2\Big) \Big\}. 
\label{dgamadsdcosthetav}
\end{eqnarray}
\begin{figure}[t]
\psfrag{c}{\hskip 0.3cm $s\ (\textrm{GeV}^2)$}
\psfrag{b}{$ \textrm{cos}\,\theta_{+}$}
\psfrag{a}{\hskip -3cm \Large {${d^2{\cal B} \over ds\ d\cos\theta_{+} |V_{ub}|^2 }$}}
\begin{center}
\includegraphics[width=12cm,height=9cm]{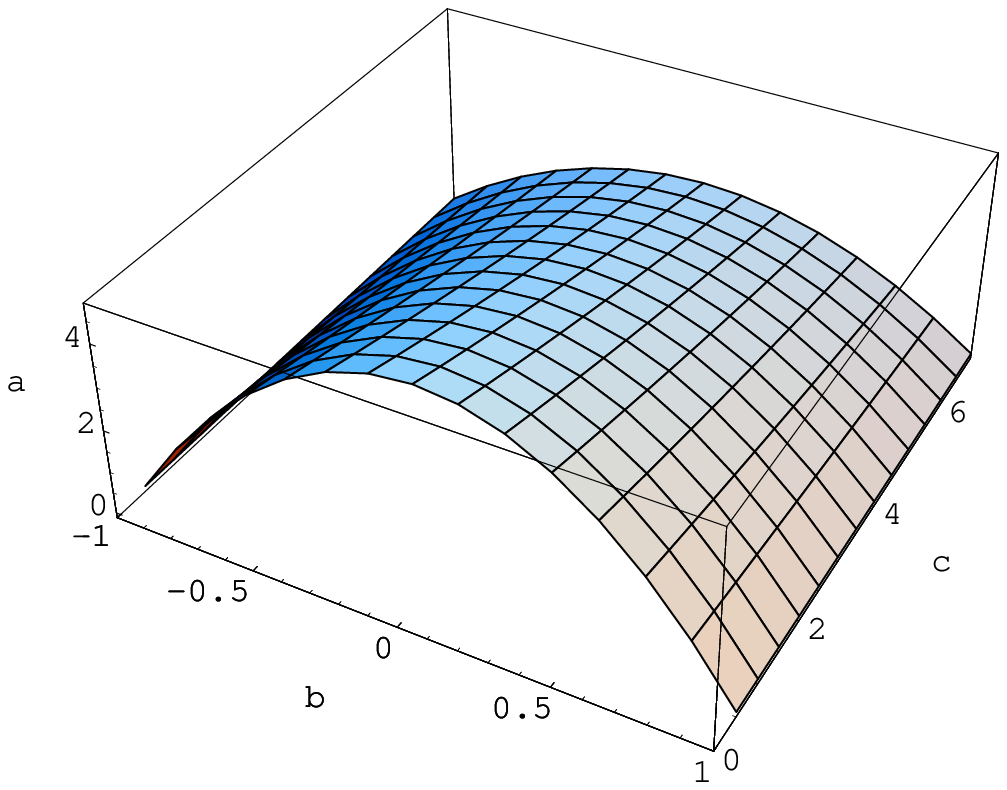}
\caption{\it Dalitz distribution {\Large${d^2{\cal B}\over
\d\cos\theta_{+}\ ds}$} for $ B \rightarrow  \rho \ell \nu_{\ell}$~\cite{Ali:2002qc}.}
\label{dalitzdsdcostheta+}
\hskip 0.5 cm 
\end{center}
\end{figure}
Similarly, We give here the $\theta_{+}$ double angular distribution
as following:
\begin{eqnarray}
{d^2{\cal B}\over d\cos\theta_{+}\ ds} &=& \tau_B {G_{F}^2\ s\ \sqrt{\lambda}
\over 256 m_{B}^3 \pi^3} |V_{ub}|^2 ({\it{{\cal B}(\rho \rightarrow \pi^+ \pi^-)}})\nn \\ 
&& \Big \{2\  sin^2\theta_{+}\  |H_0(s)|^2+(1 -  \ cos\theta_{+})^2\
 |H_{+}(s)|^2 + (1 + cos\theta_{+})^2\ |H_{-}(s)|^2 \Big\}\nn \\
{} & = &{d^2{\cal B}_{|H_{0}|^2} \over d\cos\theta_{+}\ ds}+{d^2{\cal
B}_{|H_{+}|^2} \over d\cos\theta_{+}\ ds}+{d^2{\cal B}_{|H_{-}|^2} \over
d\cos\theta_{+}\ ds},
\label{dgamadsdcos+}
\end{eqnarray}
where the corresponding partial double angular distributions read as:
\begin{eqnarray}
{d^2{\cal B}_{|H_{+}|^2} \over d\cos\theta_{+}\ ds} &=&  \tau_B {G_{F}^2\ s\
\sqrt{\lambda} \over 256 m_{B}^3 \pi^3} |V_{ub}|^2
 ({\it{{\cal B}(\rho \rightarrow \pi^+ \pi^-)}})\Big
 \{(1 - cos\theta_{+})^2\ |H_{+}(s)|^2 \Big\}\nn \\
{d^2{\cal B}_{|H_{-}|^2} \over d\cos\theta_{+}\ ds} &=&  \tau_B {G_{F}^2\ s\
\sqrt{\lambda} \over 256 m_{B}^3 \pi^3} |V_{ub}|^2
 ({\it{{\cal B}(\rho \rightarrow \pi^+ \pi^-)}})\Big
 \{(1 + cos\theta_{+})^2\ |H_{-}(s)|^2 \Big\}\nn \\
{d^2{\cal B}_{|H_{0}|^2} \over d\cos\theta_{+}\ ds} &=&  \tau_B {G_{F}^2\ s\
\sqrt{\lambda} \over 256 m_{B}^3 \pi^3} |V_{ub}|^2 ({\it{{\cal B}(\rho \rightarrow \pi^+ \pi^-)}}) \Big \{2\  sin^2\theta_{+}\  |H_0(s)|^2\Big\}.
\end{eqnarray}
%
\begin{figure}[t]
\psfrag{c}{\hskip 0.3cm $s\ (\textrm{GeV}^2)$}
\begin{center}
\psfrag{a}{\hskip -3cm \Large {${d^2{\cal B}\over ds\ d\cos\theta_{\rho}|V_{ub}|^2}$}}
\psfrag{b}{$ \textrm{cos}\,\theta_{\rho}$}
\includegraphics[width=12cm,height=9cm]{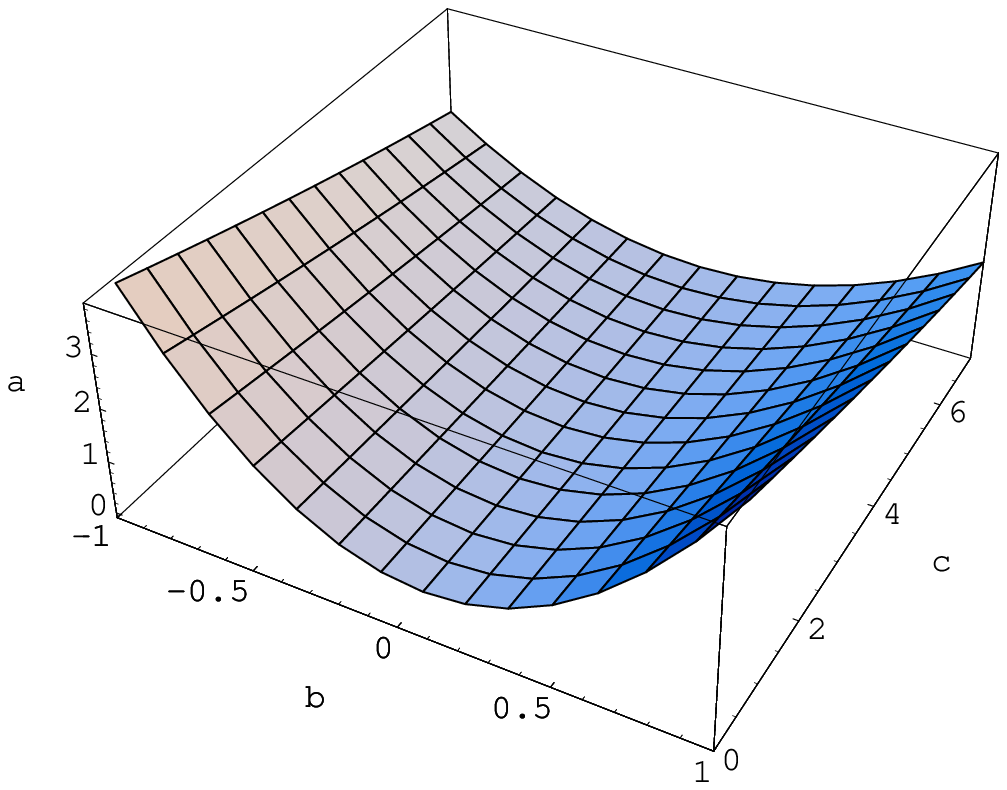}
\caption{\it Dalitz distribution   {\Large${d^2{\cal B} \over \d\cos\theta_{\rho}\ ds}$} for $ B \rightarrow  \rho \ell \nu_{\ell}$~\cite{Ali:2002qc}.}
\label{dalitzdsdcosthetav}
\hskip 0.5 cm 
\end{center}
\end{figure}

\begin{figure}[H]\vspace{1.cm}
\psfrag{c}{\hskip 0.3cm $s\ (\textrm{GeV}^2)$}
\begin{center}
\psfrag{b}{$ \textrm{cos}\,\theta_{\rho}$}
\psfrag{a}{\hskip -3cm \Large{${d^2{\cal B}_{|H_{-}^2|}\over ds\ d\cos\theta_{+}|V_{ub}|^2} $}}
\includegraphics[width=12cm,height=9cm]{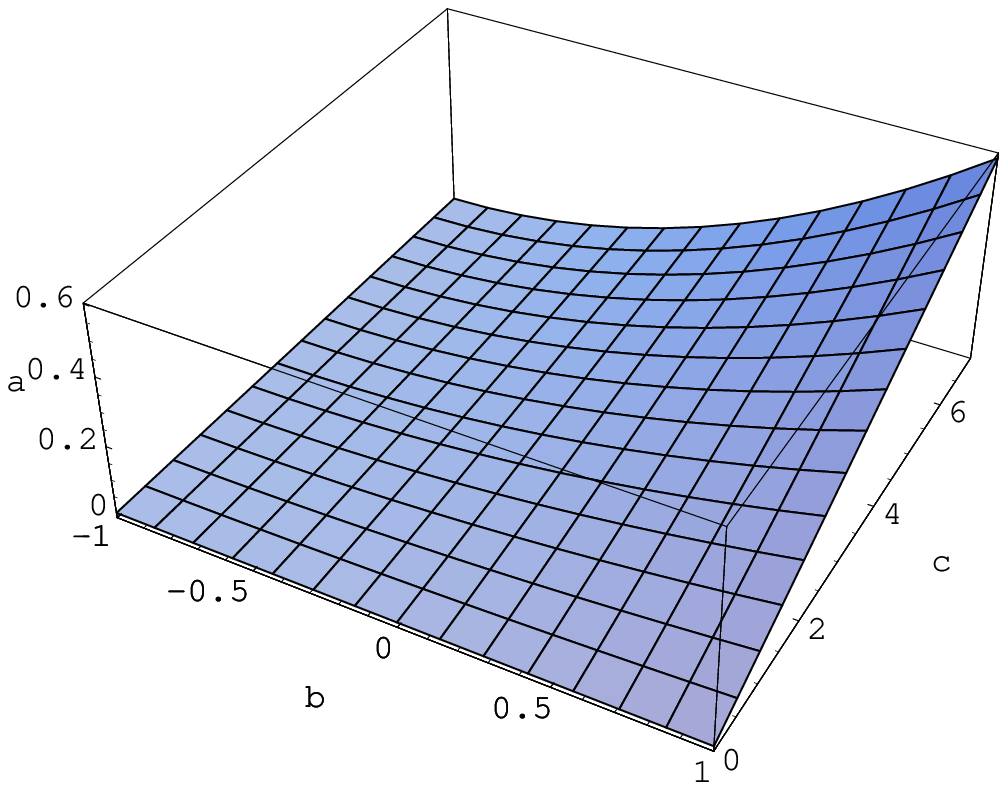}
\caption{\it Partial Dalitz distribution {\Large ${d^2{\cal B}_{|H_{-}|^2}\over
d\cos\theta_{+}\ ds}$} for $ B \rightarrow  \rho \ell
\nu_{\ell}$~\cite{Ali:2002qc}.} 
\label{dalitzdsdcosthetaH-}
\vspace{2cm}
\psfrag{a}{\hskip -3cm \Large {${d^2{\cal B}_{|H_{0}^2|}\over ds\ d\cos\theta_{+}|V_{ub}|^2}$}}
\includegraphics[width=12cm,height=9cm]{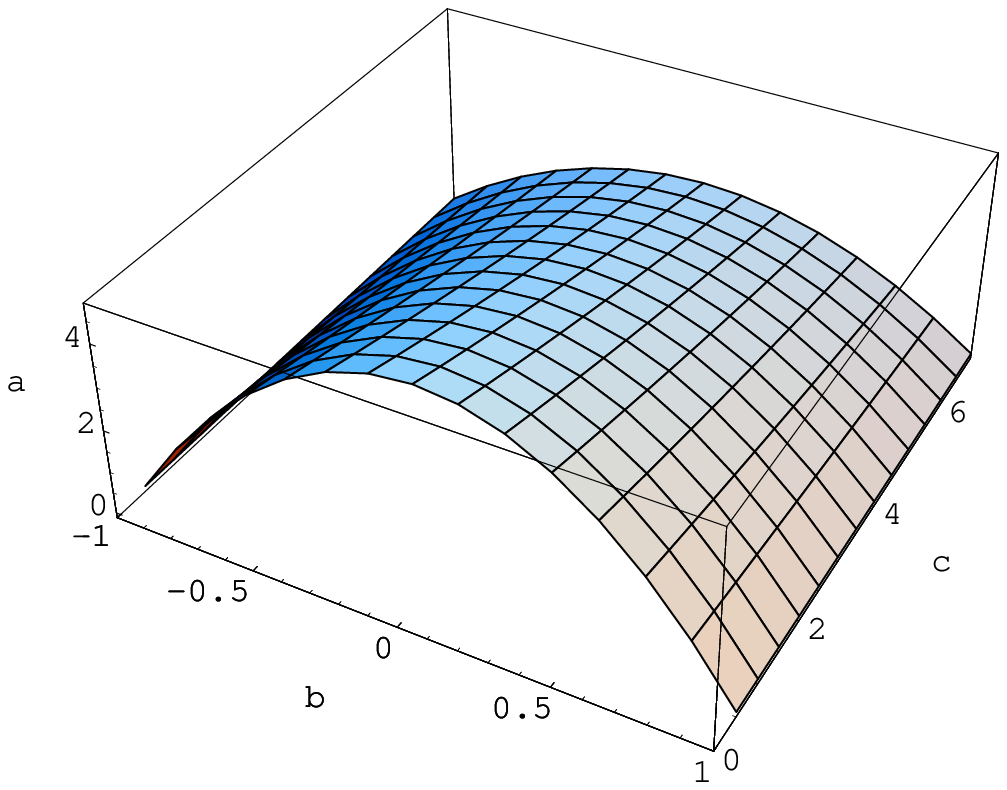}
\caption{\it Partial Dalitz distribution {\Large${d^2{\cal B}_{|H_{0}|^2} \over
d\cos\theta_{+}\ ds}$} for $ B \rightarrow  \rho \ell
\nu_{\ell}$~\cite{Ali:2002qc}.} 
\label{dalitzdsdcosthetaH0}
\end{center}
\end{figure}

Implementing the $O(\alpha_s)$-improvements in the various helicity
amplitudes above, we have shown respectively in
Figs.~(\ref{dalitzdsdcosthetav}) and (\ref{dalitzdsdcostheta+}) the
explicit behavior of the $\theta_{\rho}$ and $\theta_{+}$ angular
distributions. Wheras in Figs.~(\ref{dalitzdsdcosthetaH-}) and
(\ref{dalitzdsdcosthetaH0}) we have presented the $\theta_{+}$ partial angular
distributions: ($d^2{\cal B}_{|H_{-}|^2}/ d\cos\theta_{+}\ ds$) and
($d^2{\cal B}_{|H_{0}|^2}/ d\cos\theta_{+}\ ds$).
%
\setcounter{footnote}{0}
To project out experimentally the various helicity components namely
$|H_{0}|^2$ and $|H_{-}|^2$,  one can use the $\theta_{\rho}$ and
$\theta_{+}$ Dalitz distributions respectively. Extracting the
$|H_{-}|^2$ ones, which is the more easiest one$^\star$\footnote{$^\star$ this is due to the fact that $|H_{-}|\approx \xi^{(\rho)}_{\perp}$, see
Eq.~(\ref{H-}).}, requires a precise measurement at $\theta_{+}=0$. This
means that the charged lepton should be back to back to the recoiling
$\rho$-meson. However measuring $|H_{0}|^2$ imply a higher efficiency at
$\theta_{\rho}=0$ or $\pi$, by means that the $\pi$-meson should be
measured in the same axe as the recoiling $\rho$-meson.
%
%
        \subsection{ Dilepton mass spectrum}
Finally, integrating out the polar angle $\theta_+$, $\theta_{\rho}$
and $\phi$ from Eq.~(\ref{diffrho}), we obtain the total branching decay
rate for the $B \rightarrow  \rho \ell \nu_{\ell}$ transition: 

\begin{eqnarray}
{d{\cal B} \over ds} &=& \tau_B {G_F^2\ s\ \sqrt{\lambda} \over 96 m_{B}^3 \pi^4} |V_{ub}|^2 ({\it{{\cal B}(\rho \rightarrow \pi^+ \pi^-)}}) \Big\{|H_0(s)|^2 + |H_{+}(s)|^2 + |H_{-}(s)|^2 \Big\} \nn\\ 
{} &=& {d{\cal B}_{|H_{0}|^2} \over ds}+{d{\cal B}_{|H_{+}|^2} \over
ds}+{d{\cal B}_{|H_{-}|^2} \over ds}.
\label{dgamads}
\end{eqnarray}

The partial distribution amplitudes defined in Eq.~(\ref{dgamads}), read as follow:

\begin{eqnarray}
{d{\cal B}_{|H_{-}|^2} \over ds} &=& \tau_B {G_F^2\ s\ \sqrt{\lambda} \over 96 m_{B}^3 \pi^4} |V_{ub}|^2 ({\it{{\cal B}(\rho \rightarrow \pi^+ \pi^-)}}) \Big\{|H_{-}(s)|^2 \Big\},\nn\\ 
{d{\cal B}_{|H_{+}|^2} \over ds} &=& \tau_B {G_F^2\ s\ \sqrt{\lambda} \over 96 m_{B}^3 \pi^4} |V_{ub}|^2 ({\it{{\cal B}(\rho \rightarrow \pi^+ \pi^-)}}) \Big\{|H_{+}(s)|^2 \Big\},\nn\\  
{d{\cal B}_{|H_{0}|^2} \over ds} &=& \tau_B {G_F^2\ s\ \sqrt{\lambda}
\over 96 m_{B}^3 \pi^4} |V_{ub}|^2 ({\it{{\cal B}(\rho \rightarrow
\pi^+ \pi^-)}}) \Big\{|H_{0}(s)|^2 \Big\}.
\label{dgamadspartial}
\end{eqnarray}

\begin{figure}[H]\vspace{-0.5cm}
\psfrag{b}{\hskip -3.cm \Large{${d{\cal B}_{|H_{-}|^2}\over ds\ |V_{ub}|^2 }$}}
\psfrag{c}{}
\psfrag{a}{}
\begin{center}
\includegraphics[width=10cm,height=7.5cm]{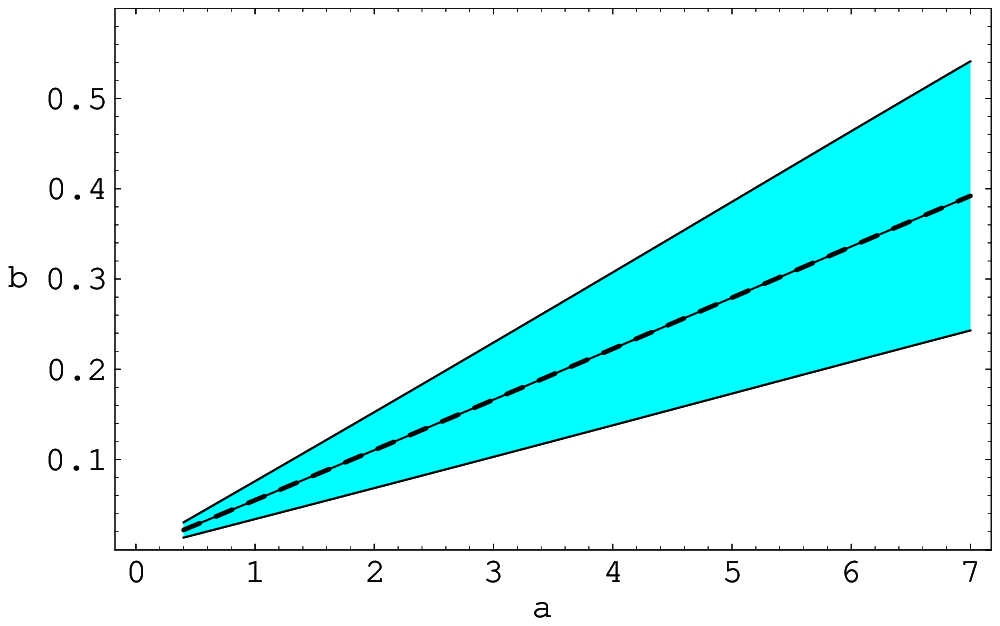}
\\
\psfrag{b}{\hskip -3.cm \Large{${d{\cal B}_{|H_{+}|^2}\over ds\ |V_{ub}|^2 10^{-4} }$}}
\includegraphics[width=10cm,height=7.5cm]{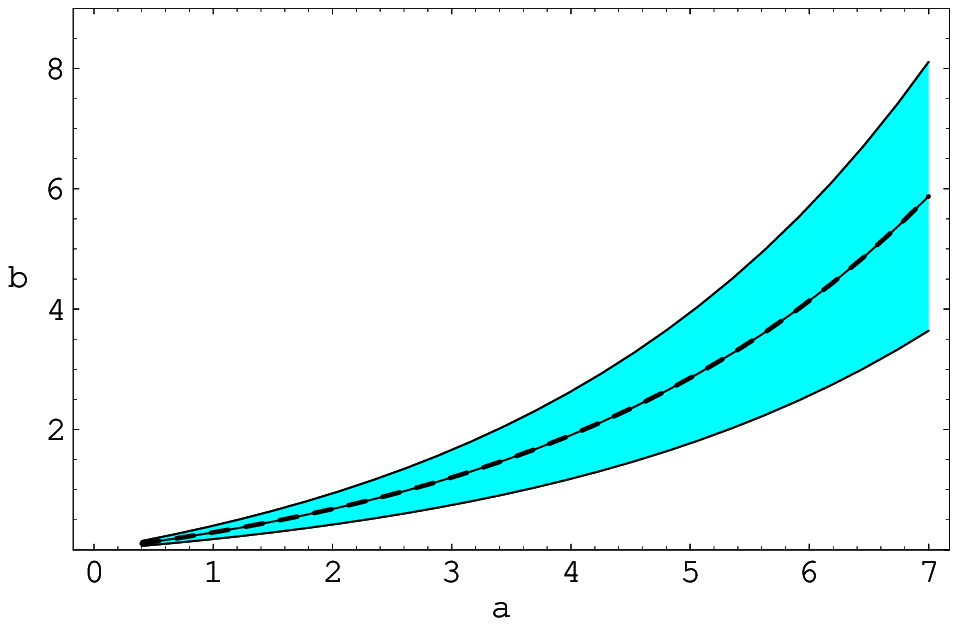}
\\
\psfrag{a}{\hskip 0.cm $\mathrm{s\ (GeV^2)}$}
\psfrag{b}{\hskip -3.cm \Large{${d{\cal B}_{|H_{0}|^2}\over ds\
|V_{ub}|^2 }$}}
\includegraphics[width=10cm,height=7.5cm]{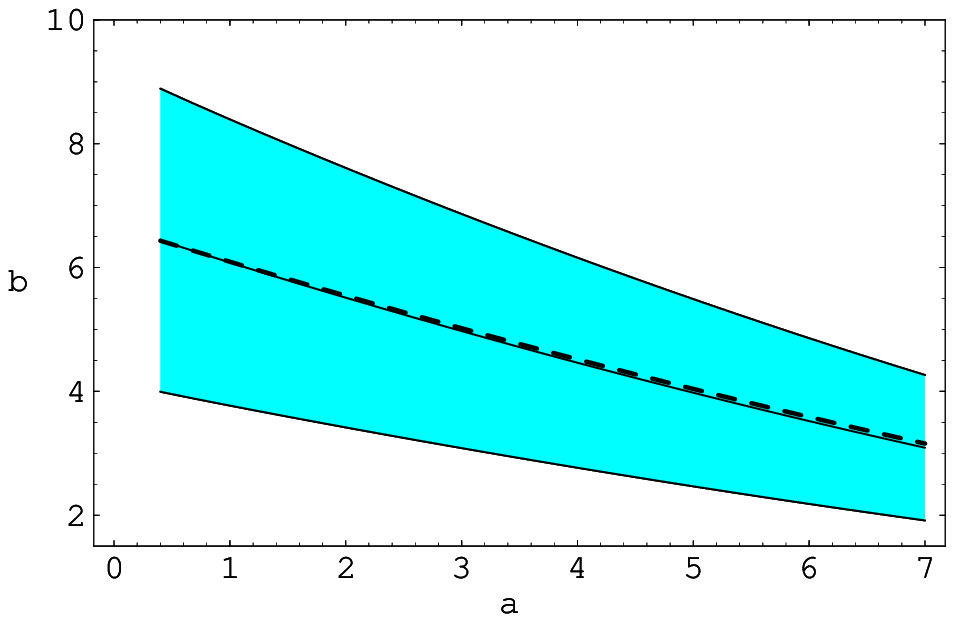}
\caption{\it The dilepton invariant mass distributions
{\large${d{\cal B}_{|H_{-}|^2}\over ds}$} (upper-plot),
{\large${d{\cal B}_{|H_{+}|^2}\over ds}$}(middle-plot) and {\large${d{\cal
B}_{|H_{0}|^2}\over ds}$} (lower-plot) for $B \rightarrow \rho \ell
\nu_{\ell}$ at NLO (solid center line) and LO (dashed).The band reflects the theoretical uncertainties from
input parameters~\cite{Ali:2002qc,Safir:SUSY02}.} 
\label{dBHrho}
\end{center}
\end{figure}

\begin{figure}[t]\vspace{-0.5cm}
\psfrag{a}{$s\ (GeV^2)$}
\psfrag{b}{\hskip -2.5cm \Large{${d{\cal B}\over ds\ |V_{ub}|^2 }$}}
\psfrag{c}{}
\begin{center}
\includegraphics[width=14cm,height=10cm]{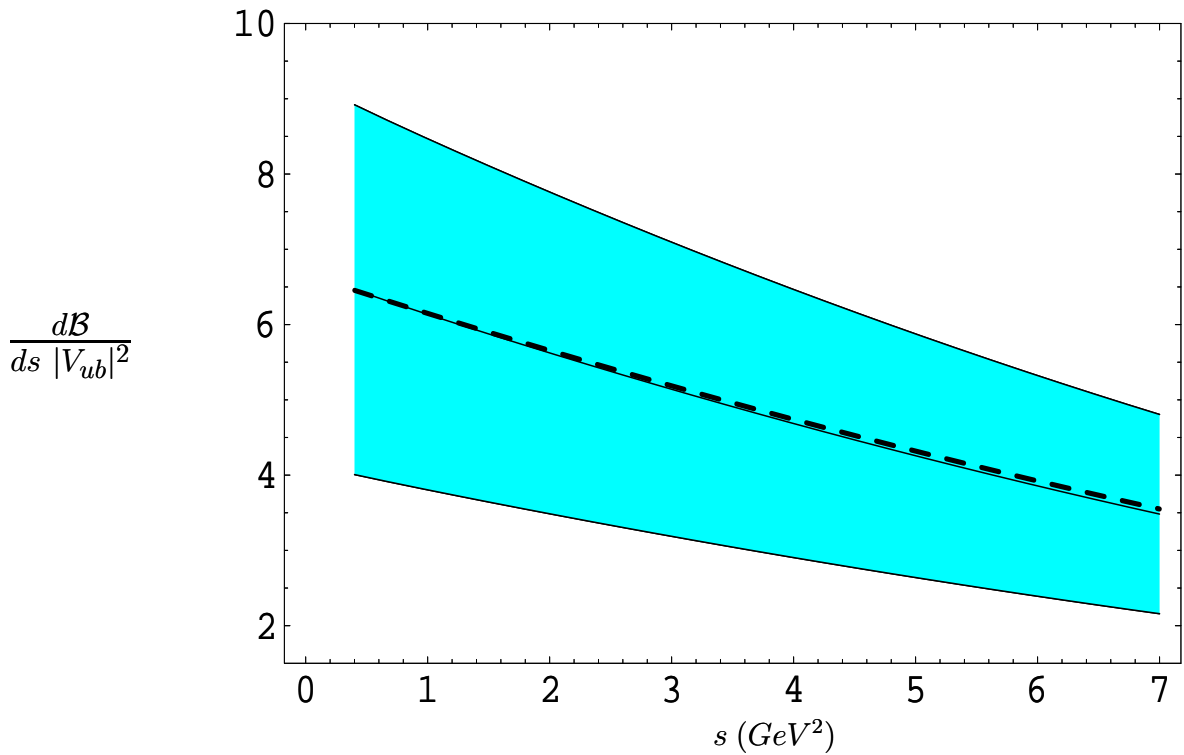}
\caption{\it The total dilepton invariant mass distribution for $ B \rightarrow
\rho \ell \nu_{\ell}$ at NLO (solid center line) and LO
(dashed). The band reflects theoretical uncertainties from 
input parameters~\cite{Ali:2002qc}.}
\label{dBrrho}
\end{center}
\end{figure}

In Figs.~(\ref{dBHrho}) and (\ref{dBrrho}), we have plotted
respectively the various partial dilepton invariant mass distributions
and the total  one.

The observed helicity $|H_{+}|$ component$^\circ$\footnote{$^\circ$ 
Note that from Eq.~(\ref{dgamadspartial}) ${d{\cal B}_{|H_{+}|^2}
\over ds} \approx  |H_{+}(s)|^2$.}, in Fig.~(\ref{dBHrho}-middle 
plot), is completeley negligeable comparing to the two others
helicities. This result is just a direct consequence of the $(V-A)$
coupling. The semileptonic process$^\ast$\footnote{$^\ast$ the same
argument holds for 
$b\to c~ \ell^- \bar{\nu}$ process.} $b\to u \ell^- \bar{\nu}$
produce a $u$ quark which is predominately helicity $\lambda=-1/2$. In
a $B \rightarrow X_{\{u\bar{q}\}} \ell \bar{\nu}$ process, the
helicity of the $X$ meson is then determined by whether the $u$ quark
combines with a spectator quark that has helicity $\lambda=+1/2$ or
$\lambda=-1/2$. If $X$ is a spin-zero meson, only $\lambda=+1/2$ 
spectator quark contributes. However, if $X$ has spin 1, both
helicities of the spectator quark contribute, leading to $X$
helicities of $\lambda=0$ and $\lambda=-1$, but not $\lambda=+1$. 

Consequently,
the $B \rightarrow \rho~ \ell^- \bar{\nu}_{\ell}$ transition should be completely
dominated by the two helicity components $|H_{0}|$ and $|H_{-}|$, in a
good agreement with what we have observed in Fig.(\ref{dBHrho}
upper-plot and lower-one). Contrary to the $B \rightarrow K^* \ell^+
\ell^- $ decay rate, the $B \rightarrow \rho \ell \nu_{\ell}$ one is
totally dominated by the 
helicity $|H_{0}|$ component, as it is shown Figs.~(\ref{dBHrho}-lower
plot) and (\ref{dBrrho}). 

Note that the impact of the NLO correction on the
total branching ratio and the partial ones are less significant as for
the $B \rightarrow K^*  \ell^+  \ell^- $ decay, and these is simply
due to the absent of the penguin form factor corrections in the $B
\rightarrow \rho \ell \nu_{\ell}$ decay. However the large systematic
error of our prediction comes from the uncertainty in the form factors with
their current large uncertainty and to the SU(3)-breaking
effects. There is no doubt that a precise measurement of the
long-distance effect, will reduce considerably our uncertainty on $B
\rightarrow \rho \ell \nu_{\ell}$ decay. 
%
\section{Phenomenological Discussion on $R_b$ 
\label{chap:Vub}}
In the Standard Model, the charged current weak interactions of three 
generations of quarks are governed by a Lagrangian which contains a 
transformation from the mass eigenbasis to the flavour (generation) 
eigenbasis~\cite{PDG2000,bib:BABPHYS,bib:BURAS2}.
This flavor-mixing is expressed as a 3$\times$3 complex matrix
$V_{\mbox{\footnotesize CKM}}$ known as the Cabibbo-Kobayashi-Maskawa
(CKM) matrix \cite{CKM} \big(see Eq.~(\ref{eq:ckm})\big).

The Unitarity of this  matrix reduces the number of independent
parameters to nine, which can be chosen as three real mixing angles
and six imaginary phases. Five of the phases are removable. The four
remaining parameters are fundamental constants of nature, to be
determined by experiment since the SM itself gives no guidance as to their
values. Therefeore, an important target of particle physics is
the determination of the CKM matrix \cite{CKM}. 

\begin{figure}[htp]
\vspace{0.10in}
\centerline{
\epsfysize=5.0truecm
\epsffile{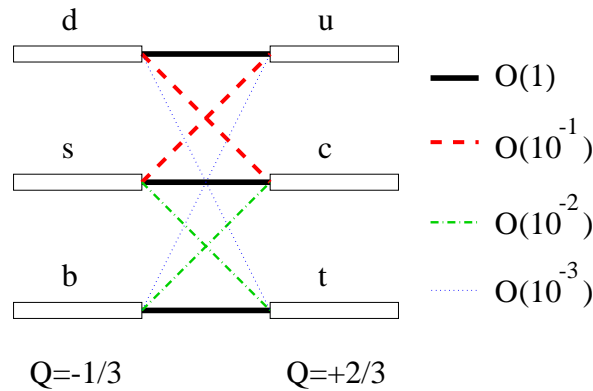}}
\caption[]{\it 
Hierarchy of the quark transitions mediated through charged 
currents~\cite{Fleischer:2002ys}.}\label{fig:term}
\end{figure}

\noi Fig.~(\ref{fig:term}) illustrates the hierarchy of the
strengths of the quark transitions mediated through charged-current
interactions$^\diamond$\footnote{$^\diamond$ Transitions within the
same generation are governed by CKM  
elements of ${\cal O}(1)$, those between the first and the second 
generation are suppressed by CKM factors of ${\cal O}(10^{-1})$, those 
between the second and the third generation are suppressed by 
${\cal O}(10^{-2})$, and the transitions between the first and the third 
generation are even suppressed by CKM factors of ${\cal
O}(10^{-3})$.}, and Table~(\ref{tab:Ros1}) their present
experimental source of information~\cite{Rosner:Scottland}.
It it known that the only CP violation source in the SM is supposed to
arise from a single phase in the CKM matrix. This is a very remarkable
property of the Kobayashi-Maskawa picture of CP violation: quark
mixing and CP violation are closely related to each other. This
property is often used to determine the angles of the unitarity
triangle without the study of CP violating quantities. 
\vspace*{-.5cm}
\begin{figure}[H]
\begin{center}
\psfrag{a}{ $\alpha$}
\psfrag{b}{ $\beta$}
\psfrag{g}{ $\gamma$}
\psfrag{Rb}{ $R_u$}
\psfrag{Rt}{ $R_t$}
\epsfig{file=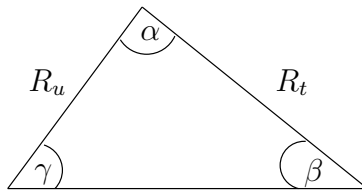,width=0.3\linewidth}
\caption{ \it Unitarity triangle.}
\label{fig:UT}
\end{center}
\end{figure}
The measurement of $R_b=|V_{ub}/V_{cb}|$ $(=|V_{ub}/V_{tb}V_{ts}^*|)$
constraints the length $R_u$ of the unitarity triangle \big(see
Fig~.(\ref{fig:UT})\big) through the
relation~\cite{bib:BABPHYS}:
\begin{eqnarray}
R_u=\Big|{V_{ub}^* V_{ud}\over V_{cb}^* V_{cd}}\Big|.
\label{eq:Ru}
\end{eqnarray}
While the two elements $V_{ud}$ and $V_{cd}$ are known with high
accuracy~\cite{PDG2000,bib:BABPHYS}, the two left in $R_u$, namely
$|V_{ub}|$ and $|V_{cb}|$, are under extensive discussion at present,
especially $|V_{ub}|$.
Their values are measured mainly in
semileptonic $B$-decays using two independent methods. An endpoint
analysis in inclusive semileptonic $B$-decays yields a direct 
determination of $|V_{ub}/V_{cb}|$~\cite{Athanas:1994fk}, while
measurements of branching fractions of exclusive final states such as
$B\to (\pi, \rho) \ell \nu_{\ell}$ measure
$|V_{ub}|$~\cite{Alexander:1996qu}. The model-dependence in either
method is quite substantial. Since we have analyzed the exclusive
decays $B\to K^* \ell^+\ell^-$ and $B\to \rho \ell \nu_{\ell}$, we
will propose a model-independent analysis of this ratio. Before doing
that, let's have some inside about their experimental status. 
\begin{table}
\begin{center}
\begin{tabular}{|l| l| l|} \hline \hline
Relative amplitude & Transition & Source of information \\ \hline
$\sim$ 1 & $u \leftrightarrow d$ & Nuclear $\beta$-decay \\
$\sim$ 1 & $c \leftrightarrow s$ & Charmed particle decays \\
$\sim 0.22$ & $u \leftrightarrow s$ & Strange particle decays \\
$\sim 0.22$ & $c \leftrightarrow d$ & Neutrino charm prod. \\
$\sim 0.04$ & $c \leftrightarrow b$ & $b$ decays \\
$\sim 0.003-0.004$ & $u \leftrightarrow b$ & Charmless $b$ decays \\
$\sim$ 1 & $t \leftrightarrow b$ & Dominance of $t \to W b$ \\
$\sim 0.04$ & $t \leftrightarrow s$ & Only indirect evidence \\
$\sim 0.01$ & $t \leftrightarrow d$ & Only indirect evidence \\ 
\hline\hline
\end{tabular}
\caption{\it Relative strengths of charge-changing weak
transitions~\cite{Rosner:Scottland}.} 
\label{tab:Ros1}
\end{center}
\end{table}
\subsection{$R_b$ Phenomenology}
Exclusive semileptonic $b \rightarrow u \ell \nu_{\ell}$ decays
are an active area of experimental and theoretical 
study \citer{lkgprl,boyd}.
These rare processes can be used to extract the magnitude of 
$V_{ub}$, one of the smallest and least well
known elements of the Cabibbo-Kobayashi-Maskawa (CKM) 
quark-mixing matrix~\cite{CKM}.  
Because $\vert V_{ub} / V_{cb} \vert \approx 0.08$,
the branching fractions for exclusive $b \rightarrow u \ell \nu_{\ell}$
processes are small, of order $10^{-4}$, and they have
only recently become experimentally accessible.  

Extracting $\vert V_{ub} \vert$ 
from a measured decay rate
requires significant theoretical input because the matrix
elements for such processes involve complex strong-interaction
dynamics.  Although the underlying
$b\to u\ell \nu_{\ell}$ decay is a relatively simple weak 
process, it is difficult to calculate
the strong-interaction effects involved in the
transition from the heavy $B$ meson to
the light daughter meson.
Because of these theoretical uncertainties,
even a perfectly measured $B\to\rho\ell\nu_{\ell}$
branching fraction would not at present lead to a precise value
of $|V_{ub}|$. 

The dynamics in $B \to \rho \ell \nu_{\ell}$ decay are in
contrast with
$b\to c\ell \nu_{\ell}$ decays, such as $B \to D^* \ell \nu_{\ell}$,
where a heavy quark
is present both in initial and final states. In this case,
techniques based on HQET can be used to calculate the decay amplitude
with good precision, 
particularly for the kinematic configuration in which the 
charm hadron has zero recoil velocity.
The zero-recoil point in $B\to\rho\ell\nu_{\ell}$ cannot be
treated with similar techniques, however, because the daughter $u$ quark
is not heavy compared to the scale of hadronic energy transfers.
Nevertheless, substantial progress has
been made using a variety of theoretical methods, including
quark models~\citer{wsb,demchuk}, 
lattice QCD~\citer{elc,ukqcd}, 
QCD sum rules~\citer{narison,ruckl},
and models relating form factors measured in
$D \rightarrow K^* \ell \nu_{\ell}$ decay~\cite{e791}
to those in $B \to \rho \ell \nu_{\ell}$ decay. 

Experimentally, the main difficulty in
observing signals from $b\to u\ell \nu_{\ell}$ processes is the very large background 
due to $b\to c\ell \nu_{\ell}$. Because a significant fraction of $B \to \rho
\ell \nu_{\ell}$ events have lepton energy beyond the endpoint for 
$b\to c\ell \nu_{\ell}$ decay, lepton-energy requirements
provide a powerful tool for background
suppression. However, extrapolation of the decay rate measured
in this portion of phase space to the full rate again requires the
use of theoretical models, and it introduces model 
dependence beyond that associated with simply extracting
the value of $\vert V_{ub} \vert$ from the branching fraction. 

The BABAR collaboration has recently presented a preliminary
measurement of the CKM matrix elements $|V_{ub}|$ with the charmless
exclusive semileptonic $B^0 \to \rho^- e^+ \nu_e$ decay. Their result
is \cite{Aubert:2002ph}:
\begin{eqnarray}
{\cal B}(B^0\to \rho^-~e^+~\nu_e) &=& (3.39\pm 0.44\pm 0.52\pm0.60)
\times 10^{-4}\nn\\ 
|V_{ub}|&=&(3.69 \pm 0.23 \pm 0.27 ^{+0.40}_{-0.59}) \times 10^{-3},
\end{eqnarray}
where the quoted errors are statistical, systematic, and theoretical
respectively. To extract $|V_{ub}|$, they have used different
form-factor calculation. 

In order to reduce the large theoretical errors on the form-factor, it
is more convenient to study the 
distribution of $s$ is reflected in the $\rho$ momentum 
spectrum.  Eventually, studies of the $s$ distribution, as well as of the 
angular distributions of the decay products,
should reduce the model dependence on
$\vert V_{ub} \vert$ by constraining theoretical models
for the decay form factors. In the next subsection, we propose a
model-independent analysis of the ratio $R_b$ using the $s$
Helicity distribution.  
\subsection{Model-independent analysis of $R_b$
\label{sec:Vub}}
To reduce the non-perturbative uncertainty in the extraction of
$|V_{ub}|$, we propose to study the ratios of the differential decay
rates in $B \to \rho \ell \nu_\ell$ and $B \to K^* \ell^+ \ell^-$
involving definite helicity states. These $s$-dependent ratios
$R_{i}(s)$, $(i=0,-1,+1)$ are defined as follows:  
\begin{eqnarray}
R_{i}(s) = {{d\Gamma_{H_i}^{B \rightarrow  K^{*} \ \ell^{+}  \ell^{-}}/ds}
\over{d\Gamma_{H_i}^{B \rightarrow  \rho \ \ell \nu_{\ell}}/ds}}.
\label{Ratio}  
\end{eqnarray}

From Eqs.~(\ref{dBrdl2}) and (\ref{dgamads}), one obtains straightforwardly:
\begin{eqnarray}
R_{i}(s) = { \alpha_{em}^2 m_{b}^2 \over 4 \pi s} {1 \over R_b^2 }
{|H_i^{(K^{*})}(s)|^2\over|H_i^{(\rho)}(s)|^2},  \label{Ratio1} 
\end{eqnarray}

\noi where $R_b = \vert V_{ub}\vert/\vert V_{tb} V_{ts}^* \vert$ and
the term $H_i^{(K^{*})}(s)$ \big($H_i^{(\rho)}(s)$\big) should be understood as
the helicty amplitudes for the decay $B \rightarrow \
K^{*}(\rightarrow \ K + \pi) \ \l^{+} \ \l^{-}$ \big($B\rightarrow \rho
(\rightarrow \pi^{+} \pi^{-})\ \l \nu_\l$\big), defined in section~3.3
(3.4).  

The ratio $R_{-}(s)$ suggests itself as the most interesting one, as the
form factor dependence essentially cancels. It is easy to see this
cancellation especially in the LEET approach as the function
$R_{-}(s)$ scales in term of form factor as:
\begin{eqnarray}
R_{-}(s) \propto  {1 \over R_b^2}\, 
\Big({\xi^{(K^*)}_{\perp}(s)\over \xi^{(\rho)}_{\perp}(s)}\Big)^2\,
{\cal F}(s,m_{K^*}, m_{\rho}, ...),  
\label{R-} 
\end{eqnarray}
\begin{figure}[t]
\begin{center}
\includegraphics[width=12cm,height=9cm]{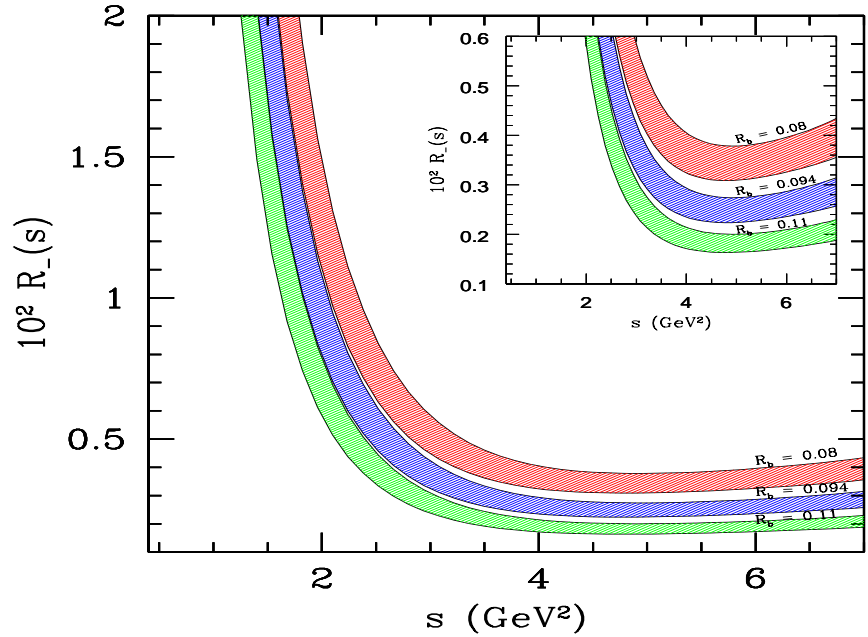}
\caption{ \it
The Ratio $R_{-}(s)$ with three indicated values of the CKM ratio $R_{b} \equiv
|V_{ub}|/|V_{tb}V_{ts}^*|$. The bands reflect the theoretical
uncertainty from $\zeta_{SU(3)}=1.3 \pm 0.06$ and
$\xi^{(K^*)}_{{\perp}}(0)=0.28\pm 0.04$~\cite{Ali:2002qc}.} 
\label{R-Vub}
\end{center}
\end{figure}
\begin{figure}[t]
\begin{center}
\includegraphics[width=12cm,height=9cm]{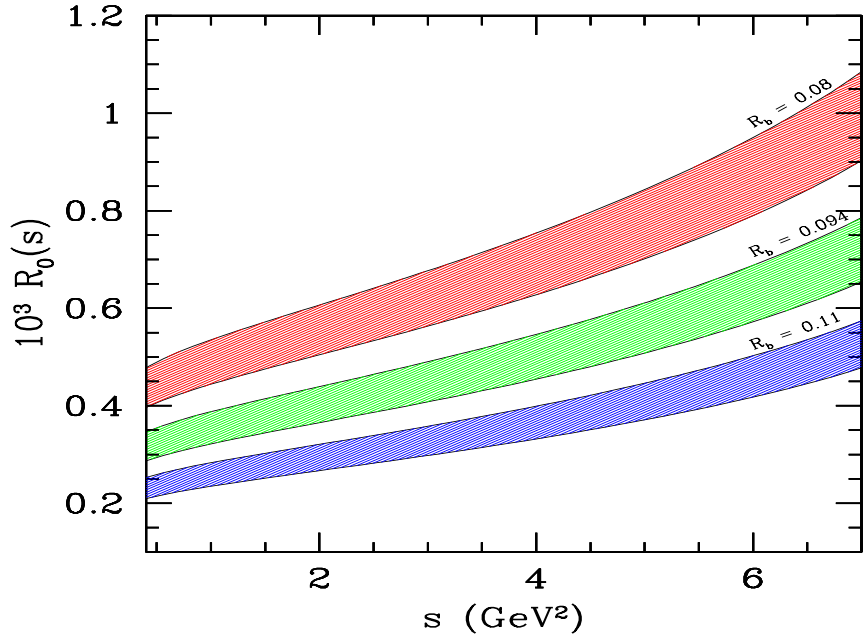}
\caption{ \it
The Ratio $R_{0}(s)$ with three indicated values of the CKM ratio $R_{b} \equiv
|V_{ub}|/|V_{tb}V_{ts}^*|$. The bands reflect the theoretical
uncertainty from $\zeta_{SU(3)}=1.3 \pm 0.06$ and
$\xi^{(K^*)}_{{\perp}}(0)=0.28\pm 0.04$~\cite{Ali:2002qc}.} 
\label{R0Vub}
\end{center}
\end{figure}

\noi where the function ${\cal F}(s,m_{K^*}, m_{\rho}, ...)$ denotes
the kinematical contribution extracted from the ratio
$|H_i^{(K^{*})}(s)|^2/ |H_i^{(\rho)}(s)|^2$. Since the helicty
amplitudes $H_{-}^{(K^{*})}(s)$ and $H_{-}^{(\rho)}(s)$ depend just on one 
universal form factor, respectively $\xi^{(K^*)}_{\perp}$ and
$\xi^{(\rho)}_{\perp}$, one can easily factorize {\it apriori} the ratio
$R_{-}(s)$ as a kinematic contribution \big(expressed in ${\cal
F}(s,m_{K^*}, m_{\rho}, ...)$\big) and a dynamical contribution 
\big(encoded in the ratio of $\xi^{(K^*)}_{\perp}/\xi^{(\rho)}_{\perp}$\big).

There is no doubt that in the SU(3) symmetry limit, the function
$R_{-}(s)$ will be defined as a kinematical function and thus no
uncertainty from the non-perturbative regime. Unhappily, the reality
is far from being that, and one has to incorporate the SU(3)-breaking
effects. Then the only source of uncertainty coming from the long
distance contribution, will be translated in the ratio
$\zeta_{SU(3)}=\xi^{(K^*)}_{\perp}/\xi^{(\rho)}_{\perp}$, which turns
out to be $\zeta_{SU(3)}= 1.3 \pm 0.06$, already defined in
section~3.4.

On the other hand, the ratio $R_{0}(s)$ could bring a certain hint on
the the structure 
of $R_b$. However its dependence on the form factors is more involved and
cann't be fudged away, as one can see:
\begin{eqnarray}
R_{0}(s) \propto  {1 \over R_b^2}\,
{|\xi^{(K^*)}_{\perp}(s)\,\, {\cal F}_1^{(K^*)}(s,m_{K^*}, ...) +
\xi^{(K^*)}_{||}(s)\,\, {\cal F}_2^{(K^*)}(s,m_{K^*}, ...)|^2\over 
|\xi^{(\rho)}_{\perp}(s)\,\, {\cal F}_1^{(\rho)}(s,m_{\rho}, ...) +
\xi^{(\rho)}_{||}(s)\,\, {\cal F}_2^{(\rho)}(s,m_{\rho}, ...)|^2},
\label{R0} 
\end{eqnarray}

\noi where ${\cal F}_{1,2}^{(K^*)}(s,m_{K^*}, ...)$ \big(${\cal
F}_{1,2}^{(\rho)}(s,m_{K^*}, ...)$\big) are a certain dynamical
function$^\wr$\footnote{$^\wr$ To extract the functions ${\cal F}_{1,2}^{(K^*)}$ (${\cal F}_{1,2}^{(\rho)}$) , one has just to write
$|H_0^{K^{*}}(s)|^2$ ($|H_0^{\rho}(s)|^2$), defined in
Eqs.~(\ref{H2def}) (\ref{H0}), as (\ref{R0}) and then identify them.} 
obtained from $|H_0^{K^{*}}(s)|^2$ \big($|H_0^{\rho}(s)|^2$\big).
Thus, we see that the ratio $R_{0}(s)$ is less attractive
than the $R_{-}(s)$ ones for the extraction of the CKM ratio $R_b$, since its
uncertainty is more poluated with the long distance contribution.
%

In Figs.~(\ref{R-Vub}) and (\ref{R0Vub})  we plot $R_{-}(s)$ and
$R_{0}(s)$ respectively,  for three representative
values of the CKM ratio $R_b = 0.08$, $0.094$, and
$0.11$. However, as we noticed earlier, the ratio $R_{-}(s)$ may be
statistically limited due to the dominance of the decay $B \to \rho
\ell \nu_\ell$ by the Helicity-$0$ component. For the LEET form factors used here,
the compounded theoretical uncertainty is shown by the shaded regions.  
This figure suggests that high statistics experiments may be able to
determine the CKM-ratio from measuring $R_{0}(s)$ at a competitive level
compared to the other methods {\it en vogue} in experimental studies.
\section{Summary and Outlook}
In this Chapter we have investigated an $O(\alpha_s)$-improved
analysis of the various helicity components in the decays $B \to K^*
\ell^+ \ell^-$ and $B \to \rho \ell \nu_\ell$, carried out in the
context of the Large-Energy-Effective-Theory. Using that and borrowing
the corresponding form factors from QCD sum rule, we have
investigated the corresponding distributions, decay rates and
Forward-bckward asymmetry\footnote{the FB asymmetry is investigated
just for the $B \to K^* \ell^+ \ell^-$ decay.}.

In the first part, we have concentrated mainly on the study of double
and single angular distributions, and the FB asymmetry.
Our findings can be summarized as follows \cite{Ali:2002qc}:
\begin{itemize}  
\item We have calculated the helicity components implementing the
$O(\alpha_s)$ corrections and shown that the $+1$-helicity component is
completely negligeable compared to the two other components, namely
$H_{0}$ and $H_{-}$.
\item The total dilepton invariant mass distribution {\large${d{\cal
B}\over ds}$} is dominated by the partial single distribution 
{\large${d{\cal B}_{|H_{-}|^2}\over ds}$}. The nex-to-leading order
correction to the total dilepton invariant mass distribution is
significant in the low dilepton mass region $(s\leq
2(\textrm{GeV}^2))$.
\item We have shown that the $O(\alpha_s)$ effects on the forward-backward
asymmetry shifts the predicted location of its zero by $\sim 1~
\textrm{GeV}^2$, confirming essentially the earlier work of {\it Beneke,
Feldmann and Seidel}~\cite{Beneke:2001at}.
\item We have carried out the $B \to K^* \ell^+ \ell^-$ decay analysis
in the so-called transversity basis. We have compared the LEET-based
amplitudes in this basis with the data currently available on $B \to
K^* J/\psi(\to \ell^+ \ell^-)$ and find that the short-distance based
transversity amplitudes are very similar to their long-distance counterparts.
\end{itemize}
In the same spirit we have studied the $B\to \rho \ell \nu_{\ell}$
decay in the second part of this chapter, using the helicity analysis
in the large energy effective theory . After presenting various double
and single angular distributions, we summarize \cite{Ali:2002qc}:
\begin{itemize}  
\item Considering the SU(3)-breaking effects, we have related the $B
\rightarrow \rho \ell \nu_{\ell}$ LEET-form factors, namely
$\xi^{(\rho)}_{\perp}(s)$ and $\xi^{(\rho)}_{||}(s)$, to the
corresponding form factors in $B \to K^* \ell^+ \ell^-$. Our numerical
estimates on $\xi^{(\rho)}_{\perp}(s)$ and $\xi^{(\rho)}_{||}(s)$ are
in agreement with the ones worked out for the full QCD form
factors in the QCD sum-rule approach in \cite{Ball:1998kk}.

\item Implementing the $O(\alpha_s)$ corrections to the $B\to \rho
\ell \nu_{\ell}$ helicity components, we have shown that the
$+1$-helicity component is completely negligeable compared to the two
other components, namely $H_{0}$ and $H_{-}$.
\item The total dilepton invariant mass distribution {\large${d{\cal
B}\over ds}$} is dominated by the partial single distribution 
{\large${d{\cal B}_{|H_{0}|^2}\over ds}$}. The nex-to-leading order
correction to the total dilepton invariant mass distribution is
completely negligeable.
\end{itemize}

Finally, combining the analysis of the decay modes $B \to K^* \ell^+
\ell^-$ and $B \to \rho \ell \nu_\ell$, we have shown that the ratios
of differential decay rates involving definite helicity states,
$R_{-}(s)$ and $R_{0}(s)$, can be used for extracting the CKM matrix
elements $\vert V_{ub}\vert/\vert V_{ts}\vert$ in a model-independent way.

\clearpage
\chapter{Exclusive $B \to \ K^{*} \ell^{+}  \ell^{-}$ Decay in SUSY
\label{chap:susy}}
This chapter is devoted to the semileptonic rare $B \to \ K^{*} \ell^{+}
\ell^{-}$ decay, by contrasting its anticipated phenomenological
profile in some variants of supersymmetric models. We discuss the
constraints on the Wilson coefficients $C_7$, $C_8$, $C_9$ and
$C_{10}$, that the current data on rare $B$ decays implies in the
context of minimal flavour violating model and in more general
scenarios admitting additional flavour changing mechanisms. As probes
of new physics effects in $B \to \ K^{*} \ell^{+} \ell^{-}$, we
propose to study the ratios $R_0(s)$ and $R_-(s)$ (introduced in the
previous chapter) using some generic SUSY effects. 
\section{Introduction}
Although the Standard Model (SM) of the elementary particle physics is
successful in explaining almost all experimental results,
it is possible that physics beyond the SM exists just above the
presently available energy scale. Since new physics  may affect
various processes at low energy such as the flavor changing neutral
current (FCNC) processes of $K$-mesons and $B$-mesons, new 
physics searches in these processes are as important as direct
particle searches at collider experiments. A prime example is the
$b\to s~\gamma$, process. Experimentally the current world average
based on the improved measurements by the BABAR~\cite{Aubert:2002pd},
CLEO~\cite{Chen:2001fj}, ALEPH~\cite{alephbsg} and
BELLE~\cite{bellebsg} collaborations, ${\cal B}(B \to X_s
\gamma)=(3.43^{+0.42}_{-0.37} ) \times 10^{-4}$. It is 
known that this process puts very strong constraints on various new
physics beyond the SM, for example two Higgs 
doublet model and supersymmetric (SUSY) extension of the SM. 

Along with the $b\to s~\gamma$ process, another important rare $b$ decay
process is the $b\to s~\ell^+ \ell^-$ decay. In
particular, after the first measurements of the semileptonic
rare $B$-decays reported in the inclusive $B \to X_s \ell^+ \ell^-$ 
$(\ell^{\pm}= e^{\pm}, \mu^{\pm})$ mode, with ${\cal B}(B \to X_s
\ell^+ \ell^-)=(6.1 \pm 1.4 ^{+1.3}_{-1.1})\times 10^{-6}$, by the BELLE
collaboration~\cite{bellebsg} as well as the exclusive $B \to K^*
\ell^+ \ell^-$ $(\ell^{\pm}= e^{\pm}, \mu^{\pm})$ one, typically
${\cal B}(B \to K^* \ell^+ \ell^-)=(1.68^{+0.68}_{-0.58}\pm
0.28)\times 10^{-6}$, by the BABAR 
collaboration~\cite{babarbsll}.   

With increased statistical power of experiments at the $B$-factories
in the next several years, the decays discussed above  and 
related rare $B$ decays will be measured very precisely. On the
theoretical side, impressive progress in the theoretical precision has been
achieved concerning the exclusive as well as the inclusive
semileptonic (and radiative) rare $B$-decays, with the completion of
NNLO (NLO) QCD calculations~\citer{BMU,Beneke:2001at}. Although the
theoretical uncertainties can be addressed only with a complete 
NLO, as $b\to s\gamma$ is calculated so far only in NLO
, the SM value for the branching ratio is in agreement with
the experimental measurement within the $1-2\sigma$ level.

Since the abovementioned FCNC rare $B$-processes are forbidden in the
Born approximation any significant deviation from the SM would imply
strongly the existence of new physics, such as Supersymmetry (SUSY). The
reason is simply that, while in the SM the $b \to s$ transition are
dominated by one loop contributions with the exchange of a virtual $W$
and the top quark, in SUSY \cite{Nilles:1983ge,Wess:1982ng} several
competing sources of FCNC are present. To begin with, in SUSY models
the Higgs sector is richer than in the SM, since at least two Higgs
doublets must be present. Consequently, there exists at least one
physical charged scalar $H^{\pm}$ which can be exchanged in the one-loop
contribution to  $b \to s$, together with an up quark. The second
obvious source of FCNC comes from the supersymmetrization of the W and
the charged Higgs contributions, where the up quark is replaced by an
up squark and $W$.

In this chapter we present a SUSY analysis of the semileptonic rare $B
\to \ K^{*} \ell^{+} \ell^{-}$ decay, in the so-called minimal
supersymmetric standard model (MSSM). After a brief review of the MSSM
in section~\ref{sec:MSSM}, we present in section~\ref{sec:MSSMparameter} 
the allowed region of the SUSY parameter space. Section~\ref{sec:bsllMSSM} 
shows the various supersymmetric contribution to the $b \to s~\ell^+\ell^-$ 
transition, while in section~\ref{sec:MSSM-models} we
study the $B \to K^{*} \ell^+ \ell^-$ decay in some specific
SUSY-models, such as Minimal supergravity (mSUGRA) model, Minimal
flavor violating supersymmetric model (MFV) and Extended Minimal
flavor violating supersymmetric model (EMFV). As probes
of new physics effects in $B \to \ K^{*} \ell^{+} \ell^{-}$, we
propose in section~\ref{sec:ratio-MSSM} to study the ratios $R_0(s)$
and $R_-(s)$ (introduced and calculated in the SM in Chapter~3) in the generic mSUGRA model.
Finally, we summarized our analysis in section~\ref{sec:MSSM-Summary}.
\section{The Minimal Supersymmetric Standard Model \label{sec:MSSM}} 
The minimal supersymmetric standard model  (MSSM) is the by far most
widely studied potentially realistic SUSY model. It owes its
popularity mostly to its simplicity, being essentially a
straightforward supersymmetrization of the Standard Model (SM), where
one introduces only those couplings and fields that are necessary for
consistency~\citer{Nilles:1983ge,Haber:1984rc}. 

The single-particle states of the MSSM fall naturally into irreducible
representations of the corresponding algebra which are called {\it
supermultiplets}. Each supermultiplet contains both fermion and boson
states with the same electric charge, weak isospin, and color degrees
of freedom, which are commonly known as {\it superpartners} of each
other.
All of the Standard Model fermions (the known quarks and leptons) 
are members of chiral supermultiplets. The names for the spin-0
partners of the quarks and leptons are constructed by prepending an
``s", which is short for scalar. It seems clear that the Higgs scalar
boson must reside in a chiral supermultiplet, since it has spin 0.
Actually, it turns out that one chiral supermultiplet is not
enough. One way to see this is to note that if there were only one
Higgs chiral supermultiplet, the electroweak gauge symmetry would
suffer a triangle gauge anomaly, and would be inconsistent as a quantum theory.

The basic structure of the MSSM is well-known and has been thoroughly 
discussed in the literature \cite{HaberKane,Derendinger}.  We therefore 
recall just those aspects of the theory which are pertinent to $b \to
s~\ell^+ \ell^-$ transitions.  We first display  nomenclature conventions for matter 
superfields and their left handed fermion and scalar components in
Table~\ref{tab:chiral}, classified according to their transformation properties under the SM gauge group $SU(3)_C\times SU(2)_L \times U(1)_Y$.   
\renewcommand{\arraystretch}{0.8}
\begin{table}[t]
\vspace{0.4cm}
\begin{center}
\begin{tabular}{|c|c|c|c|}
\hline\hline
\mco{2}{|c|}{Names} & Fermions & Scalars \\
\hline\hline
& & &\\
squarks, quarks & $Q_i=\pmatrix{U_i \cr D_i \cr}$  &  $q_i = \pmatrix{u_i \cr d_i \cr}$ & 
 $\tilde{q}_i = \pmatrix{\tilde{u}_i \cr \tilde{d}_i \cr}$ 
\\ 
($\times 3$ families)& \quad $U_i^c$ \quad & \quad $u_i^c$ \quad & \quad $\tilde{u}_i^c$
\\
& \quad $D_i^c$ \quad & 
\quad $d_i^c$ \quad & \quad $\tilde{d}_i^c$ 
\\
& & &\\
\hline
& & &\\
sleptons, leptons & \quad $L_i=\pmatrix{N_i \cr E_i \cr}$ \quad &
\quad $\ell_i = \pmatrix{\nu_i \cr e_i \cr}$ \quad & 
\quad $\tilde{\ell}_i = \pmatrix{\tilde{\nu}_i \cr \tilde{e}_i \cr}$ \quad
\\
($\times 3$ families) &\quad $E_i^c$ \quad & 
\quad $e_i^c$ \quad & \quad $\tilde{e}_i^c$ 
\\
& & &\\
\hline
& & &\\
Higgs, higgsinos &\quad $H_1=\pmatrix{H^0_1 \cr H^-_1 \cr}$ \quad & 
\quad $\tilde{h}_1 = \pmatrix{\tilde{h}_1^0 \cr \tilde{h}_1^- \cr}$ \quad & 
\quad $h_1 = \pmatrix{{h^0_1}^* \cr -h_1^- \cr}$ 
\quad
\\
& & &\\
&\quad $H_2=\pmatrix{H^+_2 \cr H^0_2 \cr}$ \quad & 
\quad $\tilde{h}_2 = \pmatrix{\tilde{h}_2^+ \cr \tilde{h}_2^0 \cr}$ \quad & 
\quad $h_2 = \pmatrix{h_2^+ \cr h_2^0 \cr}$ 
\quad
\\
& & &\\
\hline\hline
\end{tabular}
\caption{\it 
Chiral supermultiplets in
the Minimal Supersymmetric Standard Model.\label{tab:chiral}}
\end{center}
\end{table}

The vector bosons of the SM clearly must reside
in gauge supermultiplets. Their fermionic superpartners are
generically referred to as {\it gauginos}.
The $SU(3)_C$ color gauge interactions of
QCD are mediated by the gluon, whose spin-1/2
color-octet supersymmetric partner is the {\it gluino}. As usual,
a tilde is used to denote the supersymmetric partner of
a SM state, so the symbols for the gluon and
gluino are $g$ and $\tilde{g}$ respectively. 
The electroweak gauge symmetry $SU(2)_L\times U(1)_Y$ has associated
with it spin-1 gauge bosons $W^+, W^0, W^-$ and $B^0$, with spin-1/2
superpartners $\tilde{W}^+, \tilde{ W}^0, \tilde{W}^-$ and $\tilde {B}^0$,
called {\it winos} and {\it bino}.
After electroweak symmetry breaking, the $W^0$, $B^0$ gauge
eigenstates mix to give mass eigenstates $Z^0$ and $\gamma$.
The corresponding gaugino mixtures of $\tilde{W}^0$ and $\tilde{B}^0$
are called zino ($\tilde{Z}^0$) and photino ($\tilde \gamma$); if
supersymmetry were unbroken, they would be
mass eigenstates with masses $m_Z$ and 0.
Table \ref{tab:gauge} summarizes the gauge supermultiplets of a
minimal supersymmetric extension of the SM.
\renewcommand{\arraystretch}{1.55}
\begin{table}[t]
\vspace{0.4cm}
\begin{center}
\begin{tabular}{|c|c|c|}
\hline\hline
Names & Fermions & Vectors \\
\hline\hline
gluino, gluon &$ \tilde{g}$& $g$ \\
\hline
winos, W bosons & $ \tilde{W}^\pm\>\>\> \tilde{W}^0 $& $W^\pm\>\>\> W^0$ \\
\hline
bino, B boson &$\tilde{B}^0$ & $B^0$ \\
\hline\hline
\end{tabular}
\caption{\it 
Gauge supermultiplets in
the Minimal Supersymmetric Standard Model.\label{tab:gauge}}
\end{center}
\end{table}

After this brief introduction on the field content of the
MSSM, we will review in what follows, the aspects of the theory
relevant to the $b \to s~ \ell^+ \ell^- $ transitions considering only
the case of unbroken R-parity. The superpotential which determines the
supersymmetry preserving 
interactions among matter fields is:
\begin{eqnarray}\label{superpotential}
W = \mu_{susy}~ H_1 H_2 + Y^\U_{ij} Q_i U^c_j H_2
+ Y^\D_{ij} Q_i D^c_j H_1 + Y^\E_{ij} L_i E^c_j H_1,
\end{eqnarray}
where $Q$, $U^c$, $D^c$, $L$ and $E^c$ are the superfields
corresponding to the $SU(2)$ doublets and singlets for quarks and
leptons, $H_1$ and $H_2$ are the two Higgs superfields, $Y^f$ are the
Yukawa matrices and $\mu$ is the Higgs quadratic coupling. After
vector superfield terms are included, the supersymmetric  
Lagrangian schematically appears in component form as~\cite{Cho:1996we}
\begin{eqnarray}
\CL_{\rm SUSY} &=& -\quarter {F_\G^\A}^{\u\v} {F_\G^\A}_{\u\v}
 + \bar{\lambda_\G^\A} i \Dslash_{\A\BB} \lambda_\G^\BB 
 + (D^\mu \phi)^\dagger (D_\mu \phi) + \bar{\psi} i \Dslash \psi \cr \nn\\
 && - \left[ \left( {d W \over d \Phi_i} \right)^* 
     \left( {d W \over d \Phi_i} \right) + \half \left( 
     {\partial^2 W \over \partial \Phi_i \partial \Phi_j} 
     \psi_i^\T C \psi_j + \hc \right) \right]_{\Phi \to \phi} \cr\nn\\
 && - \sqrt{2} g_\G \left[ \phi^\dagger T^\A_\G {\lambda^\A_\G}^\T C \psi 
     + \hc \right] - \half g^2_\G ( \phi^\dagger T^\A_\G \phi) 
        ( \phi^\dagger T^\A_\G \phi) .\label{LSUSY}
\end{eqnarray}
The index $G$ labels the color, weak isospin and hypercharge factors in 
the Standard Model gauge group, and indices $A$ and $B$ range over the 
nonabelian subgroups' adjoint representations.  All MSSM scalars are 
assembled into $\phi$, while matter fermions and gauginos are 
respectively contained within the four-component left handed $\psi$ and 
$\lambda$ fields. 

        Since supersymmetry is manifestly violated in the low energy world, 
the MSSM Lagrangian is supplemented with the soft supersymmetry breaking terms~\cite{Cho:1996we}
\begin{eqnarray} 
\CL_{\rm soft} &=& -\half \Bigl[ m_{\tilde{g}} \tilde{g}^{a \,\T} C \tilde{g}^a
+ m_{\tilde{\W}} \tilde{W}^{i \,\T} C \tilde{W}^i
+ m_{\tilde{\BB}} \tilde{B}^\T C \tilde{B} + \hc \Bigr]- {\Delta}_1^2 h_1^\dagger h_1 - {\Delta}_2^2 h_2^\dagger h_2
\cr \nn\\
&& 
- \tilde{q}^\dagger_i (M^2_{\tilde{q}})_{ij} \tilde{q}_j 
- \tilde{u}^{c\,\dagger}_i (M^2_{\tilde{u}^c})_{ij} {\tilde{u}}^c_j 
- \tilde{d}^{c\,\dagger}_i (M^2_{\tilde{d}^c})_{ij} {\tilde{d}}^c_j 
- \tilde{\ell}^\dagger_i (M^2_{\tilde{\ell}})_{ij} \tilde{\ell}_j 
- \tilde{e}^{c\,\dagger}_i (M^2_{\tilde{e}^c})_{ij} {\tilde{e}}^c_j
\cr \nn\\
&& 
+ \Bigl[ A^\U_{ij} \tilde{q}_i \tilde{u}^c_j h_2 
+ A^\D_{ij} \tilde{q}_i \tilde{d}^c_j h_1
+ A^\E_{ij} \tilde{\ell}_i \tilde{e}^c_j h_1
+ B \mu_{susy} h_1 h_2 + \hc \Bigr],  
\label{Lsoft}
\end{eqnarray}
where $m_{\tilde{g}}$ ($m_{\tilde{\W}}$ and $m_{\tilde{\BB}}$), 
${\Delta}_i^2$ $(i=1,2)$ and $B \mu_{susy}$ are mass terms for gluino (wino
and bino) and for the Higgs fields, respectively. 
The scalar mass terms $M^2_{\tilde{q}}$, $M^2_{\tilde{u}^c}$,
$M^2_{\tilde{d}^c}$, $M^2_{\tilde{\ell}}$ and $M^2_{\tilde{e}^c}$ are
in general hermitian $3\times 3$ matrices, while $A^\U~h_2$, $A^\D~h_1$
and $A^\E~h_1$  are general $3\times 3$ matrices. Allowing all the
parameters in (\ref{Lsoft}) to be complex, we end up with 124 masses,
phases and mixing angles as free parameters of the model.

Electroweak symmetry breaking induces mixing among MSSM fields. 
In the matter sector, primed mass eigenstates are related to unprimed 
gauge eigenstate counterparts as follows:

\begin{eqnarray} 
u' &=& {\cal S}^{U_L} u+{\cal S}^{U_R} C{\bar{u^c}}^{\>\T},~~~~~~~~~~~~~~~~~~~ \tilde{u}' = {\cal T}^{U}\pmatrix{{\cal S}^{U_L} \> \tilde{u} \cr {\cal S}^{U_R} \> \tilde{u^c}^*\cr}  \nn \\
d' &=& {\cal S}^{D_L} d - {\cal S}^{D_R} C {\bar{d^c}}^{\>\T}  \nn,
~~~~~~~~~~~~~~~~~~~  \tilde{d}' = {\cal T}^{D} \pmatrix{{\cal S}^{D_L} \> \tilde{d} \cr -{\cal S}^{D_R} \> \tilde{d^c}^* \cr} \\
\nu' &=& {\cal S}^{N_L} \nu,~~~~~~~~~~~~~~~~~~~~~~~~~~~~~~~~~~~~~~~~ 
\tilde{\nu}' = {\cal T}^{N} {\cal S}^{E_L} \tilde{\nu}  \nn \\
e' &=& {\cal S}^{E_L} e - {\cal S}^{E_R} C {\bar{e^c}}^{\>\T},
~~~~~~~~~~~~~~~~~~~ 
\tilde{e}' = {\cal T}^{E} \pmatrix{{\cal S}^{E_L} \> \tilde{e} \cr -{\cal S}^{E_R} \> \tilde{e^c}^* \cr}. \label{masseigenstates}   
\end{eqnarray}
The unitary ${\cal S}$ and ${\cal T}$ transformations rotate fermion and sfermion mass 
matrices into real and diagonal forms.   The $3\times 3$ quark and lepton mass 
matrices are simply related to the Yukawa couplings in the superpotential:

\begin{eqnarray} \label{massmatrices}
M_{U} &=& {v \sin\beta \over \sqrt{2}} {\cal S}^{U_R} {Y^U}^\T {{\cal S}^{U_L}}^\dagger \nn\\
M_{D} &=& {v \cos\beta \over \sqrt{2}} {\cal S}^{D_R} {Y^D}^\T {{\cal S}^{D_L}}^\dagger \nn\\
M_{E} &=& {v \cos\beta \over \sqrt{2}} {\cal S}^{E_R} {Y^E}^\T
{{\cal S}^{E_L}}^\dagger, 
\label{massmatrices}
\end{eqnarray}
where $v_1=v \cos\beta$ and $v_2=v \sin\beta$  are the expectation
values of the two Higgs doublets and ${\cal S}^f~ \big(f=U_R, U_L, D_R,
D_L, E_R, E_L \big)$. The $6\times 6$ squared mass matrices for the
squarks (we do not present the lepton ones since they are not relevant
for the forthcoming analysis of SUSY contributions) read:

\begin{eqnarray}
M^2_{\tilde{u}} &=&{\cal T}^{U} \left(\begin{array}{cc}
 {\cal S}^{U_L} M^2_{\tilde{q}} {{\cal S}^{U_L}}^\dagger + M_{U}^2 + \ldots 
& \mu_{susy} M_{U} \cot\beta -\displaystyle{v \sin\beta \over \sqrt{2}} 
{\cal S}^{U_L} {A^{U}}^* {{\cal S}^{U_R}}^\dagger \\
\mu_{susy}^* M_{U} \cot\beta -\displaystyle{v \sin\beta \over \sqrt{2}} 
{\cal S}^{U_R} {A^U}^\T {{\cal S}^{U_L}}^\dagger &
{\cal S}^{U_R} {M^2_{\tilde{u}^c}}^\T {{\cal S}^{U_R}}^\dagger + M_U^2
 + \ldots\end{array}\right){{\cal T}^U}^\dagger
 \nn\\[1cm] 
M_{\tilde{d}}^2 \hspace*{-0.2cm}&=&\hspace*{-0.2cm}{\cal T}^{D}\left(\begin{array}{cc}
{\cal S}^{D_L}\,M_{\tilde{q}}^2\, {{\cal S}^{D_L}}^\dagger +M_D^2+\ldots & 
\hspace*{-0.5cm}
\mu_{susy} M_D \tan\beta -\displaystyle{v \cos\beta \over \sqrt{2}} 
{\cal S}^{D_L}\,{A^{D}}^*\, {{\cal S}^{D_R}}^\dag\\
\mu^*_{susy} M_D \tan\beta - \displaystyle{v \cos \beta \over \sqrt{2}}
{\cal S}^{D_R}\,A^{D^T}\, {{\cal S}^{D_L}}^\dag& 
\hspace*{-0.5cm}
{\cal S}^{D_R}\,{M^2_{\tilde{d}^c}}^\T\, {{\cal S}^{D_R}}^\dag +M_D^2+\ldots 
\end{array}\right){{\cal T}^{D \dag}},
\nn\\
\hspace*{-3cm}
\label{eq:4.5}
\end{eqnarray}
where $M^2_{\tilde{u}}$ and  $M_{\tilde{d}}^2$ are the diagonal mass
matrices of the up and down squarks and the dots stand for terms
proportional to $m^2_Z$. Moreover, $M^2_{\tilde{u}}$ and
$M_{\tilde{d}}^2$ are matrices in the basis in which the squark fields
undergo the same rotations as the quark ones. This means that we
diagonalize the matrices $M_{U}$ and $M_{D}$ applying rigid rotations
to the quark superfields and that there is not any flavour change in
vertices with both quark and squarks. In the literature this basis is
usually referred as the {\it SuperCKM (SCKM)} one.

Mixing also takes place in the gaugino and Higgs sectors.  The 
physical Dirac chargino and Majorana neutralino eigenstates are respectively linear combinations of left handed Winos, Binos and Higgsinos.
Thus in the weak eigenstates basis, the chargino and neutalino mass
matrices are given respectively, by
\begin{eqnarray}
M_{\tilde{\chi}^\pm} &= {\cal U}^* \left(\begin{array}{cc}
m_{\tilde{\W}} & \sqrt{2} \mW \sin\beta \\
\sqrt{2} \mW \cos\beta & -\mu_{susy} \end{array}\right) 
{\cal V}^\dagger , 
\label{charginomassmatrix}
\end{eqnarray}
and
\begin{eqnarray}
M_{\tilde{\chi}^0} &= {\cal N}^* \pmatrix{
m_{\tilde{\BB}} & 0 & -\mZ \sin\th \cos\beta & \mZ \sin\th \sin\beta \cr 
0 & m_{\tilde{\W}} & \mZ\cos\th \cos\beta & -\mZ\cos\th \sin\beta \cr
-\mZ \sin\th \cos\beta & \mZ\cos\th \cos\beta & 0 & \mu_{susy} \cr 
\mZ \sin\th \sin\beta & -\mZ\cos\th \sin\beta & \mu_{susy} & 0 \cr}
{\cal N}^\dagger,\nn\\
\label{neutralinomassmatrix}
\end{eqnarray}
where, the unitary transformations ${\cal U}$, ${\cal V}$ and ${\cal
N}$ are unitary matrices, which diagonalize these fields mass
matrices. 

After the gauge eigenstate fields in the supersymmetric Lagrangian 
\ref{LSUSY} are rewritten in terms of their mass eigenstate counterparts,
\footnote{ We suppress primes on mass eigenstate fields from here on.}
it is straightforward to work out the interactions of gluinos,
charginos and neutralinos with quarks and squarks.  We list below the
resulting terms which participate at one-loop order in $d_i \to d_j
\ell^+ \ell^-$ decay:
\begin{eqnarray}
\CL_{\tilde{g},\tilde{\chi}} & = & -\sqrt{2} g_3 \sum_{a=1}^8 
\bar{\tilde{g}^a_\M} \tilde{d}^\dagger ( {\cal T}^\DL L - 
{\cal T}^\DR R) T^a d \label{Lchi}\\
& & + \sum_{\I=1}^2 \bar{\tilde{\chi}}^{\, -}_\I 
\tilde{u}^\dagger ( {\cal X}_\I^\UL L + {\cal X}_\I^\UR R) d +
\sum_{\I = 1}^4 (\bar{\tilde{\chi}}^0_\M)_\I 
\tilde{d}^\dagger ( {\cal Z}_\I^\DL L + {\cal Z}_\I^\DR R)
d + \hc, 
\label{Lgluino,chi}
\end{eqnarray}
where
\begin{eqnarray}
{\cal X}_\I^\UL &=& g_2 \Bigl( -{\cal V}^*_{\I1} {\cal T}^\UL + {\cal V}^*_{\I2} {\cal T}^\UR
{M_\U \over \sqrt{2} \mW \sin\beta} \Bigr) V_{\mathrm CKM}, \nn\\
{\cal X}_\I^\UR &=& g_2 \Bigl( {\cal U}_{\I2} {\cal T}^\UL V_{\mathrm CKM} 
{M_\D \over \sqrt{2} \mW \cos\beta} \Bigr), \nn\\
{\cal Z}_\I^\DL &=& -{g_2 \over \sqrt{2}} \Bigl[ ( -{\cal N}^*_{\I2} + \third \tan\th 
{\cal N}^*_{\I1}) {\cal T}^\DL +{\cal N}^*_{\I3} {\cal T}^\DR {M_\D \over \mW \cos\beta} \Bigr], 
\nn\\
{\cal Z}_\I^\DR &=& -{g_2 \over \sqrt{2}} \Bigl( \twothirds \tan\th {\cal N}_{\I1} {\cal T}^\DR
+{\cal N}_{\I3} {\cal T}^\DL {M_\D \over \mW \cos\beta} \Bigr).  
\label{XandZ}
\end{eqnarray}
We see clearly that the flavor mixing enters into these interactions through the Kobayashi-Maskawa matrix $V_{\tiny{ CKM}}={\cal S}^\UL {{\cal S}^\DL}^\dagger$ and the $6\times 3$ block components of ${\cal T}^\U$
and ${\cal T}^\D$: 

\begin{eqnarray}
{\mathcal T}^{\U(D)}_{6 \times 6} = \Big(
{\cal T}^{\UL(\DL)}_{6
\times 3} , {\cal T}^{\UR(\DR)}_{6 \times 3} \Big).
\label{Gammamatrices}
\end{eqnarray}
The Feynman rules for all these interactions may be found in the
literature \cite{HaberKane,Rosiek}. Having set up the basic MSSM framework, we are now ready to explore its large parameter space.  We take up this topic in the following section.
        \section{MSSM parameter space \label{sec:MSSMparameter}} 
Before predictions can be derived from the Minimal Supersymmetric 
Standard Model, explicit values for the parameters in the superpotential 
(\ref{superpotential}) and soft supersymmetry breaking Lagrangian
(\ref{Lsoft}) must be specified.  

In order to determine the allowed region of the SUSY parameter space, we 
require the following phenomenological constraints\cite{PDG2000}

\renewcommand{\theenumi}{(\arabic{enumi})}
\renewcommand{\labelenumi}{\theenumi}
\begin{enumerate}
\item
  \label{item:bsg}
$b \to s \gamma$ constraint from BABAR~\cite{Aubert:2002pd},
CLEO~\cite{Chen:2001fj}, ALEPH~\cite{alephbsg} and
BELLE~\cite{bellebsg} collaborations, i.e., $3.06 \times 10^{-4}<{\cal B}(B \to X_s \gamma)<3.85 \times 10^{-4}$.
\item
  \label{item:lightsp}
From the recent experiment at LEP 2\cite{PDG2000}, we impose that all
the charged SUSY particles are heavier than 70 GeV
\item
The neutarlino ${\tilde \chi}_1^0$ mass is larger than 32 GeV.
\item
  All sneutrino masses are larger than 43 GeV \cite{PDG2000}.
\item
The gluino and squark mass bounds from Fermilab Tevatron
experiments~\cite{TEV}. The precise bounds on the gluino mass and the
averaged squark mass except for the top squark is restricted to be
larger than about 180 GeV.
\item
The stop ${\tilde t}_1$ mass is larger than 87 GeV.
\end{enumerate}

Having these phenomenological constraints at hand, let's explore the
SUSY contributions to the $b\to s \ell^+ \ell^-$ transition.
\section{Supersymmetric contribution to $ b \to s~ \ell^+\ell^-$ 
\label{sec:bsllMSSM}}
The Major theoretical breakthrough in the analysis of the FCNC tests
in SUSY models came in 1983 when {\it Duncan}~\cite{Duncan:1983iq} and,
independentely, {\it Donoghue}, {\it Nilles and Wyler}~\cite{Donoghue:1983mx}
noticed that FC transitions could occur at the $\tilde{g}-\tilde{q}-q$
vertices. The replacement of the weak coupling of the W with the strong
coupling of the gluino, and the presence of approximately the same CKM
mixing angles, raised the hope of possible SUSY enhancements of FCNC
processes through one-loop gluino exchanges. Indeed, in the general
context of MSSMs, it was found that these gluino mediated FCNC
contributions could play a relevant role in CP violation of the
K-system \cite{Duncanetal} and, more recently, in the radiative $b\to
s \gamma$ \cite{Bertolini:1986tg} and  $b\to s g$
\cite{Bertolini:1987pk} decays. 

Within the MSSM,  the $b\to s~ \ell^+ \ell^-$ transition is governed by
five possible classes of contributions. They correspond to four
classes of one loop SUSY diagrams that produce $b\to s~ \ell^+ \ell^-$
transition, in additions to W-exchange. We will classify them
according to the particles running in the loop:

\renewcommand{\theenumi}{(\arabic{enumi})}
\renewcommand{\labelenumi}{\theenumi}

\begin{enumerate}
\item
\label{item:W-up}
$W^{-}$ and up quarks (SM),
\item\label{item:H-up}
$ H^{-}$ and up quarks,
\item\label{item:chi-sup}
$\tilde{\chi}^{-}$  and up squarks,
\item\label{item:g-sdown}
$\tilde{g}$ and down squarks,
\item\label{item:chi0-sdown}
 $ \tilde{\chi}^{0}$  and down squarks.
\end{enumerate}

\begin{figure}[t]
\begin{center}
\psfrag{b}{ $b$}
\psfrag{s}{ $s$}
\psfrag{gma}{ $\gamma,~Z$}
\psfrag{Z}{ $$}
\psfrag{g}{ $$}
\psfrag{Hmns}{ $H^{-}$}
\psfrag{chit}{ $\tilde{\chi}^-$}
\psfrag{chio}{ $\tilde{\chi}^0$}
\psfrag{gluino}{ $\tilde{g}$}
\psfrag{d}{ $d$}
\psfrag{l}{ $l$}
\psfrag{u}{ $u,c,t$}
\psfrag{ut}{ $\tilde{u},\tilde{c},\tilde{t}$}
\psfrag{dt}{ $\tilde{d},\tilde{s},\tilde{b}$}
\psfrag{af}{ $(\textrm{a})$}
\psfrag{aaf}{ $(\textrm{b})$}
\psfrag{bf}{ $(\textrm{b})$}
\psfrag{bbf}{ $(\textrm{d})$}
\psfrag{boson}{$$}
\psfrag{cf}{ $(\textrm{b})$}
\psfrag{df}{ $(\textrm{d})$}
\epsfig{file=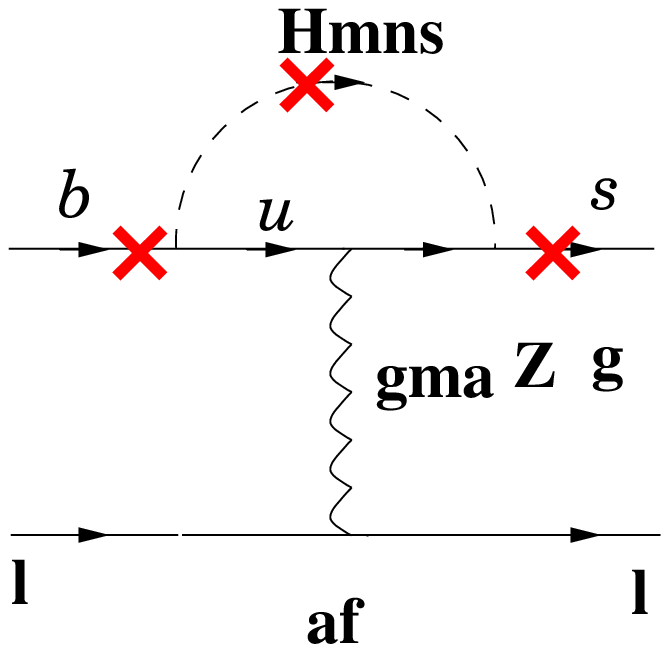,width=0.3\linewidth}
\hspace*{4cm}
\epsfig{file=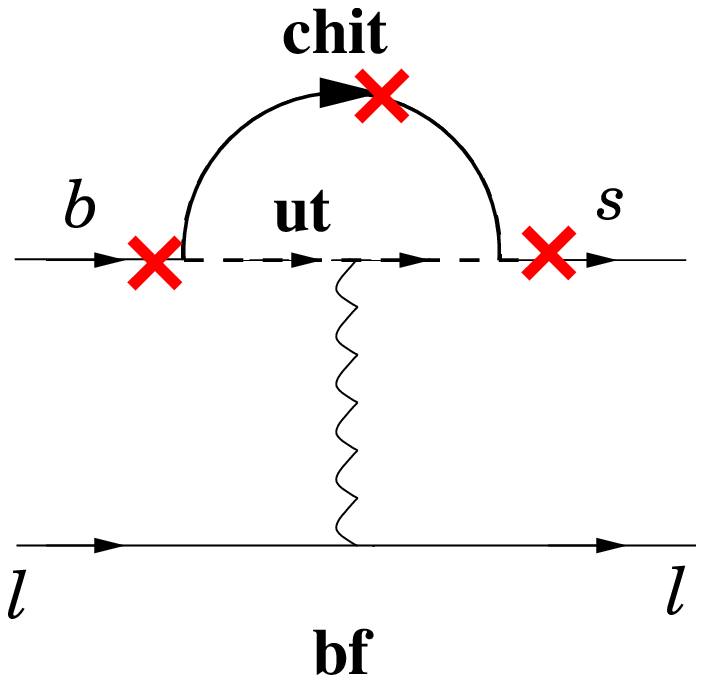,width=0.3\linewidth}
\\
\vspace*{1cm}
\psfrag{b}{ $b$}
\psfrag{s}{ $s$}
\psfrag{gma}{ $\gamma,~Z$}
\psfrag{Z}{ $$}
\psfrag{g}{ $$}
\psfrag{Hmns}{ $H^{-}$}
\psfrag{chit}{ $\tilde{\chi}^0$}
\psfrag{chio}{ $\tilde{\chi}^0,~\tilde{g}$}
\psfrag{d}{ $d$}
\psfrag{l}{ $l$}
\psfrag{u}{ $d$}
\psfrag{ut}{ $\tilde{d},\tilde{s},\tilde{b}$}
\psfrag{boson}{$Z$}
\psfrag{dt}{ $\tilde{d},\tilde{s},\tilde{b}$}
\psfrag{af}{ $(\textrm{a})$}
\psfrag{af1}{ $(\textrm{b})$}
\psfrag{bf}{ $(\textrm{c})$}
\psfrag{bf1}{ $(\textrm{d})$}
\psfrag{cf}{ $(\textrm{c, d})$}
\psfrag{bbf}{ $(\textrm{e})$}
\psfrag{boson}{$Z$}
\epsfig{file=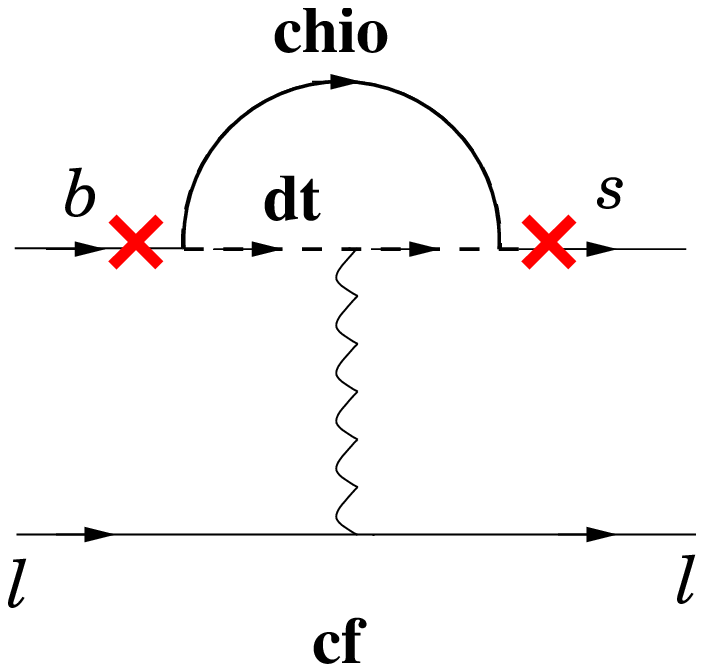,width=0.3\linewidth}
\hspace*{4cm}
\epsfig{file=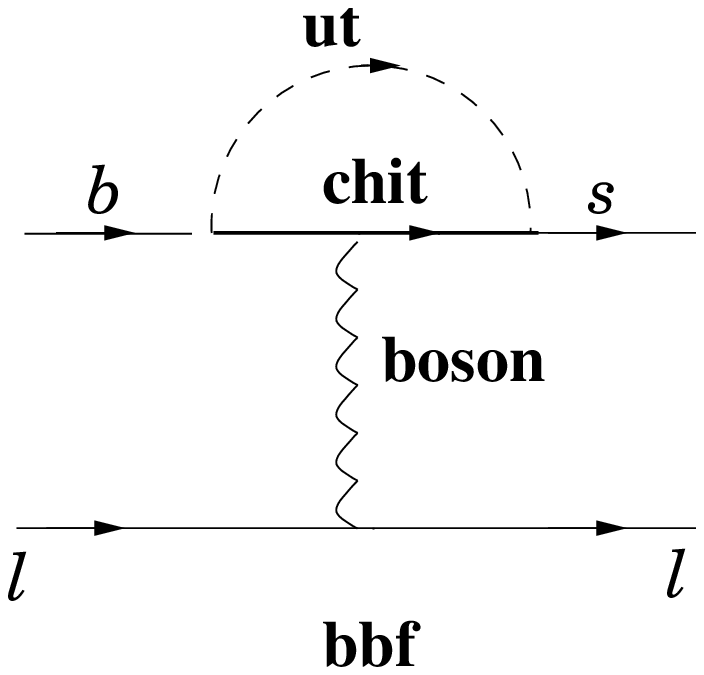,width=0.3\linewidth}
\caption{ \it
SUSY-Penguin diagrams relevant to the semileptonic $b \to s~ \ell^+
\ell^-$ transition in the MSSM. The cross denotes a possibility to
attach the photon or the $Z$-boson. 
}
\label{fig:MSSM-Peng}
\end{center}
\end{figure}

\begin{figure}[t]
\begin{center}
\psfrag{Hmns}{ $H^{-}$}
\psfrag{nul}{ $\nu_{l}$}
\psfrag{b}{ $b$}
\psfrag{s}{ $s$}
\psfrag{u}{ $u,c,t$}
\psfrag{l}{ $l$}
\psfrag{af}{ $(a)$}
\epsfig{file=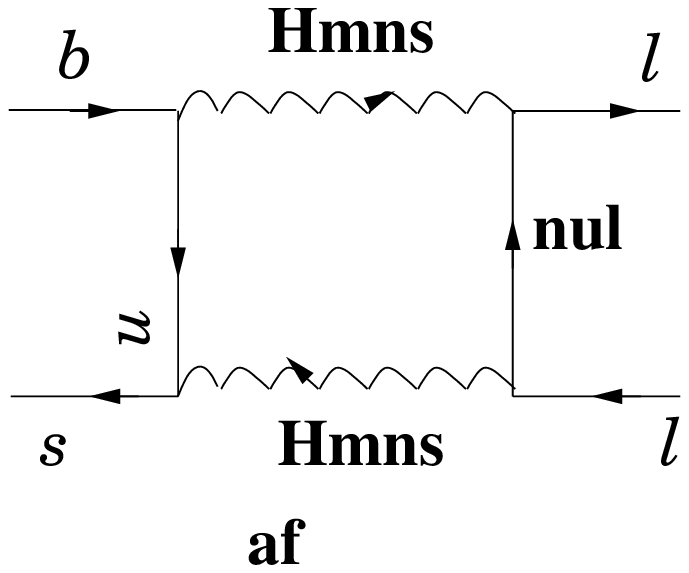,width=0.3\linewidth}
\hspace*{4cm}
\psfrag{chit}{ $\tilde{\chi}^-$}
\psfrag{nult}{ $\tilde{\nu}_{l}$}
\psfrag{ut}{ $\tilde{u},\tilde{c},\tilde{t}$}
\psfrag{bf}{ $(b)$}
\epsfig{file=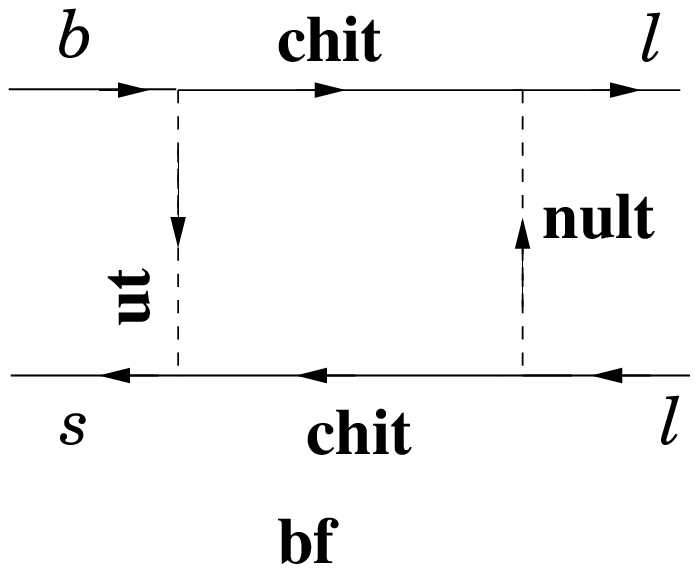,width=0.3\linewidth}
\hspace*{0.2cm}
\\ 
\vspace*{1cm}
\psfrag{chio}{ $\tilde{\chi}^o$}
\psfrag{lt}{ $\tilde{l}$}
\psfrag{dt}{ $\tilde{d},\tilde{s},\tilde{b}$}
\psfrag{cf}{ $(c)$}
\epsfig{file=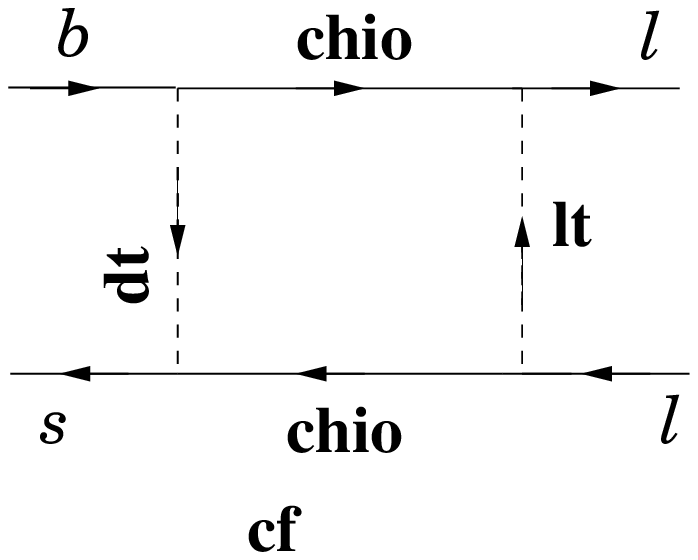,width=0.3\linewidth}
\hspace*{4cm}
\psfrag{df}{ $(d)$}
\epsfig{file=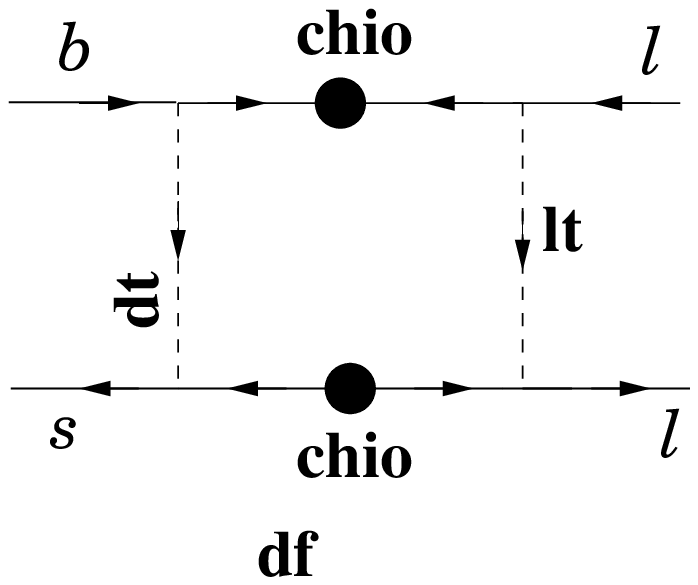,width=0.3\linewidth}
\caption{ \it
SUSY-Box diagrams relevant to the semileptonic $b \to s~ \ell^+
\ell^-$ transition in the MSSM. The bubble in figure $(d)$ reflect the
Majorana nature of the neutralinos.}
\label{fig:MSSM-Box}
\end{center}
\end{figure}

We list below the $W$-scale matching contributions to the Wilson 
coefficients $C_7$, $C_9$ and $C_{10}$ which arise from one-loop MSSM diagrams,
presented in Fig.~(\ref{fig:MSSM-Peng}) and (\ref{fig:MSSM-Box}). 
The total contribution to $C_i$ $(i=7,9,10)$ and the normalized ratio
can thus be written respectively as  
\begin{eqnarray}
C_i^{tot}(m_W) &=& C_i^{SM}(m_W)+ C_i^{NP}(m_W),\label{eq:CiTot}
\end{eqnarray}
and
\begin{eqnarray}
R_i = {C_i^{tot}(m_W)\over  C_i^{SM}(m_W)}.\label{eq:Ri}
\end{eqnarray}

We start with the $W$-scale matching contributions to the 
coefficient $C_7$ of the magnetic moment operator in the $\Delta B=1$ 
effective Hamiltonian which arise from one-loop MSSM diagrams,
presented in Fig.~(\ref{fig:MSSM-Peng}) (only the $\gamma$-penguin
diagrams). Thus its total contribution reads as 
\begin{eqnarray}
C_7^{tot}(m_W) &=& C_7^{SM}(m_W)+ C_7^{NP}(m_W),\label{eq:C7Tot}
\end{eqnarray}
\noi where $C_7^{NP}(m_W)$ represent the new physics contribution at
the scale $m_W$ to the $C_7$, defined as:
\begin{eqnarray}
C_7^{NP}(m_W)= \delta C_7^{h_{\pm}}(m_W)+
                \delta C_7^{\tilde{\chi}^{\pm}}(m_W)+
                \delta C_7^{\tilde{\chi}^{0}}(m_W)+
                \delta C_7^{\tilde{g}}(m_W). 
\label{eq:C7NP}
\end{eqnarray}

\newpage
\noindent The explicit expressions for the various terms are~\cite{Cho:1996we}:

\begin{itemize}
\item
Standard Model graphs: 
\begin{eqnarray}\label{Bsgmatchone}
C_7^{SM} = {\xt \over 4} f_1(\xt)
\end{eqnarray}

\item
Graphs with charged Higgs loops:
\begin{eqnarray}\label{Bsgmatchtwo}
\delta C_7^{h_{\pm}} &= {1 \over 6} \left\{ \half {\mtsq \over \mhsq}
\cot^2 \beta \, f_1\bigl({\mtsq \over \mhsq} \bigr) + f_2 \bigl({\mtsq \over 
\mhsq} \bigr) \right\} 
\end{eqnarray}

\item
Graphs with chargino loops:
\begin{eqnarray}
\delta C_7^{\tilde{\chi}^{\pm}} &=& {1 \over 3 g_2^2 \KManglesb} \sum_{\A = 1}^6 \sum_{\I = 1}^2 
{\mW^2 \over \mchI^2} \label{Bsgmatchthree}\\
&& \qquad \times \Bigl\{ -\half ({\cal X}_\I^\UL)^\dagger_\twoA 
 ({\cal X}_\I^\UL)_\Athree f_1 \Bigl({\musqA^2 \over \mchI^2} \Bigr) 
+ ({\cal X}_\I^\UL)^\dagger_\twoA ({\cal X}_\I^\UR)_\Athree 
  {\mchI \over m_b} f_2 \Bigl({\musqA^2 \over \mchI^2} \Bigr) \Bigr\} \nn
\end{eqnarray}

\item
Graphs with neutralino loops:
\begin{eqnarray}
\delta C_7^{\tilde{\chi}^{0}} &=& - {1 \over 3 g_2^2 \KManglesb} \sum_{\A = 1}^6 \sum_{\I = 1}^4 
{\mW^2 \over \mneI^2} \label{Bsgmatchfour}\\
&& \qquad \times \Bigl\{ \half ({\cal Z}_\I^\DL)^\dagger_\twoA 
 ({\cal Z}_\I^\DL)_\Athree f_3 \Bigl({\mdsqA^2 \over \mneI^2} \Bigr) 
+ ({\cal Z}_\I^\DL)^\dagger_\twoA ({\cal Z}_\I^\DR)_\Athree 
  {\mneI \over m_b} f_4\Bigl({\mdsqA^2 \over \mneI^2} \Bigr) \Bigr\}
\nn
\end{eqnarray}

\item
Graphs with gluino loops:
\begin{eqnarray}
\delta C_7^{\tilde{g}} &=& {4 g_3^2 \over 9 g_2^2 \KManglesb} 
\sum_{\A = 1}^6 {\mW^2 \over \mgluino^2} \label{Bsgmatchfive}\\
&& \qquad \times \Bigl\{ - ({\cal T}^\DL)^\dagger_\twoA 
 ({\cal T}^\DL)_\Athree f_3 \Bigl({\mdsqA^2 \over \mgluino^2} \Bigr) 
+ 2 ({\cal T}^\DL)^\dagger_\twoA ({\cal T}^\DR)_\Athree 
  {\mgluino \over m_b} f_4\Bigl({\mdsqA^2 \over \mgluino^2} \Bigr)
\Bigr\}\nn 
\end{eqnarray}

\end{itemize}
\noindent
The one-loop integral functions which enter into these matching conditions 
are given in Appendix~\ref{app:susy}. Concerning the SUSY contributions to the
chromo-magnetic coefficient $C_8$, it has the same structure as the
$C_7$ ones, with different colour factors and loop-functions.
The $W$-scale matching contributions to the semileptonic
coefficients , namely $C_9^{eff}$ and $C_{10}$, which arise from
one-loop MSSM diagrams, presented in Figs.~(\ref{fig:MSSM-Peng}) and
(\ref{fig:MSSM-Box}), and can be written respectively as
\begin{eqnarray} 
C_{9,10}^{tot}(m_W)= C_{9,10}^{SM}(m_W)+C_{9,10}^{NP}(m_W),
\end{eqnarray}

\noi where the SUSY contributuions to the semileptonic coefficients,
are defined as:
\begin{eqnarray}
C_{9}^{NP}(m_W) &=& \delta
C_{9}^{(Z,\gamma)-peng,(H,\tilde{\chi},\tilde{\chi}^0,\tilde{g})} +
\delta
C_{9}^{'(Z,\gamma)-peng,(H,\tilde{\chi},\tilde{\chi}^0,\tilde{g})}+
\delta C_{9}^{(\tilde{\chi},\tilde{\chi}^0)-box}+
\delta C_{9}^{'(\tilde{\chi},\tilde{\chi}^0)-box}, 
\nn\\
C_{10}^{NP}(M_W) &=& C_{10}^{(Z,\gamma)-peng,(H,\tilde{\chi},\tilde{\chi}^0,\tilde{g})}+\delta C_{10}^{(\tilde{\chi},\tilde{\chi}^0)-box}. \label{eq:defc9c10}
\end{eqnarray}

\newpage
\noi The various terms are given as follow~\cite{Cho:1996we}: 
\begin{itemize}
\item
Z-penguin graphs with charged Higgs loops:
\begin{eqnarray}
\delta C_{9,10}^{Z-peng,H} &=& \mp{1 \over 8~ \sin^2 \theta_W} 
\cot^2 \beta \, \xt f_5\bigl({\mtsq \over \mhsq} \bigr)\nn\\
\delta C_{9}^{' Z-peng,H} &=& {1 \over 2} 
\cot^2 \beta \, \xt f_5\bigl({\mtsq \over \mhsq} \bigr)
\label{kpieematchone}
\end{eqnarray} 

\item 
$\gamma$-penguin graphs with charged Higgs loops:
\begin{eqnarray}
\delta C_{9,10}^{\gamma-peng,H} &=& 0 \nn\\
\delta C_{9}^{' \gamma-peng,H} &=& {1 \over 18} \cot^2 \beta \, f_6\bigl({\mtsq \over \mhsq} 
\bigr) \label{kpieematchtwo}
\end{eqnarray}

\item 
Z-penguin graphs with chargino loops:
\begin{eqnarray}
&&\delta C_{9,10}^{Z-peng,\tilde{\chi}} = \pm {1 \over 2 g_2^2 ~ \sin^2 \theta_W \KMangles}
\sum_{\A,\BB = 1}^6 \sum_{\I,\J = 1}^2 ({\cal X}_\I^\UL)^\dagger_\twoA
({\cal X}_\J^\UL)_\Bi \nn\\   
&& \qquad \times \Bigl\{ c_2(\mchI^2, \musqA^2, \musqB^2)
({\cal T}^\UL {{\cal T}^\UL}^\dagger)_{\A\BB} \, \dIJ
- c_2(\musqA^2, \mchI^2, \mchJ^2) ~ \dAB~ {\cal V}^*_{\I1}~ {\cal
V}_{\J1} \nn\\  
&& \qquad + \half~ \mchI~ \mchJ~ c_0(\musqA^2, \mchI^2, \mchJ^2)~ \dAB
~{\cal U}_{\I1}~ {\cal U}^*_{\J1} \Bigr\} \nn\\
&&\delta C_{9}^{' Z-peng,\tilde{\chi}} = -{2 \over  g_2^2  \KMangles}
\sum_{\A,\BB = 1}^6 \sum_{\I,\J = 1}^2 ({\cal X}_\I^\UL)^\dagger_\twoA
({\cal X}_\J^\UL)_\Bi \nn\\ 
&& \qquad \times \Bigl\{ c_2(\mchI^2, \musqA^2, \musqB^2)~
({\cal T}^\UL {{\cal T}^\UL}^\dagger)_{\A\BB} \, \dIJ
- c_2(\musqA^2, \mchI^2, \mchJ^2)~  \dAB~ {\cal V}^*_{\I1}~ {\cal
V}_{\J1} \nn\\  
&& \qquad + \half~ \mchI~ \mchJ~ c_0(\musqA^2, \mchI^2, \mchJ^2)~ \dAB~
{\cal U}_{\I1}~ {\cal U}^*_{\J1} \Bigr\} \label{kpieematchthree}
\end{eqnarray}

\item 
$\gamma$-penguin graphs with chargino loops:
\begin{eqnarray}
\delta C_{9,10}^{\gamma-peng,\tilde{\chi}} &=& 0 \nn\\ 
\delta C_{9}^{' \gamma-peng,\tilde{\chi}} &=& -{1 \over 4 g_2^2
\KMangles} \sum_{\A=1}^6 \sum_{\I=1}^2 
{\mWsq \over \musqA^2} ({\cal X}_\I^\UL)^\dagger_\twoA ({\cal X}_\I^\UL)_\Ai 
f_7\bigl( {\mchI^2 \over \musqA^2} \bigr)\label{kpieematchfour} 
\end{eqnarray}

\item 
Z-penguin graphs with neutralino loops:
\begin{eqnarray}
&& \delta C_{9,10}^{Z-peng,\tilde{\chi}^0} =\pm{1 \over 2 g_2^2 ~
\sin^2 \theta_W \KMangles}  
\sum_{\A,\BB=1}^6 \sum_{\I,\J=1}^4
({\cal Z}_\I^\DL)^\dagger_\twoA ({\cal Z}_\J^\DL)_\Bi \nn\\  
&& \quad \times \Bigl\{ c_2(\mneI^2, \mdsqA^2, \mdsqB^2) 
({\cal T}^\DR {{\cal T}^\DR}^\dagger)_{\A\BB} \, \dIJ 
- c_2(\mdsqA^2, \mneI^2, \mneJ^2)~ 
\dAB~ ({\cal N}^*_{\I3} {\cal N}_{\J3} - {\cal N}^*_{\I4} {\cal
N}_{\J4} ) \nn\\  
&& \quad - \half~ \mneI~ \mneJ~ c_0(\mdsqA^2, \mneI^2, \mneJ^2)~ 
\dAB~ ({\cal N}_{\I3} {\cal N}^*_{\J3} - {\cal N}_{\I4} {\cal
N}^*_{\J4} ) \Bigl\} 
\nn \\
&& \delta  C_{9}^{' Z-peng,\tilde{\chi}^0} = -{2 \over  g_2^2  \KMangles} 
\sum_{\A,\BB=1}^6 \sum_{\I,\J=1}^4
({\cal Z}_\I^\DL)^\dagger_\twoA ({\cal Z}_\J^\DL)_\Bi \nn\\  
&& \quad \times \Bigl\{ c_2(\mneI^2, \mdsqA^2, \mdsqB^2) 
({\cal T}^\DR {{\cal T}^\DR}^\dagger)_{\A\BB} \, \dIJ 
- c_2(\mdsqA^2, \mneI^2, \mneJ^2)~ 
\dAB~ ({\cal N}^*_{\I3} {\cal N}_{\J3} - {\cal N}^*_{\I4} {\cal
N}_{\J4} ) \nn\\  
&& \quad - \half~ \mneI~ \mneJ~ c_0(\mdsqA^2, \mneI^2, \mneJ^2)~ 
\dAB~ ({\cal N}_{\I3} {\cal N}^*_{\J3} - {\cal N}_{\I4} {\cal
N}^*_{\J4} ) \Bigl\}\label{kpieematchfive} 
\end{eqnarray}

\item 
$\gamma$-penguin graphs with neutralino loops:
\begin{eqnarray}
\delta  C_{9,10}^{\gamma-peng,\tilde{\chi}^0} &=& 0 \nn\\ 
\delta C_{9}^{' \gamma-peng,\tilde{\chi}^0} &=&  {1 \over 54 g_2^2 \KMangles} \sum_{\A=1}^6 \sum_{\I=1}^4
{\mWsq \over \mdsqA^2} ({\cal Z}_\I^\DL)^\dagger_\twoA ({\cal Z}_\I^\DL)_\Ai 
f_8\bigl( {\mneI^2 \over \mdsqA^2} \bigr) 
\label{kpieematchsix}
\end{eqnarray}

\item 
Z-penguin graphs with gluino loops:
\begin{eqnarray}
\delta C_{9,10}^{Z-peng,\tilde{g}} &=& \pm{4 g_3^2 \over 3 g_2^2 ~
\sin^2 \theta_W \KMangles}  
\sum_{\A,\BB=1}^6
({\cal T}^\DL)^\dagger_\twoA ({\cal T}^\DL)_\Bi~ c_2(\mgluino^2, \mdsqA^2, 
\mdsqB^2) ({\cal T}^\DR {{\cal T}^\DR}^\dagger)_{\A\BB} \nn\\ 
\delta C_{9}^{' Z-peng,\tilde{g}} &=& -{16 g_3^2 \over 3 g_2^2 \KMangles} 
\sum_{\A,\BB=1}^6 
({\cal T}^\DL)^\dagger_\twoA ({\cal T}^\DL)_\Bi~ c_2(\mgluino^2, \mdsqA^2, 
\mdsqB^2) ({\cal T}^\DR {{\cal T}^\DR}^\dagger)_{\A\BB} 
\label{kpieematchseven}
\end{eqnarray}

\item 
$\gamma$-penguin graphs with gluino loops:
\begin{eqnarray}
\delta C_{9,10}^{\gamma-peng,\tilde{g}} &=& 0 \nn\\
\delta C_{9}^{' \gamma-peng,\tilde{g}} &=&  {4 g_3^2 \over 81 g_2^2
\KMangles} \sum_{\A=1}^6 
{\mWsq \over \mdsqA^2} 
({\cal T}^\DL)^\dagger_\twoA ({\cal T}^\DL)_\Ai f_8\bigl( {\mgluino^2 \over 
\mdsqA^2} \bigr)
\label{kpieematcheight} 
\end{eqnarray}

\item 
Chargino box graph: 
\begin{eqnarray}
\delta C_{9,10}^{\tilde{\chi}-box} &=& \pm {\mWsq \over g_2^2 ~ \sin^2 \theta_W \KMangles} 
\sum_{\A=1}^6 \sum_{\I,\J=1}^2
({\cal X}_\I^\UL)^\dagger_\twoA ({\cal X}_\J^\UL)_\Ai d_2(\mchI^2,
\mchJ^2, \musqA^2,  
\msneutrinoone^2) {\cal V}^*_{\I1} {\cal V}_{\J1} \nn\\
\delta C_{9}^{' \tilde{\chi}-box} &=& 0 \label{kpieematchnine}
\end{eqnarray}

\item 
Neutralino box graphs: 
\begin{eqnarray}
&& \delta C_{9,10}^{\tilde{\chi}^{0}-box} = \pm 2 \;\delta
C_{9}^{' \tilde{\chi}^0-box} 
\pm {\mWsq \over 2 g_2^2 ~ \sin^2 \theta_W \KMangles} 
\sum_{\A=1}^6 \sum_{\I,\J=1}^4
({\cal Z}_\I^\DL)^\dagger_\twoA ({\cal Z}_\J^\DL)_\Ai \label{kpieematchten} \\
&& \quad \times \Bigl\{ d_2(\mneI^2, \mneJ^2, \mdsqA^2, \meslone^2) 
  ({\cal N}^*_{\I2} + \tan\theta_W {\cal N}^*_{\I1})({\cal N}_{\J2} + \tan\theta_W {\cal N}_{\J1}) \cr 
&& \qquad + \half~ \mneI~ \mneJ ~ d_0(\mneI^2, \mneJ^2, \mdsqA^2, \meslone^2) 
  ({\cal N}_{\I2} + \tan\theta_W {\cal N}_{\I1})({\cal N}^*_{\J2} + \tan\theta_W {\cal N}^*_{\J1}) \Bigr\} \nn\\
&& \delta C_{9}^{' \tilde{\chi}^0-box} =- {4 \mWsq \over g_2^2 \cos^2\theta_W \KMangles} 
\sum_{\A=1}^6 \sum_{\I,\J=1}^4
({\cal Z}_\I^\DL)^\dagger_\twoA ({\cal Z}_\J^\DL)_\Ai  \nn\\
&& \quad \times \Bigl\{ d_2(\mneI^2, \mneJ^2, \mdsqA^2, \meslfour^2) 
  {\cal N}^*_{\I1} {\cal N}_{\J1} 
  + \half~\mneI ~\mneJ~ d_0(\mneI^2,\mneJ^2,\mdsqA^2,\meslfour^2) {\cal
N}_{\I1} {\cal N}^*_{\J1} 
  \Bigr\}
\nn
\end{eqnarray}
\end{itemize} 
The one-loop integral functions which appear within these MSSM matching 
conditions are given in Appendix~\ref{app:susy}. The total Wilson
coefficients given above, namely $C_7^{tot}$ and
$C_{9,10}^{tot}$, are evaluated perturbatively at the $W$ scale
and then evaluated down to the renormalization scale $\mu\sim b_b$ by
the renormalization group equation (RGE). The details of this strong
interaction running are quit cumbersome, and we will not present them
here. However, the details of this RGE can be found in \cite{Ali:2002jg}.
%
\section{Analysis in supersymmetry \label{sec:MSSM-models}}
Having now at hand the general MSSM contribution to the $b\to s~
\ell^+ \ell^-$ transition, we will turn now to more restricted
framework of the MSSM.
We employ the following models to study the rare 
$B \to K^{*} \ell^+ \ell^-$ decays: 
\renewcommand{\theenumi}{(\arabic{enumi})}
\renewcommand{\labelenumi}{\theenumi}
\begin{enumerate}
\item
\label{item:Sugra}
Minimal supergravity (mSUGRA)~\citer{hw97,goto99},
\item\label{item:MFV}
Minimal flavor violating supersymmetric model (MFV) 
~\cite{giudice},
\item\label{item:EMFV}
Extended Minimal flavor violating supersymmetric model
(EMFV)~\cite{Ali:2001ej}. 
\end{enumerate}
The last of these models serves as a generic supersymmetric extension of 
the SM having non-CKM flavor violations.

             \subsection{SUGRA model}
%
In supergravity (SUGRA) model~\cite{Freedman:1976xh,Freedman:1976py}, 
which is a result of the unification of the supersymmetry transformations
with the space-time symmetries of general relativity, the soft SUSY
breaking terms are supposed to arise from a hidden sector of the
theory  which can only  communicate with the ordinary matter fields
through gravitational interactions.

Since gravity is flavour-blind, the breaking terms can be realized in
a minimal version~\cite{SUGRA}, introducing a common scalar mass
parameter $m_0$ and trilinear coupling $A_X$, a universal gaugino mass
parameter $M_{1/2}$ and the bilinear Higgs parameter $b$. This reduced
set of breaking parameters is called {\it minimal supergravity} (mSUGRA).

In the minimal SUGRA model, the soft SUSY breaking terms are assumed
to take the following universal structures at the Planck or GUT scale:

\begin{eqnarray}
(M^2_{\tilde{q}})_i^j &=&(M^2_{\tilde{u}^c})_i^j=(M^2_{\tilde{d}^c})_i^j
=(M^2_{\tilde{\ell}})_i^j =(M^2_{\tilde{e}^c})_i^j= m_0^2\,\delta_i^{~j} ~,
 \nn\\
  \Delta_1^2 &=& \Delta_2^2 ~=~ \Delta_0^2 ~,
\nn\\
  A_\D^{ij} &=& f_{D}^{ij} A_X m_0 ~, ~~ 
  A_\E^{ij} ~=~ f_{L}^{ij} A_X m_0 ~, ~~ 
  A_\U^{ij} ~=~ f_{U}^{ij} A_X m_0 ~,
\nn\\
m_{\tilde{g}} &=& m_{\tilde{\W}} ~=~ m_{\tilde{\BB}}~=~ M_{gX} ~.  
\label{softGUT}
\end{eqnarray}

In the minimal supergravity model the soft breaking parameters $m_0$ and 
$\Delta_0$ are assumed to be equal\footnote{ whereas in the nonminimal case we
treat the two as independent parameters}. 
With the above initial conditions we can solve the one-loop
RGEs for the SUSY breaking parameters and determine these parameters at
the electroweak scale \cite{RGE}. We also require that the electroweak
symmetry breaking occurs properly to give the correct $Z^0$ boson mass.

Scanning over the soft SUSY breaking parameter space in the range
$m_0\leq 600~{\textrm GeV}$, ${\Delta}_0\leq 600~{\textrm
GeV}$, $M_{gX}\leq 600~{\textrm GeV}$ and $|A_X|\leq 5$ for each fixed
value of $\tan\beta$, it turns out that 
the parameter space of this model may be decomposed into two qualitatively 
different regions:~\cite{Ali:1999mm,goto96}

\begin{itemize}  
\item
For small $\tan\beta$, say $\tan\beta \sim 3$, the sign of $\cse$ is the same as in the SM. Hence, no spectacular deviations from the 
SM can be expected in the $B \to K^* \ell^+ \ell^-$ decay mode. Given
the theoretical uncertainties in the SM estimates, it would be very
difficult to disentangle any SUSY effects for this scenario in this
decay~\cite{Ali:1999mm}.  
\item
For large $\tan\beta$, the situation is more interesting due to correlations
involving the branching ratio for $B \to X_s \gamma$, the mass of the 
lightest CP-even Higgs boson, $m_h$, and sign$(\mu_{susy})$. 
In this case, there are two branches for the solutions for $m_h$
and ${\cal B}(B \to X_s \gamma)$. The interesting scenario for
SUSY searches in $B \to K^{*} \ell^+ \ell^-$ is the one in which 
sign$(\mu_{susy})$ and $m_h$ admit $\cse$ to be
positive\footnote{ For example,  
this happens for $\tan\beta \geq 10$, in which case $m_h=(115$-$125)$ GeV and 
$\cse$ is positive and obeys the $B \to X_s \gamma$ bounds 
\cite{goto99}. Following the generic case shown earlier, one expects 
a constructive interference of the terms depending on $\cse$ and $\cn$ in 
the dilepton invariant mass spectra.}. 
\end{itemize} 
However, a large\footnote{ $\tan\beta= 30$ corresponds to\cite{goto99}: $r_7=-1.2,~~r_9=1.03,~~r_{10}=1.0 ~$.} $\tan\beta$ values can get a considerable impact on
the $b\to s~ \ell^+ \ell^-$ branching ratio:
\begin{itemize}  
\item
At the low-$s$ region the branching ratio $B\to K \mu^+ \mu^-$ could
be enhanced by about $30 \% $ compared to the SM one, nevertheless
this enhancement is difficult to disentangle from the non-perturbative
uncertainties attendant with the SM-distributions~\cite{Ali:1999mm}. 
\item
The $B \to K^* \mu^+ \mu^-$ dilepton mass distribution could
be enhanced by about $100 \%$  and this
is distinguishable from the SM-related theoretical
uncertainties~\cite{Ali:1999mm}. 
A very similar supersymmetric effects have been worked out for the
inclusive decays $B \to X_s \ell^+ \ell^-$ \cite{goto96}, where enhancements 
of ($50$-$100$)\% were predicted in the low-$s$ branching ratios.  
\end{itemize} 
Summarizing for the SUGRA theories, small $\tan\beta$ implies the sign
of $\cse$ being the same as in the SM. Hence, no spectacular
deviations from the SM can be expected in the $B \to K^* \ell^+
\ell^-$ decay mode. However, large $\tan\beta$ solutions lead to
$\cse$ being positive, and one expects an enhancement up to a factor
two in the dilepton mass distribution in $B \to K^* \ell^+
\ell^-$. This would be a drastic deviation from the SM, which cannot be fudged away due to non-perturbative effects.   

               \subsection{The MFV model\label{MFV}}
The minimal flavour violating (MFV) SUSY  model~\cite{giudice} is
based on the assumption of minimal flavor violation, which means that
all the genuine new sources of flavour changing transitions other than
the CKM matrix are switched off. Here, quarks and squarks are aligned
so there is no  flavor-changing
$q-\tilde{q}^{\prime}-(\tilde{Z},\tilde{\gamma},\tilde{g})$ vertex and
the charged one, $d-\tilde{u}-\tilde{\chi}^\pm$, is governed by the CKM matrix.
As a consequence, in this model neutralino-down-squark and gluino-down-squark
graphs do not contribute to either $b \to s \gamma$ or $b \to s \ell^+ 
\ell^-$ transitions. In addition to the charged Higgs-top graphs,
chargino-up type squarks loops with a light stop $\tilde{t}_1$, and
the $W^\pm$-top quark loops, present in the SM, give the dominant
contribution. While not holding generally, the assumptions in the
MFV-SUSY model are valid over an important part of the MSSM parameter space. 
%

In order to derive the new physics range contributions to $C_9$ and
$C_{10}$, we scan over the parameter space in the range $78.6~{\textrm
GeV} \leq m_{H^{\pm}}\leq 1~{\textrm TeV}$, $0 \leq m_{\tilde W},
|\mu_{susy}|\leq 1~{\textrm TeV}$ and $m_{\tilde q}= m_{\tilde t_2}
\geq 90~{\textrm GeV}$, where $m_{\tilde q}$ denotes the (degenerate)
masses of other than top squarks, and $m_{\tilde \n} \geq 50~{\textrm GeV}$. 
We reject too light charginos, demanding $m_{\chi_i^\pm}\geq
90~{\textrm GeV}$ and also taking the $B \to X_s \gamma$ experimental
constraint. We have chosen a stop mixing angle
$|\theta_{\tilde{t}}|\leq \pi/2$, i.e. the light stop $\tilde{t}_2=
-\sin\theta_{\tilde{t}} \tilde{t}_L+ \cos\theta_{\tilde{t}} \tilde{t}_R$ is  almost left handed. 
\begin{figure}[t]
\begin{center}
\psfrag{R}{ $r$}
\epsfig{file=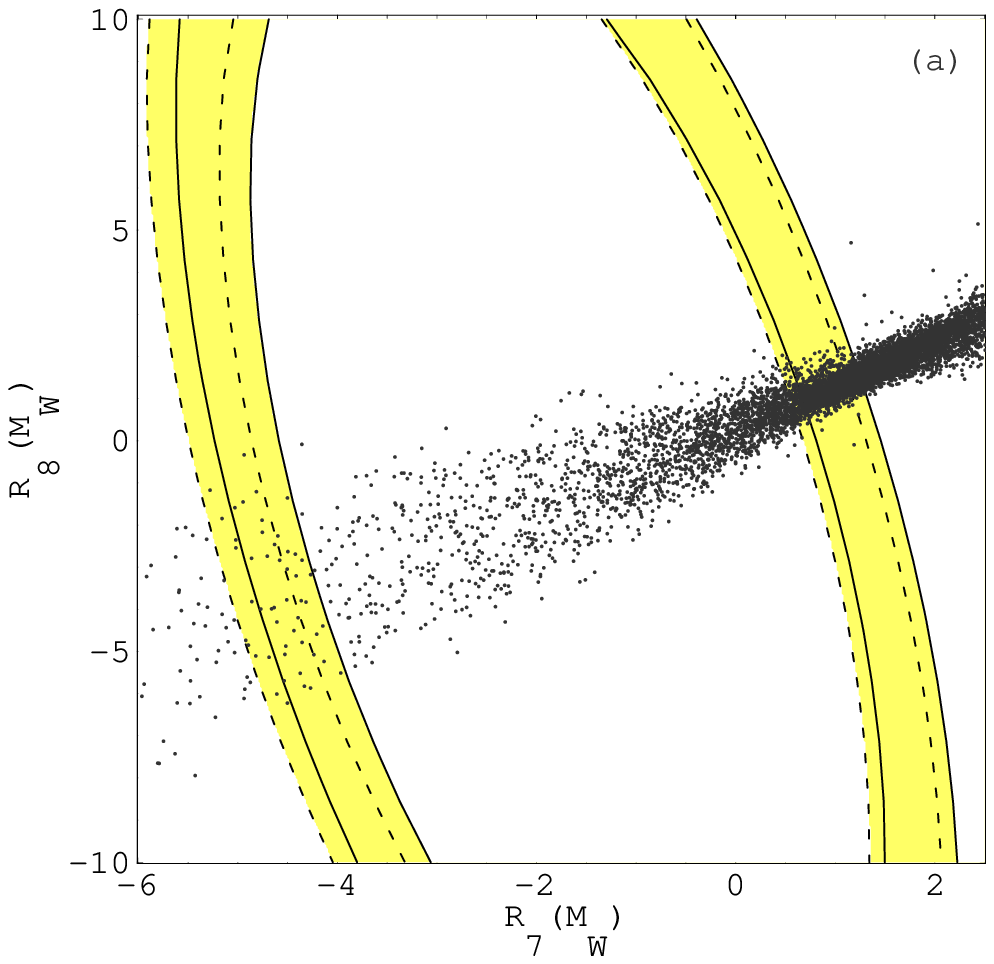,width=0.48\linewidth}
\hspace*{.2cm}
\epsfig{file=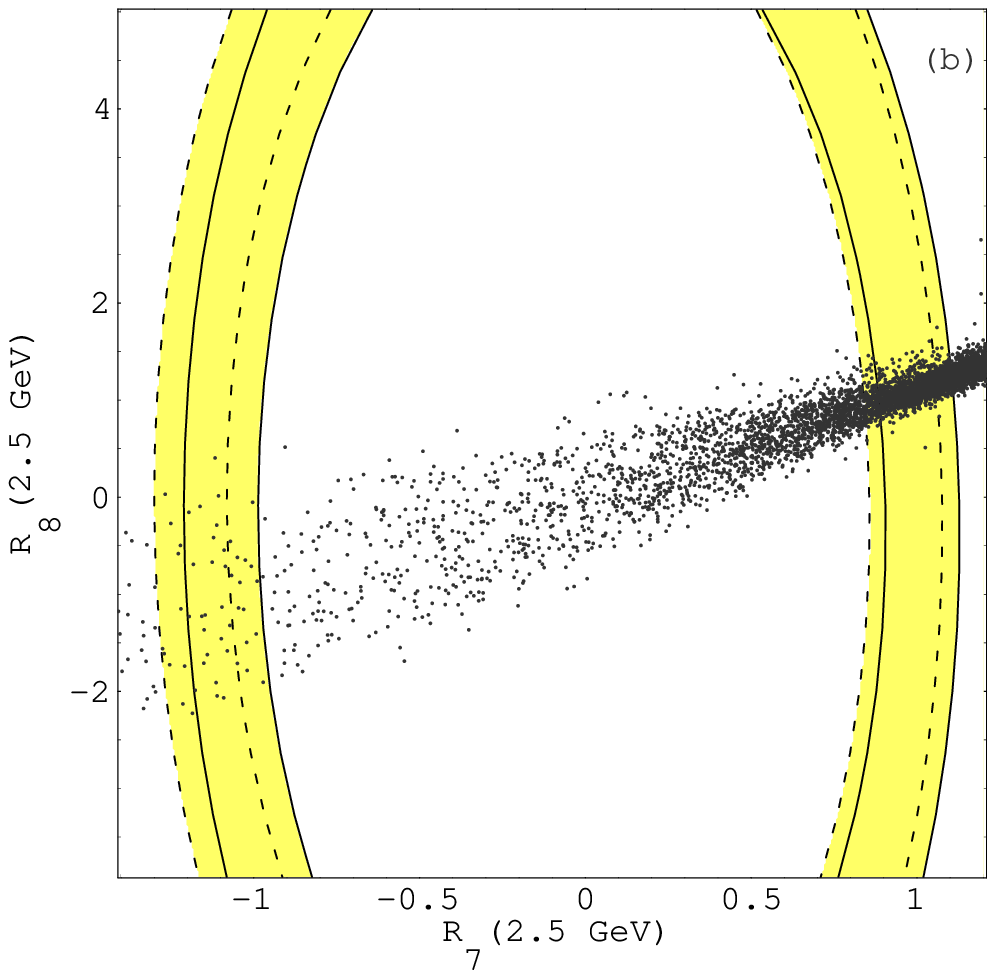,width=0.48\linewidth}
\caption{\it $\cl{90}$ bounds in the $[R_7 (\m), R_8(\m)]$ plane
following from the world average $B\to X_s \g$ branching ratio for
$\mu=m_W$ (left-hand plot) and $\mu=2.5$ GeV (right-hand plot).
Theoretical uncertainties are taken into account. The solid and dashed
lines correspond to the $m_c = m_{c,pole}$ and $m_c =
m_{c}^{\overline{MS}} (\m_b)$ cases respectively. The scatter points
correspond to the expectation in MFV models (the ranges of the SUSY
parameters are specified in the text)~\cite{Ali:2002jg}.}
\label{bsg}
\end{center}
\end{figure}
In order to produce bounds that can be compared with the model
independent allowed regions plotted in Fig~(\ref{fig:total}), it turns
out that the surviving SUSY points can be divided to two
sets~\cite{Ali:2002jg}: 
\begin{itemize}  
\item
For small $\tan\beta$, for which we again take $\tan\beta=2.3$, we find that 
the ratio $R_7$ remains positive, i.e. $\cse <0$, and lies within the 
experimentally allowed 
bounds from $B \to X_s \gamma$, and the corresponding bounds on the
semileptonic coefficients are in the range 
$-0.2< C_9^{MFV}(\m_W)<0.4$ and $-1.0<C_{10}^{MFV}(\m_W)<0.7$.
\item
For large $\tan\beta$, taken to be 50, $\cse$ changes 
sign ($R_7 <0$). The corresponding bounds on $C_9^{MFV}$ and $C_{10}^{MFV}$
tend to be in the range $-0.2< C_9^{MFV}(\m_W)<0.3$ and
$-0.8<C_{10}^{MFV}(\m_W)<0.5$. 
\end{itemize}  
The above discussion applies to any supersymmetric model
with flavour universal soft-breaking terms, such as
mSUGRA MSSM and gauge-mediated supersymmetry
breaking models. Beyond-the-SM flavour violations in such models are
induced only via renormalization group running, and are tiny.
Before ending this subsection, let us discuss the impact of $b\to s
\gamma $ on MFV models in varying the MFV SUSY parameters: 
\begin{itemize}  
\item
The strong correlation between the values of the Wilson coefficients
$C_7$ and $C_8$, shown in Fig.~\ref{bsg}. In fact, the
SUSY contributions to the magnetic and chromo--magnetic coefficients
differ only because of colour factors and loop-functions. 
\item
The dependence of the charged Higgs contribution to magnetic
coefficient at the scale $\m_b$, can be seen from Fig.~12 in
ref~\cite{Ali:2002jg}. It turns out that with a specific scenario, it
is possible to obtain lower bounds on some SUSY particles. 
\item
The chargino contributions on $C_7(\m_b)$, show very strong
consequences\footnote{ As it has been argued in\cite{Ali:2002jg}, 
one can exploit the $\theta_{\tilde t}$ and $\tan \beta$ dependence
since (for non 
negligible values of the stop mixing angle) the chargino contribution
is essentially proportional to $\sin \theta_{\tilde t} \tan \beta$.} 
on the $b\to s\g$ transition as it has been shown in~\cite{Ali:2002jg}.
Assuming for instance $m_{\tilde t_2} = m_{\chi}= 500\; \gev$ the chargino
contributions (normalized to $(\sin \theta_{\tilde t} \tan \beta)$) 
is of order 0.2. If one then allows for larger values of the stop
mixing angle and of $\tan \beta$, the contribution can easily 
violate the $b\to s\g $ constraint by more than one order of magnitude
(e.g. for $\sin \theta_{\tilde t} = 0.5 $ and $\tan \beta=50$ one obtains
something of order 6 that is orders of magnitude above the current
limit)~\cite{Ali:2002jg}.
\end{itemize}
\begin{figure}[t]
\begin{center}
\epsfig{file=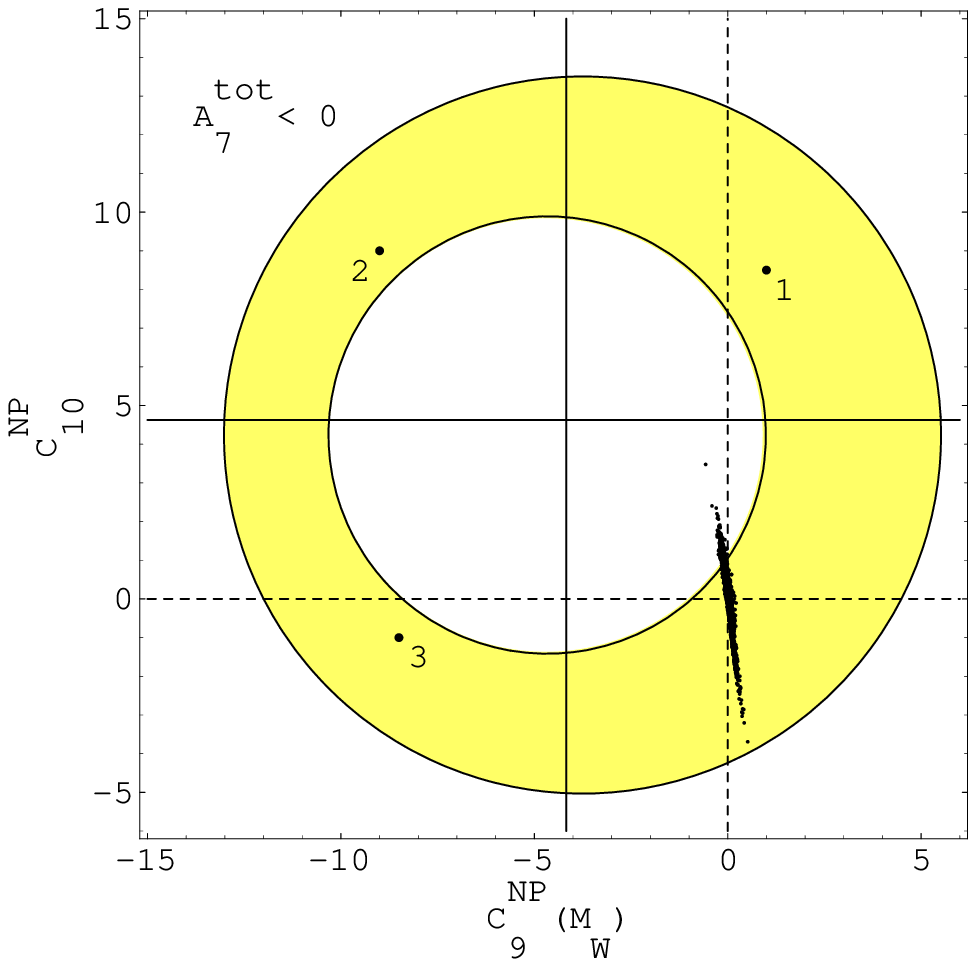,width=0.4\linewidth}
\hskip 0.5cm 
\epsfig{file=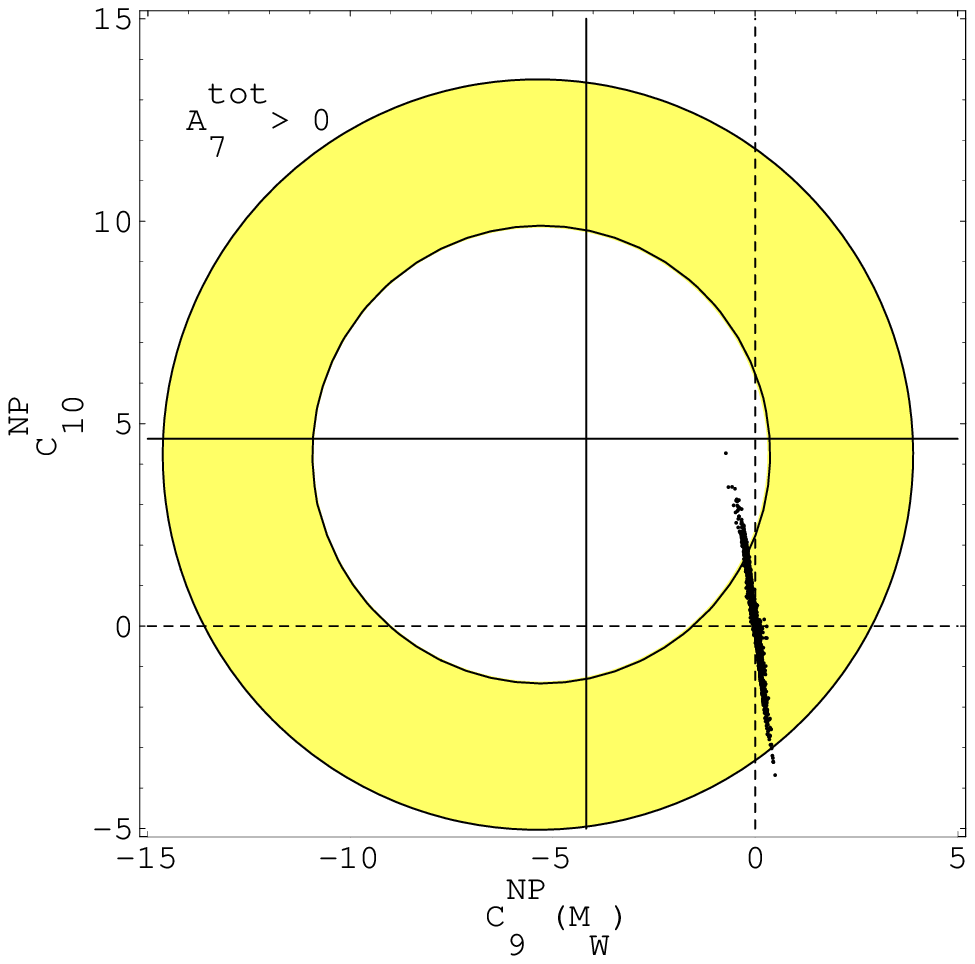,width=0.4\linewidth}
\caption{\it {\bf NNLO Case.} Superposition of all the
constraints. The plots correspond to the $A_7^{\rm tot}(2.5 \;
\gev)<0$ and $A_7^{\rm tot}(2.5 \; \gev) >0$ case, respectively. The
points are obtained by means of a scanning over the EMFV parameter
space and requiring the experimental bound from $B\to X_s \g$ to be
satisfied~\cite{Ali:2002jg}.}
\label{fig:total}
\end{center}
\end{figure}
Up to this point we have been looking at the minimal case where all
genuine sources of flavour changing transitions in the MSSM are
attributed to the CKM matrix elements. However, an interesting question
to ask is whether there might be other, non-universal scenarios, where
new flavour changing transitions (other than the CKM mixing elements)
could occur. A model which incorporates these features is the
so-called Extended-MFV (EMFV) model~\cite{Ali:2001ej}, discussed in
the following subsection. 
        \subsection{The Extended-MFV model\label{EMFV}}
The Extended-MFV (EMFV) model which is a generalization of the
MFV-model using the language of minimal insertion approximation (MIA)
\cite{hkr86} in a supersymmetric context. Its main assumption lies in
the fact that new sources of flavour changing transitions other than
the CKM mixing elements could occur.

EMFV-models are based on the heavy squarks and gluino assumption.  
Moreover, the charged Higgs and the lightest chargino and stop
masses are required to be heavier than $100 \; \gev$ in order to
satisfy the lower bounds from direct searches.  The rest of the SUSY
spectrum is assumed to be almost degenerate and heavier than $1
\;\textrm{TeV}$. The lightest stop is almost right-handed and the
stop mixing angle (which parameterizes the amount of the left-handed
stop $\tilde t_L$ present in the lighter mass eigenstate) turns out to
be of order $\sim 10\%$; for definiteness we will take
$|\theta_{\tilde t}| \leq \pi/10$. 

The assumption of a heavy ($\ge 1$ TeV) gluino totally suppresses any
possible gluino--mediated SUSY 
contribution to low energy observables. Note that even in the presence
of a light gluino (i.e. $m_{\tilde g} \simeq O(300 \; \gev)$) these
penguin diagrams remain suppressed due to the heavy down squarks
present in the loop. On the other hand, the presence of only a single light
squark mass eigenstate (out of twelve) has strong consequences due to
the rich flavour structure which emerges from the squark mass
matrices. Adopting the MIA-framework~\cite{hkr86}, all FC
transitions which are not generated by the CKM mixing matrix are
proportional to the properly normalized off-diagonal elements of the
squark mass matrices:
\begin{eqnarray}
(\delta_{ij})^{U,D}_{AB} \equiv {(M^2_{ij})^{\U,\D}_{AB} \over
M_{\tilde q_i} M_{\tilde q_j}}
\label{miadef}
\end{eqnarray}
where $i,j=1,2,3$ and $A,B=L,R$. In this approach, some remarks are in order:
\begin{itemize}  
\item
The only sizable contributions arise from the inserted mass insertions involving the light stop.
\item
All the other diagrams require necessarily a loop with at least
two heavy ($\geq 1 \; \textrm{TeV}$) squarks and are therefore automatically
suppressed. 
\end{itemize}
This leaves us with only two unsuppressed flavour changing
sources other than the CKM matrix, namely the mixings $\tilde u_L -
\tilde t_2$ (denoted by $\delta_{\tilde u_L \tilde t_2}$) and $\tilde c_L
- \tilde t_2$ (denoted by $\delta_{\tilde c_L \tilde t_2}$).  We note that
$\delta_{\tilde u_L \tilde t_2}$ and $\delta_{\tilde c_L \tilde t_2}$ are mass
insertions extracted from the up-squarks mass matrix after the
diagonalization of the stop system and are therefore linear
combinations of $(\delta_{13})^U_{LR}$, $(\delta_{13})^U_{LL}$ and of
$(\delta_{23})^U_{LR}$, $(\delta_{23})^U_{LL}$, respectively. The insertions
relevant to our discussion are normalized as follows:
\beq
\label{delta}
\delta_{\tilde u(\tilde c)_L \tilde t_2} 
\equiv {M^2_{\tilde u(\tilde c)_L \tilde t_2}
\over M_{\tilde t_2} M_{\tilde q}} { |V_{td(s)}| \over V_{td(s)}^*} 
\; .  
\eeq 
The phenomenological impact of $\delta_{\tilde t_2 \tilde u_L}$ has
been studied in~\cite{Ali:2001ej} and its impact on the $b\to s \g$ and $b\to s
\ell^+ \ell^-$ transitions is indeed negligible. Therefore, we are
left with the MIA parameter $\delta_{\tilde t_2 \tilde c_L}$ only.
Thus, the SUSY parameter space that we have to deal with are the same
as the MFV-model with  MIA parameter $\delta_{\tilde t_2 \tilde c_L}$.
In order to explore the region in the $[C_9^{NP},C_{10}^{NP}]$ plane
(where $C_{9,10}^{NP}$ are the sum of MFV and MI contributions) that
is accessible to these models, one has to apply the $b\to s \g$
constraint on the EMFV parameter space, which are the same as the
MFV-ones apart from  $|\theta_{\tilde t}| \leq \pi/10$ and
$|\delta_{\tilde t_2 \tilde c_L}| \leq1$.
The surviving points are shown in Fig.~\ref{fig:total} together with the
model independent constraints. To get them, one has to use the integrated branching ratios to put constraints on the effective coefficients. This procedure allows multiple solutions, which can be disentangled from each other only
with the help of both the dilepton mass spectrum and the
forward-backward asymmetry. Only such measurements would allow us to
determine the exact values and signs of the Wilson coefficients $C_7$,
$C_9$ and $C_{10}$. 
\section{The ratios $R_-(s)$ and $R_0(s)$ as probes of New Physics 
in $B \rightarrow  K^{*} \ell^{+}  \ell^{-}$ 
\label{sec:ratio-MSSM}}
In order to look for new physics in $B \rightarrow  K^{*} \ell^{+}
\ell^{-}$,  we propose to study  the ratios $R_{0}(s)$ and $R_{-}(s)$,
introduced in the previous section. As well known, new physics can distort
the dilepton invariant mass spectrum and the forward-backward asymmetry 
in a non-trivial way. 
\begin{figure}[t]
\begin{center}
\psfrag{a}{ $s\ (GeV^2)$}
\psfrag{b}{\hskip -2.5cm $R_{-}(s)/ 10^{-2}$}
\epsfig{file=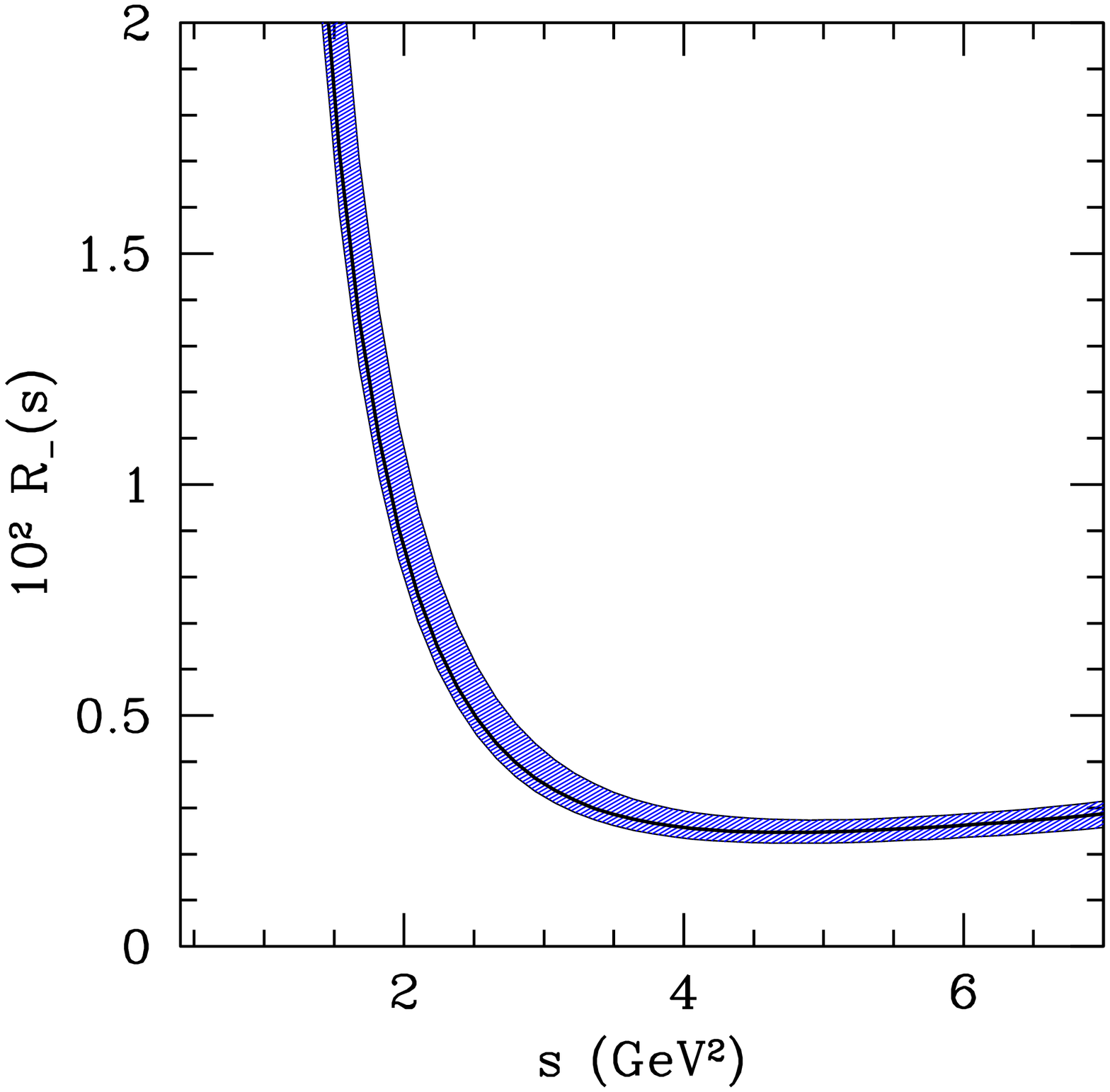,width=0.48\linewidth}
\hspace*{.2cm}
\epsfig{file=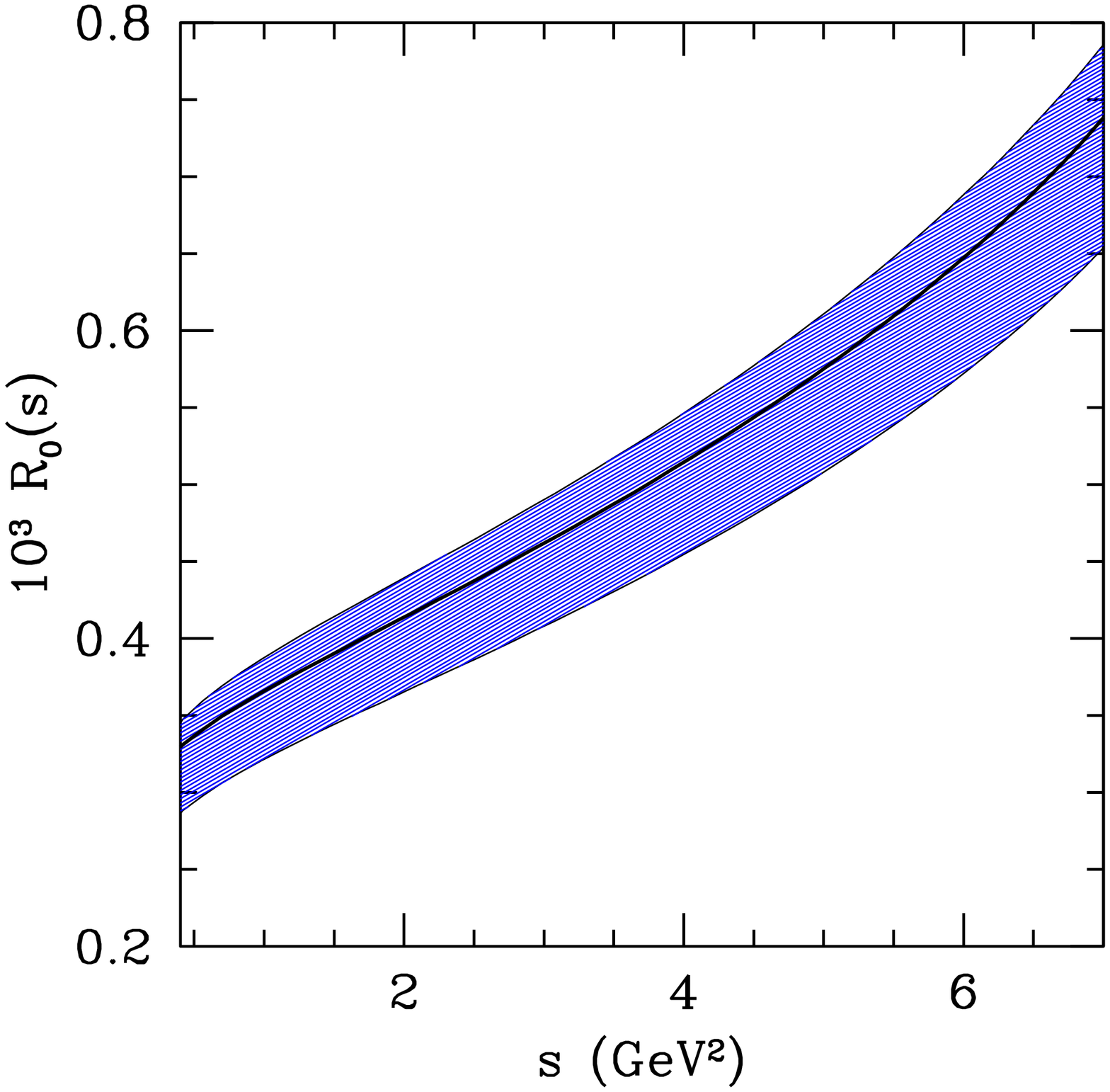,width=0.48\linewidth}
\caption{\it 
The Ratios $R_{-}(s)$ (left-hand plot) and $R_{0}(s)$ (right-hand
plot) in the Standard Model with
$R_b=0.094$, $\zeta_{SU(3)}=1.3$, 
$\xi^{(K^*)}_{{\perp}}(0)=0.28$ and in SUGRA, with $(r_7,\ r_8)=(1.1,\
1.4)$.
The SM and the SUGRA contriburions are  represented respectively by the shaded
area and the solid curve. The shaded area depicts the theoretical
uncertainty on $\zeta_{SU(3)}=1.3 \pm 0.06$  and on
$\xi^{(K^*)}_{{\perp}}=0.28\pm 0.04$~\cite{Safir:SUSY02}.}
\label{R-SMetSUGRA}
\end{center}
\end{figure}

To illustrate generic SUSY effects in ${B \rightarrow  K^{*}
\ell^{+} \ell^{-}}$, 
we note that the Wilson coefficients $C^{\bf eff}_7$, $C^{\bf eff}_8$,
$C_9$ and $C_{10}$ receive additional contributions from the supersymmetric
particles. We incorporate these effects by assuming that the ratios of the
Wilson coefficients in these theories and the SM deviate from 1. These ratios
for $k= 7, 8, 9, 10$ are defined as follows\footnote{ These ratios
have been introduced in Eq.~(\ref{eq:Ri}), and we change slightly their
notations by $r_i$ instead of $R_i$ and keep this later notation for
our helicity ratios.}:
\begin{eqnarray}
r_{k}(\mu) = {C_{k}^{SUSY}\over C_{k}^{SM}}~.
\label{rk} 
\end{eqnarray}
%

\begin{figure}[t]
\begin{center}
\psfrag{a}{ $s\ (GeV^2)$}
\psfrag{b}{\hskip -2.5cm $R_{0}(s)/ 10^{-3}$}
\epsfig{file=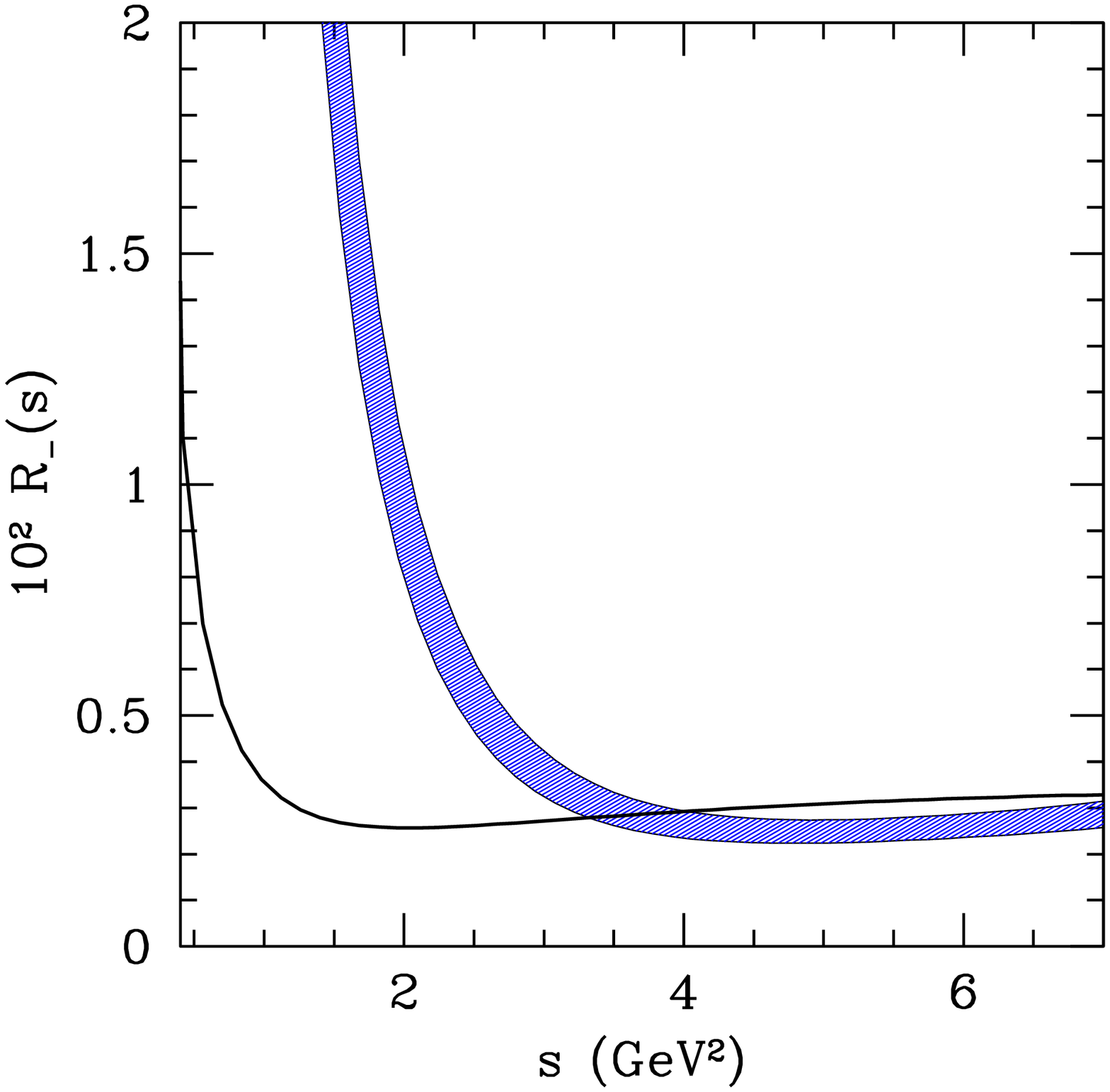,width=0.48\linewidth}
\hspace*{.2cm}
\epsfig{file=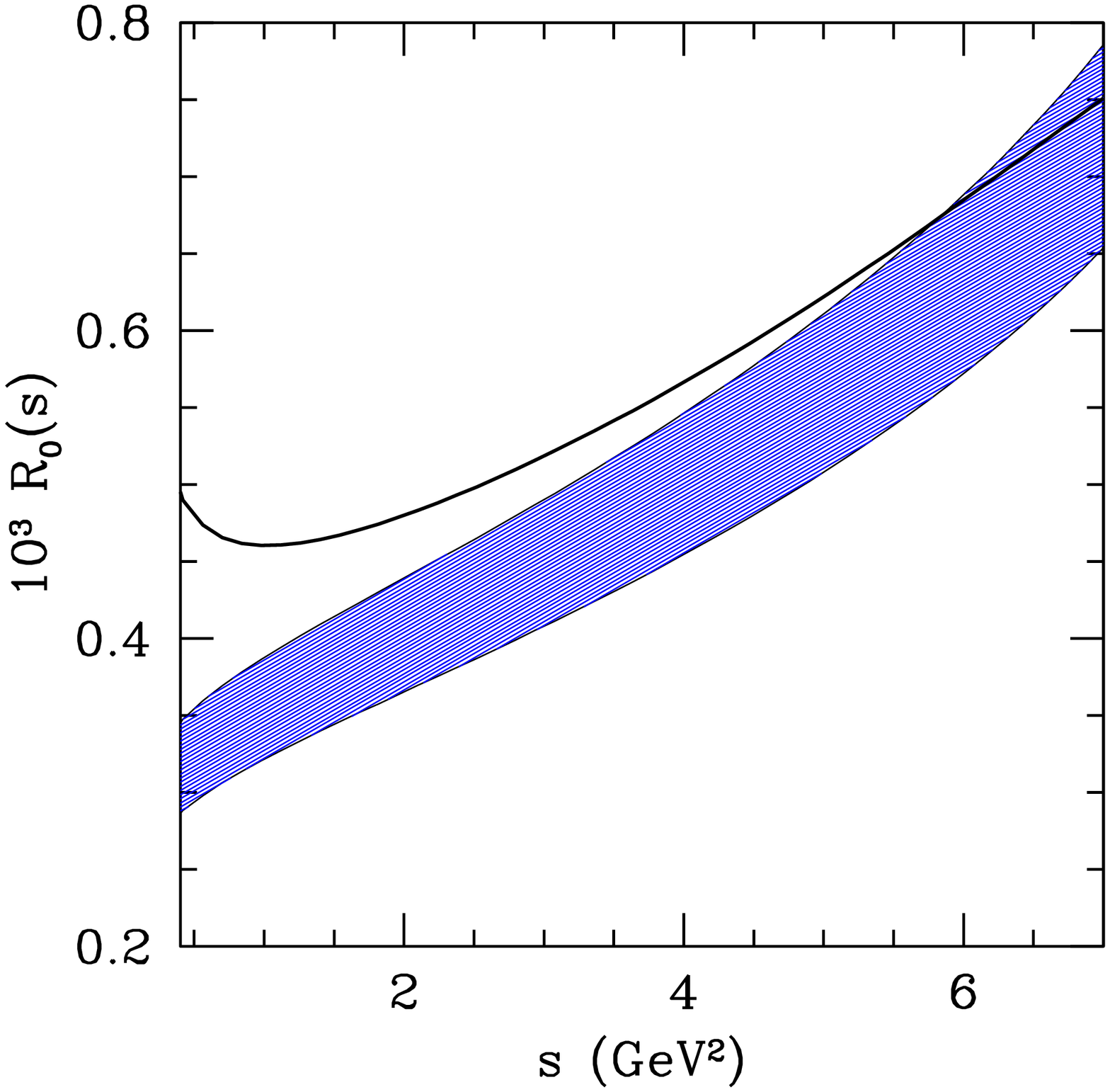,width=0.48\linewidth}
\caption{\it The Ratios $R_{-}(s)$ (left-hand plot) and $R_{0}(s)$ (right-hand
plot) in the Standard Model with
$R_b=0.094$, $\zeta_{SU(3)}=1.3$, 
$\xi^{(K^*)}_{{\perp}}(0)=0.28$ and in SUGRA, 
with $(r_7,\ r_8)=( -1.2,\ -1)$.
The SM and the SUGRA contributions are  represented respectively by the shaded
area and the solid curve. The shaded area depicts the theoretical
uncertainty on $\zeta_{SU(3)}=1.3 \pm 0.06$  and on
$\xi^{(K^*)}_{{\perp}}=0.28\pm 0.04$~\cite{Ali:2002qc}.}
\label{R0SMetSUGRA}
\end{center}
\end{figure}
\noi They depend on the renormalization scale (except for $C_{10}$),
for which we take $\mu =m_{b,pole}$. For the sake of illustration,
we use representative values for the large-$\tan \beta$  SUGRA model,
in which the ratios $r_7$ and $r_8$ actually keep or change their signs. The
supersymmetric effects on the other two Wilson coefficients $C_9$ and
$C_{10}$ are generally small in the SUGRA models, leaving $r_9$ and $r_{10}$
practically unchanged from their SM value. To be specific, we take the
two allowed scenarios\footnote{ We thank Enrico Lunghi for providing us with these numbers.}
\begin{eqnarray}
r_{7} = -1.2,\ \ r_{8} = -1,\ \    \ r_{9} = 1.03,\ \ r_{10} = 1.0,
\label{rnegativ}
\end{eqnarray} 
\noi and 
\begin{eqnarray}
r_{7} = 1.1,\ \ r_{8} = 1.4,\ \   \ r_{9} = 1.03,\ \ r_{10} = 1.0.
\label{rpositiv} 
\end{eqnarray} 
In Figs.~(\ref{R-Vub}), (\ref{R0Vub}), (\ref{R-SMetSUGRA}) and
(\ref{R0SMetSUGRA}), we present a comparative study of the SM and
SUGRA partial distribution for $H_{-}$ and $H_{0}$, respectively. In
doing this, we also show the attendant theoretical uncertainties for
the SM, worked out in the LEET approach~\cite{Ali:2002qc}. 
For these distributions, we have used the form factors from 
\cite{Ali:1999mm} with the SU(3)-symmetry breaking parameter taken in the
range $\zeta_{SU(3)}=1.3\pm 0.06$.

From Fig.~(\ref{R-SMetSUGRA}) 
, where
$r_{k}>0,~~ (k= 7,8, 9, 10)$, it is difficult to work out a signal of new
physics from the SM picture. There is no surprise to be expected, due to the
fact that in these scenario the corresponding ratio $r_k$ is
approximatively one, which makes the SUGRA picture closer to the SM one.
However, Fig.~(\ref{R0SMetSUGRA}) with $(r_7,\ r_8)<0$ illustrate
clearly that despite non-perturbative uncertainties, it is possible,
in principle,  in the low $s$ region  to 
distinguish between the SM and a SUGRA-type models, provided the ratios
$r_k$ differ sufficiently from 1. 
%
\section{Summary and Outlook \label{sec:MSSM-Summary}}
In this chapter we have presented a phenomenological profile of the
semileptonic rare $B \to K^* \ell^+ \ell^-$ decay in the context of
supersymmetric theories. Considering the straightforward
supersymmetrization version of the SM, the so-called MSSM, we have
reviewed in details its contribution to the $b \to s \ell^+ \ell^-$
process. We illustrate the constraints on the Wilson coefficients
$C_7$, $C_8$, $C_9$ and $C_{10}$ appearing in the effective
Hamiltonian formalism, that the current data on rare $B$ decays
implies in the context of minimal flavour violating model and in more
general scenarios admitting additional flavour changing
mechanisms. Finally, incorporating these supersymmetric effects on the
corresponding Wilson coefficients, we have shown their
phenomenological impact on the ratios $R_{0}(s)$ and $R_{-}(s)$ as
probes of new physics.

Our studies can be reported as follows~\cite{Ali:2002qc}~:
\begin{itemize}  
\item We have shown within the MSSM framework the complete SUSY
contributions to the $b \to s \ell^+ \ell^-$ decay. Beyond the
$W$-exchange in the SM, four other classes contribute to this process,
namely the $Higgs$-exchange, $Chargino$-exchange, $gluino$-exchange and the $Neutralino$-ones.
\item Using the current data constraints on rare $B$ decays, we have
discussed their phenomenological impacts on the Wilson coefficients
$C_7$, $C_8$, $C_9$ and $C_{10}$. It turns out that present
experimental measurements leave considerable room for beyond-the-SM
contributions , especially, to the allowed region in the $[C_9,C_{10}]$
plane. However, in the SUGRA models they are practically unchanged from their SM value.
\item 
In order to look for new physics in $B \rightarrow  K^{*} \ell^{+}
\ell^{-}$,  we propose to study  the ratios $R_{0}(s)$ and $R_{-}(s)$.
For the sake of illustration, we use representative values for the
large-$\tan \beta$  SUGRA model with two scenarios, namely $(r_7,\ r_8)<0$
and $(r_7,\ r_8)>0$ . We have noticed that despite non-perturbative
uncertainties, it is possible when $(r_7,\ r_8)<0$ , in principle,  in
the low $s$ region  to distinguish between the SM and a SUGRA-type models, provided the ratios $r_k$ differ sufficiently from 1. 
\end{itemize}

\chapter{Summary \& Future \label{chap:out}}
While waiting for the completion of the {\it second-generation} 
experiments at hadron colliders, $\mathrm{BTeV}$ (Fermilab) and LHCb
(CERN), $B$-physics is among the most active and promising fields in
recent particle physics. Its importance lies in the deeper
understanding of the Standard Model and particle physics in general.
The futur investigation of $b$ decays (such as rare $B$-decays, two
leptonic $B$-decays, $\cdots$ ) at the $B$-factories, namely
BABAR and BELLE, and at hadron colliders will probe the flavour sector
of the SM with unprecedented precision, {\it and may be} reveal new
physics effects. 

Rare $B$ decays involving flavour-changing-neutral-current (FCNC)
transitions, such as $b\to s \gamma$ and $b \to s \ell^+ \ell^-$, have
received a lot of theoretical interest \cite{Greub:1999sv}. 
Especially, after the first measurements of the radiative decay $B \to
X_s \gamma$ were reported by the CLEO collaboration \cite{Alam:1995aw} in 1995 
and recently the first measurements of the semileptonic
rare $B$-decays reported in the inclusive (exclusive) $B \to X_s
(K, K^*) \ell^+ \ell^-$  by the BELLE collaboration~\cite{bellebsg}
(BABAR collaboration~\cite{babarbsll}). With increased statistical
power of experiments at the $B$-factories 
in the next several years, the decays discussed above  and 
related rare $B$ decays will be measured very precisely. We summarize 
the projections for improvement in the experimental contributions to
the precision of CKM matrix elements $V_{ub}$, $V_{cb}$ and
$V_{ts}$~\cite{SuperBaBar} in Table~\ref{tab:CKMexp} while in 
Table~\ref{tab:exp}~\cite{SuperBaBar} the decay reach of the rare
$B$-decays is given.

\begin{table}[t]
\renewcommand{\arraystretch}{1}
\begin{center}
\begin{tabular}{|c||c|c|c|c|}
\hline\hline
$V_{ij}$ & Experimental & $\sigma~(\%)$ & $\sigma~(\%)$ & $\sigma~(\%)$   \\ 
         & Measurement  & 2001     & 2006     & 2011        \\
         &              & stat/sys & stat/sys & stat/sys    \\
\hline\hline
$V_{ub}$& ${\mathcal B}(B\to \rho~\ell \nu_{\ell})$ &$4.3/8$
  &$8.6/2.4$ &$1.4/2.4$\\
& ${\mathcal B}(B\to u~\ell \nu_{\ell})$ &$3.4/16$
  &$4.0/2.4$ &$2.8/2.4$\\
\hline
$V_{cb}$& ${\mathcal B}(B\to D~\ell \nu_{\ell})$ &$3.1/4$
  &$0.4/2$ &$0.10/1$\\
& ${\mathcal B}(B\to c~\ell \nu_{\ell})$ &$2.5/2$
  &$0.3/1$ &$0.07/0.5$\\
\hline
$V_{ts}$& $\Delta m_{s}$ & & &\\
\hline\hline
\end{tabular}
\end{center}
\caption{\it Projections for improvement in the experimental
  contributions to the precision of CKM matrix elements $V_{ub}$,
  $V_{cb}$ and $V_{ts}$ \cite{SuperBaBar}.
\label{tab:CKMexp}}
\end{table}
%

Within this thesis,  we have reported an
$O(\alpha_s)$-improved analysis of the various helicity components
in the decays $B \to K^* \ell^+ \ell^-$ and $B \to \rho \ell \nu_\ell$,
carried out in the context of the Large-Energy-Effective-Theory. Our studies can be summarized as follows~\cite{Ali:2002qc,Safir:SUSY02}~:
\begin{itemize}  
\item
The underlying symmetries in the large energy limit lead to an enormous
simplification as they reduce the number of independent form factors
in these decays. The LEET-symmetries are broken by QCD
corrections, and we have calculated the helicity components
implementing the $O(\alpha_s)$ corrections. 
\item
The results presented here
make use of the form factors calculated in the QCD sum rule approach. 
The LEET form factor $\xi_\perp^{(K^*)}(0)$ is constrained by current data on $B
\to K^* \gamma$. 
\item
As the theoretical analysis is restricted to the lower
part of the dilepton invariant mass region in $B \to K^* \ell^+ \ell^-$,
typically $s < 8$ GeV$^2$, errors in this form factor are
not expected to severely limit theoretical precision. This implies that
distributions involving the $H_{-}(s)$ helicity component can be
calculated reliably. Precise measurements of the two LEET form factors
$\xi_{\perp}^{(\rho)}(s)$ and $\xi_{\parallel}^{(\rho)}(s)$ in
the decays $B \to \rho \ell \nu_\ell$ can be used to largely reduce the
residual model dependence. 
\item
With the assumed form factors, we have worked
out a number of single and double (Dalitz) distributions in $B \to \rho
\ell \nu_\ell$, which need to be confronted with data. 
\item
An analysis of the
decays $B \to K^* \ell^+ \ell^-$ is also carried out in the so-called
transversity basis. We have compared the LEET-based amplitudes in this
basis with the data currently available on $B \to K^* J/\psi(\to \ell^+
\ell^-)$ and find that the short-distance based transversity amplitudes
are very similar to their long-distance counterparts.  
\item
We also show the
$O(\alpha_s)$ effects on the forward-backward asymmetry, confirming
essentially the earlier work of {\it Beneke, Feldmann and Seidel}
\cite{Beneke:2001at}. 
\item
Combining the analysis of the decay modes $B \to K^* \ell^+ \ell^-$
and $B \to \rho \ell \nu_\ell$, we show that 
the ratios of differential decay rates involving definite helicity states, 
$R_{-}(s)$ and $R_{0}(s)$, can be used for testing the SM precisely.
We work out the dependence of these ratios on the CKM matrix elements
$\vert V_{ub}\vert/\vert V_{ts}\vert$. We have also analyzed possible
effects on these 
ratios from New Physics contributions, exemplified by representative
values for the effective Wilson coefficients in SUGRA models. 
\end{itemize}
 
The main thrust of this work lies, however, on showing that the currently
prevailing theoretical uncertainties on the SM distributions in $B \to K^*
\ell^+ \ell^-$ can be largely reduced by using the LEET approach and data
on $B \to K^* \gamma$ and $B \to \rho \ell \nu_\ell$ decays. 

Finally, we remark that the current experimental limits on $B \to
(X_s, K, K^*) \ell^+ \ell^-$ decays~\citer{bellebsll,Anderson:2001nt}
are already probing the SM-sensitivity. With the integrated luminosities   
over the next couple of years at the $B$-factories, the
helicity analysis in  $B \to \rho \ell \nu_\ell$
and $B \to K^* \ell^+ \ell^-$ decays presented here can be
carried out experimentally. This work will help the search for flavour
changing neutral current $B 
\to K^* \ell^+ \ell^-$ and in particular, will contribute
to precise determinations of the LEET form factors, and the CKM matrix
elements $\vert V_{ub} \vert/\vert V_{ts} \vert$ using as well the $B
\to \rho \ell \nu_{\ell}$ decay in forthcoming $B$-facilities.

\renewcommand{\arraystretch}{0.8}
\begin{table}[t]
\renewcommand{\arraystretch}{1}
\begin{center}
\begin{tabular}{|c||c||c|c|c||c|c|}
\hline\hline
&& \multicolumn{3}{c||}{Hadron Collider Experiments} & \multicolumn{2}{|c|}{$e^+e^-$ $B$ Factories}\\ 
\hline
Decay Mode& Branching & {\bf CDF} & {\bf BTeV} & {\bf ATLAS} & {\sc BaBar} & Super-\\
  & Fractions & {\bf D0}  & {\bf LHCb} & {\bf CMS} & {\bf Belle} &{\sc BaBar}\\
 & & $(2\,\mathrm{fb}^{-1})$ & $(10^7 \,\mathrm{s})$ & $(1\,\mathrm{Year})$ & $(0.5\,\mathrm{ab}^{-1})$ & $(10\,\mathrm{ab}^{-1})$\\
\hline\hline
$B\to X_{s}\gamma$    & $(3.3\pm 0.3)$ &  &  &  & 11K & 220K\\
                      &  $\times 10^{-4}$        &  &  &  & 1.7K & 34K\\
                      & & & & & $(\mathrm{B~ Tagged})$ & $(\mathrm{B~ Tagged})$
\\\hline
$B\to K^*\gamma$    & $5\times 10^{-6}$ & 170 & 25K &  & 6K & 120K\\
$B\to \rho(\omega)\gamma$   & $2\times 10^{-6}$ & &  & &300 & 6K\\ 
\hline
$B\to X_{s}\mu^+ \mu^-$  &$(6.0 \pm 1.5)\times 10^{-6}$ & &3.6K &  & 300 & 6K\\
$B\to X_{s} e^+ e^-$  & & & &  & 350 & 7K\\
$B\to K^*\mu^+ \mu^-$   &$(2 \pm 1)\times 10^{-6}$ & 60-150 &
2.2K/4.5K & 665/4.2K & 120 & 2.4K\\
$B\to K^* e^+ e^-$      &  & &  & &150 & 3K\\     
\hline\hline
\end{tabular}
\end{center}
\caption{\it Decay reach of $B$ experiments for rare decays \cite{SuperBaBar}.\label{tab:exp}}
\end{table}

\newpage
\chapter*{ Acknowledgments}

I am extremely grateful to my supervisor Prof. Ahmed Ali, who suggested
this work. I have benefited from his strong interest in this work and advice in
numerous clarifying discussions. 

Further the DESY theory group members are gratefully 
acknowledged for instructive discussions and practical help, 
especially  Markus Diehl, Claus Gebert, L.T. Handoko and Enrico Lunghi.
My gratitude also goes to Thorsten Feldmann and Jerome Charles for 
stimulating discussions in the early stage of this work. 

My warmest thanks go to Olivier P\'ene for several helpful and
interesting discussions as well as for his invaluable encouragements
especially when they were missing.

I gratefully acknowledge the German Academic Exchange Service
(DAAD) and DESY for financial support.
Finally, many thanks go to my parents for their continuous
encouragement and support over the years.

%
\begin{appendix}
\chapter{Generalities \label{app:generalities}}
\setcounter{equation}{0}
\section{Input Parameters \label{app:input}}
\renewcommand{\arraystretch}{1.5}
\begin{table}[h] 
   \begin{center} 
     \begin{tabular}{| l l| l l |} 
\hline\hline
\rule[-2mm]{0mm}{7mm}
     $M_W$                           & $80.4$~GeV & 
        $f_B$                           & $200 \pm 20$~MeV\\ 
     
        $\hat m_t(\hat m_t)$            & $167 \pm 5$~GeV & 
         $f_{K^*,\parallel}$             & $225 \pm 30$~MeV \\

    $m_{b,pole}(2\,\mbox{GeV})$   & $4.6 \pm 0.1$~GeV & 
        $f_{K^*,\perp}(1\,\mbox{GeV})$  & $185 \pm 10$~MeV \\

         $m_c$                           & $1.4 \pm 0.2$~GeV &
       $f_{\rho}$~(1 GeV)             & $198\pm 7 $~MeV \\

      $\alpha_{em}$               & $ 1/137$ & 
        $\lambda_{B,+}^{-1}$            & $(3 \pm 1)$~GeV \\
        
        $\tau_B$                & 1.65 ps &
       $a_1( K^*)_{\perp,\,\parallel}$    & $0.2 \pm 0.1$ \\ 

      $|V_{ts}^* V_{tb}|$                      & $0.041 \pm 0.003$ &
       $a_2( K^*)_{\perp,\,\parallel}$    & $0.05 \pm 0.1$  \\

 $R_{b}=|V_{ub}|/|V_{ts}^* V_{tb}|$      & $0.094\pm 0.014$&
      $\xi^{(K^*)}_\perp(0)$                & $ 0.28 \pm 0.04$ \\

        $\Lambda_{\rm QCD}^{(n_f=5)}$   & $220 \pm 40$~MeV &
        $\xi^{(\rho)}_\perp(0)$                & $ 0.22 \pm 0.04$ \\
        
        $\langle \ell_{+}^{-1}\rangle^{(\rho)}_{+}$     & $0.3 \pm 0.2$~
        $(GeV)^{-1}$&
        $\langle \bar{u}^{-1}\rangle^{(\rho)}_{||}$            & 3.48 
        \\[0.15cm]
\hline \hline
       \end{tabular} 
\end{center} 
\vspace{0.1cm}
\centerline{\parbox{14cm}{\caption{\label{parameters} 
\it Input parameters and their uncertainties used in the 
calculations of the decay rates for $B \to K^* \ell^+ \ell^-$ and $B
\to \rho \ell \nu_\ell$ in the LEET approach.}}} \vspace{0.1cm}
\end{table} 
%
\section{QCD \label{app:qcd}}
The QCD Lagrangian reads in covariant gauge \cite{tarrachbook}
($A^a_{\mu}$: gluon field) 
\begin{eqnarray}
{\cal L}_{QCD}=\sum_{q=u,d,s,c,b,t} \bar{q} (i \slash{D}-m_q) q-\frac{1}{4} 
G_{\mu \nu}^a G^{a \, \mu \nu} + {\cal L}_{fix} +{\cal L}_{ghosts} \;
\label{Lqcd} ,
\end{eqnarray}
where the gauge fixing term and the one for ghosts $c^a,\bar{c}^a$ are given as
\begin{eqnarray}
{\cal L}_{fix} &= &-\frac{1}{2 \xi} (\partial \cdot A^a)^2 \; , \\
{\cal L}_{ghosts} &= &\bar{c}^a \partial \cdot D^{ab} c^b \; .
\end{eqnarray}
The chromomagnetic field strength tensor and covariant derivative 
are written as
\begin{eqnarray}
G^a_{\mu \nu} &=& \partial_\mu A_\nu^a-\partial_\nu A_\mu^a+g f^{axy} A_\mu^x
A_\nu^y \; , \\
D_\mu&=& \partial_\mu -i g T^x A_\mu^x \; , 
\label{eq:textbookDmu} \\
D_\mu^{ab}&=&\delta^{ab} \partial_\mu-g f^{abx} A_\mu^x \; ,
\end{eqnarray}
where $f^{abx}$ are the structure constants of $SU(3)$, defined by
\begin{eqnarray}
[ T^a, T^b ]=i f^{abx} T^x \; .
\end{eqnarray}
We have the identities
\begin{eqnarray}
[D_\mu, D_\nu] &=&-i g T^x G^x_{\mu \nu} \; , \\
D^{ax}_\mu D^{xb}_\nu-D^{ax}_\nu D^{xb}_\mu & =& -g f^{abx} G^x_{\mu \nu} \; .
\end{eqnarray}
Often the abbreviation is used 
\begin{eqnarray}G_{\mu \nu}=G_{\mu \nu}^a T^a \; .
\label{eq:Gqcd}
\end{eqnarray}
$T^a, \, a=1, \dots ,8$ are the generators of QCD.
They are related to the Gell-Mann $(3\times3)$ matrices $\lambda^a$ through
$T^a=\frac{\lambda^a}{2}$.
The $T^a$ obey the following relations ($ i,j,k=1,2,3$)
\begin {eqnarray}
Sp(T_a)&=&0 \; , \\
Sp(T_a T_b)&=&\delta_{a b}/2 \; , \\
\label{eq:TaTa}
T^a_{i j} T^a_{k l}&=&-\frac{1}{2 N_c} \delta_{i j} \delta_{k l}+\frac{1}{2}
 \delta_{i k} \delta_{j l} \; , \\
T^a_{i l}   T^a_{l k}&=& \delta_{i k} C_F  \; ,
\end{eqnarray}
with the invariant $C_F$ in an arbitrary $SU(N_c)$ given as
\begin{equation}
C_F=\frac{N_c^2-1}{2 N_c} \; \; (= \frac{4}{3} \; \; {\mbox{for}} \; N_c=3) 
\; .
\end{equation}
The coefficients of the QCD beta function (see Eq.~(\ref{eq:qcdbeta}))
are written as:
\begin{eqnarray}
\beta_0&=&\frac{11 N_c-2 N_f}{3} \; , \\
\beta_1&=&\frac{34 N_c^2-10 N_c N_f-6 C_F N_f}{3} \; .
\end{eqnarray}
Here, $N_c$ denotes the number of colours ($N_c=3$ for QCD) and
$N_f$ denotes the number of {\it active} flavours ($N_f=5$ for the effective
Hamiltonian theory relevant for $b$ decays).
\section{Feynman Rules \label{app:feynrules}}
The covariant derivative consistent with our definition of the operator basis 
and the corresponding Wilson coefficients given in section \ref{sec:effham} is 
\cite{grinstein90}
\begin{eqnarray}
\label{eq:HeffDmu}
D_{\mu} \equiv \partial_{\mu}+i g T^a A_{\mu}^a+i e Q A_{\mu} \; ,
\end{eqnarray}
where $A^a_\mu, A-\mu$ denote the polarization four-vectors of the gluon, photon
respectively.
Note that the sign convention of the strong coupling here is opposite to the 
usual one appearing in QCD text books \cite{tarrachbook,Yndurainbook} given in
Eq.~(\ref{eq:textbookDmu}), but can be made consistent with the substitution 
$g \to -g$.
The Feynman rules consistent with eq.~(\ref{eq:HeffDmu}) are given here with
boson propagators in Feynman gauge. In a general 
gauge with gauge parameter $\xi$ they are written as:
\begin{eqnarray}
-i \frac{g_{\mu \nu}+ (\xi-1) k_{\mu} k_{\nu}/(k^2+i \epsilon)}{k^2+i \epsilon}
\; ,
\end{eqnarray} 
with $\xi=1,0$ corresponding to Feynman, Landau gauge, respectively. The Feynman rules are~:
\begin{figure}[H]
\vskip -1.0truein
\centerline{\epsfysize=10in
{\epsffile{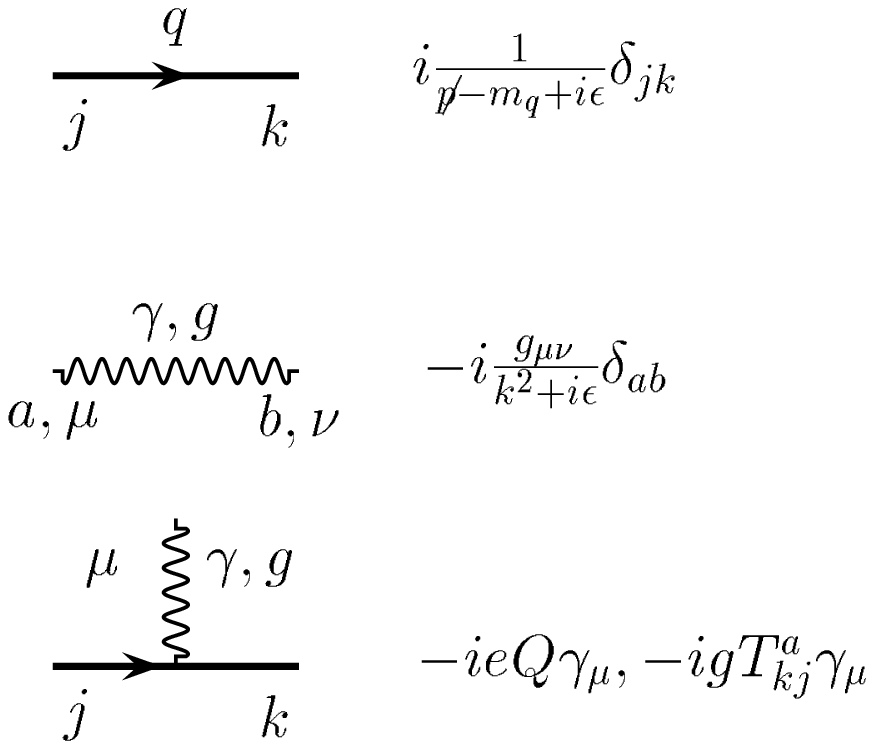}}}
\vskip -5.9truein
\label{fig:frules}
\end{figure}  
\noi complemented by the rules~:
\begin{itemize}
\item evaluate fermion lines {\bf against} the momentum flow
\item add a $(-1)$ for a closed fermion loop and perform the trace over the 
string of $\gamma$ matrices
\end{itemize}
The rule for an $O_7$ operator insertion is, using $\partial_{\mu}=i q_{\mu}$
for an {\bf out going} photon and further $\epsilon \cdot q=0$ for a real
photon, and $F^{\mu \nu}=\partial^\mu A^\nu-\partial^\nu A^\mu$, 
\begin{eqnarray}
\sigma F =\sigma_{\mu \nu} F^{\mu \nu}=i [\slash{\partial},\slash{A}]=
2 \gamma_{\mu} \slash{q} \epsilon^{\mu} \; .
\end{eqnarray}
The Fierz transformation in $d=4$ dimensions is defined as:
\begin{eqnarray}
\label{eq:fierzLL}
(\bar{q}_1 \gamma_{\mu} L q_2)(\bar{q}_3 \gamma_{\mu} L q_4)&=&
+(\bar{q}_1 \gamma_{\mu} L q_4)(\bar{q}_3 \gamma_{\mu} L q_2) \; , \\
(\bar{q}_1 \gamma_{\mu} L(R) q_2)(\bar{q}_3 \gamma_{\mu} R(L) q_4)&=&
(-2)(\bar{q}_1 R(L) q_4)(\bar{q}_3  L(R) q_2) \; .
\label{eq:fierzLR}
\end{eqnarray}
\section{Utilities}
A variety of tools for 1-loop calculations is collected in the 
appendix of ref.~\cite{Yndurainbook}. \\
{\it Distributions}~:
\begin{eqnarray}
\delta(x)&=&\frac{1}{2 \pi} \int_{\mbox{R}} dq e^{i q x} dq \; , \\
\theta(x)&=& \lim_{\epsilon \to 0} \frac{-i}{2 \pi} 
\int_{\mbox{R}} dq \frac{e^{i q x}}{q-i \epsilon} dq\; , \\
\frac{d \theta}{dx}(x)&=&\delta(x) \; . 
\end{eqnarray}
{\it Geometrical series}~:
\begin{eqnarray}
\frac{1}{1 \pm x}=1 \mp x + x^2  +\sum_{n=3}^{\infty}(\mp x)^n \; .
\end{eqnarray}

Special Functions useful for loops\\
{\it Poly-logarithms}~:
\begin{eqnarray}
Li_n(z)&=&\sum_{k=1}^{\infty} \frac{z^k}{k^n} ; \; |z| <1  \; , \\
Li_2(z)&=&- \int_0^z \frac{dt}{t} \ln(1-t)  \; .
\end{eqnarray}
{\it Spence function}~: 
\begin{eqnarray}
Sp(z) &\equiv& Li_2(z)=-\int_0^1 \frac{dt}{t} \ln(1-z t)  \; , \\
Sp(0) & = &0 \; \; , \; \;   Sp(1)=\frac{\pi^2}{6} \; \; , 
\; \;   Sp(-1)=\frac{\pi^2}{12}  \;  \; , \\
Sp(z)&=&-Sp(1-z) +\frac{\pi^2}{6}-\ln(z) \ln(1-z) \; , \\
Sp(z)&=&-Sp(\frac{1}{z}) -\frac{\pi^2}{6}-\frac{1}{2} \ln^2(-z) \; .
\end{eqnarray}
Useful identities for loops:
\begin{eqnarray}
\arctan(z) =\frac{1}{2i} \ln\frac{1+i z}{1-i z} \; , 
~~{\mbox{arctanh}}(z)=\frac{1}{2} \ln\frac{1+z}{1-z} \; .
\end{eqnarray}
Phase space element, 
$d^3 \vec{p}=|\vec{p}|^2~ d|\vec{p}|~ d \cos{\theta}~ d\phi \; ,\cos{\theta} \in [-1,+1] \; ,
\phi \in [0,2 \pi[ $.
\begin{eqnarray}
\frac{d^3 \vec{p}}{2 E}=\int d^4 p~ \delta(p^2-m^2)~ \theta(E) \; \; ;~~
E=\sqrt{{\vec{p}}^2+m^2}\; .
\end{eqnarray}
Dirac algebra identities, for more see ref.~\cite{itzykson}, 
especially the appendix.
\begin{eqnarray}
\{ \gamma_{\mu}, \gamma_\nu \}=2 g_{\mu \nu} \; ,
~~\sigma_{\mu \nu}=\frac{i}{2}[\gamma_{\mu}, \gamma_\nu] \; ,
~~\gamma_{\mu} \gamma_\nu=g_{\mu \nu}-i \sigma_{\mu \nu} \; .
\end{eqnarray}
A useful tool within this context is the TRACER routine 
\cite{tracer} running under the symbolic algebra program  mathematica.

Chiral projectors $L(R)\equiv(1 \mp \gamma_5)/2$:
\begin{eqnarray}
\gamma_5^2&=&1 \; ,~~ \gamma_5^\dagger=\gamma_5 \; , \\
(L(R))^2&=&L(R)  \; ,~~ L R=R L=0 \; , ~~(L(R))^\dagger=L(R)  \; .
\end{eqnarray}
Further we have
\begin{eqnarray}
\gamma_0^2&=&1 \; ,~~ \gamma_0^\dagger=\gamma_0 \; .
\end{eqnarray}
Fermion fields~: 
\begin{eqnarray}
\bar{\psi}& \equiv &\psi^{\dagger} \gamma_0=(\psi^{*})^T \gamma_0\; ,
~~ \psi_{L(R)} \equiv L(R) \psi \; , \nonumber \\
\bar{\psi_{L(R)}}&=&\bar{L(R) \psi}=(L(R) \psi)^\dagger \gamma_0=
\psi^\dagger (L(R))^\dagger \gamma_0= \psi^\dagger \gamma_0 R(L)
=\bar{\psi} R(L) \; .
\label{eq:chiralfields}
\end{eqnarray}
\chapter{The large energy expansion \label{app:leet}}
\setcounter{equation}{0}
%
\section{Feynman Rules in the Large Energy Limit of QCD
\label{app:leetrules} }
In the hadron limit of an infinitely heavy meson $M$ with mass $m_M \to
\infty$ and large energy for the final one $E_{P(V)}\to \infty$, the
QCD Lagrangian defined in (\ref{Lqcd}) is just:
\begin{eqnarray}
 {\mathcal L}_{\rm QCD}={\mathcal L}_{\rm LEET}+{\cal O}(1/E)\;,
\end{eqnarray}

\noi where the effective Lagrangian is given by
\begin{eqnarray}
\label{leet}
  {\mathcal L}_{\rm LEET} &=& 
    \bar q_n \, \frac{\slash{n}_+}{2} \,i n_{-} \cdot D\, q_n\;.
\end{eqnarray}
\noi where $q_n(x)=e^{i E_q n_- \cdot x} \, \frac{\slash{n}_- \slash
{n}_+}{4} \,q(x)$ are the large components of the light quark spinor field.  
The $n_+=2 v-n_-$ is a light-like vector with $n_+ \cdot n_- =
2$ and $E_q\approx E$ is the energy of the light quark.
Here $v$  and $n$ denote respectively the four-velocity ($v^2=1$) of
the heavy quark $Q$ with momentum $p_Q=m_Q v+k$ and the four
light-like vector ($n_-^2= 0$) parallel to the the light
quark $p_q=E n_-+k'$, where the small residual momentum $k$ (and
$k'$ )of order $\Lambda_{QCD}$. 

It follows immediately, the Feynman rules of the LEET formalism:
\\
\\
\begin{figure}[H]
\begin{center}
\psfrag{mu}{  $\mu$}
\psfrag{j}{ $j$}
\psfrag{k}{$k$}
\psfrag{vertex}{\large $-i~g~{\slash v} T^a_{kj}~n_{\mu}$}
\psfrag{propagator}{\Large {${i~{\slash  v}\over n\cdot k +i~\epsilon} 
\frac{\slash n~ \slash v}{ 2}$}$\delta_{kj}$}
\epsfig{file=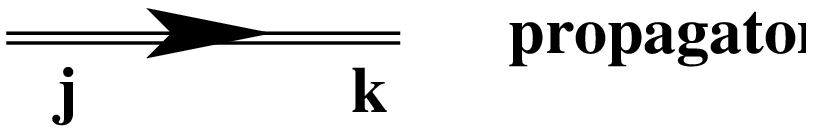,width=0.3\linewidth}
\vskip 0.3truein
\epsfig{file=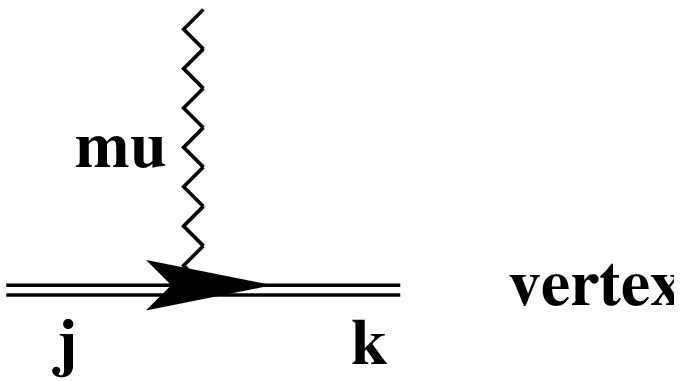,width=0.3\linewidth}
\end{center}
\label{fig:hqetrules}
\end{figure} 
where the light quark $q$ is represented by a double line.
\section{The Factorizabe corrections $\Delta F_i$ 
\label{app:leetcorr}}
We give here the ${\cal O}(\alpha_s)$- factorizable corrections 
to the LEET form factors (defined in
Eqs.~(\ref{add+})-(\ref{vertexcorr2})). 
We now present the result for the hard scattering correction to 
$B \to P$ form factors, as defined by $\Delta f_{+,0,T}$ in
Eqs.~(\ref{add+}), (\ref{vertexcorr1}). The renormalisation 
convention\footnote{
In ref.~\cite{Beneke:2000wa}, it was convenient to define the
factorisation scheme (or renormalisation conventions for the ``soft form factors'') by imposing the condition that 
$ f_+ \equiv  \xi^{(P)},\;\; V   \equiv  \frac{m_B+m_V}{m_B} \,
  \xi^{(V)}_\perp, \;\; A_0 \equiv  \frac{E_V}{m_V} \, \xi^{(V)}_\parallel$, 
hold exactly to all orders in perturbation theory.} 
implies $\Delta f_+ \equiv  0$ by definition. 
Thus, they  are then given by \cite{Beneke:2000wa} 
\begin{eqnarray}
\Delta f_+ &=&  0, \\
\Delta f_0  &=&  \frac{m_B-2 E_P}{2 E_P} \, \Delta F_P, \\ 
\Delta f_T  &=& - \frac{m_B+m_P}{2 E_P}  \, \Delta F_P,
\label{eq:pi_final}
\end{eqnarray}
with the quantity
\begin{eqnarray}
\Delta F_P
    &=& \frac{8 \pi^2  f_B f_P}{N_C m_B}
          \, \langle l_+^{-1} \rangle_+ \,
          \langle \bar u^{-1} \rangle_{P}, 
\label{pihard}
\end{eqnarray}
is defined in terms of moment of the leading twist distribution 
amplitude (as usual $\bar u = 1-u$)
\begin{equation}
\langle \bar{u}^{-1}\rangle_P = \int du \frac{\phi(u)}{\bar u}.
\end{equation}
For the $B$ meson, the quantity $\langle l_+^{-1}\rangle_+$ reads
\begin{equation}
\langle l_+^{-1}\rangle_+=\int dl_+ \frac{\phi^B_+(l_+)}{l_+}. 
\label{bwfmom}
\end{equation} 
Note that $\Delta f_0$ vanishes at $s=0$ ($E_P=m_B/2$) as required on 
general grounds. 
Following the renormalisation convention,  the hard
correction to the other $B\to V$ form factors, defined in Eqs.~(\ref{A0cor})-(\ref{vertexcorr2}), reads \cite{Beneke:2000wa}
\begin{eqnarray}
\Delta A_1 &=&\Delta V= 0 , \nn\\
\Delta A_2 &=& \frac{m_V}{E_{V}}\,\frac{m_B}{m_B-m_V} \, \frac{m_B(m_B-2 E_{V})}{4
    E_{V}^2} \, \Delta F_\parallel \ ,
\label{hardrho0}\\
\Delta T_1 &=& \frac{m_B}{4 E_{V}}\,\Delta F_\perp,  \nn\\
\Delta T_2 &=& \frac{1}{2} \, \Delta F_\perp, \nn\\ 
\Delta T_3 &=& \frac{m_B}{4 E_{V}}\,\Delta F_\perp + \frac{m_V}{E_{V}}\, 
\left(\frac{m_B}{2 E_{V}}\right)^2\,\Delta F_\parallel.
\label{hardrho}
\end{eqnarray}
We introduce the quantities 
\begin{eqnarray}
\Delta F_\parallel
    &=& \frac{8 \pi^2  f_B f_V}{N_C m_B} 
          \, \langle l_+^{-1} \rangle_+ \,
          \langle \bar u^{-1} \rangle_{\parallel},\\
\Delta F_\perp
    &=& \frac{8 \pi^2  f_B f_{\perp}}{N_C m_B} 
          \, \langle l_+^{-1} \rangle_+ \,
          \langle \bar u^{-1} \rangle_{\perp},
\label{rhohard}
\end{eqnarray}
where the two terms, namely $\langle \bar{u}^{-1}\rangle_\parallel$
and $\langle \bar{u}^{-1}\rangle_\perp$ are given by
\begin{eqnarray} 
\langle \bar{u}^{-1}\rangle_\parallel = \int du \,
\phi_\parallel(u)/\bar u, 
\end{eqnarray}
and
\begin{eqnarray}
\langle \bar{u}^{-1}\rangle_\perp = \int du\, 
\phi_\perp(u)/\bar u.
\end{eqnarray}

\section{The Functions $F_{i}^{(j)}$
\label{app:leetcorr2}} 
In this appendix we list the functions $F_{8}^{(7,9)}$, 
$F_1^{(9)}$ and $F_2^{(7, 9)}$  (defined in Eqs.~(\ref{start3}) and
(\ref{end3})), representing the power correction to the matrix
elements of the operators ${\cal O}_{8}$, ${\cal O}_{1}$ and ${\cal O}_{2}$ respectively. The corrected one loop matrix elements of ${\cal O}_8$ read \cite{AAGW}:
\begin{eqnarray}
  F_8^{(7)} &=& -\frac{32}{9} \, \ln \frac{\mu}{m_b} 
  - \frac 89 \, \frac{\tilde s}{1- \tilde s} \, \ln \tilde s- \frac89 \, i\pi 
    - \frac 49 \, \frac{ 11 - 16 \tilde s + 8 \tilde s^2 }{(1-\tilde s)^2}
\cr && \quad +
    \frac 49 \, \frac{1}{(1-\tilde s)^3} \,
   \left[ 
     (9 \tilde s - 5 \tilde s^2 + 2 \tilde s^3) \, B_0(\tilde s)
    - (4 + 2 \tilde s) \, C_0(\tilde s) \right],
\end{eqnarray}
and
\begin{eqnarray}
  F_8^{(9)} &=& 
    \frac {16}{9} \, \frac{1}{1- \tilde s}\, \ln \tilde s
             + \frac{8}{9}  \, \frac{5 - 2 \tilde s}{(1-\tilde s)^2}
 -
    \frac 89 \, \frac{4 - \tilde s}{(1-\tilde s)^3} 
   \left[  (1 + \tilde s) \, B_0(\tilde s)
         - 2 \, C_0(\tilde s)
   \right],       
\end{eqnarray}

\noi where $\tilde s = s/m_b^2$, $B_0(\tilde s)=B_0(q^2,m_b^2)$ is given 
in (\ref{b0def}), and the integral 

\begin{eqnarray}
 C_0(\tilde s) &=& \int_0^1 dx \, \frac{1}{x \, (1-\tilde s) + 1} \, 
                 \ln \frac{x^2}{1- x \, (1-x) \, \tilde s}
\end{eqnarray}

\noi can be expressed in terms of dilogarithms. The corresponding power
correction to the matrix elements of the operators ${\cal O}_{1}$ and
${\cal O}_{2}$ are given respectively (using $L_\mu = \ln(\mu/m_b)$,
$L_s = \ln (\tilde{s})$ and $\tilde{m}_c =m_c/m_b$ ) as \cite{AAGW}:

\begin{eqnarray}
{}
    F_1^{(9)} &=&
    \left (-{\tfrac {1424}{729}}+{\tfrac {16}{243}}\,i\pi +
    {\tfrac {64}{27}}\,{ L_c}\right ){ L_\mu}
    -{\tfrac {16}{243}}\,{ L_\mu}\,{ L_s}+
    \left ({\tfrac {16}{1215}}-{\tfrac {32}{135}}\,{{ \tilde{m}_c}}^{-2}\right )
    { L_\mu}\,{ \tilde{s}}
    \nn \\
    &+&
    \left ({\tfrac {4}{2835}}-{\tfrac {8}{315}}\,{{ \tilde{m}_c}}^{-4}\right )
    { L_\mu}\,{{ \tilde{s}}}^{2}+
    \left ({\tfrac {16}{76545}}-{\tfrac {32}{8505}}\,
    {{ \tilde{m}_c}}^{-6}\right ){ L_\mu}\,{{ \tilde{s}}}^{3}\nn\\
    &-&{\tfrac {256}{243}}\,{{ L_\mu}}^{2} + f_1^{(9)} \, ,
\end{eqnarray}
\renewcommand{\arraystretch}{1}
\begin{table}[t]
\begin{center}
\begin{tabular}{| c | rcl | rcl | rcl |}
\hline\hline \ & \multicolumn{3}{c|}{$\tilde{m}_c=0.27$} & \multicolumn{3}{c|}{$\tilde{m}_
c=0.29$} & \multicolumn{3}{c|}{$\tilde{m}_c=0.31$} \\ \hline \hline
$k_1^{(9)}(0,0)$  & $ -12.327$&$+$&$0.13512\,i $  
& $ -11.973$&$+$&$0.16371\,i $
  & $ -11.65$&$+$&$0.18223\,i $\\
$k_1^{(9)}(0,1)$  & $ -0.080505$&$-$&$0.067181\,i $  & $ 
-0.081271$&$-$&$0.059691\,i $  & $
-0.080959$&$-$&$0.051864\,i $  \\
$k_1^{(9)}(1,0)$  & $ -33.015$&$-$&$0.42492\,i $  & $ 
-28.432$&$-$&$0.25044\,i $
  & $ -24.709$&$-$&$0.13474\,i $\\
$k_1^{(9)}(1,1)$  & $ -0.041008$&$+$&$0.0078685\,i $  & 
$ -0.040243$&$+$&$0.016442\,i $  & $
-0.036585$&$+$&$0.024753\,i $  \\
$k_1^{(9)}(2,0)$  & $ -76.2$&$-$&$1.5067\,i $  & $ 
-57.114$&$-$&$0.86486\,i $  &
 $ -43.588$&$-$&$0.4738\,i $  \\
$k_1^{(9)}(2,1)$  & $ -0.042685$&$+$&$0.015754\,i $  & $ -0.035191$&$+$&$
0.027909\,i $  & $
-0.021692$&$+$&$0.036925\,i $  \\
$k_1^{(9)}(3,0)$  & $ -197.81$&$-$&$4.6389\,i $  & $ 
-128.8$&$-$&$2.5243\,i $  &
 $ -86.22$&$-$&$1.3542\,i $  \\
$k_1^{(9)}(3,1)$  & $ -0.039021$&$+$&$0.039384\,i $  & $ 
-0.017587$&$+$&$0.050639\,i $  & $
0.013282$&$+$&$0.052023\,i $  \\
\hline\hline
\end{tabular}
\caption{\it Coefficients in the decomposition of $f_1^{(9)}$  for three values of $\tilde{m}_c$ \cite{AAGW}.}
\label{coeff1}
%
%
%
%
\vspace*{2cm}
\begin{tabular}{| c | rcl | rcl | rcl |}
\hline\hline \ & \multicolumn{3}{c|}{$\tilde{m}_c=0.27$} & \multicolumn{3}{c|}{$\tilde{m}_
c=0.29$} & \multicolumn{3}{c|}{$\tilde{m}_c=0.31$} \\ \hline \hline
$k_2^{(9)}(0,0)$  & $ 7.9938$&$-$&$0.81071\,i $  & $ 6.6338$&$-$&$0.98225\,i $  & $ 5.4082$&$-$&$1.0934\,i $  \\
$k_2^{(9)}(0,1)$  & $ 0.48303$&$+$&$0.40309\,i $  & $
0.48763$&$+$&$0.35815\,i $ & $ 0.48576$&$+$&$0.31119\,i $
\\
$k_2^{(9)}(1,0)$  & $ 5.1651$&$+$&$2.5495\,i $  & $ 3.3585$&$+$&$1.5026\,i $  & $ 1.9061$&$+$&$0.80843\,i $  \\
$k_2^{(9)}(1,1)$  & $ 0.24605$&$-$&$0.047211\,i $  & $
0.24146$&$-$&$0.098649\,i $  & $ 0.21951$&$-$&$0.14852\,i $
\\
$k_2^{(9)}(2,0)$  & $ -0.45653$&$+$&$9.0402\,i $  & $ -1.1906$&$+$&$5.1892\,i $ & $ -1.8286$&$+$&$2.8428\,i $  \\
$k_2^{(9)}(2,1)$  & $ 0.25611$&$-$&$0.094525\,i $  & $
0.21115$&$-$&$0.16745\,i $  & $ 0.13015$&$-$&$0.22155\,i $
\\
$k_2^{(9)}(3,0)$  & $ -25.981$&$+$&$27.833\,i $  & $ -17.12$&$+$&$15.146\,i $  &
 $ -12.113$&$+$&$8.1251\,i $  \\
$k_2^{(9)}(3,1)$  & $ 0.23413$&$-$&$0.2363\,i $  & $ 0.10552$&$-$&$0.30383\,i $ 
 & $ -0.079692$&$-$&$0.31214\,i $
\\
\hline \
$k_2^{(7)}(0,0)$  & $ 4.3477$&$+$&$0.56054\,i $  & $ 4.0915$&$+$&$0.44999\,i $  
& $ 3.8367$&$+$&$0.3531\,i $  \\
$k_2^{(7)}(0,1)$  & & 0 & & & 0 & & & 0 &  \\
$k_2^{(7)}(1,0)$  & $ 1.5694$&$+$&$0.9005\,i $  & $ 1.4361$&$+$&$0.73732\,i $  &
 $ 1.3098$&$+$&$0.60185\,i $  \\
$k_2^{(7)}(1,1)$  & $ 0.0010623$&$-$&$0.12324\,i $  & 
$ -0.016454$&$-$&$0.11806\, i $  & $
-0.031936$&$-$&$0.10981\,i $  \\
$k_2^{(7)}(2,0)$  & $ -0.14311$&$+$&$1.2188\,i $  & $ 0.011133$&$+$&$1.05\,i $  
& $ 0.13507$&$+$&$0.89014\,i $  \\
$k_2^{(7)}(2,1)$  & $ -0.12196$&$-$&$0.099636\,i $  & 
$ -0.13718$&$-$&$0.068733\, i $  & $
-0.14169$&$-$&$0.035553\,i $  \\
$k_2^{(7)}(3,0)$  & $ -2.5739$&$+$&$0.59521\,i $  & $ -1.6949$&$+$&$0.76698\,i $
  & $ -1.0271$&$+$&$0.77168\,i $
\\
$k_2^{(7)}(3,1)$  & $ -0.18904$&$-$&$0.0025554\,i $  & $ -0.17416$&$+$&$0.049359
\,i $  & $ -0.13592$&$+$&$0.093\,i
$  \\
\hline\hline
\end{tabular}
\end{center}
\caption{\it Coefficients in the decomposition of $f_2^{(9)}$
and $f_2^{(7)}$ for three values of $\tilde{m}_c$ \cite{AAGW}.}
\label{coeff2}
\end{table}
and
\begin{eqnarray}
{}
    F_2^{(9)} &=&
    \left ({\tfrac {256}{243}}-{\tfrac {32}{81}}\,i\pi -
    {\tfrac {128}{9}}\,{ L_c}\right ){ L_\mu}+
    {\tfrac {32}{81}}\,{ L_\mu}\,{ L_s}+
    \left (-{\tfrac {32}{405}}+{\tfrac {64}{45}}\,
    {{ \tilde{m}_c}}^{-2}\right ){ L_\mu}\,{ \tilde{s}}
    \nn \\ &+&
    \left (-{\tfrac {8}{945}}+{\tfrac {16}{105}}\,
    {{ \tilde{m}_c}}^{-4}\right ){ L_\mu}\,{{ \tilde{s}}}^{2}+
    \left (-{\tfrac {32}{25515}}+{\tfrac {64}{2835}}\,
    {{ \tilde{m}_c}}^{-6}\right ){ L_\mu}\,{{ \tilde{s}}}^{3}\nn\\
    &+&{\tfrac {512}{81}}\,{{ L_\mu}}^{2}
    +f_2^{(9)} \, ,
\\
\nn \\
{}
    F_2^{(7)} &=& {\tfrac {416}{81}}\,{ L_\mu}
    +f_2^{(7)} \, .
\end{eqnarray}

The analytic results for $f_1^{(9)}$, $f_1^{(7)}$, $f_2^{(9)}$, and
$f_2^{(7)}$ (expanded up to $\tilde{s}^3$ and $(\tilde{m}_c^2)^3$) are 
rather lengthy.
The formulas become relatively short,
however, if we give the charm quark mass dependence in numerical
form (for the characteristic values of 
$\tilde{m}_c$=0.27, 0.29 and 0.31).

We write the functions $f_a^{(b)}$ as

\begin{eqnarray}
\label{f1decomp}
    f_a^{(b)} = \sum_{i,j} \, k_a^{(b)}(i,j) \,
    \hat{s}^i \, L_s^j  \quad \quad
    \quad (a=1,2;\,b=7,9;\,i=0,...,3;\,j=0,1).
\end{eqnarray}

\noi The numerical values for the quantities $k_a^{(b)}(i,j)$
are given in Tables~(\ref{coeff1}) and (\ref{coeff2}).

%
\chapter{$B\to K^* \ell^+ \ell^-$ in SUSY \label{app:susy}}
\setcounter{equation}{0}
\section{The functions $f_i(x)$ 
\label{loopfct:fi}}
In this appendix we give the various one-loop integral functions,
namely $f_i(x)$ $(i=1,\cdots, 8)$, which
appear within the MSSM contributions to the magnetic dipole moment
and to the semileptonic coefficients. They are given respectively by~\cite{Cho:1996we}:
\begin{eqnarray}
f_1(x) &=& {-7 + 5 x + 8x^2  \over 6(1-x)^3} - {2 x - 3 x^2 \over (1-x)^4} 
  \log x, \\
f_2(x) &=& {3x-5x^2 \over 2(1-x)^2} + {2x-3x^2 \over (1-x)^3} 
  \log x, \\
f_3(x) &=& {2+ 5 x - x^2 \over 6(1-x)^3} + {x \over (1-x)^4} \log x, \\
f_4(x) &=& {1 + x \over 2 (1-x)^2} + {x \over (1-x)^3} \log x,\\ 
f_5(x) &=& {x \over 1-x} + {x \over (1-x)^2} \log x, \\
f_6(x) &=& {38x-79x^2+47 x^3 \over 6(1-x)^3} + {4x-6x^2+3x^4 \over (1-x)^4} 
\log x, \\
f_7(x) &=& {52-101x+43x^2 \over 6(1-x)^3} + {6-9x+2x^3 \over (1-x)^4} 
\log x, \\
f_8(x) &=& {2-7x+11x^2 \over (1-x)^3} + {6x^3 \over (1-x)^4} \log x. 
\label{gfunctions}
\end{eqnarray}
\section{The Auxiliary Functions 
$c_i(m_1^2, m_2^2, m_3^2)$, $d_i(m_1^2, m_2^2, m_3^2, m_4^2)$
\label{loopfct:cdi}}
%
The various auxiliary functions $c_i(m_1^2, m_2^2, m_3^2)$ and
$d_i(m_1^2, m_2^2, m_3^2, m_4^2)$ are listed below~\cite{Cho:1996we}:
\begin{eqnarray}
c_0(m_1^2, m_2^2, m_3^2) &=& -\Bigl[
  {m_1^2 \, \displaystyle{\log{m_1^2 \over \mu^2}} \over
  (m_1^2-m_2^2)(m_1^2-m_3^2)}
+ (m_1 \leftrightarrow m_2)
+ (m_1 \leftrightarrow m_3) \Bigr], \\ 
c_2(m_1^2, m_2^2, m_3^2) &=& - {1 \over 4} \Bigl[
  {m_1^2 \, \displaystyle{\log{m_1^4 \over \mu^2}} \over
  (m_1^2-m_2^2)(m_1^2-m_3^2)}
+ (m_1 \leftrightarrow m_2)
+ (m_1 \leftrightarrow m_3) \Bigr] \nn\\
&&+{3 \over 8},\\
%
d_0(m_1^2, m_2^2, m_3^2, m_4^2) &=& 
- \Bigl[ 
{m_1^2 \, \displaystyle{\log{m_1^2 \over \u^2}} \over
  (m_1^2-m_2^2)(m_1^2-m_3^2)(m_1^2-m_4^2)} 
+ (m_1 \leftrightarrow m_2) \nn\\
&&+ (m_1 \leftrightarrow m_3) 
+ (m_1 \leftrightarrow m_4) \Bigr],  \\
d_2(m_1^2, m_2^2, m_3^2, m_4^2) &=& 
- {1 \over 4} \Bigl[  
{m_1^4 \, \displaystyle{\log{m_1^2 \over \u^2}} \over
  (m_1^2-m_2^2)(m_1^2-m_3^2)(m_1^2-m_4^2)} 
+ (m_1 \leftrightarrow m_2) \nn\\
&&+ (m_1 \leftrightarrow m_3)  
+ (m_1 \leftrightarrow m_4) \Bigr]. \label{functions}
\end{eqnarray}
%
%
\end{appendix}

\end{document}